\documentclass[resetfootnote,trackchanges,twocolumn,twocolappendix]{aastex701}

\usepackage{tabularx}
\usepackage{amsmath}
\usepackage{tablefootnote}
\usepackage{placeins}
\usepackage{enumitem}      
\usepackage{array}         
\usepackage{booktabs}        
\usepackage{xcolor}
\usepackage{colortbl}
\usepackage{multirow}
\newcommand{\lya}{Ly$\alpha$}

\newcommand{\HI}{\rm H\,{\textsc {i}}}
\newcommand{\HII}{\rm H\,{\textsc {ii}}}
\newcommand{\SiII}{\rm Si\,{\textsc {ii}}}
\newcommand{\SiIII}{\rm Si\,{\textsc {iii}}}
\newcommand{\SiIV}{\rm Si\,{\textsc {iv}}}
\newcommand{\CII}{\rm C\,{\textsc {ii}}}
\newcommand{\CIV}{\rm C\,{\textsc {iv}}}

\definecolor{gold}{RGB}{255,166,0}   
\definecolor{silver}{RGB}{138,138,138}
\definecolor{bronze}{RGB}{150,75,0}

\begin{document}

\title{Kinematically Coherent Multiphase Galactic Winds in Star-Forming Galaxies Revealed by Unified Radiative Transfer Modeling of UV Emission and Absorption Lines}

\author[0000-0001-5113-7558]{Zhihui Li}
\affiliation{Center for Astrophysical Sciences, Department of Physics \& Astronomy, Johns Hopkins University, Baltimore, MD 21218, USA}
\email[show]{zli367@jh.edu}

\author[0000-0001-6670-6370]{Timothy Heckman}
\affiliation{Center for Astrophysical Sciences, Department of Physics \& Astronomy, Johns Hopkins University, Baltimore, MD 21218, USA}
\email{theckma1@jhu.edu}

\author[0000-0003-2491-060X]{Max Gronke}
\affiliation{Max-Planck Institute for Astrophysics, Karl-Schwarzschild-Str. 1, D-85741 Garching, Germany}
\affiliation{Astronomisches Rechen-Institut, Zentrum f\"{u}r Astronomie, Universit\"{a}t Heidelberg, Mönchhofstra$\beta$e 12-14, 69120 Heidelberg, Germany}
\email{max.gronke@uni-heidelberg.de}

\author[0000-0002-9217-7051]{Xinfeng Xu}
\affiliation{Department of Physics and Astronomy, Northwestern University, 2145 Sheridan Road, Evanston, IL, 60208, USA}
\affiliation{Center for Interdisciplinary Exploration and Research in Astrophysics (CIERA), Northwestern University, 1800 Sherman Avenue, Evanston, IL, 60201, USA}
\email{xinfeng.xu@northwestern.edu}

\author[0000-0002-6586-4446]{Alaina Henry}
\affiliation{Center for Astrophysical Sciences, Department of Physics \& Astronomy, Johns Hopkins University, Baltimore, MD 21218, USA}
\affiliation{Space Telescope Science Institute, 3700 San Martin Drive, Baltimore, MD 21218, USA}
\email{ahenry@stsci.edu}

\author[0000-0001-9735-7484]{Evan Schneider}
\affiliation{Department of Physics \& Astronomy and PITT-PACC, University of Pittsburgh, 100 Allen Hall, 3941 O’Hara Street, Pittsburgh, 15260, PA, USA.}
\email{eschneider@pitt.edu}

\author[0000-0002-7918-3086]{Matthew Abruzzo}
\affiliation{Department of Physics \& Astronomy and PITT-PACC, University of Pittsburgh, 100 Allen Hall, 3941 O’Hara Street, Pittsburgh, 15260, PA, USA.}
\email{mwa2113@columbia.edu}

\author[0000-0002-4153-053X]{Danielle Berg}
\affiliation{Department of Astronomy, The University of Texas at Austin, 2515 Speedway, Stop C1400, Austin, TX 78712, USA}
\email{daberg@austin.utexas.edu}

\author[0000-0003-4372-2006]{Bethan James}
\affiliation{Space Telescope Science Institute, 3700 San Martin Drive, Baltimore, MD 21218, USA}
\email{bjames@stsci.edu}

\author[0000-0001-9189-7818]{Crystal Martin}
\affiliation{Department of Physics, University of California, Santa Barbara, CA 93106, USA}
\email{cmartin@physics.ucsb.edu}

\author[0000-0002-0302-2577]{John Chisholm}
\affiliation{Department of Astronomy, The University of Texas at Austin, 2515 Speedway, Stop C1400, Austin, TX 78712, USA}
\email{chisholm@austin.utexas.edu}

\begin{abstract}

We present \texttt{PEACOCK}, a three-dimensional Monte Carlo radiative transfer (RT) framework designed to self-consistently model rest-frame ultraviolet emission and absorption lines arising from multiphase, clumpy galactic winds. Applied to deep \textit{HST}/COS spectra of 50 nearby star-forming galaxies, \texttt{PEACOCK} reproduces 220 observed profiles of \lya, \SiII, \CII, \SiIII, \SiIV, and \CIV\ spanning pure absorption, pure emission, and P-Cygni–like morphologies within a single physically motivated CGM model. By combining Monte Carlo RT with deep-learning acceleration and nested sampling, the framework enables fully converged multi-line inference at a small fraction of the computational cost of traditional RT grids. Systematic parameter experiments show that ion column densities, bulk outflow velocities, and turbulent motions leave distinct imprints on line profiles with minimal degeneracy, allowing the underlying gas properties to be independently constrained. Purely radial accelerating flows often fail to reproduce the observed absorption morphologies, whereas introducing macroscopic velocity dispersion naturally produces the broad asymmetric troughs commonly seen in the data, indicating that turbulent motions are a necessary component of outflow kinematics. The inferred kinematics reveal strong coherence among low- and high-ionization metal lines in both bulk and turbulent velocities, consistent with a dynamically coupled multiphase wind. In contrast, neutral hydrogen shows weaker correspondence with metals, suggesting incomplete mixing and a distinct kinematic structure. Joint multi-line fitting further tightens constraints on the bulk outflow parameters while preserving the quality of individual line fits. By unifying emission and absorption diagnostics across multiple ions, \texttt{PEACOCK} provides a physically grounded bridge between UV observations and theoretical models of galactic winds from the local universe to the epoch of reionization.

\end{abstract}



\section{Introduction} 

Galactic-scale outflows (or galactic winds) are a ubiquitous feature of star-forming galaxies across cosmic time and play a central role in regulating baryon cycling, quenching star formation, enriching the circumgalactic medium (CGM), and shaping the growth of galaxies and their halos \citep{Veilleux2005, Tumlinson2017, Thompson2024}. These outflows are thought to be driven by stellar feedback, supernovae, and radiation pressure, and they operate across a wide range of physical scales and phases. Observationally, galactic winds are commonly probed through optical and ultraviolet (UV) emission and absorption lines, which trace multiphase gas spanning a broad range of densities, temperatures, and ionization states \citep[e.g.,][]{Heckman2000, Steidel2010, Rubin2014}. Among the most informative diagnostics are resonant lines such as \lya\ emission and low- to high-ionization metal lines including \CII, \SiII, \SiIII, \CIV, and \SiIV, which collectively provide a powerful multi-phase diagnostic of outflow kinematics, ionization structure, and mass-loading across the ISM and CGM \citep[e.g.,][]{Erb2014, Bordoloi2014, Chisholm2016, Xu2022}.

Observational surveys over the past decade have provided compelling evidence for the ubiquity of galactic-scale outflows across cosmic time. In star-forming galaxies at $z \gtrsim 2$, deep rest-frame UV spectroscopy frequently reveals blue-shifted interstellar absorption lines and redshifted \lya\ emission relative to systemic velocity -- features widely interpreted as signatures of large-scale outflows. The blueshifted absorption lines trace cool gas flowing toward the observer, while the redshifted \lya\ emission arises from resonant scattering in an expanding medium. These characteristics have been consistently observed in both gravitationally lensed galaxies and large spectroscopic samples, reinforcing the view that feedback-driven winds are a generic property of high-redshift star-forming systems \citep[e.g.,][]{Shapley2003, Steidel2010, Jones2012, Kornei2012, Erb2014, Trainor2015, Patricio2016, Rigby2018}.

At lower redshifts, surveys with the Cosmic Origins Spectrograph (COS) on board the Hubble Space Telescope have mapped the CGM in absorption against background quasars, enabling systematic studies of gas kinematics and ionization structure in the halos of $z \lesssim 0.5$ galaxies. The COS-Halos program \citep{Werk2013, Werk2014, Prochaska2017} has revealed that cool and ionized gas, traced by a wide range of metal absorption lines, is ubiquitous in galaxy halos out to impact parameters of $\sim$\,150 kpc. These findings indicate that outflows are not only common but also capable of entraining multiphase material and sustaining extended metal-enriched reservoirs over gigayear timescales. Complementary observations of metal-line emission in nearby galaxies reinforce this picture, revealing kinematically broad and spatially extended outflow components traced by \SiII\ and other far-UV transitions \citep[e.g.,][]{Chisholm2015, Chisholm2016, Xu2022, Xu2023}. These emission-line studies offer a spatially resolved, in situ view of the inner regions of galactic winds, complementing absorption line measurements along off-center lines of sight.

Despite significant observational progress, the physical interpretation of UV emission and absorption line profiles remains challenging due to the combined effects of resonant scattering and the complex, multiphase structure of the ISM and CGM. Previous modeling efforts have often focused on individual transitions — particularly \lya\ — using Monte Carlo radiative transfer (RT) simulations with idealized assumptions about the geometry and kinematics of neutral hydrogen \citep[e.g.,][]{Richling03, Ahn03, Ahn04, Dijkstra06a, Dijkstra06b, Verhamme2006, Gronke2016}. Metal lines, on the other hand, are frequently analyzed using simplified empirical or semi-analytic models. Two widely adopted approaches — the partial covering model (PCM) and the semi-analytical line transfer (SALT) framework — highlight the limitations of such techniques \citep[e.g.,][]{Scarlata2015, Carr2018, Xu2022}. The PCM phenomenologically assumes partial coverage of the central source by the absorbing material and infers the covering fraction and optical depth as functions of the observed velocity. Since it does not account for complex RT effects, this approach may substantially underestimate the gas column densities and outflow rates.\citep{Huberty2024, Carr2025, Jennings2025}. In contrast, SALT improves upon purely empirical models by incorporating radial velocity gradients and solving the RT equation semi-analytically. However, SALT also relies on a number of idealized assumptions. In particular, it is built upon the Sobolev approximation, which requires large velocity gradients and therefore limits its applicability to outflows with nearly flat or non-monotonic velocity profiles or significant velocity dispersion. As a result, SALT cannot fully capture RT effects associated with non-monotonic velocity fields, multiple scatterings, or significant turbulent and thermal line broadening. In addition, while SALT allows for flexible outflow geometries, it assumes a continuous, homogeneous wind rather than a clumpy, multiphase medium, which may not fully capture the physical complexity of galactic outflows. These simplifications restrict SALT’s ability to accurately recover outflow properties in galaxies with complex kinematics or narrow observational apertures that fail to capture the full extent of the outflow \citep{Carr2025}. More sophisticated 3D Monte Carlo RT simulations have also been applied to metal lines (e.g., \citealt{Michel-Dansac2020, Garel2024}). However, these studies are typically performed in idealized homogeneous gaseous media and have not yet been systematically applied to large observational galaxy samples.

In addition, several studies have employed radiation-hydrodynamical cosmological zoom-in simulations combined with RT post-processing to generate synthetic line profiles for comparison with observations \citep[e.g.,][]{Mauerhofer2021, Katz2022, Gazagnes2023, Gazagnes2024, Jennings2025}. While these efforts have achieved reasonable agreement between simulated and observed spectra, they are subject to several limitations. In particular, the resulting synthetic spectra are frequently generated from large ensembles of sightlines through a single simulated galaxy, whose global properties and outflow kinematics may differ substantially from those of the observed systems. This mismatch complicates the physical interpretation of best-fit parameters derived from such comparisons and limits their applicability to individual galaxies. In addition, current simulations often lack the spatial and mass resolution required to fully capture the multiphase morphology and small-scale dynamics of galactic winds and the CGM, and RT is typically performed at resolutions much coarser than the photon mean free path.

All of the aforementioned limitations highlight the need for a physically motivated, unified modeling framework that can self-consistently handle multiple spectral lines, incorporate complex gas kinematics, and connect absorption and emission signatures across a broad range of ionization states. Previous analyses based on a single transition are often subject to strong parameter degeneracies, allowing very different physical conditions to reproduce similar line profiles (e.g., \citealt{Li22}). Moreover, parameters inferred independently from \lya\ frequently appear inconsistent with those derived from low-ionization metal lines, underscoring the limitations of single-line modeling \citep[e.g.,][]{Orlitova18}. Joint modeling of multiple ions therefore represents an important avenue for mitigating these degeneracies and achieving a physically consistent interpretation of galactic outflows. With recent advances in computational RT methods and the growing availability of high-quality, multi-line UV spectra, such comprehensive modeling is now becoming feasible, opening new windows into the structure and energetics of galactic winds.

In this work, we present the first joint RT modeling analysis of \lya, \CII, \SiII, \SiIII, \CIV, and \SiIV\ line profiles for a sample of 45 low-redshift star-forming galaxies from the COS Legacy Archive Spectroscopic SurveY\footnote{https://archive.stsci.edu/hlsp/classy} (CLASSY, \citealt{Berg2022}), supplemented by five additional starbursts with archival HST/COS spectroscopy \citep{Heckman2015}. While not all six transitions are available in every galaxy, our analysis leverages the available subset of lines in each target to jointly constrain the galactic outflow properties. We employ a self-consistent, multiphase outflow model that incorporates a clumpy gas structure and simultaneously accounts for resonant scattering in strictly resonant lines (such as \lya, \SiIII, \SiIV, and \CIV) as well as fluorescent re-emission in transitions with non-resonant decay channels (such as \SiII\ and \CII). This approach enables us to probe both the neutral and ionized components of galactic winds within a coherent physical framework. By jointly fitting emission and absorption features across multiple transitions, we break key degeneracies inherent in single-line modeling and evaluate whether a single, radially varying outflow can consistently reproduce the observed line profiles across a broad ionization range.

This paper is organized as follows. In Section~\ref{sec:RT_lines}, we introduce the \texttt{PEACOCK} RT framework and describe its physical assumptions and numerical implementation for modeling UV emission and absorption lines arising from clumpy, multiphase galactic winds. Section~\ref{sec:data} summarizes the \textit{HST}/COS observations from the CLASSY survey and the additional archival datasets used in this work. In Section~\ref{sec:individual_modeling}, we present the results of line-by-line RT modeling for \lya, \CII, \SiII, \SiIII, \SiIV, and \CIV, demonstrating how the framework reproduces the diverse observed spectral morphologies. Section~\ref{sec:physical_basis} examines the physical origin of these successful fits, emphasizing how bulk outflow motions, turbulence, ion column densities, and dust extinction each leave distinct and non-degenerate imprints on the emergent line profiles. In Section~\ref{sec:param_compare}, we present a cross-transition comparison of the best-fit model parameters, examining the physical consistency of the inferred bulk outflow velocities, turbulent motions, and ion column densities across different ions. In Section~\ref{sec:multi_line}, we extend the analysis from single-line modeling to joint multi-line spectral fitting and demonstrate how shared kinematic constraints improve the robustness of the inferred outflow properties. We then present a discussion in Section~\ref{sec:discussion}, including the key advances of this work and its caveats and limitations. Finally, we summarize our main conclusions in Section~\ref{sec:conclusion}. A companion paper (Paper II) will build upon the modeling framework and results presented here to connect the inferred galactic wind properties to global host-galaxy properties, quantify the kinetic energy and pressure budget of the cool-to-warm CGM wind, and investigate the physical origin of the large turbulent velocities, including the role of stellar feedback as a dominant energy source.


\section{Radiative Transfer Modeling of UV Emission and Absorption Lines}\label{sec:RT_lines}

We introduce \texttt{PEACOCK} (Profiles of Emission and Absorption from Clumpy Outflows with Complex Kinematics), a physically motivated radiative transfer (RT) framework developed to model UV absorption and emission line profiles emerging from a multiphase, clumpy medium. \texttt{PEACOCK} is built on the 3D \lya\ Monte Carlo RT code \texttt{tlac} \citep{Gronke14}, and in particular inherits the structural framework for \lya\ propagation in a clumpy medium \citep{Gronke2016,Gronke16_model}. It extends our previous modeling approach by enabling ion-specific RT calculations and incorporating both turbulent and radially varying velocity fields. The framework supports the joint modeling of multiple emission and absorption lines from various species (e.g., \lya, \CII, \SiII, \SiIII, \CIV, \SiIV), thereby helping to break degeneracies and place significantly tighter constraints on key model parameters.

\begin{figure*}
\centering
\includegraphics[width=\textwidth]{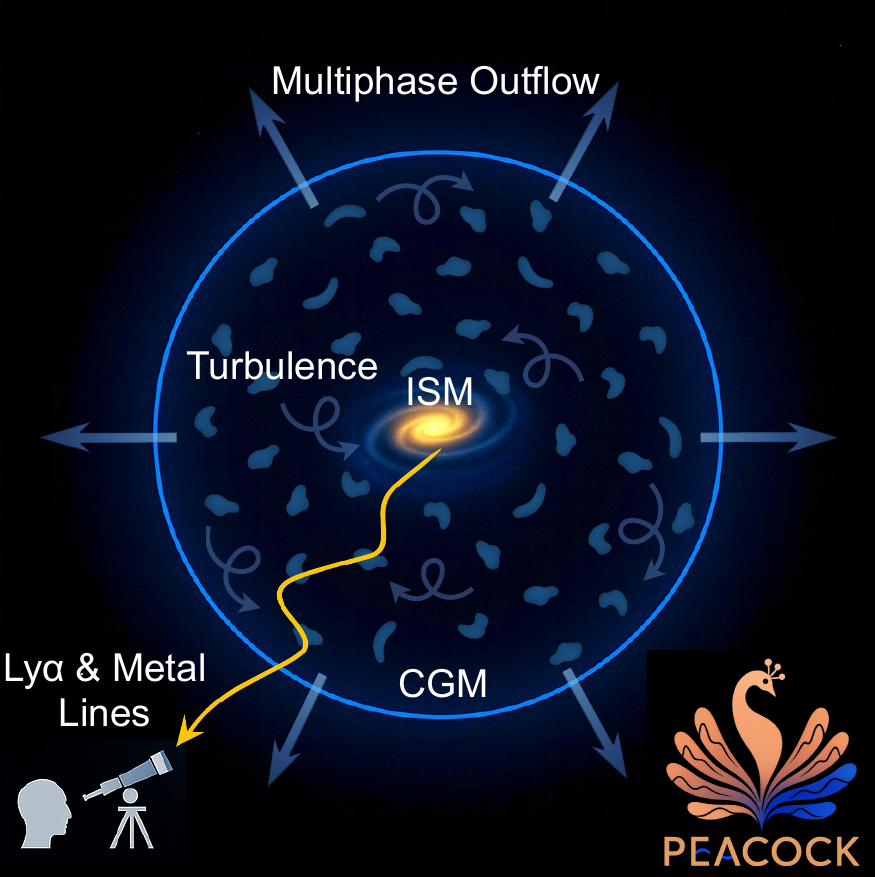}
    \caption{\textbf{Schematic of the multiphase, clumpy RT model \texttt{PEACOCK} used in this work. } A central star-forming galaxy is embedded in a multiphase, clumpy CGM halo. Cool and warm ($T\!\sim\!10^{4}$ – $10^{5}\,$K) gas clumps populate the spherical halo with a radial number density profile. Each clump exhibits intrinsic microscopic turbulence ($b_{\rm D,\,cl}$), macroscopic turbulent motions ($\sigma_{\rm cl}$), and participate in a large-scale radial outflow ($v_{\rm cl,\,out}(r)$). Photons at the wavelengths of \lya\ and UV metal lines (e.g., \CII, \SiII, \SiIII, \CIV, \SiIV) are assumed to be emitted by the central galaxy and propagate through the clumpy medium, undergoing resonant scattering (and, for partially resonant transitions like \CII\ and \SiII, fluorescence), as well as dust absorption and scattering. Note that in the model, clumps are idealized as spheres rather than irregular shapes shown in the schematic.}
    \label{fig:schematic}
\end{figure*}

\subsection{Basic Physics and Model Configuration}
We consider RT through galactic winds in star-forming galaxies, where photons emitted from the central star-forming regions propagate through a multiphase medium, undergoing resonant scattering, absorption, and re-emission. Motivated by this physical picture and by the increasingly favored view that galactic winds in low-$z$ star-forming galaxies are multiphase and clumpy in nature \citep[see e.g.,][]{Fielding2022, Faucher2023, Thompson2024}, as illustrated in Figure~\ref{fig:schematic}, we adopt a minimal geometric realization of the outflow. Following Occam’s razor, we begin with the simplest configuration: the cool-to-warm ($\sim10^{4} - 10^{5}\,\rm K$) gas is represented by a population of identical, spherical clumps distributed within a spherical halo bounded by an inner and an outer radius ($R_{\rm in}$ and $R_{\rm out}$, respectively)\footnote{The RT model is intrinsically scale-free and can be rescaled to different physical sizes without loss of generality. The specific physical scales adopted in the numerical implementation are chosen for convenience and do not affect the dimensionless RT behavior or the inferred kinematic and column density parameters. In Paper~II, when converting the model results to physical quantities such as the mass outflow rate, an absolute spatial scale must be specified. We therefore adopt $R_{\rm in} = 2\,r_{50}$ as the clump launch radius, where $r_{50}$ is the galaxy half-light radius measured by \citet{Xu2022}, and assume $R_{\rm out} = 10\,R_{\rm in}$.}. Each clump is characterized by a single radius $R_{\rm cl}$, and the clump number density decreases with galactocentric distance according to a power-law profile, $n_{\rm cl}(r)\propto r^{-2}$.

The clumps may contain a range of ionic species in various ionization states, including \HI, C$^{+}$, Si$^{+}$, Si$^{2+}$, C$^{3+}$, and Si$^{3+}$. Unlike the two-phase model adopted in our previous work \citep[e.g.,][]{Erb23, Li2024}, which incorporated a diffuse, volume-filling hot inter-clump medium, the present model assumes that all neutral and low- to intermediate-ionization gas relevant for the UV metal-line transitions is assumed to reside within the clumps. This change is motivated by the expectation that the volume-filling hot phase has a very low neutral fraction, while the corresponding metal-ion column densities are also sufficiently small that their contribution to the metal-line RT is negligible. We therefore exclude a hot inter-clump medium in the current model to reduce complexity while retaining the dominant contributors to the observed line profiles. We further assume that \lya\ and metal-line photons, whether originating from continuum or line emission, are produced in the central galaxy and propagate through the clumpy medium. This assumption effectively treats the emission as arising from a compact central source and can be rescaled to account for more extended emission regions. The RT of each transition is computed independently, taking into account the relevant ions and atomic transitions, as discussed in the next section.

The emergent spectra are derived from the frequency distribution of photons escaping the spherical CGM halo. Owing to the intrinsically isotropic geometry of the model, the emergent photon frequency distribution is independent of viewing direction. We therefore include escaping photons over all directions to maximize the signal-to-noise ratio of the model spectra, subject to a maximum impact parameter $b_{\rm max}$ that mimics the effective aperture of the observation (see discussion below). In subsequent spectral fitting, this procedure implicitly assumes that the observed down-the-barrel spectra sample a representative subset of the overall distribution of escaping photons from galaxies.

We then assign several modes of motion\footnote{In this work, we assume both the coherent bulk outflow and the macroscopic random motions to be purely radial, since down-the-barrel spectra are most sensitive to radial velocity components.} to the clumps. First, each clump is assumed to possess an intrinsic line broadening that includes both thermal and non-thermal contributions. This small-scale, internal broadening is characterized by the clump Doppler parameter, $b_{\rm D,\,cl}$. 
Second, we assume that the clumps have large-scale macroscopic clump-to-clump random motions, parameterized by a Gaussian velocity distribution with a standard deviation of $\sigma_{\rm cl}$. Lastly, we assign a large-scale galactic outflow velocity to the clumps. The outflow follows a physically motivated radial velocity profile in which clumps are initially accelerated by a force that scales with radius as $r^{-2}$ and are subsequently decelerated by the gravitational pull of an isothermal\footnote{In our previous work \citep[e.g.,][]{Erb23, Li2024}, we adopted NFW-type dark matter halo density profiles. In this study, we instead employ a slightly simpler isothermal profile to ensure that the clump outflow velocity is independent of halo mass, thereby enabling consistent modeling across a galaxy sample that spans a broad range of masses and redshifts. We find that the resulting outflow velocity profiles under the two assumptions exhibit similar shapes.} dark matter halo. The motion of each outflowing clump is governed by the following kinematic equation \citep[e.g.,][]{Murray05, Martin2012, Dijkstra2012}:

\begin{equation}
\frac{\mathrm{d}v_{\rm cl,\,out}(r)}{\mathrm{d}t}=-\frac{GM(r)}{r^2} + Ar^{-2}
\label{eq:clump_momentum}
\end{equation}
where $M(r)$ scales linearly with $r$. After reparameterization, the solution to this differential equation can be written as:

\begin{align}
v_{\rm cl,\,out}(r)  = v_0 \sqrt{\left(1 - \frac{r_{\min}}{r}\right) R - \ln\left(\frac{r}{r_{\min}}\right)}
\label{eq:v}
\end{align}
where $r_{\min}$ denotes the clump launching radius. As a fiducial choice, we set $r_{\min} = R_{\rm in}$, corresponding to the inner boundary of the halo. Eq. (\ref{eq:v}) contains two free parameters $v_0$ and $R$, where $v_0$ sets the overall velocity normalization and and $R$ is a dimensionless parameter that characterizes the ratio of outward acceleration to gravitational deceleration (see e.g., \citealt{Heckman2015}). It is relatively straightforward to show that $v_{\rm cl,out}(r)$ increases with radius before declining, and reaches its maximum at $r/r_{\min} = R$. The corresponding maximum outflow velocity is
\begin{equation}
v_{\rm cl,\,out,\,max} = v_0 \sqrt{(R - 1) - \ln R}
\label{eq:vcl_max}
\end{equation}

In the RT model, the total number of clumps in the halo is determined by the volume filling factor, $F_{\rm V}$, which quantifies the fractional volume occupied by all clumps within the spherical halo. If $N_{\rm cl,\,tot}$ is the total number of clumps, each of radius $R_{\rm cl}$, then
\begin{equation}
F_{\rm V} 
=
\frac{N_{\rm cl,\,tot}\,R_{\rm cl}^3}
     {R_{\rm out}^3 - R_{\rm in}^3}
\end{equation}
where $R_{\rm in}$ and $R_{\rm out}$ denote the inner and outer boundaries of the clump distribution, respectively. In this sense, $F_{\rm V}$ is a global, halo-integrated quantity that characterizes the overall clump population\footnote{One may also define a \emph{local} or radius-dependent volume filling factor $F_{\rm V}(r)$, which expresses the fractional volume contributed by clumps within a thin spherical shell at radius $r$. In that case, the local volume filling factor is related to the clump number density by $F_{\rm V}(r) = \frac{4\pi}{3} n_{\rm cl}(r){R_{\rm cl}^3}$.}.

Assuming the clump number density follows a radially decreasing power law, $n_{\rm cl}(r) = Cr^{-2}$, the normalization constant $C$ of the clump distribution can be expressed as
\begin{equation}
C \;=\; \frac{F_{\rm V}\,\left(R_{\rm out}^{3}-R_{\rm in}^{3}\right)}{4\pi R_{\rm cl}^{3}\,\left(R_{\rm out}-R_{\rm in}\right)}
\;=\; \frac{F_{\rm V}\,\left(R_{\rm out}^{2}+R_{\rm out}R_{\rm in}+R_{\rm in}^{2}\right)}{4\pi R_{\rm cl}^{3}}
\end{equation}
\label{eq:C_ncl}

The clump covering factor, defined as the average number of clumps intercepted along a sightline from the galaxy to the observer, can then be written as (see e.g., \citealt{Hansen06})
\begin{equation}
\begin{split}
f_{\rm cl}
&= \int_{R_{\rm in}}^{R_{\rm out}} n_{\rm cl}(r)\,\pi R_{\rm cl}^{2}\,{\rm d}r \\
&= \frac{F_{\rm V}}{4R_{\rm cl}}\,
   \frac{\big(R_{\rm out}^{2}+R_{\rm out}R_{\rm in}+R_{\rm in}^{2}\big)
         \big(R_{\rm out}-R_{\rm in}\big)}{R_{\rm in}R_{\rm out}} 
\end{split}
\label{eq:fcl_r2}
\end{equation}

$f_{\rm cl}$ is the key parameter characterizing the degree of ``clumpiness'' in a clumpy medium \citep{Gronke2017}. In the limit $R_{\rm out} \gg R_{\rm in}$ (we adopt $R_{\rm out}/R_{\rm in} = 10$ in our model), this equation simplifies to
\begin{equation}
f_{\rm cl} \;\approx\; \frac{F_{\rm V}}{4R_{\rm cl}}\,
\frac{R_{\rm out}^2}{R_{\rm in}}
\end{equation}

Assuming each clump has a constant ionic column density $N_{\rm ion,\,cl}$, the average total column density along a given sightline can be expressed as
\begin{equation}
N_{\rm ion,\,LOS} \;=\; \frac{4}{3}\,f_{\rm cl}\,N_{\rm ion,\,cl}
\label{eq:NionLOS}
\end{equation}
where the factor of $4/3$ arises from the mean chord length through a sphere.

As we will show in the later sections, our spectral fitting experiments have shown that a simple, single-population model is generally sufficient to reproduce nearly all of the observed line profiles for \CII, \SiII, \SiIII, \CIV, and \SiIV. For \lya, however, the situation is more complex: in many cases, the absorption trough near the line center cannot be well reproduced by a single outflowing clump population. In such cases, we find that introducing an additional \HI\ component that is characterized by random motions and a small bulk velocity (could be either inflow or outflow) successfully resolves this issue. This secondary clump population shares the same geometric set-up as the outflowing one (i.e., the same clump radius and power-law radial distribution) but is assigned its own set of free parameters. Physically, this additional component may represent recycled material that has decelerated and re-virialized in the CGM, becoming dynamically decoupled from the bulk large-scale outflow, or a semi-static, high-column-density \HI\ component located near the ISM. We find that this component is frequently required to reproduce the \lya\ profiles but is rarely needed for the metal lines, likely because the much larger \HI\ column densities make recycled neutral gas much easier to detect in \lya\ spectra than the corresponding metal species. We will return to this issue in more detail in Section \ref{sec:lya}.

\subsection{Radiative Transfer of Individual Transitions}\label{sec:individual_RT}

In this section, we describe the RT treatment for the for the individual transitions relevant to this work — \lya, \CII, \SiII, \SiIII, \CIV, and \SiIV. These lines fall into two classes: strictly resonant (\lya, \SiIII, \CIV, \SiIV) and partially resonant with fluorescent channels (\CII, \SiII).

\begin{figure*}
\centering
\includegraphics[width=\textwidth]{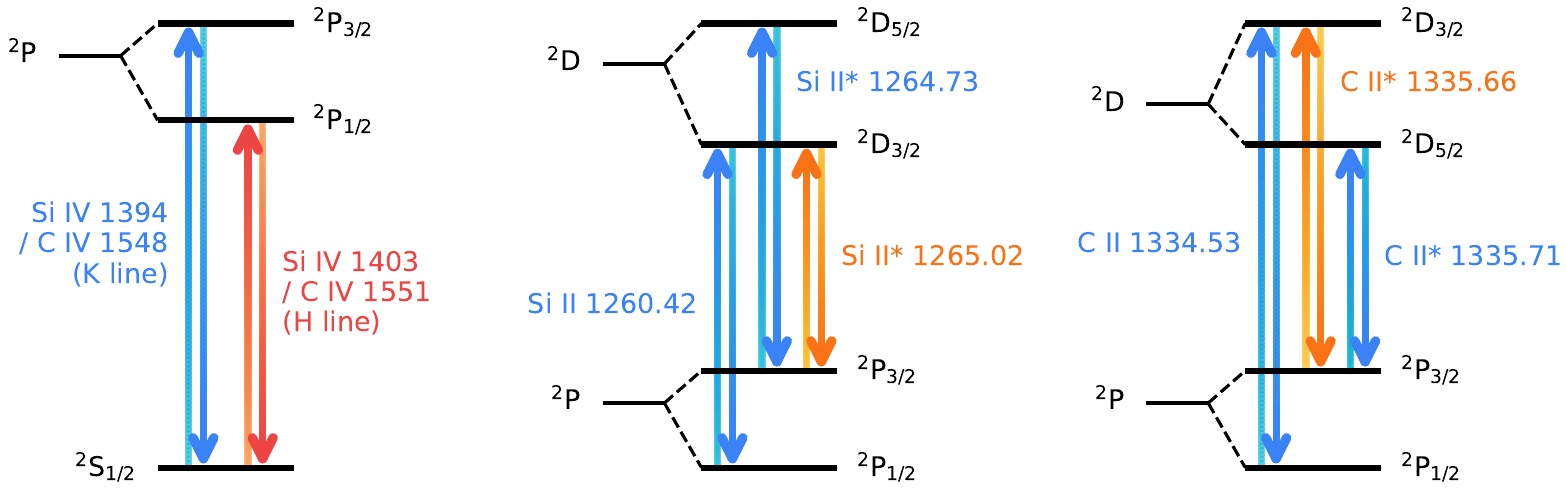}
    \caption{\textbf{Energy level diagrams that illustrate fine-structure splitting of ground and excited states for \CII, \SiII, \CIV, and \SiIV\ ions.} Starred labels (e.g., \SiII*, \CII*) indicate fluorescent transitions.  \emph{Left:} strictly resonant doublets like \CIV\ and \SiIV, which arise from transitions between the ground state $^2S_{1/2}$ and the fine-structure–split excited states $^2P_{1/2}$ and $^2P_{3/2}$. The oscillator strength of the higher-frequency component (the “K” line, shown in blue) is typically twice that of the lower-frequency “H” line (shown in red). \emph{Middle and right:} partially resonant systems like \SiII\ and \CII, where the ground term $^2P_{1/2,\,3/2}$ is fine-structure split, allowing both resonant and fluorescent channels to the excited $^2D_{3/2,\,5/2}$ states. In each set of three allowed transitions, one fluorescent line (shown in orange) has a much lower oscillator strength and effectively serves as a bridge between the other two transitions (one resonant and one fluorescent, both shown in blue).}
    \label{fig:energy_level}
\end{figure*}

\subsubsection{Strictly Resonant Transitions}
Ions such as C$^{3+}$ and Si$^{3+}$ have hydrogen-like electron configurations, each with a single valence electron. These ions often exhibit UV doublets that arise from transitions between the ground state $S_{1/2}$ and two fine-structure split excited states: $P_{1/2}$ and $P_{3/2}$. Since the ground state has no fine-structure splitting, both transitions are {\it{strictly resonant}}, without any fluorescent emission channels. For \lya, the wavelength separation between the two fine-structure transitions is negligible (only $\sim$\,0.006\,\AA), so it is typically treated as a single line. For \CIV\ and \SiIV, the wavelength separation between the two components is significantly larger (2.57\,\AA\ for \CIV\ and 9.02\,\AA\ for \SiIV), making them clearly distinguishable in velocity space as doublets. The oscillator strength ($f$) of the lower-wavelength component (often referred to as the “K” line) is typically twice that of the higher-wavelength “H” line. As for \SiIII, the 1206.5\,\AA\ transition occurs strictly between the ${}^{1}S_{0}$ and ${}^{1}P_{1}$ levels and has no allowed fluorescent channels.

\subsubsection{Partially Resonant Transitions}
Unlike highly ionized species such as C$^{3+}$ and Si$^{3+}$, ions like C$^{+}$ and Si$^{+}$ have more than one valence electron outside a closed shell, resulting in a richer energy level structure with multiple low-lying excited states. Transitions from the fine-structure-split ground levels $^2P_{1/2}$ and $^2P_{3/2}$ to excited $^2D_{3/2}$ and $^2D_{5/2}$ states give rise to multiple UV lines. The presence of fine-structure splitting in the ground state enables {\it{fluorescent absorption and emission}} pathways, in addition to strictly resonant scattering, as illustrated in Figure~\ref{fig:energy_level}. Previous studies have shown that incorporating these fluorescent transitions into modeling provides valuable constraints on the geometry and kinematics of galactic outflows (e.g., \citealt{Prochaska2011, Scarlata2015, Xu2023}).

For \SiII, three allowed transitions connect the ground-state sublevels ($^2P_{1/2,\,3/2}$) to the excited states ($^2D_{3/2,\,5/2}$). The strongest line at 1260.42\,\AA\ corresponds to the resonant transition $^2P_{1/2} \rightarrow {^2}D_{3/2}$, while two weaker fluorescent lines at 1264.73 and 1265.02\,\AA\ arise from decays of the upper states back to the $^2P_{3/2}$ ground sublevel. Since the two ground-state fine-structure levels are separated by only $\sim$\,0.036\,eV, both can be thermally populated, allowing all three transitions to be observed. The relative populations of the $^2P_{1/2}$ and $^2P_{3/2}$ levels are sensitive to local excitation conditions, particularly the electron density and the intensity of the UV radiation field (see e.g., \citealt{Xu2023}). For \CII, the energy level configuration and corresponding transitions are similar, except that the $^2D_{3/2}$ and $^2D_{5/2}$ excited states are inverted in energy due to a negative spin–orbit coupling constant.

Low-ionization species such as \SiII\ and \CII\ exhibit a set of three allowed transitions that are not of equal strength. The $^2P_{3/2} \rightarrow {^2}D_{3/2}$ transition (1265.02\,\AA\ for \SiII\ and 1335.66\,\AA\ for \CII; both shown in orange in Figure~\ref{fig:energy_level}) generally has the weakest oscillator strength — typically an order of magnitude lower than that of the other two transitions. This transition thereby effectively acts as a bridge through which photons originating from the $^2P_{1/2}$ ground sublevel can interact with the $^2D_{5/2}$ upper state. Although the probability for a photon to traverse a fluorescent channel via the $^2P_{3/2}$ sublevel is relatively small ($\sim$15\% for \SiII\ and $\sim$10\% for \CII, depending on the Einstein coefficients), once a photon reaches the $^2P_{3/2}$ level, it has a high probability of being excited into the $^2D_{5/2}$ state and subsequently escaping as fluorescent emission, since the optical depth of ions in the $^2P_{3/2}$ level is generally much lower than that in the $^2P_{1/2}$ level.

\subsubsection{Improvements to the RT Code}
To accommodate the metal lines, several major modifications to the original \lya\ RT code are required (see e.g., \citealt{Seon24, Chang24}).

(1) \textbf{Line cross section:} the cross-section for an atom with multiple transitions is given by a weighted sum of the cross-sections for each individual transition:

\begin{equation}
\sigma_\nu = \frac{1}{\sqrt{\pi} \Delta \nu_D} \sum_{i=1}^{N} \chi_i\, H(x + x_i, a_i)
\end{equation}
where \( \chi_i = f_i \left( \frac{\pi e^2}{m_e c} \right) \) is proportional to the oscillator strength \( f_i \) of the $i$-th transition, $x$ is the dimensionless frequency (see definition below), \( x_i \) is the offset of the \( i \)-th line center from the reference frequency, and \( H(x, a) \) is the Voigt-Hjerting function, defined as:

\begin{equation}
H(x, a) = \frac{a}{\pi} \int_{-\infty}^{\infty} \frac{e^{-y^2}}{(x - y)^2 + a^2} \, dy
\end{equation}
where $a = {\Gamma}/({4 \pi \Delta \nu_{\rm D}})$ is the natural line width, with $\Gamma$ denoting the Einstein A coefficient of the transition. The dimensionless frequency variable \( x \) is defined as:

\begin{equation}
x = \frac{\nu - \nu_0}{\Delta \nu_{\rm D}}
\end{equation}
where \( \Delta \nu_{\rm D} = \nu_0 (b_{\rm D}/c) \) is the Doppler broadening, and $b_{\rm D} = \sqrt{v_{\rm th}^{2}+v_{\rm turb}^{2}}$ is the Doppler parameter, which accounts for both thermal and turbulent motions. In practice, $\nu_0$ is often chosen to be the line-center frequency of the highest-frequency component in the multiplet (e.g., the K line of a resonant doublet), since the frequency separations between multiplet components are generally very small compared to their absolute frequencies. This choice simplifies the calculation without introducing significant inaccuracies.

(2) \textbf{Atom velocity sampling:} the parallel component of the normalized atomic velocity, defined as $u_\parallel = v_\parallel / b_{\rm D}$, depends on the frequency of the incident photon. For a single line, $u_\parallel$ should be sampled from the following distribution (e.g., \citealt{Zheng2002}):

\begin{equation}
f(u_\parallel | x) = \frac{a}{\pi H(x, a)} \cdot \frac{e^{-u_\parallel^2}}{(x - u_\parallel)^2 + a^2}
\label{eq:vpar}
\end{equation}
which can be understood as the (normalized) product of two factors: the Maxwellian (Gaussian) distribution of atomic velocities, \( e^{-u_\parallel^2} \), and the Lorentzian line profile, \( 1 / \left[ (x - u_\parallel)^2 + a^2 \right] \), which characterizes the probability of the atom being in resonance with the photon.

In the case of a multiplet, the probability distribution from which \( u_\parallel \) of the scattering atom should be sampled, is given by a weighted sum over the individual line contributions:

\begin{equation}
f_{\rm multi}(u_\parallel | x) = \sum_{i=1}^N w_i\, f(u_\parallel | x + x_i, a_i)
\end{equation}
where \( w_i\) is the relative probability that the photon at frequency \( x \) interacts with the \( i \)-th transition. This weight is given by:

\begin{equation}
w_i = \frac{f_i\, H(x + x_i, a_i)}{\sum_{j=1}^{N} f_j\, H(x + x_j, a_j)}
\end{equation}
where \( f_i \) is the oscillator strength of the \( i \)-th line. This formulation ensures that the contribution from each transition is properly weighted by its absorption cross section at the frequency of the incoming photon. The perpendicular component of the normalized atomic velocity is then drawn from a Gaussian distribution, as is standard in \lya\ RT modeling.

(3) \textbf{Fluorescent absorption and re-emission:} for lines such as \SiII\ and \CII\ that possess fluorescent channels, resonant photons can be converted into fluorescent photons (and vice versa) through fluorescent channels. For example, a \SiII\ photon can be transformed into a \SiII*\ photon via the pathway $^2P_{1/2} \rightarrow ^2D_{3/2} \rightarrow ^2P_{3/2}$. In our Monte Carlo RT scheme, both resonantly scattered and fluorescently re-emitted photons are explicitly tracked. Following each absorption event, the decay channel is selected probabilistically based on the Einstein $A$ coefficients of the competing transitions, and the frequency of the re-emitted photon is adjusted accordingly to reflect the chosen transition. A detailed description of how these probabilities are computed, along with the corresponding frequency updates, is provided in Appendix~\ref{sec:fluorescent_frequency}.

\subsubsection{Photon Sources and Emission Mechanisms}
For all six transitions discussed above, the photons entering the RT calculation can originate from either line emission or continuum emission produced in the central galaxy. Line emission may be generated through nebular processes within \HII\ regions, including photoionization followed by recombination as well as collisional excitation and subsequent radiative de-excitation. For resonant doublets such as \CIV\ and \SiIV, differences in oscillator strengths give rise to a characteristic doublet ratio ($F_{\rm K} / F_{\rm H}$) of approximately 2:1 in the emitted line fluxes.

In addition to line emission, UV continuum photons from the central galaxy near the line centers can resonantly excite ions, which subsequently decay and re-emit photons -- a mechanism commonly referred to as continuum pumping. In this work, we do not include potential in-situ emission from the clumps in the CGM, as the observed spectra can be well reproduced by photons originating from the central galaxy alone.

In our \lya\ modeling, the net observed equivalent width
($\mathrm{EW_{net}} = \mathrm{EW_{em}} - \mathrm{EW_{abs}}$, with positive values indicating net emission) is always greater than zero. This implies that the emergent \lya\ profiles necessarily include a line-emission component in addition to the continuum emission. We therefore adopt a composite photon source that includes both nebular line emission and continuum emission, with their relative contributions parameterized by the intrinsic equivalent width ($\mathrm{EW_{int}}$; see details below). For high-ionization resonant doublets such as \CIV\ and \SiIV, the observed profiles can exhibit pure emission, pure absorption, or P-Cygni--like combinations of both (e.g., \citealt{Jaskot2017, Berg2019, Schaerer2022, Izotov2024}). In these cases, the presence of net emission ($\mathrm{EW_{net}} > 0$) is interpreted as evidence for an underlying nebular line-emission component originating in the central galaxy, in addition to resonantly processed continuum photons. Accordingly, when $\mathrm{EW_{net}} > 0$, both line and continuum emission are included in the photon source, whereas for $\mathrm{EW_{net}} \le 0$, we assume a purely continuum source. By contrast, for lower-ionization species such as \CII, \SiII, and \SiIII, all objects in our sample exhibit $\mathrm{EW_{net}} < 0$, indicating that the observed profiles are dominated by absorption. In these cases, assuming a purely continuum photon source provides satisfactory fits to all objects.

\subsection{Additional Parameters}

Before fitting the observed UV spectra with \texttt{PEACOCK}, we introduce several supplementary model parameters.

\renewcommand{\arraystretch}{1.15}
\begin{table*}
\footnotesize
\centering
\caption{Summary of free parameters used in the \texttt{PEACOCK} radiative transfer framework}
\label{tab:model_parameters}
\begin{tabular}{lll}
\hline
\hline
\textbf{Symbol} & \textbf{Parameter} & \textbf{Physical Meaning} \\
\hline
$F_{\rm V}$ & Clump volume filling factor & Fraction of the halo volume occupied by clumps \\
$N_{\rm ion,\,cl}$ & Clump ion column density & Column density of a given ionic species in each clump \\
$b_{\rm D,\,cl}$ & Clump Doppler parameter & Intrinsic (thermal + microturbulent) velocity dispersion within a single clump \\
$\sigma_{\rm cl}$ & Macroscopic random velocity & Random (turbulent) velocity dispersion among clumps \\
$v_0$ & Outflow velocity normalization & Characteristic velocity scale of the clump radial outflow (Eq.~\ref{eq:v}) \\
$R$ & Acceleration parameter & Dimensionless ratio of outward acceleration to gravitational deceleration (Eq.~\ref{eq:v}) \\
$\tau_{\rm d,\,cl}$ & Dust optical depth & Effective dust absorption optical depth of each clump \\
$b_{\rm max}$ & Maximum photon impact parameter & Maximum impact parameter of photons included in the model spectra\\
$\Delta v$ & Velocity shift & Systemic velocity offset between the best-fit model and the observed spectra \\
$f_{\rm scale}$ & Continuum scaling factor & Scaling factor used to match the continuum level of the model and observed spectra \\
$v_{\rm bulk,\,sec}$ & Bulk velocity of secondary clump population & A small net inflow/outflow velocity of the secondary clump population (if included)\\
$R_{\rm line}$ & Line-to-continuum photon ratio & Ratio of line to continuum photons; only used for models with $\mathrm{EW_{net}}>0$ \\
$r_{\rm pop}$ & Fine-structure population ratio & Relative population of the ground-state fine-structure levels for \SiII\ and \CII\ \\
\hline
\end{tabular}
\end{table*}

For \lya, \SiIV, and \CIV, in objects that have $\mathrm{EW_{net}} > 0 $, a composite photon source function that includes both line and continuum emission is required. To quantify their relative contributions, we define the ratio of line to continuum photons as $R_{\rm line} = N_{\rm line}/N_{\rm cont}$. For each model, we perform two separate RT simulations: one using 40,000 line photons (with a 2:1 split between the K and H transitions for \SiIV\ and \CIV), and another using 40,000 continuum photons centered around the line center(s). In the line-emission simulation, the intrinsic emission is modeled as one or two Gaussian profiles centered at the line center(s) with a intrinsic velocity dispersion\footnote{The intrinsic velocity dispersion $\sigma_{\rm int}$ is fixed to a typical value as it has to be specified prior to the RT simulations. The exact choice of $\sigma_{\rm int}$ has little effect on the emergent line profiles provided it is reasonably small, since the final line shape is dominated by RT effects and by broadening from the much larger clump velocity dispersion (see e.g., \citealt{Li22, Erb23}).} of $\sigma_{\rm int} = 50\,\rm km\,s^{-1}$. In the continuum-emission simulation, the intrinsic spectrum is assumed to be flat around the mean of the K and H line centers, over a broad velocity range $\Delta V$ (typically $2500$ – $3000\,\mathrm{km\,s^{-1}}$, depending on the line separation) that encompasses both transitions. Since the RT of each photon is independent, a composite model spectrum can be constructed by combining continuum photons with a fraction of the line photons (or vice versa), modulated by the parameter $R_{\rm line}$. The corresponding intrinsic EW can then be expressed as ${\rm EW}_{\rm int,\,K+H} = R_{\rm line} \Delta V {\lambda_0} / c$, where ${\lambda_0}$ is the average wavelength of the K and H lines (see also \citealt{Li2025}).

We also account for the effects of dust absorption and scattering within the clumps. The amount of dust absorption in each clump is characterized by the dust optical depth, $\tau_{\rm d,\,cl}$, assuming a fiducial Small Magellanic Cloud (SMC) dust model following \citet{Draine03, Draine03b}. Dust scattering is described by the scattering albedo, $\alpha_{\rm d}$, and the asymmetry factor, $g$, which specify the efficiency and angular distribution of scattering in the Henyey–Greenstein phase function. Both $\alpha_{\rm d}$ and $g$ are wavelength dependent and are adopted from \citet{Draine03} for the relevant transitions. The dust treatment adopted here is consistent with that used in our previous work (e.g., \citealt{Li2025}).

To account for aperture losses in real observations, i.e., the incomplete collection of emission or absorption due to the finite slit size of spectrographs, we introduce an aperture correction factor. In our model, this effect is represented by the parameter $b_{\rm max}$, which specifies the maximum impact parameter within which photons are included when constructing the model spectra. This parameter effectively mimics the observational limitation imposed by finite spectrograph apertures. For instruments such as the \emph{HST}/COS, whose primary science aperture is circular, this treatment directly reflects the geometry of the observational beam. The observational interpretation of $b_{\rm max}$ is discussed in Appendix~\ref{sec:compare_impact_param}.

In addition, we include a velocity shift parameter, $\Delta v$, to account for any potential offset between the systemic redshift of the modeled line profile and that inferred from observations. To provide flexibility in matching the continuum level of the data, we also include a scaling factor, $f_{\rm scale}$\footnote{This scaling factor is introduced to account for possible contamination from nearby absorption or emission features that may artificially affect the observed continuum level.}. In practice, we find that $\Delta v$ is typically small and well within the observational uncertainties, while $f_{\rm scale}$ remains close to unity, as both the model and observed spectra are normalized prior to comparison\footnote{To facilitate DNN training and ensure consistent comparison between model predictions and observational data, all spectra are normalized such that the integral of the flux density over velocity is unity. We note that this normalization differs from the conventional observational practice of continuum normalization, in which the continuum level is generally set to unity.}.

As a result, our modeling includes the following free parameters for fitting individual line profiles: the clump volume filling factor ($F_{\rm V}$), the clump ion column density ($N_{\rm ion,\,cl}$), the clump Doppler parameter ($b_{\rm D,\,cl}$), the macroscopic random velocity  ($\sigma_{\rm cl}$), the clump outflow velocity normalization ($v_0$), the acceleration parameter ($R$), the dust optical depth ($\tau_{\rm d,\,cl}$), the maximum photon impact parameter\footnote{The parameter $b_{\rm max}$ sets the maximum impact parameter of photons included in the construction of the model spectra and serves as a proxy for the effective spatial extent probed by the observations. Since the model is scale-free, $b_{\rm max}$ does not uniquely correspond to the actual physical COS aperture size, but is expected to correlate with it qualitatively.} ($b_{\rm max}$), the velocity shift ($\Delta v$), and the continuum scaling factor ($f_{\rm scale}$). For spectra that require an secondary clump population (primarily for \lya), a separate set of \{${F_{\rm V}, N_{\rm ion,\,cl}, b_{\rm D,\,cl}, \sigma_{\rm cl}, \tau_{\rm d,\,cl}}$\}, along with a small, constant bulk velocity $v_{\rm bulk,\,sec}$, is assigned. For \lya, \SiIV, and \CIV\ spectra that have $\mathrm{EW_{net}} > 0 $, an additional parameter, the line-to-continuum photon ratio $R_{\rm line}$ is introduced. For \SiII\ and \CII, we instead include a fine-structure population parameter, $r_{\rm pop} = n_{^2P_{3/2}} / n_{^2P_{1/2}}$, to characterize the relative level populations in the ground-state doublet. A summary of all parameters and their physical meanings is provided in Table \ref{tab:model_parameters}.

\subsection{Fitting Pipeline}
In this section, we outline the main steps of our spectral fitting pipeline:

(1) \textbf{Mock Spectral Library Construction:} 
Our standard RT model has eight free parameters: 
$F_{\rm V}$, $N_{\rm ion,\,cl}$, $b_{\rm D,\,cl}$, $\sigma_{\rm cl}$, $v_0$, $R$, $\tau_{\rm d,\,cl}$, and $R_{\rm line}$ or $r_{\rm pop}$, along with three (or four) post-processing parameters: $b_{\rm max}$, $R_{\rm line}$ (if needed), $\Delta v$, and $f_{\rm scale}$. Properly sampling such a high-dimensional parameter space would traditionally require running a large number of RT models. In our previous works (e.g., \citealt{Erb23, Li2024}), we constructed evenly spaced multidimensional model grids and performed linear interpolation during the fitting. In this study, however, we introduce a more powerful approach based on deep neural networks (DNNs), which can efficiently and accurately learn the nonlinear mapping between model parameters and emergent spectra. This technique allows us to achieve much higher accuracy while requiring orders of magnitude fewer RT calculations compared to traditional grid interpolation.

For each observed spectrum, we begin by specifying an initial guess for the prior ranges of all model parameters\footnote{Since the spectra of our galaxies exhibit a wide range of morphologies, we find that constructing smaller, object-specific parameter sets is more efficient for both DNN training and parameter fitting than building a single, large set of models intended to capture the diversity of all galaxies. We therefore do not list fixed prior ranges for the model parameters, as these are tailored to each transition and each galaxy.} . To uniformly explore the multidimensional parameter space, we generate two hundred random parameter sets using Latin hypercube sampling (LHS), a stratified statistical method that produces near-random samples from a multidimensional parameter distribution \citep{McKay1979}. A full RT simulation is performed for each parameter set to produce synthetic spectra, which collectively form the mock spectral library. Since the variation of each parameter generally induces continuous and monotonic changes in the resulting spectra (as demonstrated later in Section \ref{sec:individual_params}), we find that a library of a few hundred models is sufficient for each DNN to learn the parameter–spectrum mapping with high accuracy in the high-dimensional parameter space. This property also motivates our adaptive expansion and targeted resampling strategy (step 4 below), whereby additional samples are iteratively added following the initial fitting run, with each expansion requiring only a few hundred new models to adequately refine the learned mapping.

(2) \textbf{DNN Training:} The synthetic spectra from the mock library are used to train the DNN. Before training, we generate twenty spectral realizations per model, each corresponding to a different value of the post-processing parameter $b_{\rm max}$ (and $R_{\rm line}$, if applicable). The corresponding spectra are then constructed for each realization. Each model spectrum is subsequently rebinned onto a uniform velocity grid and convolved with a Gaussian kernel (FWHM = $60\,\mathrm{km\,s^{-1}}$) to simulate instrumental broadening, followed by normalization. This process produces the finalized mock spectral dataset used for subsequent DNN training.

We train our DNNs using the \texttt{TensorFlow} \citep{Abadi2016} framework. The network architecture is optimized by tuning the number of hidden layers and neurons per layer to achieve a balance between accuracy and generalization. Our experiments show that a fully connected feed-forward network with four hidden layers of 512 neurons each provides a good compromise between model complexity and performance. After each training session, we perform a diagnostic check to ensure model reliability. Specifically, we identify the ten spectra with the largest deviations from their corresponding training data and visually inspect them. We confirm that the residuals are primarily caused by numerical noise rather than inadequate learning. If the discrepancies appear systematic, we adjust the training hyperparameters and retrain the network until satisfactory convergence is achieved. Once trained, the network serves as an efficient surrogate model capable of predicting spectra for arbitrary parameter combinations within the sampled range, yielding orders-of-magnitude speedup compared to direct RT simulations (see Appendix~\ref{sec:dnn_validation} for validation tests).

(3) \textbf{Fitting the Observed Spectrum:} Our fitting procedure combines the trained DNN with a nested sampling algorithm \citep{Skilling04, Skilling06} to efficiently explore the parameter space of the RT models. The DNN predicts the emergent line spectrum for any input parameter vector within the trained range. For a given parameter set $\boldsymbol{x}$, the model spectrum is generated as
\begin{equation}
F_{\rm model}(v) = {\rm DNN}\left(\boldsymbol{x}\right)
\end{equation}
where the input parameters are first standardized using pre-computed scalers, and the output fluxes are subsequently de-normalized to physical units. The predicted spectrum is then interpolated onto the observed velocity grid, shifted according to the fitted velocity offset $\Delta v$, and rescaled by the continuum correction factor $f_{\rm scale}$.

We perform parameter inference using the \texttt{dynesty} nested sampler \citep{Speagle20}, which efficiently maps the posterior probability distribution even for complex, multi-modal likelihood surfaces. The likelihood function is defined as
\begin{equation}
\begin{split}
\ln \mathcal{L}(\boldsymbol{x})
  = -\frac{1}{2}\sum_{i\in\mathcal{M}}
  \Bigg[
  &\frac{\big(F_{\rm obs}(v_i)-F_{\rm model}(v_i\mid \boldsymbol{x})\big)^2}
        {\sigma_{\rm tot}^2(v_i)}  \\
  &+\,\ln\!\big(2\pi\,\sigma_{\rm tot}^2(v_i)\big)
  \Bigg]
\end{split}
\end{equation}
where $F_{\rm obs}$ denotes the observed flux, $\sigma_{\rm tot}^2=\sigma_{\rm obs}^2+\sigma_{\rm mod}^2$ represents the total variance that includes both observational and model uncertainties\footnote{We adopt a Poisson-like model uncertainty, $\sigma_{\rm mod}(v_i)=f\sqrt{\max[F_{\rm model}(v_i),0]/N_{\rm phot}}$, where $N_{\rm phot}$ is the number of photons used in the RT simulations and $f$ is an empirical scaling factor (set to 0.1 in this work).}, and $\mathcal{M}$ denotes the set of valid spectral pixels after masking regions contaminated by nearby lines.

During each run, \texttt{dynesty} explores the multi-dimensional parameter space (including both physical and post-processing parameters) and maximizes the likelihood function to identify the global best-fit solution. After convergence, the sampler outputs the posterior-weighted parameter distributions and the corresponding best-fit spectrum. Credible intervals (e.g., 16–84\%) and covariance matrices are computed for all parameters. Each fitting run also produces trace and corner plots to visually evaluate convergence and parameter degeneracies.

\begin{figure*}
\centering
\includegraphics[width=\textwidth]{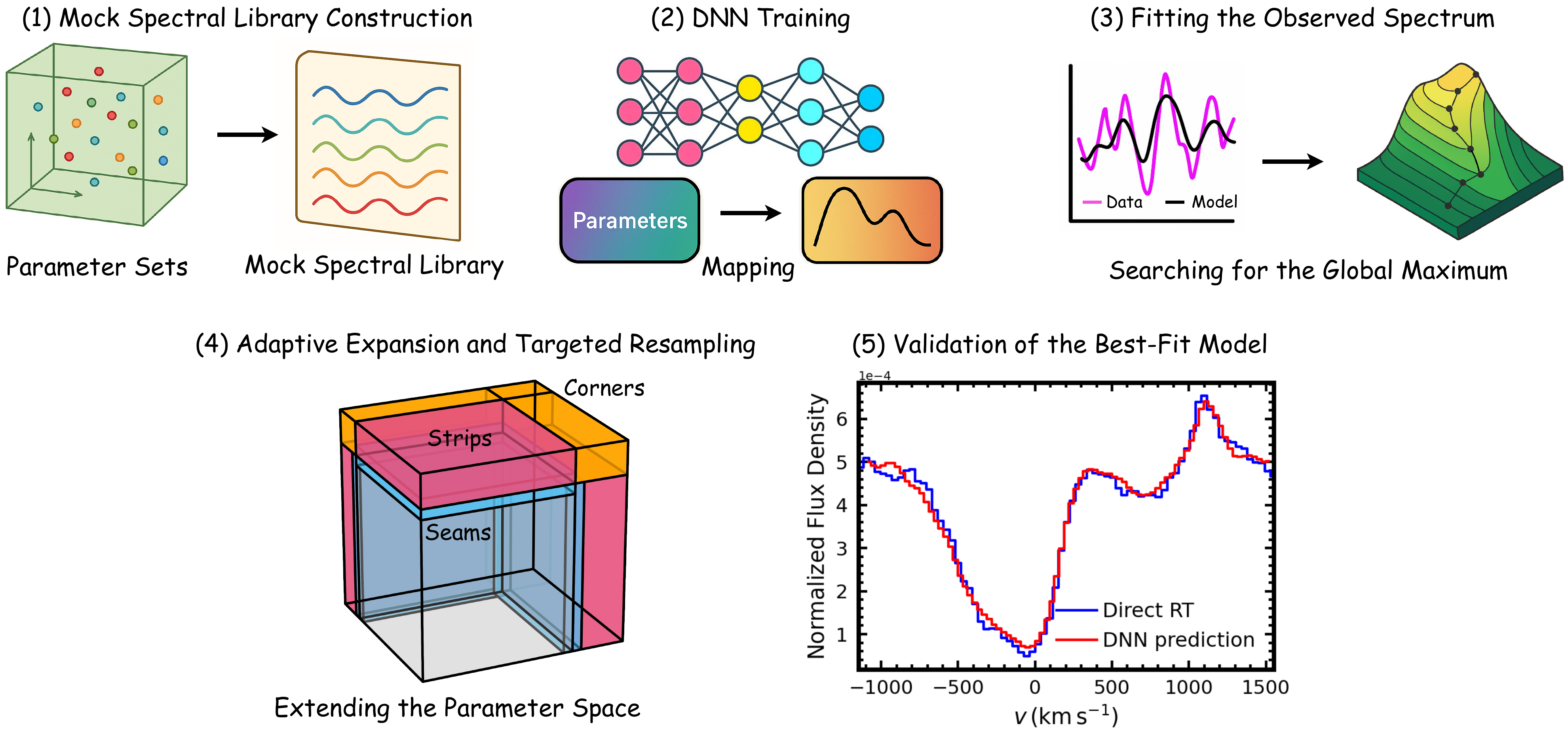}
    \caption{\textbf{Overview of the five major steps in our spectral fitting pipeline.} (1) A mock spectral library is constructed by performing RT simulations for a number of parameter sets that sample the multidimensional model space. (2) A DNN is trained to learn the nonlinear mapping between model parameters and emergent spectra. (3) The trained DNN model is coupled with a nested sampling algorithm to efficiently fit the observed spectra and locate the global likelihood maximum in parameter space. (4) Posterior-guided adaptive expansion and targeted resampling are applied when the posterior extends to the edges of the explored space, adding new training samples to refine the DNN. (5) The reliability of the best-fit solution is validated by comparing the DNN-predicted spectrum with a direct RT calculation using the same parameters, confirming close agreement between the two.}
    \label{fig:dnn_schematic}
\end{figure*}

(4) \textbf{Adaptive Expansion and Targeted Resampling:} After the initial fit, we assess whether the posterior distribution approaches the edges of the prior ranges in any dimension and, if so, adaptively expand the search domain and resample new training points near the boundaries. More specifically, we examine each parameter dimension and checks whether its posterior extends into the outer few percent of the prior range (by default, the outer 5\%). For any such dimension, we use a user-defined configuration that specifies how much the range should be expanded on each side.

Once the new parameter boundaries are defined, the code generates additional training samples concentrated in the newly opened regions of parameter space. Instead of sampling uniformly, the resampling is intentionally weighted toward the boundary regions where the posterior was previously truncated. Specifically, most of the new points are placed in three regions: (1) narrow \emph{strips} along the expanded sides, which densely populate the parameter edges; (2) \emph{corners}, where two or more parameters simultaneously approach their prior limits, capturing potential correlations between edge dimensions; and (3) thin transitional \emph{seams} — or “stitching bands” — that slightly overlap with the original parameter space to ensure a smooth connection between the old and newly extended parameter domains. Roughly half of the new samples are drawn from the edge strips, about one-third from the corners, and the remainder from these seams that bridge the old and new regions. This targeted resampling strategy ensures that the DNN receives sufficient training examples near the boundaries to learn smooth and physically consistent spectral variations across the expanded parameter space.

The expanded samples are then merged with the original training set to create an updated mock spectral library, which is then used to retrain the DNN and repeat the fitting process. This iterative refinement continues until the posterior distributions are fully contained within the explored parameter space — typically after one or two cycles. In short, we iterate through the following cycle:
\begin{align}
\text{(fit)} 
&\;\rightarrow\;
\text{(edge check)} 
\;\rightarrow\;
\text{(expand \& resample)} \nonumber\\[4pt]
&\;\rightarrow\;
\text{(RT/DNN update)} 
\;\rightarrow\;
\text{(refit)}
\end{align}
until the main body of the posterior (e.g., 16 -- 84\% quantiles) is comfortably contained away from the edges along all dimensions, indicating that no further expansion is required. This adaptive loop produces a compact and computationally efficient training set, focused on the regions of parameter space supported by the data, while maintaining smoothness across boundaries through sufficient sampling in the transitional ``corner'' and ``seam'' regions.

(5) \textbf{Validation of the Best-Fit Model:} We validate the reliability of our DNN surrogate model by comparing its predicted best-fit spectra with direct RT simulations, ensuring that the derived solutions are physically robust. For each fitting run, we take the maximum-likelihood parameter set obtained from the DNN fitting and re-run the full RT simulation using the same input parameters. The emergent spectrum from the direct RT calculation is then compared with the spectrum predicted by the DNN at the same parameter values. We consider the two spectra to be in good agreement if they are visually indistinguishable and show consistent line shapes and amplitudes across the full velocity range, with differences well below the observational uncertainties. When this criterion is satisfied, the fitting is considered successful, demonstrating that the DNN surrogate has accurately learned the nonlinear mapping between the physical parameters and the emergent spectra without introducing systematic artifacts. In cases where noticeable discrepancies arise, it indicates that the DNN has not yet been fully trained; in such instances, we refine the model by adding additional training samples and retraining the network. Once consistency between the DNN-predicted and directly computed RT spectra is achieved, we confirm that the best-fit spectra and parameter posteriors inferred from the DNN surrogate are equivalent to those obtained through full RT fitting, while achieving orders-of-magnitude gains in computational efficiency.

We provide a schematic diagram that visually summarizes the key steps of the modeling and fitting procedure in Figure \ref{fig:dnn_schematic}.

\subsection{Key Advantages of PEACOCK}

Before we finish this section, we highlight several major advantages of \texttt{PEACOCK}:

\begin{itemize}
    \item \textbf{Physically Motivated Clumpy, Multiphase Structure:} Traditional RT models often approximate galactic outflows as monolithic shells or homogeneous winds, where the gas is assumed to be smoothly distributed. While such simplified geometries facilitate computation, they fail to capture the intrinsically clumpy and inhomogeneous structure of the multiphase CGM, where cool and warm gas are distributed in discrete clouds with large density contrasts and turbulent motions. In contrast, \texttt{PEACOCK} adopts a physically motivated, clumpy multiphase structure in which the outflowing medium is composed of discrete cool-to-warm clumps. This allows for a more accurate representation of the spatial distribution and ionization structure of gas around star-forming galaxies, while providing a physically interpretable framework for how photons propagate, scatter, and escape through a clumpy, non-uniform CGM.

    \item \textbf{Incorporating Turbulence and Varying Velocity Fields:} \texttt{PEACOCK} introduces a physics-based kinematic framework that accounts for both radial velocity gradients and multi-scale turbulence. The model implements radially varying velocity profiles that capture both the acceleration and deceleration phases of galactic outflows. In addition, it includes both small-scale internal turbulence within individual clumps  and large-scale macroscopic turbulent motions among clumps. This joint treatment of outflow velocity variation and turbulence allows the model to accurately reproduce the diverse range of observed line asymmetries and widths, representing a major improvement over previous models that rely on simplistic velocity prescriptions (e.g., constant, linear, or power-law) and neglect non-thermal broadening effects.
    
    \item \textbf{Self-Consistent Treatment of Emission and Absorption:} \texttt{PEACOCK} performs fully self-consistent RT modeling of photons as they propagate through the clumpy, multiphase medium. It simultaneously accounts for photon emission, resonant scattering, and fluorescent transitions within a unified 3D framework, tracking detailed interactions between photons and gas across multiple ionic species. This holistic approach ensures that the emergent emission and absorption profiles are governed by the same underlying physical processes and geometric configuration of the media. It represents a major advancement over previous models that either treated emission and absorption separately or adopted ad hoc, oversimplified prescriptions for the emission component.

\begin{table*}
\footnotesize
\centering
\caption{Approximate computational cost (per spectrum) of each major stage in the \texttt{PEACOCK} modeling and fitting pipeline.}
\label{tab:peacock_timing}
\begin{tabular}{llll}
\hline\hline
\textbf{Stage} & \textbf{Description} & \textbf{CPU Hours} & \textbf{Wall Time} \\
\hline
Mock RT Library Construction & Monte Carlo RT for $10^2$--$10^3$ models (parallelized over 48 cores) & $\sim 10^2$--$10^3$ core hr & $\sim \mathcal{O}(10)\ \mathrm{min}$ \\
DNN Surrogate Training & Training on RT spectra & $\sim \mathcal{O}(1)$ core hr & $\sim \mathcal{O}(1)\ \mathrm{min}$ \\
Posterior Inference & Nested sampling using DNN surrogate (48 cores) & $\sim \mathcal{O}(1)$ core hr & $\sim \mathcal{O}(1)\ \mathrm{min}$ \\
Edge Detection \& Resampling & Posterior-guided adaptive expansion & $\sim 10^2$--$10^3$ core hr & $\sim \mathcal{O}(10)\ \mathrm{min}$ \\
DNN Retraining \& Refitting & Final convergence of posteriors & $\sim \mathcal{O}(1)$ core hr & $\sim \mathcal{O}(1)\ \mathrm{min}$ \\
Best-Fit Model Validation & RT–DNN consistency check & $\sim \mathcal{O}(1)$ core hr & $\sim \mathcal{O}(1)\ \mathrm{min}$ \\
\hline
Total Pipeline (Full Cycle) & From initial RT modeling to final fit & $\sim \mathcal{O}(10^3)$ core hr & $\sim \mathcal{O}(1)\ \mathrm{hr}$ \\
\hline
\end{tabular}
\end{table*}

    \item \textbf{Ion-Specific, Multi-line Radiative Transfer Modeling:} \texttt{PEACOCK} performs full RT calculations separately for each ionic transition, allowing the emergent spectra of different ions to be modeled in a self-consistent yet ion-specific manner. Each transition is characterized by its own line coefficients that determine how photons are scattered, absorbed, and re-emitted. This framework enables the model to capture the distinct radiative behaviors of both purely resonant transitions (e.g., \lya, \SiIII, \SiIV, \CIV) and fine-structure multiplets (e.g., \SiII, \CII).
    
    By simultaneously modeling multiple ions within the same underlying clumpy CGM structure, \texttt{PEACOCK} provides a unified framework that connects the observed morphologies of different spectral lines to the same physical outflow and ionization conditions. This multi-line approach not only allows the model to break degeneracies inherent to single-line analyses and place tighter constraints on key physical parameters, such as the gas kinematics, but also enables cross-ion comparisons — for example, examining the relative column densities of different species. This represents a major improvement over previous single-line modeling approaches, which often suffer from parameter degeneracies or yield weak constraints due to limited spectral information and observational uncertainties.

    \item \textbf{Flexible Treatment of Photon Sources:} 
    \texttt{PEACOCK} incorporates a flexible treatment of the photon sources, allowing it to seamlessly adapt to lines with different intrinsic emission and absorption characteristics. In our framework, the photon source can include both line and continuum components, whose relative contributions are parameterized by the intrinsic equivalent width (${\rm EW}_{\rm int}$) or the line-to-continuum photon ratio ($R_{\rm line}$). This flexibility is crucial for modeling transitions that exhibit diverse spectral morphologies — from pure absorption (e.g., \CII, \SiII, and \SiIII) to pure emission or P-Cygni–like profiles (e.g., \CIV, \SiIV, and \lya). By adjusting the balance between line and continuum emission, \texttt{PEACOCK} reproduces the full range of observed line shapes in a physically interpretable manner. This capability extends existing RT approaches by providing a unified and flexible framework for treating mixed line and continuum photon sources in star-forming galaxies.
    
    \item \textbf{Accelerated Inference via Deep Learning and Nested Sampling:} \texttt{PEACOCK} combines full RT calculations with modern machine-learning techniques to achieve orders-of-magnitude acceleration in spectral fitting. By replacing traditional grid interpolation with a DNN-based surrogate, \texttt{PEACOCK} requires far fewer RT models while maintaining high accuracy. Once trained, the network predicts spectra for arbitrary parameter sets within milliseconds, achieving $\sim10^3$ – $10^4$ speedup over direct RT calculations. This acceleration enables robust Bayesian inference with nested sampling, allowing comprehensive exploration of complex, multidimensional parameter space and obtain well-converged posterior distributions. As a result, \texttt{PEACOCK} retains the physical rigor of RT modeling while making full parameter inference computationally feasible. This synergy between deep learning and nested sampling enables fast, physically grounded exploration of the parameter space without sacrificing accuracy or completeness. It extends existing machine-learning–based approaches to RT modeling by enabling efficient, fully Bayesian inference in complex, high-dimensional parameter spaces (e.g., \citealt{Gurung2022, Gurung2025, Pengfei2025}).

\end{itemize}

To provide a quantitative perspective on the performance gains discussed above, Table~\ref{tab:peacock_timing} summarizes the approximate computational cost per spectrum for each major stage of the \texttt{PEACOCK} pipeline. These results highlight that the full modeling and fitting cycle can typically be completed within about an hour of wall-clock time per spectrum, representing a dramatic improvement in both speed and scalability compared to traditional RT analyses.

\section{Observational Data}\label{sec:data}
Our analysis is based on the COS Legacy Archive Spectroscopic SurveY \citep[CLASSY;][]{Berg2022}, an \textit{Hubble Space Telescope} (HST) treasury program (GO 15840, PI: Berg). CLASSY obtained high-signal-to-noise–ratio far-ultraviolet (FUV) spectra for 45 nearby star-forming galaxies covering $0.002 \lesssim z \lesssim 0.182$. Observations were carried out with the \textit{HST}/COS spectrograph using the G130M, G160M, and G185M/G225M gratings, providing contiguous coverage of the key resonance lines \lya\,$\lambda1216$, \SiII\,$\lambda1260$, \CII\,$\lambda1334$, \SiIII\,$\lambda1206$, \SiIV\,$\lambda\lambda1394,1403$, and \CIV\,$\lambda\lambda1548,1551$. In total, the CLASSY dataset combines 135 orbits of new HST observations with 177 archival orbits, forming the first statistically significant rest-frame FUV spectroscopic sample of low-redshift star-forming galaxies.

To extend the dynamic range at the high-SFR end, we supplement the CLASSY sample with five starburst galaxies from \citet{Heckman2015} that were not part of the original program but have archival HST/COS observations. These galaxies satisfy all CLASSY selection criteria and are therefore directly comparable to the main sample \citep{Xu2022}.

Stellar continua were removed by fitting each spectrum with linear combinations of single-age, single-metallicity templates from the \textsc{Starburst99} library \citep{Leitherer1999}, following the methodology described by \citet{Chisholm2019}. To further improve the signal-to-noise ratio, the spectra are resampled in wavelength. As described in \citet{Xu2022}, all spectra are resampled into 0.18\,\AA\ bins, corresponding to a spectral resolving power of $R \sim 6000$--$10{,}000$. The resulting continuum-normalized, wavelength-resampled spectra form the basis for all spectral line profile analyses presented in this work.

\section{Results of RT Modeling for Individual Lines}\label{sec:individual_modeling}

Our RT modeling successfully reproduces the observed UV emission and absorption line profiles for nearly all 50 galaxies in the sample, provided that the data quality is sufficient. In this section, we present the results of our line-by-line RT modeling, highlighting how different spectral morphologies trace variations in the underlying gas properties. These results form the foundation of our analysis and establish the physical interpretations discussed in the subsequent sections.

\subsection{\CII\ and \SiII}\label{sec:CII}

\begin{figure*}
\centering
\includegraphics[width=0.329\textwidth]{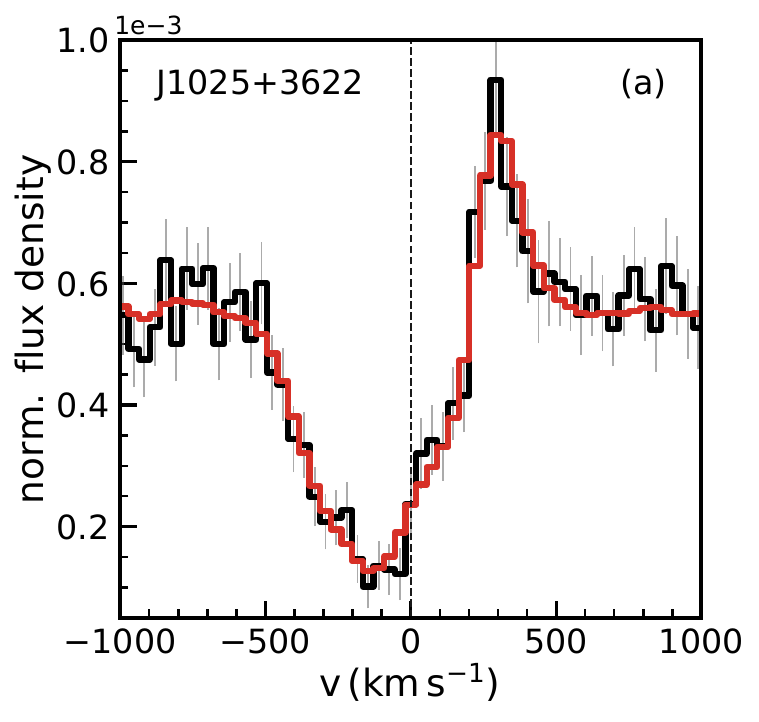}
\includegraphics[width=0.329\textwidth]{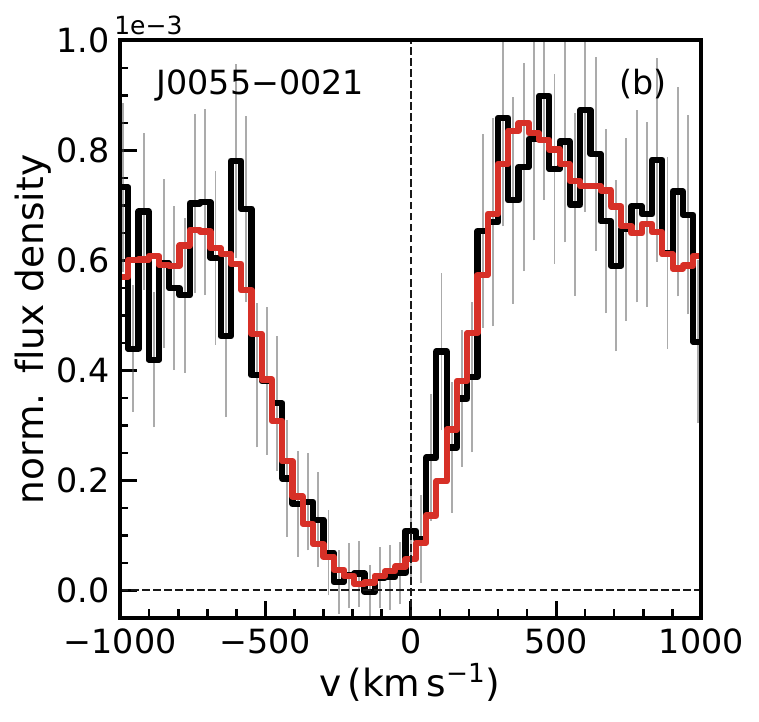}
\includegraphics[width=0.329\textwidth]{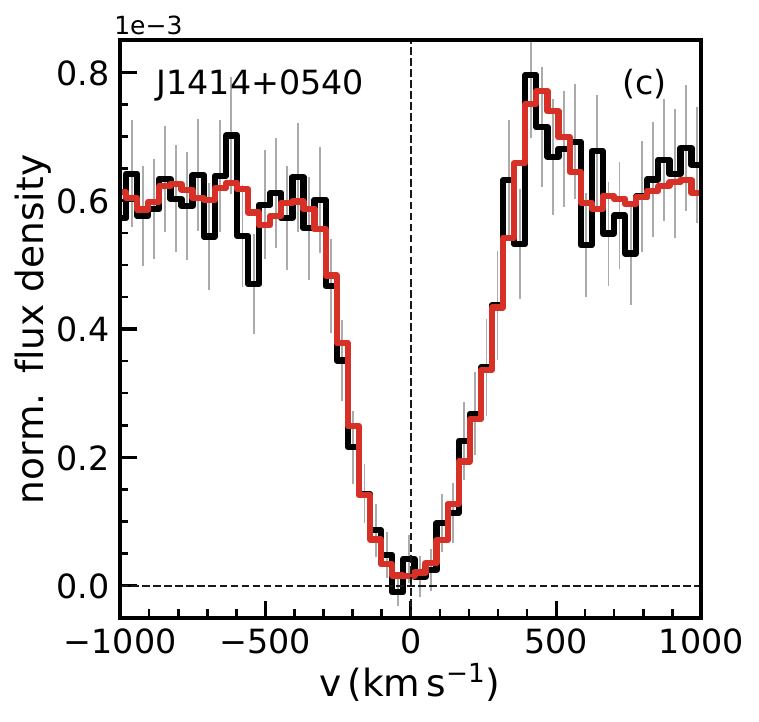}\\
\includegraphics[width=0.329\textwidth]{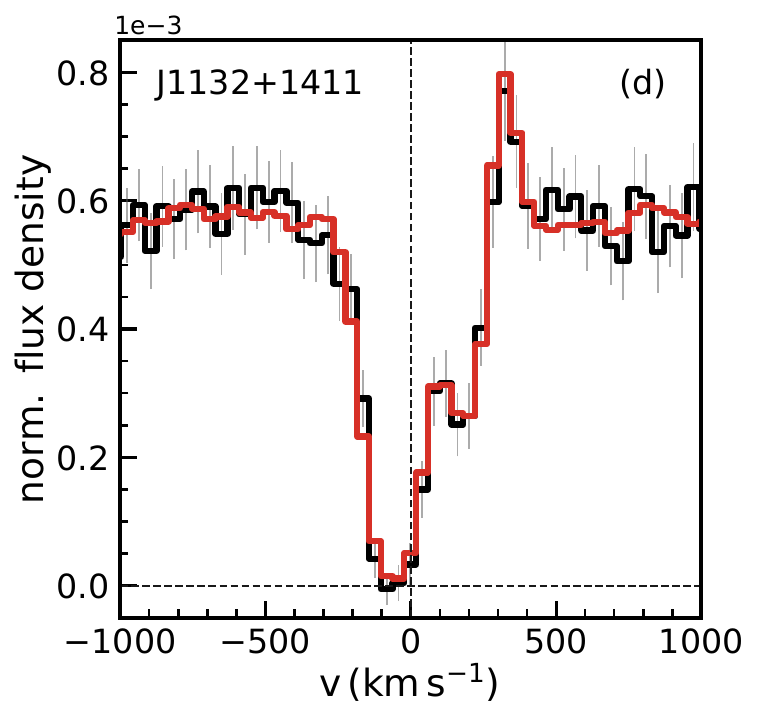}
\includegraphics[width=0.329\textwidth]{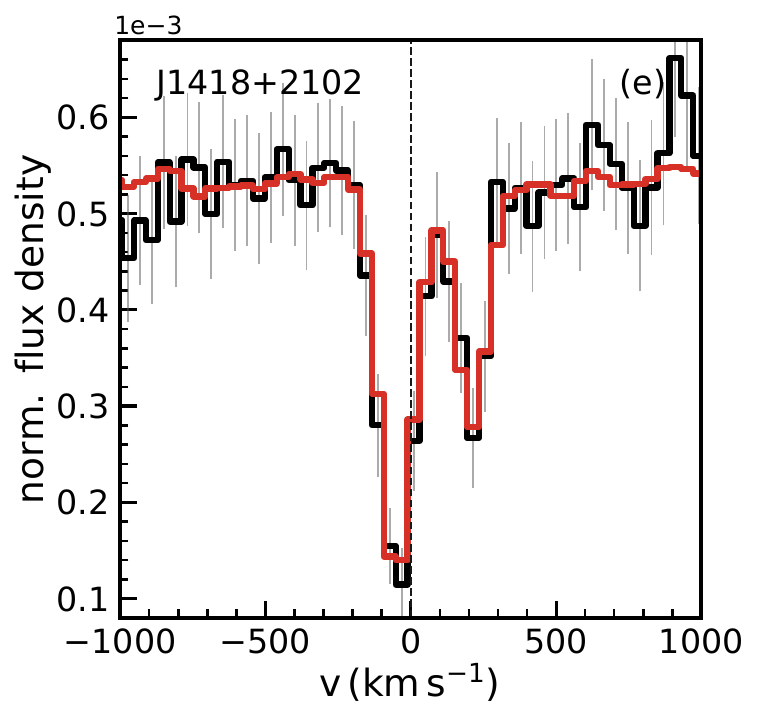}
\includegraphics[width=0.329\textwidth]{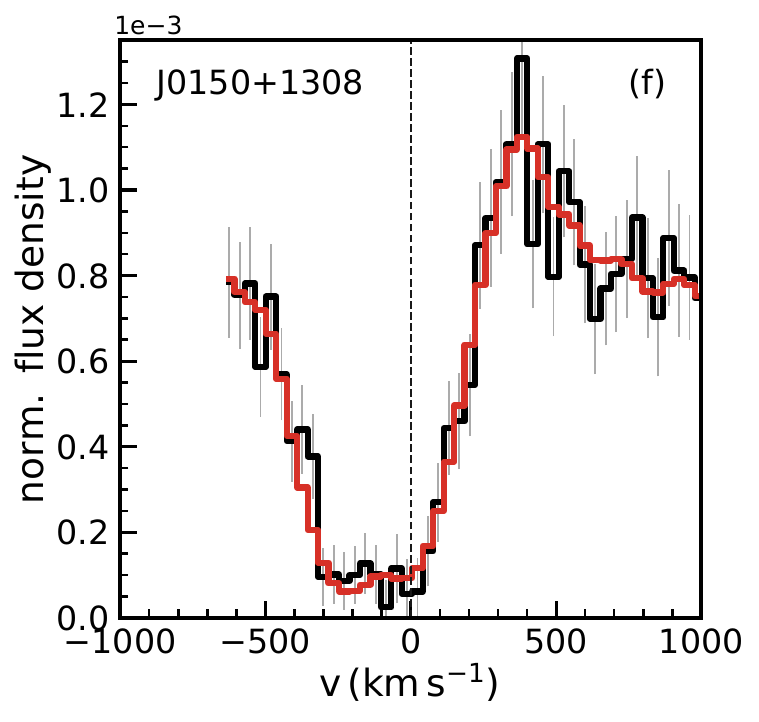}
    \caption{\textbf{Six representative C\,{\sc{ii}}\,$\lambda1334$ line profiles (black) and their best-fit RT models (red).} The galaxy names are indicated in the upper-left corners. Velocity is shown relative to the systemic redshift, and the vertical dashed line marks $v = 0$. The observed spectra are plotted with 1$\sigma$ error bars (gray). The best-fit model parameters associated with these profiles are summarized in Table \ref{tab:CII_params}.
    \label{fig:CII_fits}}
\end{figure*}

\begin{table*}
\centering
\caption{Best-fit model parameters for six representative \CII\,$\lambda1334$ profile morphologies.}
\setlength{\tabcolsep}{2.5pt}
\hspace*{-3.4cm}
\begin{tabular}{cccccccccccc}
\hline\hline
Galaxy Name & Profile Type & 
$f_{\rm cl}$ &
${\rm log}\,N_{\rm ion,\,cl}$ &
$b_{\mathrm{D,\,cl}}$ &
$\sigma_{\mathrm{cl}}$ & $r_{\rm pop}$ & $b_{\rm max}/R_{\rm out}$ & $\tau_{\rm d,\,cl}$ & 
$v_{\rm cl,\,max}$ & $\Delta v$\\
 & & & (cm$^{-2}$) & (km\,s$^{-1}$) & (km\,s$^{-1}$) & & & & (km\,s$^{-1}$) & (km\,s$^{-1}$) \\
\hline
J1025+3622 & Blue trough + red dip + red peak & $6.4_{-1.5}^{+1.7}$ & $15.0_{-0.2}^{+0.3}$ & $28_{-6}^{+11}$ & $169_{-11}^{+7}$ & $0.50_{-0.03}^{+0.05}$ & $0.47_{-0.08}^{+0.05}$ & $0.04_{-0.01}^{+0.01}$ & $140_{-21}^{+22}$ & $-31_{-12}^{+10}$ \\
J0055--0021 & Blue trough + red peak &   $10.8_{-3.5}^{+2.9}$ & $14.7_{-0.2}^{+0.2}$ & $134_{-52}^{+68}$ & $109_{-64}^{+53}$ & $0.16_{-0.10}^{+0.12}$ & $0.23_{-0.09}^{+0.11}$ & $0.08_{-0.04}^{+0.03}$ & $250_{-89}^{+82}$ & $52_{-60}^{+53}$\\
J1414+0540 & Central trough + red peak & $7.4_{-2.7}^{+3.0}$ & $14.9_{-0.3}^{+0.3}$ & $90_{-29}^{+18}$ & $74_{-37}^{+33}$ & $0.11_{-0.05}^{+0.06}$ & $0.18_{-0.07}^{+0.12}$ & $0.08_{-0.05}^{+0.05}$ & $90_{-56}^{+58}$ & $66_{-38}^{+38}$ \\
J1132+1411 & Central trough + red dip + red peak &   $8.5_{-2.4}^{+3.7}$ & $15.2_{-0.4}^{+0.3}$ & $37_{-8}^{+10}$ & $40_{-15}^{+16}$ & $0.07_{-0.05}^{+0.08}$ & $0.18_{-0.07}^{+0.17}$ & $0.06_{-0.03}^{+0.03}$ & $61_{-22}^{+32}$ & $-4_{-20}^{+26}$\\
J1418+2102 & Central trough + red dip & $11.1_{-3.5}^{+2.5}$ & $14.6_{-0.2}^{+0.3}$ & $10_{-2}^{+3}$ & $36_{-7}^{+8}$ & $0.21_{-0.08}^{+0.10}$ & $0.19_{-0.07}^{+0.08}$ & $0.08_{-0.04}^{+0.02}$ & $23_{-13}^{+14}$ & $-27_{-14}^{+14}$\\
J0150+1308 & Extended trough + red peak & $10.4_{-3.3}^{+2.8}$ & $16.0_{-0.4}^{+0.5}$ & $18_{-4}^{+6}$ & $108_{-29}^{+25}$ & $0.23_{-0.12}^{+0.15}$ & $0.31_{-0.12}^{+0.13}$ & $0.11_{-0.03}^{+0.03}$ & $274_{-49}^{+50}$ & $15_{-28}^{+32}$\\
\hline
\end{tabular}
\label{tab:CII_params}
\end{table*}

Figure \ref{fig:CII_fits} presents six representative \CII\,$\lambda1334$ spectra that span a wide range of morphologies, from blueshifted absorption troughs to redshifted absorption dips and emission peaks arising from fluorescent channels. The corresponding best-fit parameters are summarized in Table~\ref{tab:CII_params}. All listed quantities have been defined previously in Table~\ref{tab:model_parameters}, except for $v_{\rm cl,\,max}$, which represents the maximum velocity of the radial outflow profile determined by $v_0$ and $R$ (see Eq. \ref{eq:vcl_max}).

We highlight the following effects of the major model parameters:

\begin{itemize}
    \item The combination of $f_{\rm cl}$ and $N_{\rm ion,\,cl}$ determines the total line-of-sight ion column density, and hence the depth of the absorption trough. They also contribute to the overall line width, although the width is more strongly controlled by $b_{\rm D,\,cl}$ and $\sigma_{\rm D,\,cl}$, which represent the internal and macroscopic turbulent motions of the clumps, respectively, while the bulk outflow primarily governs the overall velocity extent of the absorption.

    \item Both $b_{\rm D,\,cl}$ and $\sigma_{\rm cl}$ broaden the absorption profile, but for different physical reasons -- increasing $\sigma_{\rm cl}$ expands the overall velocity dispersion of the clump ensemble, whereas increasing $b_{\rm D,\,cl}$ widens the intrinsic line width of each individual clump. As discussed later in Section \ref{sec:individual_params}, a larger Doppler parameter tends to make the absorption slightly broader and deeper without altering its overall shape. In contrast, increasing the clump random velocity redistributes absorption from the trough into the wings, producing broader wings but a shallower trough.
    
    Generally, the impact of increasing $\sigma_{\rm cl}$ is more pronounced, and models generally favor larger $\sigma_{\rm cl}$ values to reproduce the observed line widths (e.g., J1025+3622 vs.\ J1132+1411). In rare cases exhibiting exceptionally broad and smooth absorption profiles, both $\sigma_{\rm cl}$ and $b_{\rm D,\,cl}$ are required to be large ($\gtrsim 100\,\rm km\,s^{-1}$, e.g., J0055–0021).

    \item $r_{\rm pop}$ is positively correlated with the depth of the redshifted ``dip'' at the fluorescent transition at 1335.66 \AA. This feature is most pronounced when both $b_{\rm D,\,cl}$ and $\sigma_{\rm cl}$ are small (e.g., J1132+1411, J1418+2102). Otherwise, the dip tends to merge with the main absorption trough, even when $r_{\rm pop}$ is large (e.g., J1025+3622, J0150+1308).  

    \item $b_{\rm max}/R_{\rm out}$ and $\tau_{\rm d,\,cl}$ regulate the strength of the redshifted fluorescent emission peak at 1335.71 \AA. Larger values of $b_{\rm max}/R_{\rm out}$ enhance the peak by including more scattered fluorescent photons, whereas higher $\tau_{\rm d,\,cl}$ suppresses it by attenuating those photons. Galaxies with strong fluorescent emission peaks typically have large $b_{\rm max}/R_{\rm out}$ ($\gtrsim 30\%$, e.g., J1025+3622, J0150+1308), whereas galaxies lacking fluorescent emission peaks tend to have both small $b_{\rm max}/R_{\rm out}$ and large $\tau_{\rm d,\,cl}$ (e.g., J1418+2102).  

    \item $v_{\rm cl,\,max} - \Delta v$ represents the effective line-of-sight velocity (in the observed frame) where the flux reaches its minimum. Physically, this corresponds to the location in the clump outflow velocity field where $\mathrm{d}v/\mathrm{d}r \simeq 0$ in the T15 clump radial outflow velocity profile (Eq. \ref{eq:clump_momentum}) — i.e., where clumps accumulate in velocity space, maximizing their contribution to the absorption trough. We find that a radially varying outflow characterized by a maximum velocity $v_{\rm cl,\,max}$, combined with velocity broadening from $\sigma_{\rm cl}$ and internal turbulence represented by $b_{\rm D,\,cl}$, demonstrates remarkable versatility and is generally sufficient to reproduce the diverse absorption line morphologies observed in our sample (see Section~\ref{sec:physical_basis} for a detailed discussion).
    
\end{itemize}

\begin{figure*}
\centering
\includegraphics[width=0.329\textwidth]{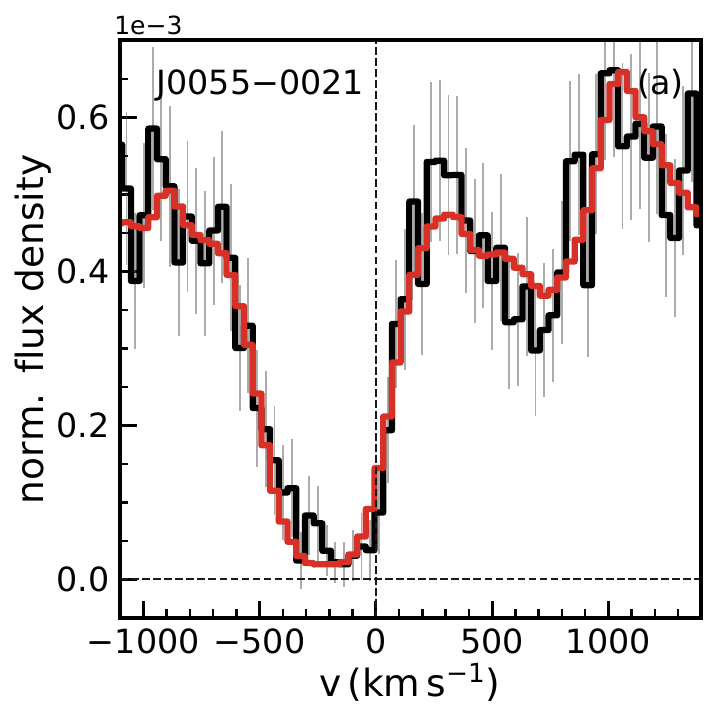}
\includegraphics[width=0.329\textwidth]{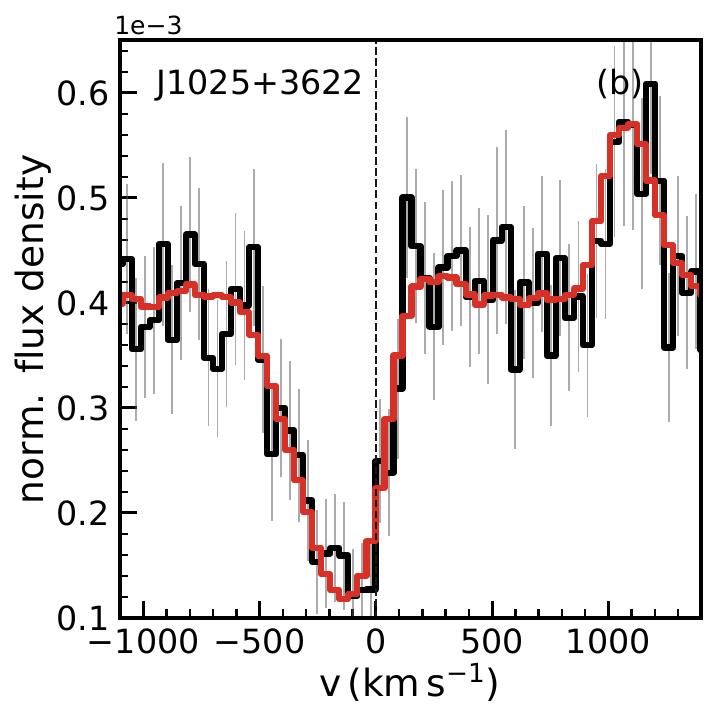}
\includegraphics[width=0.329\textwidth]{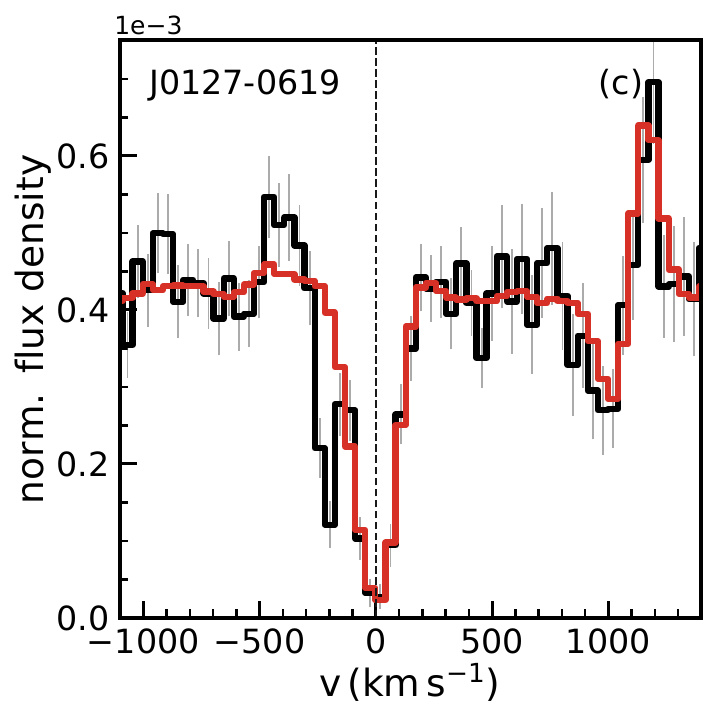}\\
    \caption{\textbf{Three representative Si\,{\sc{ii}}\,$\lambda1260$ line profiles (black) and their best-fit RT models (red).} The galaxy names are indicated in the upper-left corners. Velocity is shown relative to the systemic redshift, and the vertical dashed line marks $v = 0$. The observed spectra are plotted with 1$\sigma$ error bars (gray). The best-fit model parameters associated with these profiles are summarized in Table \ref{tab:SiII_params}. The \SiII\,$\lambda1260$ profile of galaxy J0127–0619 shows contamination from S\,{\sc{ii}} absorption at 1259.52\,\AA, located at $\sim -200\,\mathrm{km\,s^{-1}}$ on the blue side of the line center.}
    \label{fig:SiII_fits}
\end{figure*}

\begin{table*}
\centering
\caption{Best-fit model parameters for three representative \SiII\,$\lambda1260$ profile morphologies.}
\setlength{\tabcolsep}{2.5pt}
\hspace*{-3.4cm}
\begin{tabular}{cccccccccccc}
\hline\hline
Galaxy Name & Profile Type & 
$f_{\rm cl}$ &
${\rm log}N_{\rm ion,\,cl}$ &
$b_{\mathrm{D,\,cl}}$ &
$\sigma_{\mathrm{cl}}$ & $r_{\rm pop}$ & $b_{\rm max}/R_{\rm out}$ & $\tau_{\rm d,\,cl}$ & 
$v_{\rm cl,\,max}$ & $\Delta v$\\
 & & & (cm$^{-2}$) & (km\,s$^{-1}$) & (km\,s$^{-1}$) & & & & (km\,s$^{-1}$) & (km\,s$^{-1}$) \\
\hline
J0055--0021 & Blue trough + red dip + red peak & $10.1_{-2.8}^{+2.7}$ & $13.8_{-0.2}^{+0.2}$ & $115_{-47}^{+59}$ & $141_{-44}^{+46}$ & $0.07_{-0.03}^{+0.03}$ & $0.17_{-0.04}^{+0.05}$ & $0.04_{-0.03}^{+0.03}$ & $141_{-45}^{+64}$ & $-78_{-42}^{+50}$ \\
J1025+3622 & Blue trough + red peak &   $6.1_{-2.8}^{+3.6}$ & $13.4_{-0.2}^{+0.2}$ & $104_{-53}^{+62}$ & $123_{-52}^{+42}$ & $0.06_{-0.04}^{+0.04}$ & $0.29_{-0.09}^{+0.11}$ & $0.03_{-0.02}^{+0.03}$ & $135_{-39}^{+48}$ & $-3_{-35}^{+39}$\\
J0127--0619 & Central trough + red dip + red peak & $10.5_{-3.1}^{+2.7}$ & $13.6_{-0.2}^{+0.2}$ & $29_{-7}^{+11}$ & $49_{-14}^{+12}$ & $0.06_{-0.01}^{+0.02}$ & $0.17_{-0.04}^{+0.06}$ & $0.06_{-0.03}^{+0.02}$ & $73_{-19}^{+20}$ & $62_{-14}^{+16}$ \\
\hline
\end{tabular}
\label{tab:SiII_params}
\end{table*}

Having examined six representative \CII\,$\lambda1334$ profiles, we now turn to the \SiII\,$\lambda1260$ line profiles. As noted earlier, the energy-level structure of \SiII\,$\lambda1260$ closely resembles that of \CII\,$\lambda1334$, except that the separation between the resonant absorption channel and the fluorescent channels is larger and therefore more visually distinguishable. It is also important to note that the \SiII\,$\lambda1260$ profiles are often contaminated by S II absorption at 1259.52\,\AA, which typically appears at about $-200\,\mathrm{km\,s^{-1}}$ on the blue side of the \SiII\,$\lambda1260$ line center. We hereby highlight three representative \SiII\,$\lambda1260$ line profiles for brief discussion in Figure \ref{fig:SiII_fits}.

\begin{itemize}
    \item Overall, the effects of the individual parameters on the observed spectra are similar to those discussed above for \CII. One notable difference is that, for comparable absorption trough depths, the best-fit clump \SiII\ column density is typically about an order of magnitude smaller than that of \CII. This trend is consistent with the larger absorption cross section of \SiII\ relative to \CII, as reflected by their oscillator strengths ($f=1.22$ for \SiII\ $\lambda1260$ and $f=0.128$ for \CII\ $\lambda1334$), as well as with the higher intrinsic abundance of carbon compared to silicon ($12+\log{\rm (C/H)}=8.46$ versus $12+\log{\rm (Si/H)}=7.51$, \citealt{Asplund2021}).
    
    \item Another noteworthy phenomenon is that the strength of the redshifted ``dip'', which arises from the \SiII*\,$\lambda1264.73$ transition, depends on both $r_{\rm pop}$ and the aperture correction factor $b_{\rm max}/R_{\rm out}$. For example, J0055--0021 and J1025+3622 have similar values of $r_{\rm pop}$, yet the absence of a red dip in the latter galaxy can be attributed to its larger $b_{\rm max}/R_{\rm out}$. A larger aperture encompasses more fluorescent emission photons, which fill in and diminish the red dip. A similar trend is also seen in the \CII\,$\lambda1334$ line profiles shown above.
    
\end{itemize}

\subsection{\SiIII}\label{sec:SiIII}

\begin{figure*}
\centering
\includegraphics[width=0.329\textwidth]{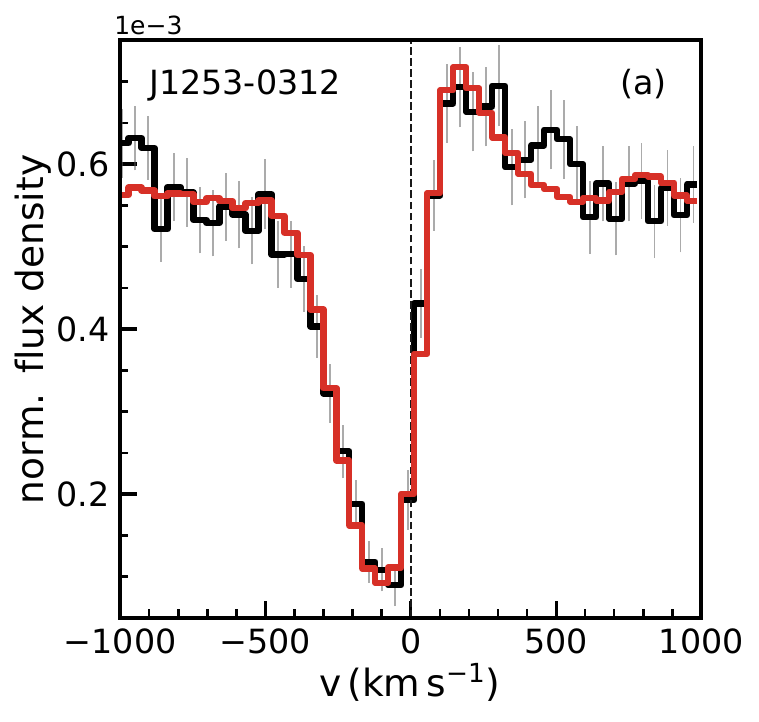}
\includegraphics[width=0.329\textwidth]{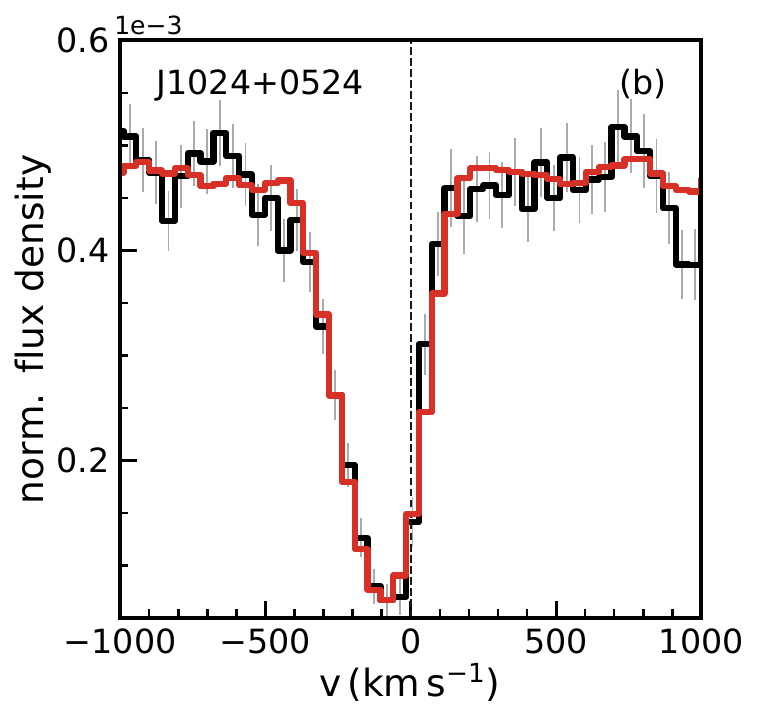}

    \caption{\textbf{Two representative Si\,{\sc{iii}}\,$\lambda$1206 line profiles (black) and their best-fit RT models (red).} The galaxy names are indicated in the upper-left corners. Velocity is shown relative to the systemic redshift, and the vertical dashed line marks $v = 0$. The observed spectra are plotted with 1$\sigma$ error bars (gray). The best-fit model parameters associated with these profiles are summarized in Table \ref{tab:SiIII_params}.
    \label{fig:SiIII_fits}}
\end{figure*}
\begin{table*}
\centering
\caption{Best-fit model parameters for two representative \SiIII\,$\lambda1206$ profile morphologies.}
\setlength{\tabcolsep}{2.5pt}
\hspace*{-2.8cm}
\begin{tabular}{ccccccccccc}
\hline\hline
Galaxy Name & Profile Type & 
$f_{\rm cl}$ &
${\rm log}N_{\rm ion,\,cl}$ &
$b_{\mathrm{D,\,cl}}$ &
$\sigma_{\mathrm{cl}}$ & $b_{\rm max}/R_{\rm out}$ & $\tau_{\rm d,\,cl}$ & 
$v_{\rm cl,\,max}$ & $\Delta v$\\
 & & & (cm$^{-2}$) & (km\,s$^{-1}$) & (km\,s$^{-1}$) & & & (km\,s$^{-1}$) & (km\,s$^{-1}$) \\
\hline
J1253--0312 & Blue trough + red peak & $10.7_{-2.6}^{+2.8}$ & $14.8_{-0.3}^{+0.4}$ & $14_{-3}^{+3}$ & $118_{-12}^{+11}$ & $0.19_{-0.05}^{+0.06}$ & $0.04_{-0.03}^{+0.01}$ & $173_{-24}^{+34}$ & $56_{-14}^{+19}$ \\
J1024+0524 & Blue trough only &   $9.9_{-1.7}^{+3.1}$ & $15.3_{-0.3}^{+0.1}$ & $12_{-2}^{+2}$ & $115_{-9}^{+9}$ & $0.11_{-0.02}^{+0.02}$ & $0.04_{-0.02}^{+0.01}$ & $120_{-20}^{+19}$ & $20_{-14}^{+14}$\\
\hline
\end{tabular}
\label{tab:SiIII_params}
\end{table*}

\begin{figure*}
\centering
\includegraphics[width=0.329\textwidth]{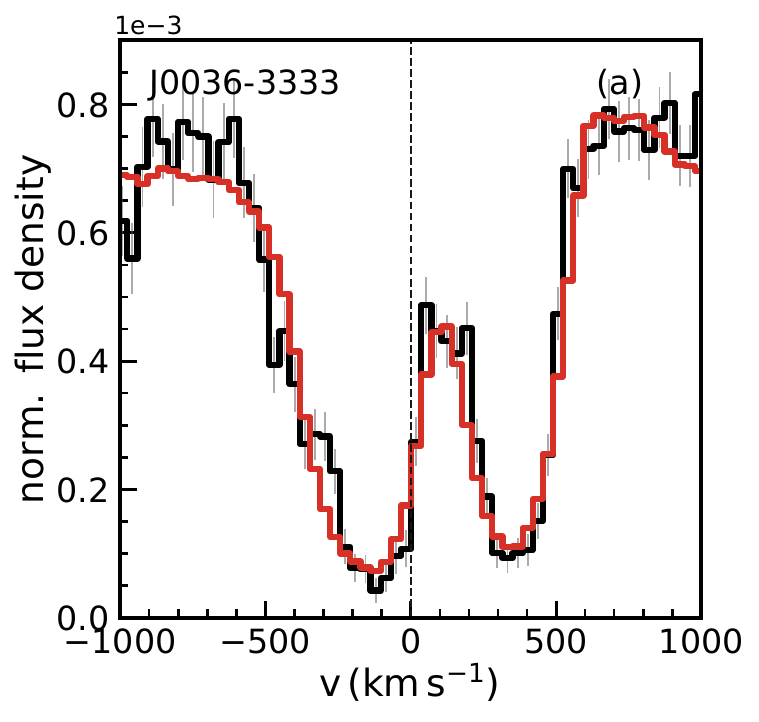}
\includegraphics[width=0.329\textwidth]{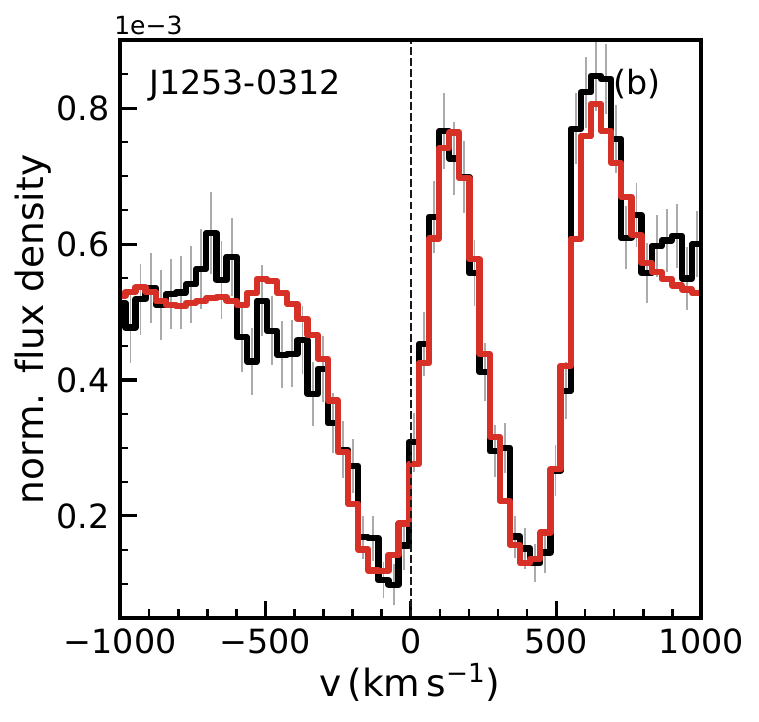}
\includegraphics[width=0.329\textwidth]{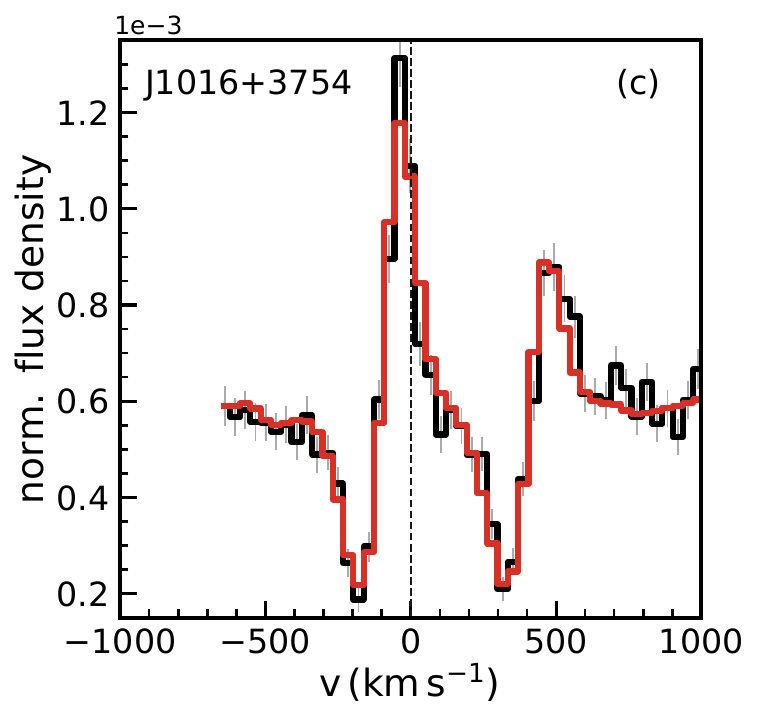}\\
\includegraphics[width=0.329\textwidth]{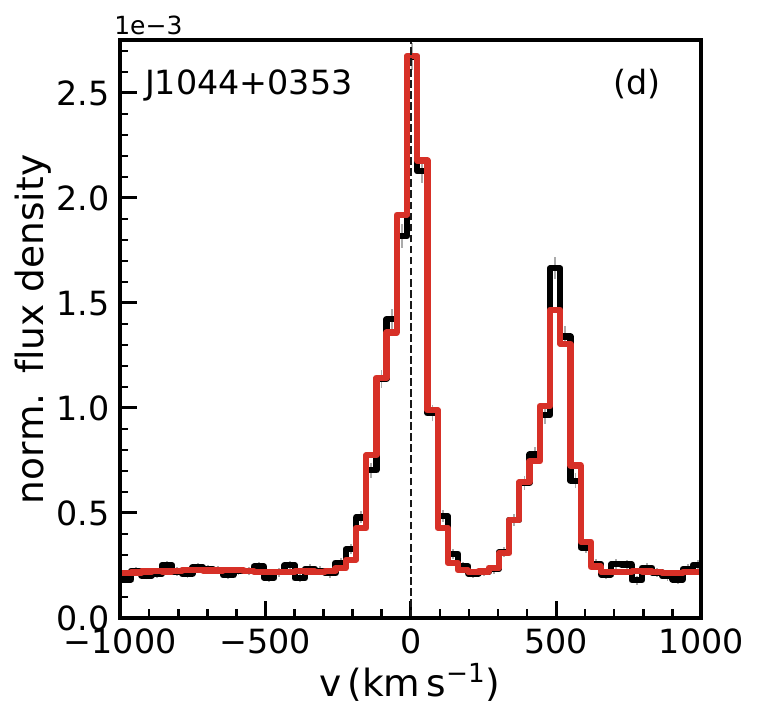}
\includegraphics[width=0.329\textwidth]{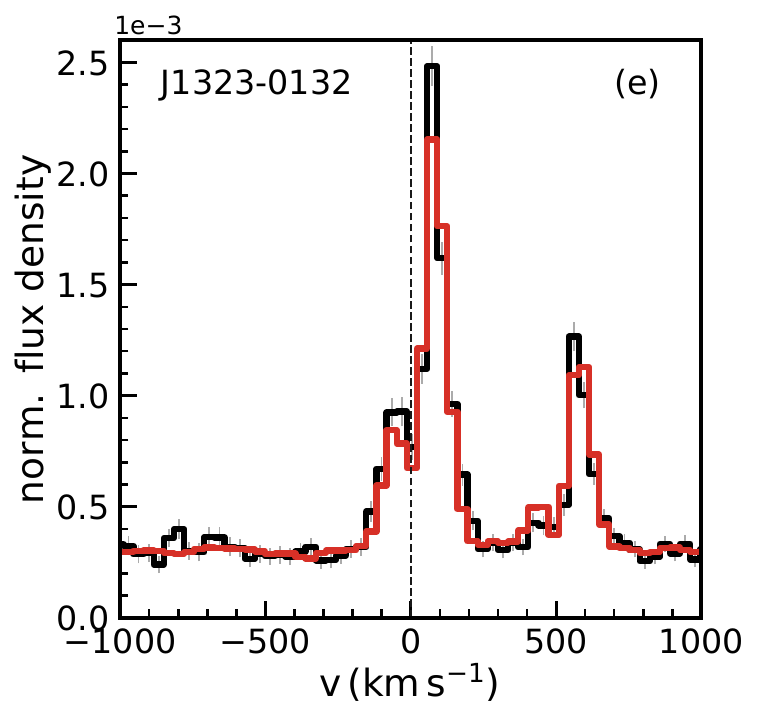}
\includegraphics[width=0.329\textwidth]{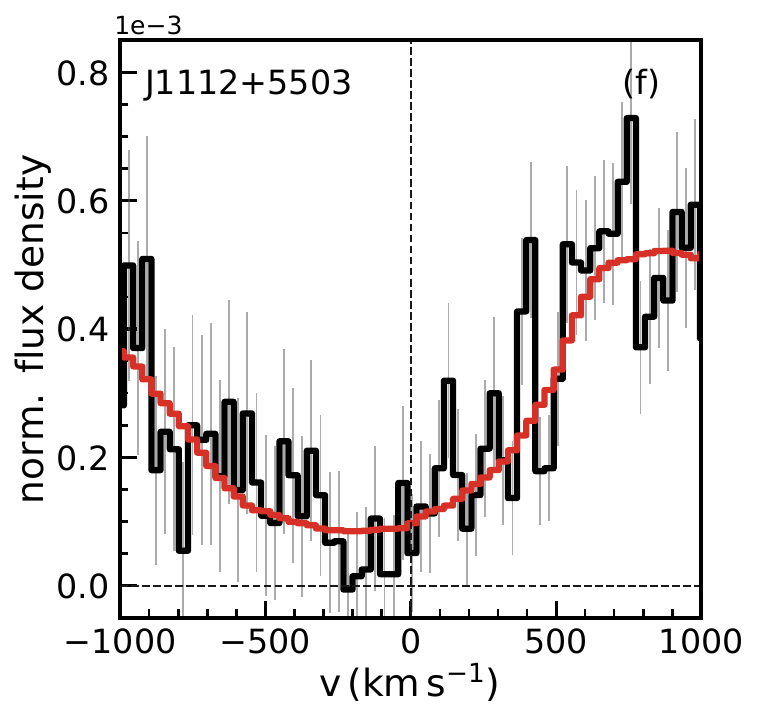}\\
    \caption{\textbf{Six representative C\,{\sc{iv}}\,$\lambda\lambda$1548,\,1550 line profiles (black) and their best-fit RT models (red).} The galaxy names are indicated in the upper-left corners. Velocity is shown relative to the systemic redshift, and the vertical dashed line marks $v = 0$. The observed spectra are plotted with 1$\sigma$ error bars (gray). The best-fit model parameters associated with these profiles are summarized in Table \ref{tab:CIV_params}.
    \label{fig:CIV_fits}}
\end{figure*}

\begin{table*}
\centering
\caption{Best-fit model parameters for six representative \CIV\,$\lambda\lambda$1548,\,1550 profile morphologies.}
\setlength{\tabcolsep}{2.5pt}
\hspace*{-2.8cm}
\begin{tabular}{cccccccccccc}
\hline\hline
Galaxy Name & Profile Type & 
$f_{\rm cl}$ &
${\rm log}N_{\rm ion,\,cl}$ &
$b_{\mathrm{D,\,cl}}$ &
$\sigma_{\mathrm{cl}}$ & $b_{\rm max}/R_{\rm out}$ & $\tau_{\rm d,\,cl}$ & 
$v_{\rm cl,\,max}$ & $\Delta v$ & $\rm EW_{\rm int}$\\
 & & & (cm$^{-2}$) & (km\,s$^{-1}$) & (km\,s$^{-1}$) & & & (km\,s$^{-1}$) & (km\,s$^{-1}$) & (\AA)\\
\hline
J0036--0033 & Double troughs & $10.3_{-2.4}^{+2.0}$ & $14.8_{-0.1}^{+0.2}$ & $36_{-4}^{+7}$ & $146_{-5}^{+5}$ & $0.18_{-0.02}^{+0.03}$ & $0.05_{-0.01}^{+0.01}$ & $192_{-21}^{+22}$ & $25_{-13}^{+14}$ & --\\
J1253--0312 & Double P-Cygni, EW$_{\rm net} < 0$ &   $6.1_{-1.7}^{+4.4}$ & $15.5_{-0.3}^{+0.2}$ & $22_{-4}^{+5}$ & $94_{-11}^{+9}$ & $0.42_{-0.10}^{+0.12}$ & $0.03_{-0.01}^{+0.02}$ & $146_{-26}^{+25}$ & $53_{-10}^{+9}$ & --\\
J1016+3754 & Double P-Cygni, EW$_{\rm net} > 0$ & $6.7_{-1.2}^{+1.7}$ & $14.5_{-0.2}^{+0.3}$ & $10_{-2}^{+2}$ & $65_{-4}^{+3}$ & $0.14_{-0.03}^{+0.04}$ & $0.008_{-0.003}^{+0.001}$ & $74_{-6}^{+7}$ & $-78_{-4}^{+4}$ & $47_{-4}^{+7}$ \\
J1044+0353 & Double peaks & $2.9_{-0.8}^{+0.9}$ & $15.2_{-0.3}^{+0.3}$ & $17_{-3}^{+3}$ & $16_{-9}^{+8}$ & $0.47_{-0.19}^{+0.28}$ & $0.05_{-0.02}^{+0.02}$ & $16_{-2}^{+2}$ & $-32_{-2}^{+2}$ & $22_{-3}^{+3}$ \\
J1323--0132 & Quadruple peaks & $8.1_{-1.3}^{+1.1}$ & $15.1_{-0.2}^{+0.2}$ & $15_{-2}^{+2}$ & $19_{-3}^{+3}$ & $0.45_{-0.15}^{+0.20}$ & $0.08_{-0.01}^{+0.00}$ & $18_{-1}^{+2}$ & $27_{-2}^{+2}$ & $16_{-1}^{+1}$ \\
J1112+5503 & Extended single trough & $9.4_{-3.5}^{+4.1}$ & $14.7_{-0.2}^{+0.3}$ & $129_{-61}^{+70}$ & $289_{-36}^{+22}$ & $0.22_{-0.08}^{+0.11}$ & $0.05_{-0.02}^{+0.02}$ & $319_{-68}^{+76}$ & $-14_{-68}^{+46}$ & --\\
\hline
\end{tabular}
\label{tab:CIV_params}
\end{table*}

Modeling \SiIII\,$\lambda1206$ is relatively straightforward compared to \CII\ and \SiII, as \SiIII\ is strictly resonant and does not possess any fluorescent channels. As a result, its line formation physics is simpler, leading to comparatively uncomplicated line profiles. In our sample, nearly all \SiIII\ profiles are characterized by a dominant blueshifted absorption trough, with a subset exhibiting an additional emission peak produced by resonant scattering.

In Figure~\ref{fig:SiIII_fits}, we present two representative examples of \SiIII\ line profiles. Galaxy J1253--0312 displays a blueshifted absorption trough accompanied by a redshifted emission peak, whereas J1024+0524 shows only a blueshifted absorption trough. The best-fitting RT model parameters, such as the amplitudes of random motions, clump outflow velocities, and column densities, are broadly similar between the two objects. The primary difference is that J1253--0312 favors a larger aperture parameter, $b_{\rm max}$, allowing a greater fraction of scattered emission photons to be captured.

\subsection{\CIV\ and \SiIV}\label{sec:CIV}

Building on our discussion of \CII, \SiII, and \SiIII, we now turn to the high-ionization transitions \CIV\ and \SiIV, which trace warmer gas phases and provide complementary insights into the multiphase structure of the outflow. As discussed in Section \ref{sec:individual_RT}, a key distinction of the \CIV\ and \SiIV\ transitions is that they lack ground-level fine-structure splitting and instead exhibit splitting only in the upper levels. Consequently, these transitions have no fluorescent channels and appear purely as resonant doublets, with the shorter-wavelength (K) component possessing roughly twice the oscillator strength of the longer-wavelength (H) component.

Another notable difference is that, unlike \CII\ and \SiII, which consistently produce net negative equivalent widths, the \CIV\ and \SiIV\ profiles exhibit a wider range of morphologies. They can appear as pure absorption, absorption accompanied by emission (i.e., P-Cygni–like), or even as pure emission. In the latter two cases, reproducing the observed spectra requires including a line-emission component in the photon source in addition to the flat continuum.

We start with \CIV\ and present six representative profiles in Figure \ref{fig:CIV_fits}, with several key features to highlight:

\begin{itemize}
    \item From J0036--0033 to J1253--0312, the spectra exhibit prominent double absorption troughs, while the emission peaks become increasingly significant, driving the net equivalent width toward zero. In this regime, the influence of individual model parameters on the spectral shape remains similar to that found for \CII\ and \SiII, with the photon source still being a flat continuum. The enhanced emission peaks are primarily explained by an increase in the aperture parameter.

    \item A notable transition occurs in J1016+3754, where the net equivalent width becomes positive, requiring the addition of a line-emission component to the photon source, parameterized by the intrinsic equivalent width $\rm EW_{\rm int}$. In J1044+0353, the profile evolves further -- from a double P-Cygni shape to an emission-only double peak -- requiring an even larger $\rm EW_{\rm int}$. Since no absorption trough is present in this case, the underlying clump ion column density and outflow velocity can no longer be inferred directly from the spectrum, making RT modeling indispensable.

    \item Interestingly, the emission peaks of J1044+0353 show a slight asymmetry toward the blue side, indicating a small clump outflow velocity. In contrast, J1323–0132 exhibits an unusual “quadruple-peak” morphology, where each emission peak is accompanied by an absorption trough (see also \citealt{Berg2019}). This feature can be explained by a scenario in which the clump outflow velocity remains low while the covering factor $f_{\rm cl}$ increases, thereby enhancing the total column density along the sightline and the optical depth at line center.

\begin{figure*}
\centering
\includegraphics[width=0.35\textwidth]{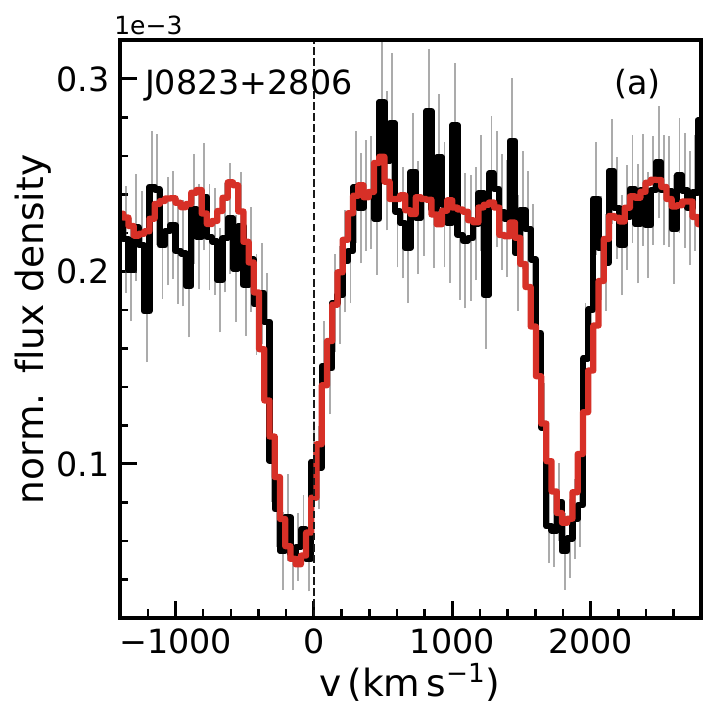}
\includegraphics[width=0.35\textwidth]{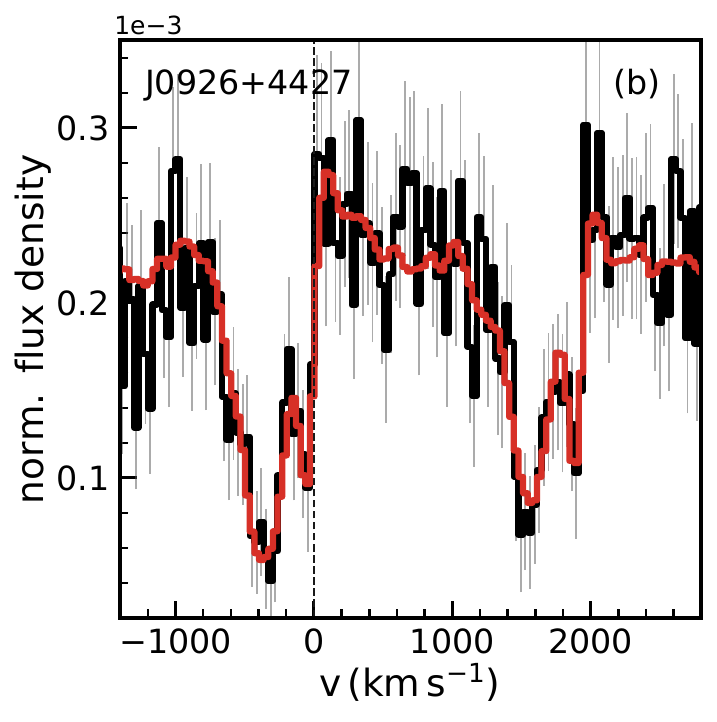}\\
\includegraphics[width=0.35\textwidth]{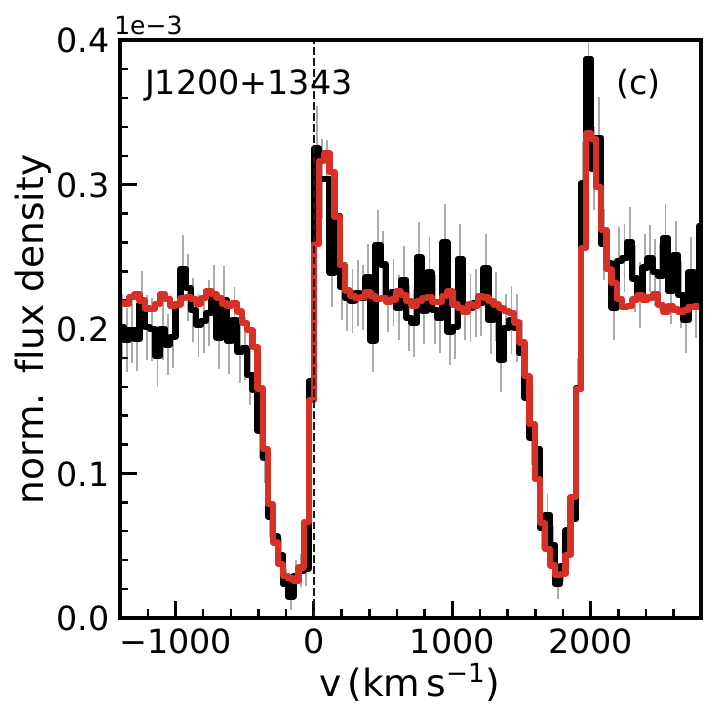}
\includegraphics[width=0.35\textwidth]{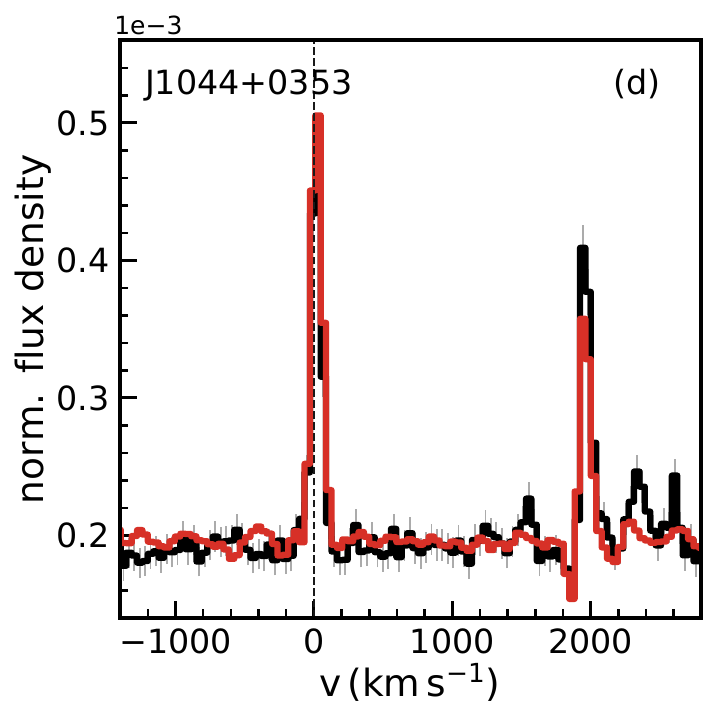}\\
    \caption{\textbf{Four representative Si\,{\sc{iv}}\,$\lambda\lambda$1393,\,1403 line profiles (black) and their best-fit RT models (red).} The galaxy names are indicated in the upper-left corners. Velocity is shown relative to the systemic redshift, and the vertical dashed line marks $v = 0$. The observed spectra are plotted with 1$\sigma$ error bars (gray). The best-fit model parameters associated with these profiles are summarized in Table \ref{tab:SiIV_params}.
    \label{fig:SiIV_fits}}
\end{figure*}

\begin{table*}
\centering
\caption{Best-fit model parameters for four representative \SiIV\,$\lambda\lambda$1393,\,1403 profile morphologies. }
\setlength{\tabcolsep}{1.5pt}
\hspace*{-3.8cm}
\begin{tabular}{cccccccccccc}
\hline\hline
Galaxy Name & Profile Type & 
$f_{\rm cl}$ &
${\rm log}N_{\rm ion,\,cl}$ &
$b_{\mathrm{D,\,cl}}$ &
$\sigma_{\mathrm{cl}}$ & $b_{\rm max}/R_{\rm out}$ & $\tau_{\rm d,\,cl}$ & 
$v_{\rm cl,\,max}$ & $\Delta v$ & $\rm EW_{\rm int}$\\
 & & & (cm$^{-2}$) & (km\,s$^{-1}$) & (km\,s$^{-1}$) & & & (km\,s$^{-1}$) & (km\,s$^{-1}$) & (\AA)\\
\hline
J0823+2806 & Double troughs & $10.7_{-2.5}^{+2.8}$ & $14.2_{-0.2}^{+0.2}$ & $34_{-7}^{+12}$ & $167_{-11}^{+16}$ & $0.24_{-0.08}^{+0.14}$ & $0.06_{-0.01}^{+0.01}$ & $183_{-20}^{+23}$ & $70_{-14}^{+18}$ & --\\
J0926+4427& Quadruple troughs & $10.4_{-3.1}^{+3.0}$ & $13.9_{-0.2}^{+0.3}$ & $53_{-26}^{+26}$ & $137_{-26}^{+42}$ & \multirow{2}{*}{$0.18_{-0.04}^{+0.06}$} & $0.03_{-0.02}^{+0.01}$ & $333_{-30}^{+27}$ & \multirow{2}{*}{$-54_{-10}^{+10}$} & \multirow{2}{*}{--}\\
\multicolumn{2}{l}{J0926+4427 -- 2nd clump population} & $8.3_{-3.6}^{+2.9}$ & $14.0_{-0.7}^{+0.7}$ & $11_{-5}^{+8}$ & $21_{-13}^{+12}$ & & $0.04_{-0.02}^{+0.01}$ & 0 & & \\
J1200+1343 & Double P-Cygni & $16.8_{-0.8}^{+0.4}$ & $14.6_{-0.2}^{+0.1}$ & $52_{-6}^{+8}$ & $75_{-7}^{+7}$ & $0.30_{-0.05}^{+0.06}$ & $0.01_{-0.00}^{+0.00}$ & $83_{-9}^{+8}$ & $-83_{-8}^{+6}$ & --\\
J1044+0353 & Double peaks & $1.4_{-0.3}^{+0.5}$ & $14.1_{-0.1}^{+0.1}$ & $29_{-4}^{+3}$ & $16_{-4}^{+3}$ & $0.16_{-0.04}^{+0.05}$ & $0.07_{-0.06}^{+0.01}$ & $30_{-3}^{+3}$ &$-34_{-4}^{+4}$ & $149_{-14}^{+23}$ \\
\hline
\end{tabular}
\label{tab:SiIV_params}
\end{table*}

    \item Finally, we note rare cases like J1112+5503, where the two absorption troughs merge into a single, extended feature. These profiles correspond to extreme conditions characterized by exceptionally large values of $b_{\mathrm{D,\,cl}} \gtrsim 100\,\rm km\,s^{-1}$ and $\sigma_{\mathrm{cl}} \gtrsim 300\,\rm km\,s^{-1}$, and are rarely observed in our sample.
\end{itemize}

Next, we turn to discussing the representative \SiIV\ line profiles, four of which are shown in Figure~\ref{fig:SiIV_fits}. The energy-level structure of \SiIV\ is similar to that of \CIV, except that the velocity splitting between its doublet components is larger and therefore more visually distinct. Overall, the \SiIV\ profiles closely resemble those of \CIV, though with several subtle differences:

\begin{figure*}
\centering
\includegraphics[width=0.329\textwidth]{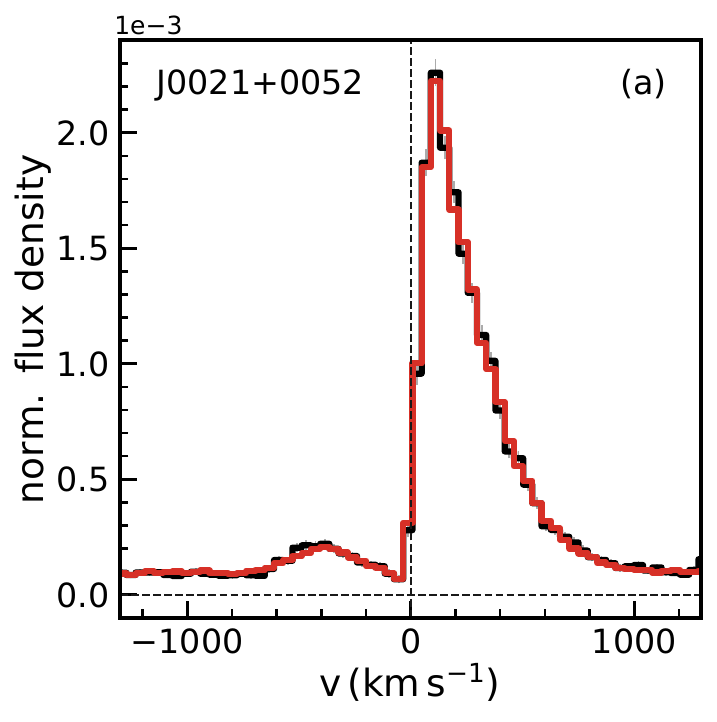}
\includegraphics[width=0.329\textwidth]{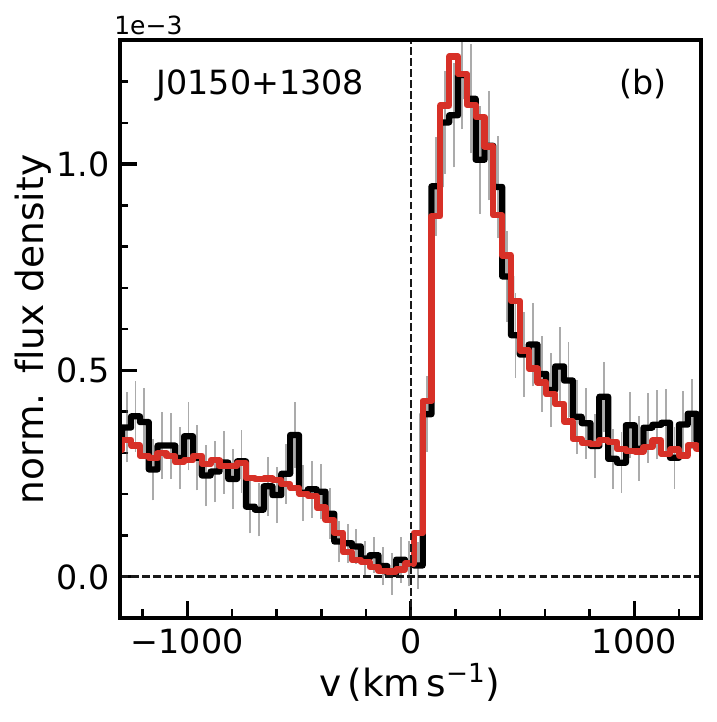}
\includegraphics[width=0.329\textwidth]{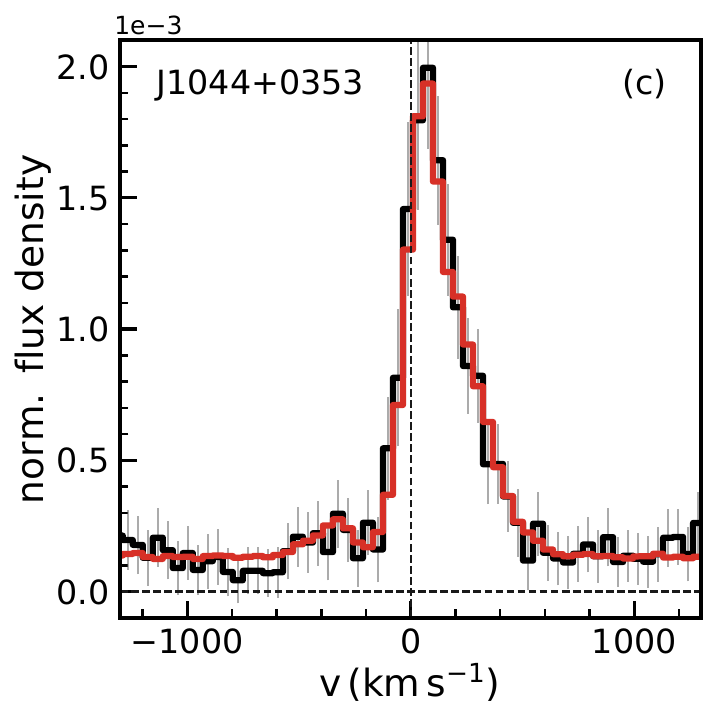}\\
\includegraphics[width=0.329\textwidth]{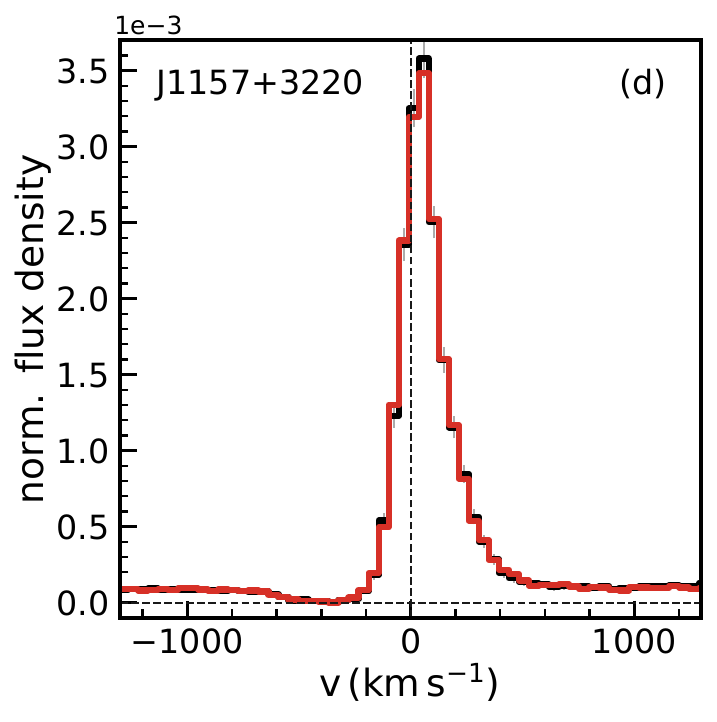}
\includegraphics[width=0.329\textwidth]{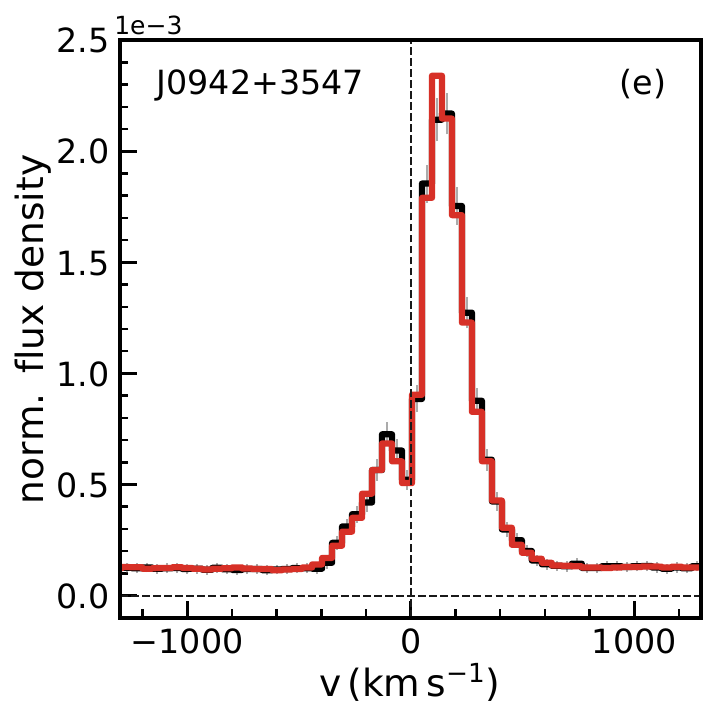}
\includegraphics[width=0.329\textwidth]{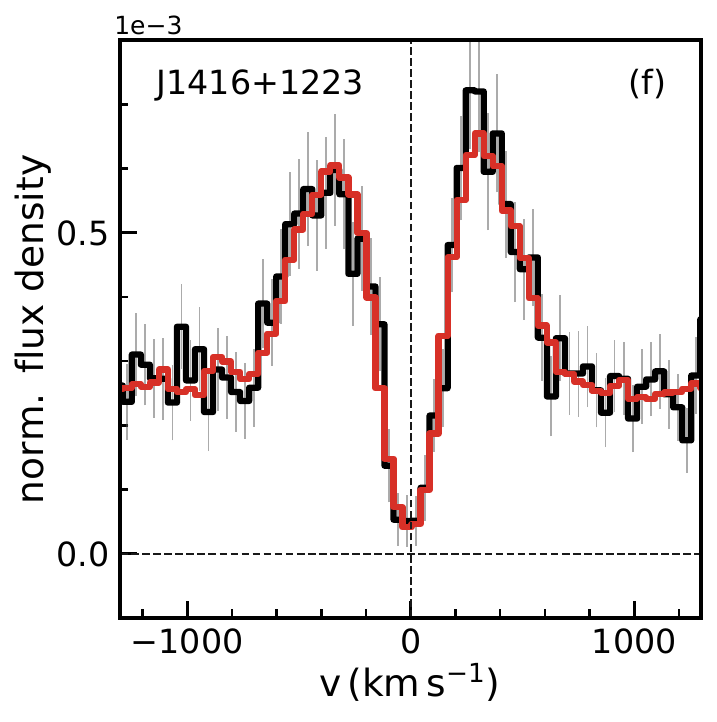}
    \caption{\textbf{Best-fit RT models (red) of the {\lya} line profiles (black) for six CLASSY galaxies.}  The observed spectra are plotted with 1$\sigma$ error bars (gray). The galaxies exhibit diverse profile morphologies -- ranging from classic double peaks (e.g., J0021+0052), to P-Cygni–like profiles (e.g., J0150+1308, J1157+3220), and asymmetric double peaks with central absorption (e.g., J1416+1223). The best-fit model parameters associated with these profiles are summarized in Table \ref{tab:lya_params}.
    \label{fig:lya_fits}}
\end{figure*}

\begin{itemize}
    \item Most of the \SiIV\ line profiles exhibit a double-trough morphology (e.g., J0823+2806), while a small fraction display an unusual “quadruple-trough” morphology (e.g., J0926+4427), in which each blueshifted trough is accompanied by an additional absorption feature located near the corresponding line center. In our sample, three out of 41 objects with well-measured \SiIV\ profiles exhibit this distinctive morphology, which can be reproduced only by including an additional non-outflowing population of clumps in the model. The best-fit parameters for both clump populations are summarized in Table~\ref{tab:SiIV_params}. In general, the two “sub-troughs” near the line center are narrower than the main troughs, corresponding to smaller values of $b_{\rm D,\,cl}$ and $\sigma_{\rm cl}$.

    \item In addition to the double- and quadruple-trough morphologies, seven objects exhibit double P-Cygni profiles (e.g., J1200+1343), while only one case of a double-peaked emission profile is observed -- compared to six such profiles for the \CIV\ doublet. This double-peaked emission profile, J1044+0353, features very narrow peaks, implying small clump $b_{\rm D,\,cl}$ and $\sigma_{\rm cl}$ values, together with low clump outflow velocities (all $\lesssim 30\,\rm km\,s^{-1}$).
\end{itemize}

\subsection{\rm Ly$\alpha$}\label{sec:lya}

\begin{table*}
\centering
\caption{Best-fit model parameters for six representative \lya\ profile morphologies. }
\setlength{\tabcolsep}{1.5pt}
\hspace*{-3.8cm}
\begin{tabular}{cccccccccccc}
\hline\hline
Galaxy Name & Profile Type & 
$f_{\rm cl}$ &
${\rm log}N_{\rm HI,\,cl}$ &
$b_{\mathrm{D,\,cl}}$ &
$\sigma_{\mathrm{cl}}$ & $b_{\rm max}/R_{\rm out}$ & $\tau_{\rm d,\,cl}$ & 
$v_{\rm cl,\,max}$\tablenotemark{a} & $\Delta v$ & $\rm EW_{\rm int}$\\
 & & & (cm$^{-2}$) & (km\,s$^{-1}$) & (km\,s$^{-1}$) & & & (km\,s$^{-1}$) & (km\,s$^{-1}$) & (\AA)\\
\hline
J0021+0052 & Classic double peaks & $13.3_{-3.9}^{+1.2}$ & $17.9_{-0.1}^{+0.1}$ & $10_{-4}^{+6}$ & $111_{-17}^{+12}$ & \multirow{2}{*}{$0.07_{-0.01}^{+0.01}$} & $0.01_{-0.01}^{+0.01}$ & $181_{-23}^{+25}$ & \multirow{2}{*}{$39_{-4}^{+4}$} & \multirow{2}{*}{$304_{-80}^{+75}$}\\
\multicolumn{2}{c}{J0021+0052 -- 2nd clump population}& $1.3_{-0.3}^{+0.3}$ & $19.4_{-0.2}^{+0.2}$ & $60_{-26}^{+44}$ & $244_{-29}^{+25}$ & & $0.02_{-0.01}^{+0.01}$ & $67_{-28}^{+21}$\\
J0150+1308 & P-Cygni & $11.2_{-2.8}^{+2.7}$ & $18.7_{-0.1}^{+0.1}$ & $55_{-16}^{+11}$ & $23_{-17}^{+29}$ & \multirow{2}{*}{$0.32_{-0.08}^{+0.09}$} & $0.006_{-0.003}^{+0.003}$ & $105_{-22}^{+29}$ & \multirow{2}{*}{$-58_{-23}^{+56}$} & \multirow{2}{*}{$27_{-8}^{+9}$} \\
\multicolumn{2}{c}{J0150+1308 -- 2nd clump population}& $3.3_{-1.9}^{+2.2}$ & $17.6_{-0.7}^{+0.6}$ & $8_{-3}^{+5}$ & $54_{-40}^{+66}$ & & $0.004_{-0.002}^{+0.003}$ & $81_{-42}^{+43}$\\
J1044+0353 & Double peaks (one near systemic) & $5.3_{-2.0}^{+2.5}$ & $17.7_{-0.3}^{+0.4}$ & $30_{-11}^{+18}$ & $229_{-58}^{+56}$ & \multirow{2}{*}{$0.66_{-0.25}^{+0.23}$} & $0.01_{-0.01}^{+0.01}$ & $222_{-63}^{+54}$ & \multirow{2}{*}{$-49_{-27}^{+26}$} & \multirow{2}{*}{$20_{-3}^{+4}$} \\
\multicolumn{2}{c}{J1044+0353 -- 2nd clump population}& $6.9_{-3.0}^{+2.9}$ & $17.6_{-0.3}^{+0.4}$ & $21_{-6}^{+11}$ & $60_{-38}^{+45}$ & & $0.01_{-0.01}^{+0.01}$ & $127_{-33}^{+22}$\\
J1157+3220 & P-Cygni (peak near systemic) & $12.1_{-3.7}^{+2.2}$ & $17.2_{-0.3}^{+0.2}$ & $45_{-3}^{+2}$ & $74_{-20}^{+23}$ & \multirow{2}{*}{$0.17_{-0.03}^{+0.04}$} & $0.006_{-0.004}^{+0.003}$ & $274_{-24}^{+27}$ & \multirow{2}{*}{$-54_{-23}^{+11}$} & \multirow{2}{*}{$45_{-4}^{+9}$} \\
\multicolumn{2}{c}{J1157+3220 -- 2nd clump population}& $1.2_{-0.4}^{+0.5}$ & $16.3_{-0.4}^{+0.5}$ & $43_{-6}^{+7}$ & $13_{-9}^{+13}$ & & $0.004_{-0.003}^{+0.004}$ & $52_{-8}^{+15}$\\
J0942+3547 & Double peaks with central flux & $5.0_{-0.7}^{+0.7}$ & $18.4_{-0.1}^{+0.1}$ & $17_{-3}^{+4}$ & $117_{-10}^{+10}$ & \multirow{2}{*}{$0.62_{-0.13}^{+0.16}$} & $0.008_{-0.004}^{+0.002}$ & $72_{-6}^{+5}$ & \multirow{2}{*}{$-5_{-11}^{+7}$} & \multirow{2}{*}{$29_{-3}^{+2}$} \\
\multicolumn{2}{c}{J0942+3547 -- 2nd clump population}& $5.8_{-1.6}^{+1.6}$ & $16.4_{-0.3}^{+0.4}$ & $14_{-2}^{+2}$ & $8_{-5}^{+8}$ & & $0.014_{-0.005}^{+0.004}$ & $-5_{-16}^{+9}$\\
J1416+1223 & Double peaks with central trough & $11.1_{-2.8}^{+2.7}$ & $17.8_{-0.3}^{+0.3}$ & $24_{-9}^{+9}$ & $40_{-26}^{+26}$ & \multirow{2}{*}{$0.22_{-0.04}^{+0.05}$} & $0.01_{-0.01}^{+0.01}$ & $7_{-5}^{+9}$ & \multirow{2}{*}{$-0_{-9}^{+10}$} & \multirow{2}{*}{$20_{-5}^{+6}$}\\
\multicolumn{2}{c}{J1416+1223 -- 2nd clump population}&$4.1_{-1.3}^{+1.5}$ & $18.9_{-0.2}^{+0.2}$ & $73_{-23}^{+26}$ & $206_{-33}^{+29}$ & & $0.01_{-0.01}^{+0.01}$ & $4_{-8}^{+7}$\\
\hline
\end{tabular}
\tablenotetext{a}{Should be interpreted as the bulk velocity of the secondary clump population ($v_{\rm bulk,\,sec}$), when applicable.} 
\label{tab:lya_params}
\end{table*}

We next examine the \lya\ line, which — unlike the metal transitions — is generally emission-dominated in our sample.
The observed profiles exhibit a rich diversity of shapes, frequently showing asymmetric peaks and an absorption trough between them.

Modeling \lya\ emission line profiles is inherently challenging because the key parameters influence multiple aspects of the line shape in a tightly coupled manner (see e.g., \citealt{Neufeld1990, Verhamme2006, Gronke16_model, Li22}). For example, increasing the total \HI\ column density widens the separation between the double peaks but also deepens the absorption trough. Likewise, increasing the clump outflow velocity enhances the peak asymmetry but also shifts the absorption trough toward bluer velocities. These intertwined effects make it difficult to match the observed \lya\ profiles with a single clump population characterized by a single velocity field and \HI\ column density, even when allowing for a global velocity shift of the model spectrum \citep[see e.g.,][]{Orlitova18}. This motivates the introduction of a secondary clump component, which provides the additional degrees of freedom required to reproduce the full complexity of the \lya\ profiles. Physically, this additional component may correspond to recycled material that has decelerated and re-virialized within the CGM without participating in the large-scale outflow, or to a semi-static, high-column-density \HI\ component located near the ISM. Evidence for such a bimodality in \HI\ column densities has been reported in previous studies. For example, \citet{Hu2023} also studied the CLASSY sample and identified relatively low-$N_{\rm HI}$ gas through \lya\ RT modeling using a shell model, alongside high-$N_{\rm HI}$, DLA-like gas inferred from fitting the extended \lya\ damping wings.

We present six representative \lya\ profiles in Figure~\ref{fig:lya_fits}, which illustrate the diversity of observed line shapes and highlight the key features reproduced by our model. The corresponding best-fit parameters for these galaxies are listed in Table~\ref{tab:lya_params}. Below, we discuss the effects of several key model parameters:

\begin{itemize}
    \item When both the blue and red peaks are present, the clump outflow velocity (characterized by $v_{\rm cl,\,max}$) is positively correlated with the flux ratio of the two peaks. Both J0021+0052 and J1044+0353 exhibit clear red-dominated double peaks that correspond to high $v_{\rm cl,\,max}$ values ($\sim170\,\mathrm{km\,s^{-1}}$ and $\sim260\,\mathrm{km\,s^{-1}}$, respectively). In contrast, J0942+3547 and J1416+1223 show much smaller red-to-blue peak flux ratios, yielding significantly lower $v_{\rm cl,\,max}$ values ($\sim90\,\mathrm{km\,s^{-1}}$ and $\sim10\,\mathrm{km\,s^{-1}}$, respectively). When only the red peak is present (J0150+1308 and J1157+3220), $v_{\rm cl,\,max}$ roughly corresponds to the velocity of the absorption trough at $\sim-100$ and $\sim-400\,\mathrm{km\,s^{-1}}$, respectively.

    \item The separation between the double peaks (or the position of the red peak in the case of a P-Cygni profile) is positively correlated with the total \HI\ column density along the sightline, $N_{\rm HI,\,LOS} = (4/3)\,f_{\rm cl}N_{\rm HI,\,cl}$, contributed by both the primary and secondary clump populations. Galaxies with broad peak separations ($\sim 600$ -- $800\,\mathrm{km\,s^{-1}}$, such as J0021+0052 and J1416+1223) exhibit substantially higher total \HI\ column densities ($\sim10^{19}$ -- $10^{20}\,\mathrm{cm^{-2}}$) than those with relatively narrow peak separations or with a red peak located close to the line center (e.g., J1044+0353 and J1157+3220), which only have total \HI\ column densities of $\sim10^{18}$ -- $10^{19}\,\mathrm{cm^{-2}}$. Furthermore, galaxies with particularly deep absorption troughs tend to require especially high values of $N_{\rm HI,\,cl}$ in the secondary component (as seen in J0021+0052 and J1416+1223, both of which have $N_{\rm HI,\,cl} \gtrsim 10^{19.0}\,\mathrm{cm^{-2}}$ for the secondary population).

    \item Similar to the metal absorption lines, the clump microscopic broadening ($b_{\rm D,\,cl}$) and macroscopic random motions ($\sigma_{\rm cl}$) contribute to the extent of the wings of the \lya\ profile, with the primary clump population providing the dominant contribution. In addition, increasing $b_{\max}/R_{\rm out}$ tends to include more emission near the line center and thereby diminishes the depth of the trough. The inferred clump optical depths are generally small ($\tau_{\rm d,\,cl} \lesssim 0.01$), indicating that dust scattering and absorption only play a minor role.
\end{itemize}

Overall, the diverse \lya\ profile morphologies observed in our sample constitute powerful diagnostics of the physical conditions of neutral gas in the CGM. Their systematic variations emphasize the necessity of detailed RT modeling to recover the underlying kinematics, column densities, and spatial distribution of the neutral gas. The emission-dominated nature of the \lya\ profiles provides a valuable complement to the metal absorption lines, offering direct access to the physical properties of the neutral gas that are otherwise difficult to infer from absorption alone.

\subsection{Statistical Distribution of Line Profile Morphologies}

\begin{figure*}
\centering
\includegraphics[width=\textwidth]{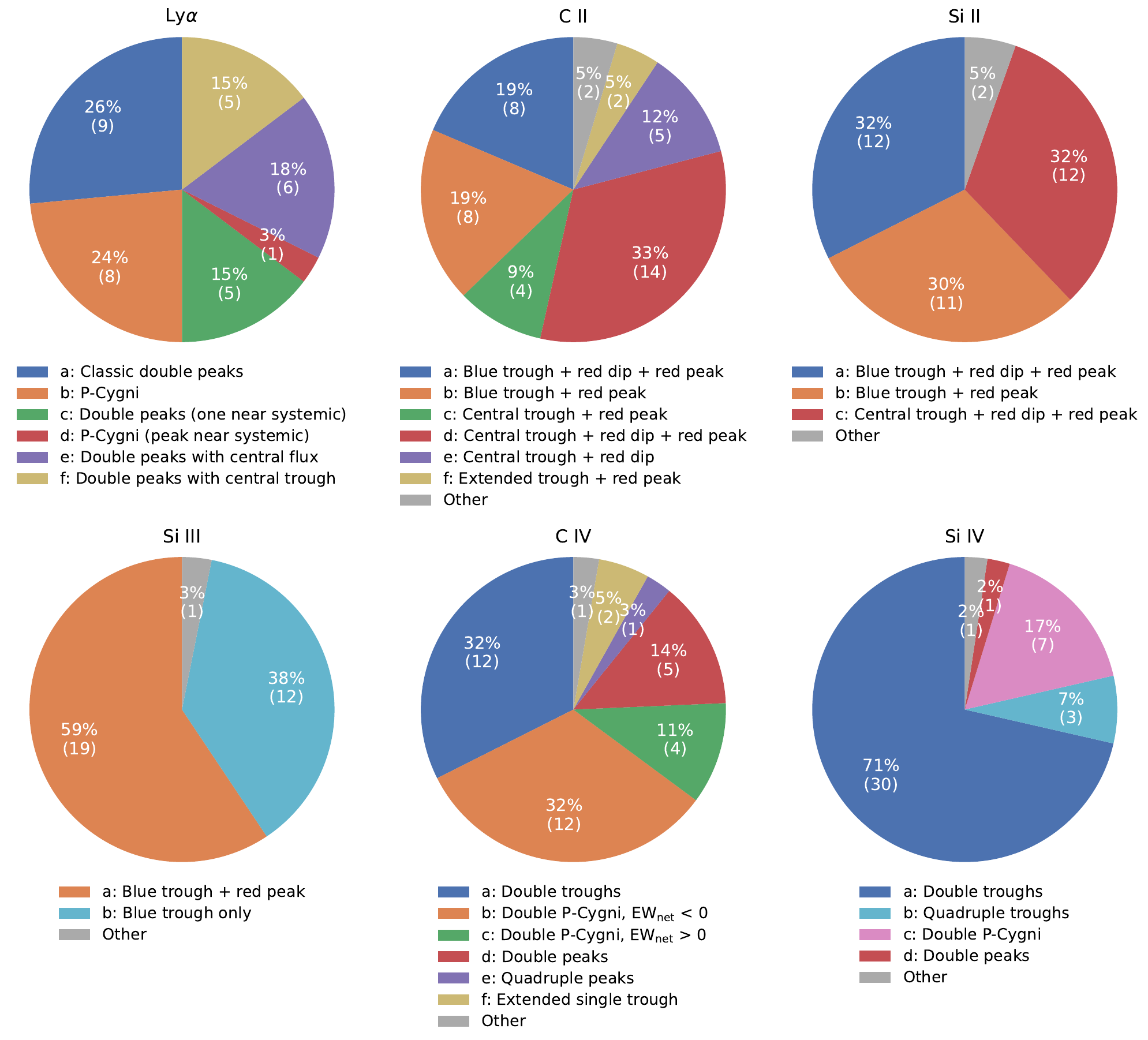}
    \caption{
    \textbf{Distribution of line profile morphology classifications for each transition in our sample.} Each pie chart shows the fraction of galaxies assigned to each morphology category, with numbers indicating the corresponding galaxy counts. The percentages are computed independently for each transition using only galaxies in which the line is detected.
    \label{fig:morph_distribution}}
\end{figure*}

After fitting all individual line profiles, we assign a morphology class to each line in each galaxy and examine the statistical distribution of line profile types. Figure~\ref{fig:morph_distribution} shows the fraction and corresponding number of galaxies in each category for every transition. For a given spectral line, the fraction is computed by counting the number of galaxies in each morphology class and dividing by the total number of galaxies in which that transition is detected.

We find that \lya\ exhibits one of the broadest spreads among morphology categories, with no single class dominating the population. The low-ionization metal lines \CII\ and \SiII\ also show relatively diverse distributions. In contrast, the intermediate- and high-ionization lines \SiIII, \SiIV, and \CIV\ are concentrated in one or two dominant categories. These differences likely reflect intrinsic variations in the RT processes of different transitions. The \lya\ line typically has the highest \HI\ optical depth, so photons undergo many scatterings and escape over a wide range of frequencies and spatial pathways, naturally producing diverse spectral shapes. The \CII\ and \SiII\ transitions include multiple fine-structure levels and fluorescent channels that redistribute photons and provide additional escape routes, further increasing profile diversity. By comparison, \SiIII, \SiIV, and \CIV\ are strictly resonant and generally have low ion column densities, leading to fewer scatterings and therefore more uniform profile morphologies.

We note that the statistical trends reported here are derived from a sample of $\sim$50 local star-forming galaxies. Larger and independent samples in the future, potentially including higher-redshift systems, will enable similar analyses and provide an important test of the robustness and generality of these trends.

\section{Understanding the Physical Basis of the Successful Spectral Fits}\label{sec:physical_basis}

\begin{figure*}
\centering
\includegraphics[width=\textwidth]{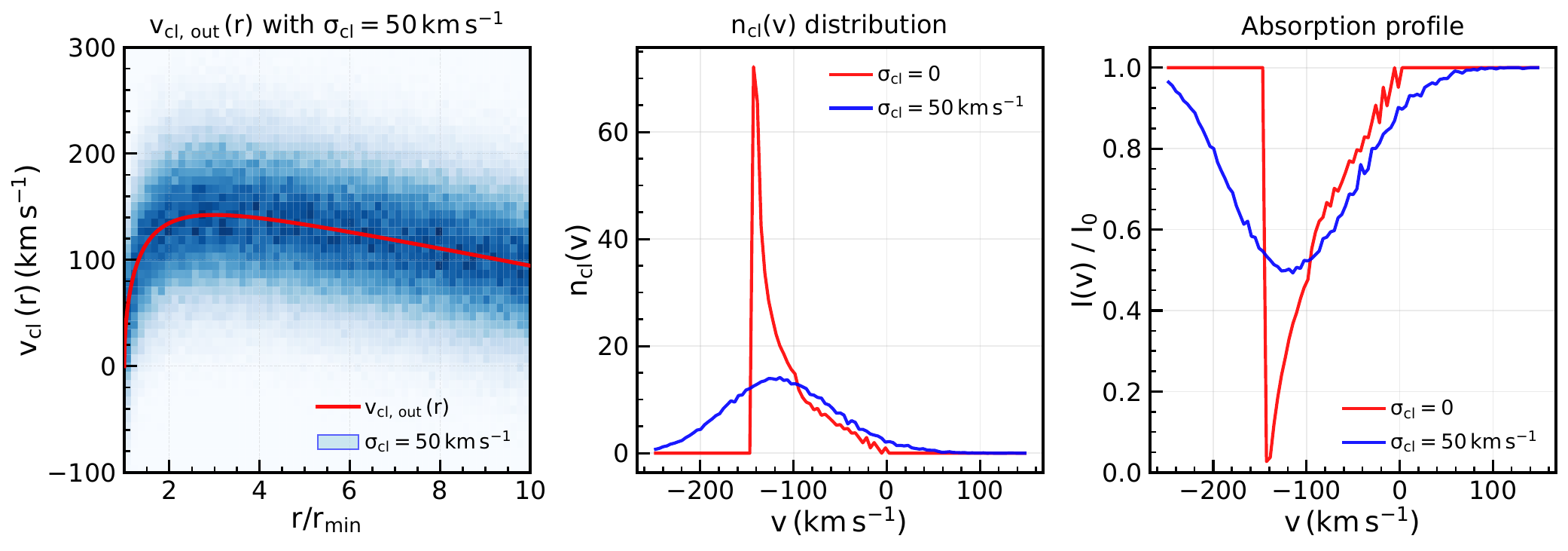}
    \caption{
    \textbf{Illustration of how macroscopic velocity dispersion modifies the mapping from a radial outflow velocity law to the clump velocity distribution and absorption profile.}
    \emph{Left:} The radial outflow velocity law $v_{\rm cl,\,out}(r)$ ($v_0 = 150\,\mathrm{km\,s^{-1}}$, $R = 3.0$, red curve) with a macroscopic velocity dispersion $\sigma_{\rm cl}=50\,\mathrm{km\,s^{-1}}$, shown as a pixelated density map in $(r,v)$ space. The dispersion broadens the velocity field around the bulk outflow profile.
    \emph{Middle:} The resulting clump velocity distribution $n_{\rm cl}(v)$, comparing the dispersion-free case ($\sigma_{\rm cl}=0$, red) to the broadened distribution ($\sigma_{\rm cl}=50\,\mathrm{km\,s^{-1}}$, blue). Velocity dispersion smooths the sharp pile-up at the maximum outflow velocity and produces a more symmetric distribution.
    \emph{Right:} The corresponding absorption profiles. In the absence of dispersion, the absorption inherits an unphysical sharp edge and strong asymmetry, whereas including $\sigma_{\rm cl}$ yields a smoother, broader, and more V-shaped profile, qualitatively resembling observed metal absorption lines. Note that the absorption is computed under a simplified prescription, $I(v) \propto \exp[-k\,n_{\rm cl}(v)]$, rather than via full RT, and is intended to illustrate the qualitative impact of velocity dispersion.
    \label{fig:ncl_v}}
\end{figure*}

In this section, we aim to elucidate the physical origin of the spectral features that enable our RT models to successfully reproduce the observed line profiles, with a primary focus on metal absorption lines, which provide the dominant constraints on the outflow kinematics compared to \lya\ emission. We first establish how the assumed kinematic structure of the outflow translates into a velocity-space distribution of absorbing clumps, thereby shaping the absorption line profiles. We then examine how individual model parameters affect the resulting spectra, highlighting the distinct signatures imprinted by variations in each parameter. Together, these analyses provide physical insight into why certain combinations of parameters yield successful fits and clarify the roles of clump properties in shaping the emergent line profiles.

\subsection{From Clump Velocity Distributions to Absorption Line Profiles}\label{sec:nv_derivation}

In this section, we demonstrate how the prescribed radially varying outflow velocity law, $v_{\rm cl,\,out}(r)$, maps onto the clump number density distribution in velocity space, $n_{\rm cl}(v)$, and reveal why the inclusion of a finite velocity dispersion $(\sigma_{\rm cl})$ is essential for producing absorption line profiles that resemble the real observations.

Since we assume that the number density of clumps follows a power-law profile $n(r)\propto r^{-2}$, the number of clumps with velocities in the interval $(v,\,v+dv)$ is
\begin{equation}
n_{\rm cl}(v)\, dv \;\propto\;
n(r)\, r^{2}\, dr \;\propto\; dr
\end{equation}
where the factor $r^2$ accounts for the spherical volume element. Thus, the velocity distribution is given by
\begin{equation}
n_{\rm cl}(v) \;\propto\;
\left|\frac{dr}{dv}\right|
 = \left|\frac{dr}{dv(r)}\right|
\end{equation}
Because $v(r)$ is steep near $r_{\min}$ but becomes flat near the turnover radius $r = R\,r_{\min}$, the resulting $n_{\rm cl}(v)$ is highly non-uniform: it exhibits a sharp spike at the maximum outflow velocity, followed by a long tail toward lower velocities. Such a characteristic skewed form of $n_{\rm cl}(v)$ results in a correspondingly unphysical absorption-line profile, as shown in the top row of Figure~\ref{fig:ncl_v}.

We now introduce a macroscopic clump velocity dispersion, $\sigma_{\rm cl}$, which modifies the mapping between radius and velocity by accounting for the intrinsically turbulent nature of galactic winds in the CGM. Both observations and simulations indicate that outflowing gas is not characterized by a single, deterministic velocity at a given radius, but instead exhibits substantial non-thermal motions driven by turbulence, instabilities, and interactions with the ambient medium (e.g., \citealt{Schneider2020, Gronke2022, HWChen2023, Schneider2024, Ghosh2025}).  

\begin{figure*}
\centering
\includegraphics[width=\textwidth]{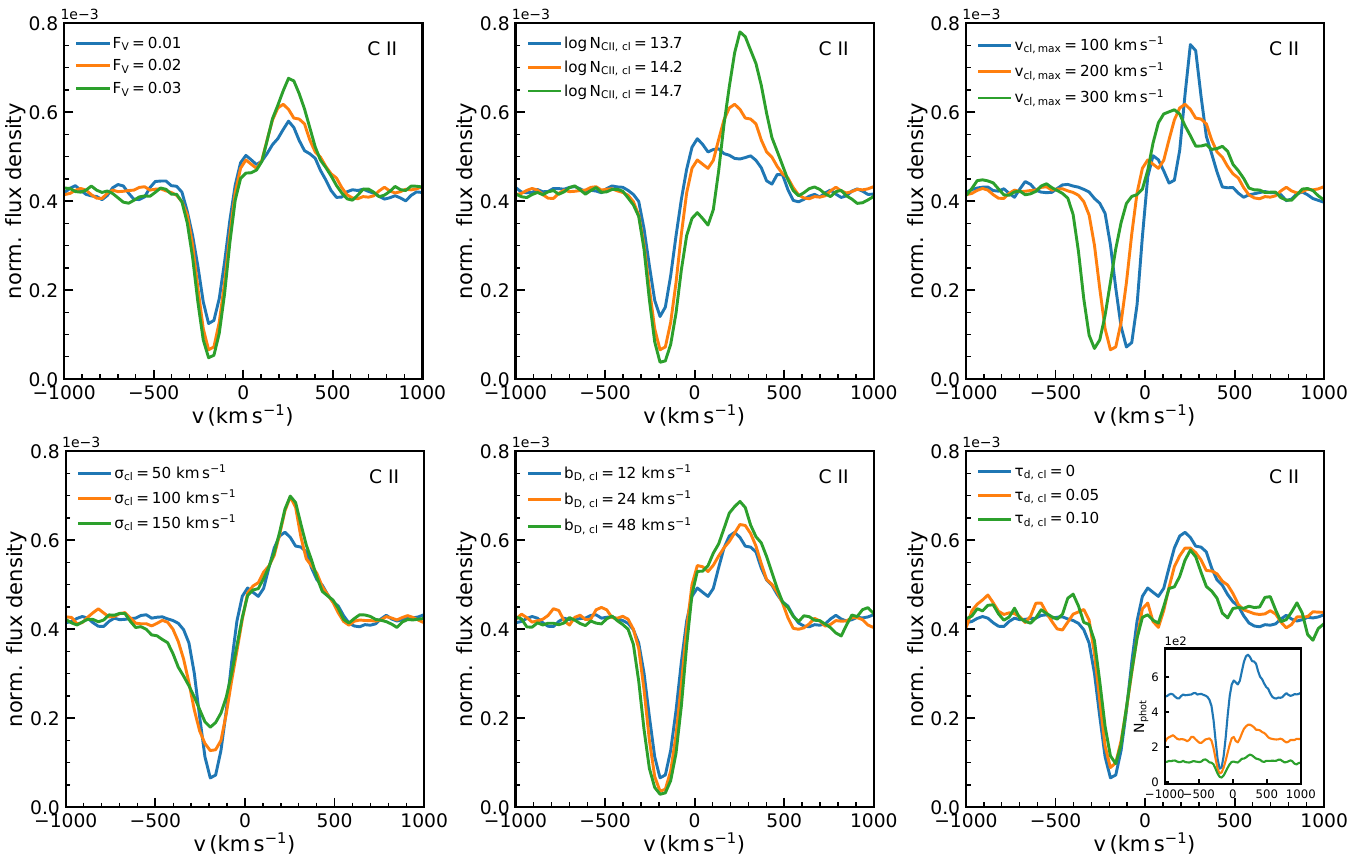}
    \caption{\textbf{Effects of varying individual model parameters on the emergent C\,\textsc{II} profile.} Each panel shows the result of changing one parameter while holding all others fixed. \emph{Top row:} increasing the volume filling factor $F_{\rm V}$ (left) strengthens both the absorption trough and red emission; increasing the ion column density $N_{\rm ion,\,cl}$ (middle) deepens and broadens the trough and enhances fluorescent absorption; increasing the maximum outflow velocity $v_{\rm cl,\,max}$ (right) shifts both absorption and emission blueward due to Doppler boosting. \emph{Bottom row:} larger macroscopic velocity dispersion $\sigma_{\rm cl}$ (left) broadens the profile and redistributes optical depth; increasing the microscopic Doppler parameter $b_{\rm D,\,cl}$ (middle) broadens the absorption while preserving its overall shape; and higher dust optical depth $\tau_{\rm d,\,cl}$ (right) suppresses the scattered emission, with stronger impact once the spectrum is unnormalized (inset). Together, these trends illustrate the distinct physical roles of clump number density, column density, velocity structure, and dust in shaping absorption profiles.
    \label{fig:varying_params}}
\end{figure*}

In this framework, the velocity of each clump at radius $r$ is drawn from a distribution centered on the mean outflow velocity,
\begin{equation}
v = v_{\rm cl,\,out}(r) + \delta v,
\qquad
\delta v \sim \mathcal{N}(0,\,\sigma_{\rm cl})
\end{equation}
which breaks the original one-to-one correspondence between $r$ and $v$. As a consequence, the sharp spike near the maximum velocity is washed out. The resulting $n_{\rm cl}(v)$ distribution becomes the convolution
\begin{equation}
n_{\rm cl}(v)
=
\int n_{\rm cl}(v')\,
\frac{1}{\sqrt{2\pi}\,\sigma_{\rm cl}}
\exp\!\left[-\frac{(v - v')^{2}}{2\sigma_{\rm cl}^{2}}\right]
\, \mathrm{d}v'
\end{equation}
so that the $\delta$-like peak present in the no-dispersion case is transformed into a much broader peak. In addition, the velocity dispersion spreads clumps to both higher and lower velocities, yielding a significantly broader and more symmetric $n_{\rm cl}(v)$ that naturally gives rise to the V-shaped absorption line profiles commonly seen in observations (see the bottom row of Figure \ref{fig:ncl_v}).

We illustrate this effect using a representative example with 
$v_0 = 150\,\mathrm{km\,s^{-1}}$ and $R = 3.0$. As shown in Figure~\ref{fig:ncl_v}, introducing a macroscopic velocity dispersion of $\sigma_{\rm cl}=50\,\mathrm{km\,s^{-1}}$ redistributes absorbing clumps over a substantially broader velocity range, smoothing the sharp pile-up present in the dispersion-free case. As a result, the corresponding absorption profile becomes broader and more symmetric. This experiment demonstrates that velocity dispersion plays a critical role in shaping the emergent line profiles and is essential for reproducing the absorption features observed in real galaxy spectra.

\subsection{Impact of Individual Model Parameters}\label{sec:individual_params}
We now turn to a more detailed examination of how each individual parameter in our RT framework influences the resulting spectra. Here we focus on the metal absorption line profiles and and explore the characteristic signatures 
imprinted by varying each model parameter independently.

We use \CII\ as an illustrative example and adopt the following set of fiducial model parameters: $F_{\rm V}=0.02$, $\log N_{\rm CII,\,cl}=14.2$, $v_{\rm cl,\,max}=200~\mathrm{km\,s^{-1}}$, $\sigma_{\rm cl}=50~\mathrm{km\,s^{-1}}$, $b_{\rm D,\,cl}=12~\mathrm{km\,s^{-1}}$, and $\tau_{\rm d,\,cl}=0$. Figure~\ref{fig:varying_params} shows how each of the six key model parameters affects the emergent C\,\textsc{ii} profile when varied individually while all others are held fixed. These trends highlight the distinct physical roles of clump number density, column density, velocity structure, and dust content in shaping the absorption line profiles, as discussed below:

\begin{enumerate}
    \item \textbf{Volume filling factor $F_{\rm V}$:}
    Increasing $F_{\rm V}$ effectively enhances the clump covering fraction as well as the number of clumps intersecting a typical sightline, thereby strengthening both the absorption trough and the re-emitted flux at positive velocities.

    \item \textbf{Ion column density $N_{\rm ion,\,cl}$:}
    Increasing $N_{\rm ion,\,cl}$ not only deepens and broadens the main absorption trough and significantly amplifies the scattered red peak, but also strengthens the fluorescent absorption when the fine-structure population ratio is fixed, since the optical depth of the fine-structure channel increases correspondingly.

    \item \textbf{Maximum outflow velocity $v_{\rm cl,\,max}$:} 
    Increasing the outflow velocity shifts both the main absorption trough and the fluorescent emission or absorption components toward the blue side, while their velocity separation remains nearly unchanged. This is because the Doppler shift is set by the bulk motion of the scattering clumps and affects all photons in the same way. In our treatment, fluorescent photons are re-emitted at the line center in the atomic rest frame, independent of the incoming photon frequency (see Appendix~\ref{sec:fluorescent_frequency}), so both components shift together while preserving their intrinsic velocity separation.
    
    \item \textbf{Macroscopic random velocity $\sigma_{\rm cl}$:} Larger $\sigma_{\rm cl}$ spreads the absorption over a wider velocity range, redistributing optical depth toward more blueshifted velocities and thereby broadening the wings while making the trough shallower.
        
    \item \textbf{Microscopic Doppler parameter $b_{\rm D,\,cl}$:} Increasing the microscopic Doppler parameter broadens each clump’s cross section, resulting in a slightly deeper and wider overall absorption feature while largely preserving the general shape of the line profile.
    
    \item \textbf{Dust optical depth $\tau_{\rm d,\,cl}$:}
    Dust primarily suppresses the scattered emission while having a comparatively mild effect on the absorption trough. Increasing $\tau_{\rm d,\,cl}$ noticeably reduces the amplitude of the red emission peak by absorbing photons during multiple scatterings. This effect is evident in the unnormalized spectra, but becomes much less pronounced once the spectra are normalized.
\end{enumerate}

\begin{figure}
\centering
\includegraphics[width=0.45\textwidth]{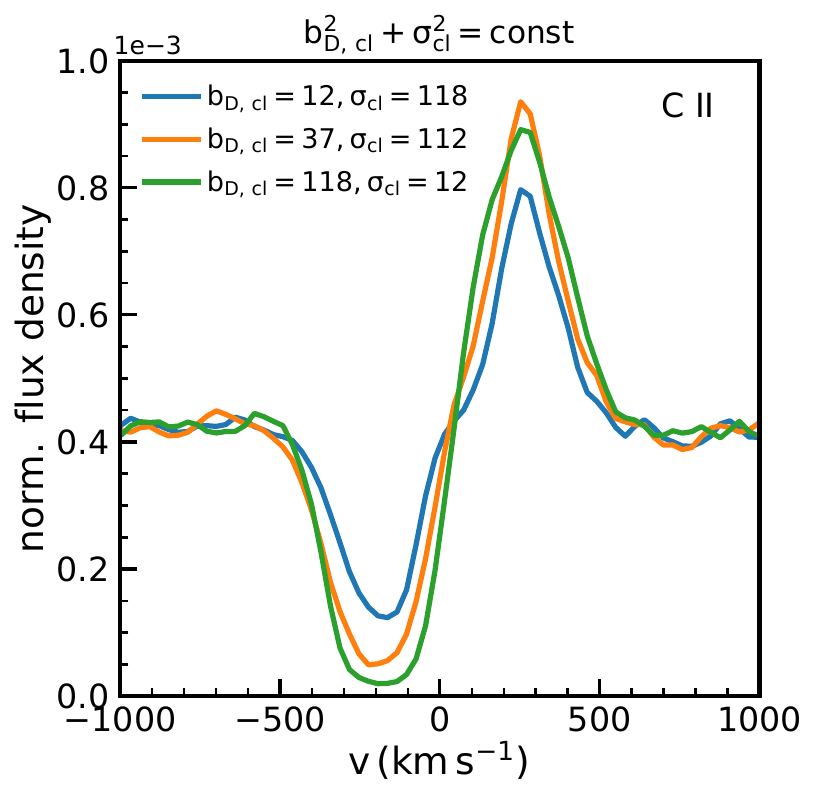}
    \caption{\textbf{Demonstration that the microscopic Doppler parameter $b_{\rm D,\,cl}$ and the macroscopic velocity dispersion $\sigma_{\rm cl}$ are not degenerate.}  We vary $b_{\rm D,\,cl}$ and $\sigma_{\rm cl}$ while keeping their quadrature sum $b_{\rm D,\,cl}^{2} + \sigma_{\rm cl}^{2}$ fixed. The resulting C\,\textsc{ii} profiles differ systematically: larger $b_{\rm D,\,cl}$ (and smaller $\sigma_{\rm cl}$) produce a deeper and wider absorption trough. These results show  that microscopic and macroscopic turbulence leave distinct observable imprints and can be independently constrained by the model.
    \label{fig:bD_sigma}}
\end{figure}

In general, since each key parameter imprints a qualitatively distinct signature on the emergent line profile, we do not find any pair of parameters to be fully degenerate. In other words, a requirement for one parameter to take a specific value cannot be compensated by adjusting other parameters. For example, reproducing a broad absorption profile with extended wings necessarily requires a sufficiently large clump velocity dispersion, $\sigma_{\rm cl}$. Increasing the volume filling factor $F_{\rm V}$ or the ion column density per clump, $N_{\rm ion,\,cl}$, can deepen the absorption but cannot substitute for the role of $\sigma_{\rm cl}$ in setting the velocity extent of the line.

In preparation for the discussion in the following sections and in Paper II, we highlight in particular how the model disentangles the microscopic Doppler parameter $b_{\rm D,\,cl}$ from the macroscopic random velocity $\sigma_{\rm cl}$. In Figure \ref{fig:bD_sigma}, we vary
$b_{\rm D,\,cl}$ and $\sigma_{\rm cl}$ while keeping their quadrature sum $b_{\rm D,\,cl}^{2}+\sigma_{\rm cl}^{2}$ fixed. The resulting profiles show that the width and
depth of the main absorption trough increase as $b_{\rm D,\,cl}$ is increased and $\sigma_{\rm cl}$ is decreased, indicating that $b_{\rm D,\,cl}$ has a more dominant effect on the line profile. This experiment demonstrates that
$b_{\rm D,\,cl}$ and $\sigma_{\rm cl}$ are not degenerate; the two types of turbulence (microscopic vs.\ macroscopic) can therefore be independently constrained by our modeling.

Taken together, these results demonstrate that our RT framework can use the detailed line shapes to disentangle the different physical components encoded in the model, such as the microscopic Doppler broadening, macroscopic random motions, and the bulk outflow. In this sense, the best-fit parameters offer a physically interpretable reconstruction of the CGM properties, rather than a purely phenomenological fit to the observational data.

\section{Cross-Transition Comparison of Model Parameters}\label{sec:param_compare}

Having completed the RT modeling for individual transitions, we now compare the best-fit parameters across multiple lines to obtain a more comprehensive view of the multiphase outflow. This cross-transition analysis allows us to identify systematic trends in clump properties — such as outflow velocity, turbulent motions, and column density — and to assess how these quantities vary across different species.

\subsection{Outflow Velocities}\label{sec:outflow_v}

\begin{figure*}
\centering
\includegraphics[width=\textwidth]{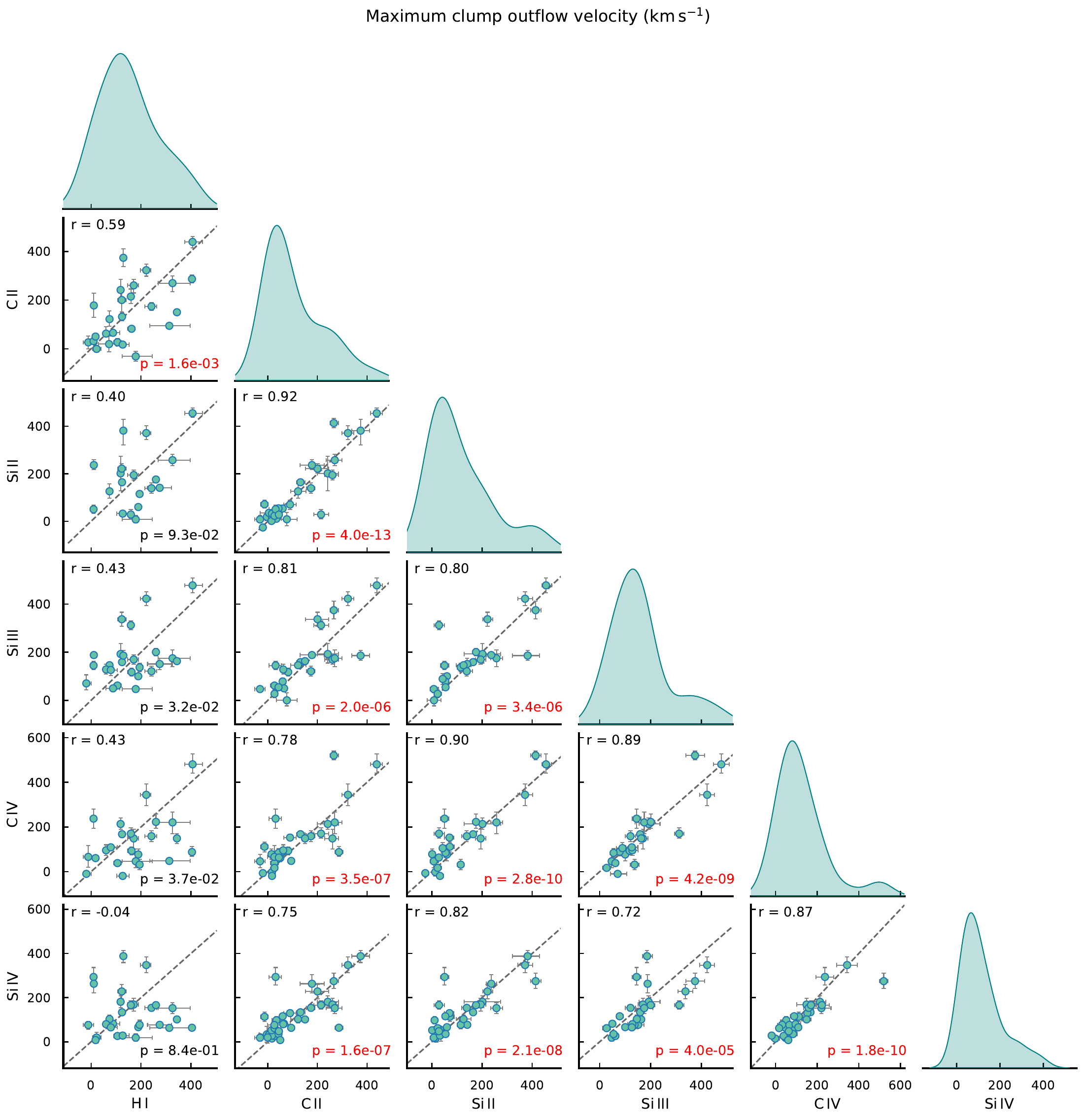}
    \caption{\textbf{Cross-transition comparison of the maximum clump outflow velocities derived from RT modeling for six individual transitions.} Each panel shows a pairwise correlation between two ions, with the Pearson correlation coefficient $r$ and $p$-value annotated. Correlations more significant than the $3\sigma$ threshold ($p < 2.7\times10^{-3}$) are highlighted in red. Histograms along the diagonal display the one-dimensional distributions of maximum clump outflow velocities for each species, and each dashed line represents a one-to-one relation. All the metal ions exhibit uniformly strong correlations ($r \simeq 0.7$--$0.9$), indicating that low- and high-ionization species share a common underlying outflow velocity structure. In contrast, neutral hydrogen shows only weak or marginal correlations with the metals. Overall, the strong ion-to-ion coherence in $v_{\rm cl,\,max}$ supports a scenario in which cool and warm gas phases form interconnected layers within a coherent, clumpy outflow.}
    \label{fig:compare_vout}
\end{figure*}

The maximum clump outflow velocity, $v_{\rm cl,\,max}$ (given by Eq. \ref{eq:vcl_max}), provides a direct measure of the energetics of the multiphase wind and therefore serves as one of the key diagnostics of the underlying outflow kinematics. Figure~\ref{fig:compare_vout} shows a cross-transition comparison of $v_{\rm cl,\,max}$ (corrected for best-fit systemic velocity offset $\Delta v$) inferred independently from the RT modeling of \HI, \CII, \SiII, \SiIII, \CIV, and \SiIV. The Spearman correlation coefficient $r$ and the corresponding $p$-value are reported in each subplot, where correlations that are more significant than the $3\sigma$ level (i.e., $p < 2.7\times10^{-3}$) are highlighted in red.

Across five metal species, the inferred maximum velocities exhibit uniformly strong correlations (typically $r \simeq 0.7-0.9$ with $p \lesssim 10^{-6}$). This tight one-to-one behavior indicates that all metal species -- despite tracing different ionization states -- share a common underlying velocity structure. The tightest correlations are found between \CII\ and \SiII\ ($p \approx 4\times10^{-13}$) and between \CIV\ and \SiIV\ ($p \approx 2\times10^{-10}$), likely because each pair of ions traces similar physical phases. The \CII\ and \SiII\ transitions both arise from the same cool ($T \sim 10^4\,\mathrm{K}$), low-ionization outflowing gas, while \CIV\ and \SiIV\ originate in warmer ($T \sim 10^{4.5}$–$10^{5}\,\mathrm{K}$) boundary layers surrounding the cool clumps, where the gas is partially mixed and ionized. The moderately strong correlations between the low- and high-ionization pairs (e.g., \CII–\CIV\ and \SiII–\SiIII-\SiIV) further suggest that the outflowing material remains dynamically coupled across phases within a common multiphase structure.

Neutral hydrogen displays only modest or marginal positive correlations with the metal ions, with the most significant case being \HI--\CII\ ($p \approx 1.6\times10^{-3}$). This behavior is unsurprising, since the neutral gas component generally is typically more extended, exhibits much higher column densities, and may not be well mixed with the metals. This is also consistent with the need for a secondary population of clumps to reproduce the \lya\ profiles, suggesting that the neutral phase does not necessarily share the same velocity structure as the metals. A plausible interpretation is that the neutral and ionized phases participate in the same global outflow but respond differently to the underlying driving mechanisms, possibly reflecting differences in their coupling efficiencies to radiation pressure, thermal pressure gradients, and hydrodynamic interactions with the surrounding hot wind.

Overall, the remarkably strong ion-to-ion correlations in $v_{\rm cl,\,max}$ indicate that the outflow maintains coherent large-scale kinematics across multiple ionization states, supporting a scenario in which cool and warm phases are not independent components but rather interconnected layers within a coherent, clumpy outflow.

\subsection{Turbulent Velocities}\label{sec:turbulent_v}
\begin{figure*}
\centering
\includegraphics[width=\textwidth]{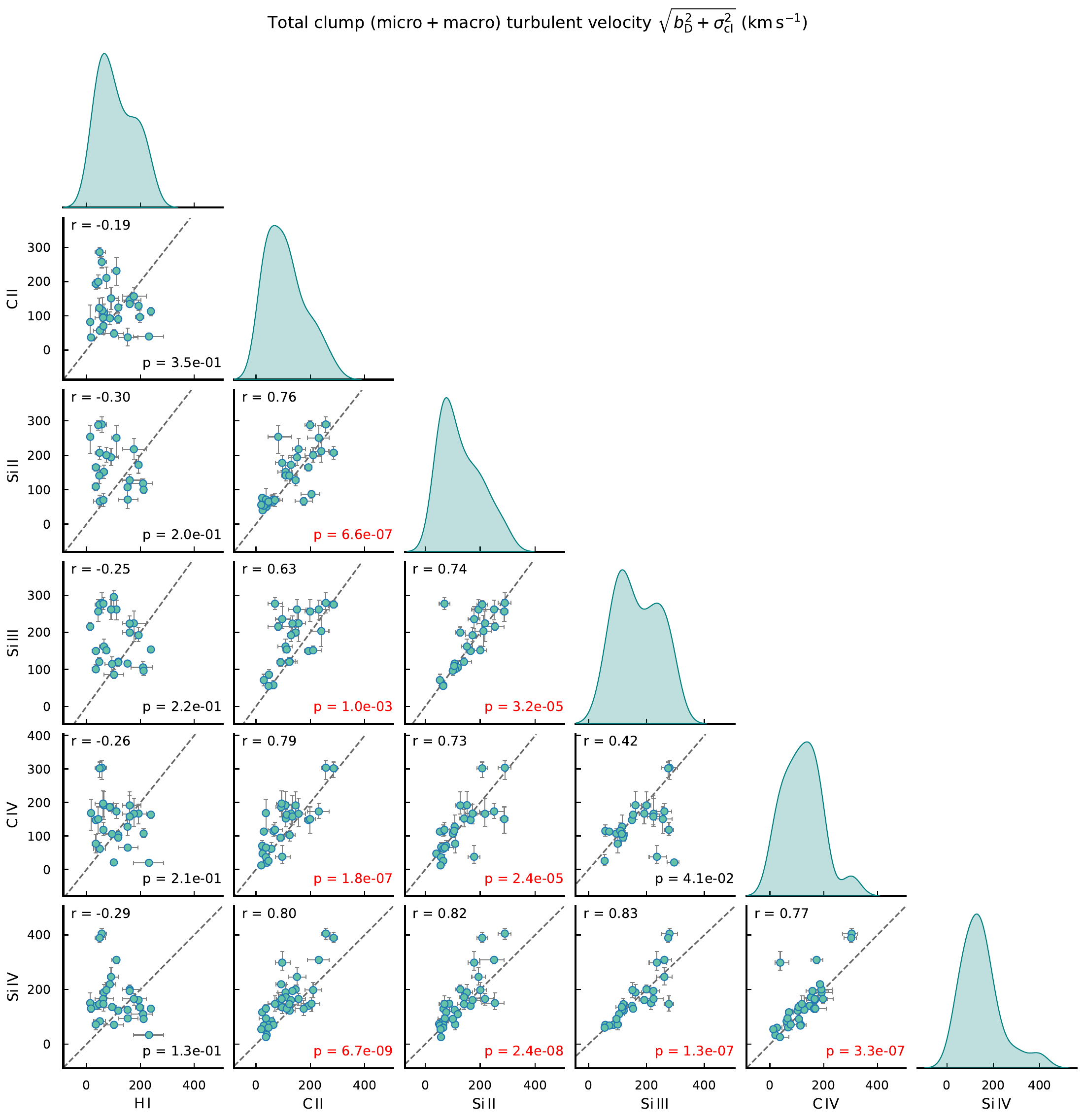}
\caption{\textbf{Cross-transition comparison of the total clump turbulent velocities $\sqrt{b_{\rm D}^2 + \sigma_{\rm cl}^2}$ inferred from RT modeling for six individual transitions.} Each panel presents a pairwise correlation between two ions, with the Pearson correlation coefficient $r$ and corresponding $p$-value annotated. Correlations more significant than the $3\sigma$ threshold are highlighted in red. Diagonal panels show the one-dimensional distributions of total turbulent velocity for each species, and dashed lines indicate a one-to-one relation. All five metal ions (except the \SiIII–\CIV\ pair due to a few outliers) exhibit exceptionally tight and monotonic correlations ($r \approx 0.7$–$0.8$), demonstrating that both low- and high-ionization species trace a common turbulent velocity field throughout the multiphase outflow. In contrast, neutral hydrogen shows only weak or marginal correlations with the metals. The strong correlations across both low- and high-ionization species indicate that the different ionization phases share comparable levels of turbulence and are likely dynamically coupled within a common, multiphase outflow structure.}\label{fig:compare_turbulence}
\end{figure*}

\begin{figure*}
\centering
\includegraphics[width=\textwidth]{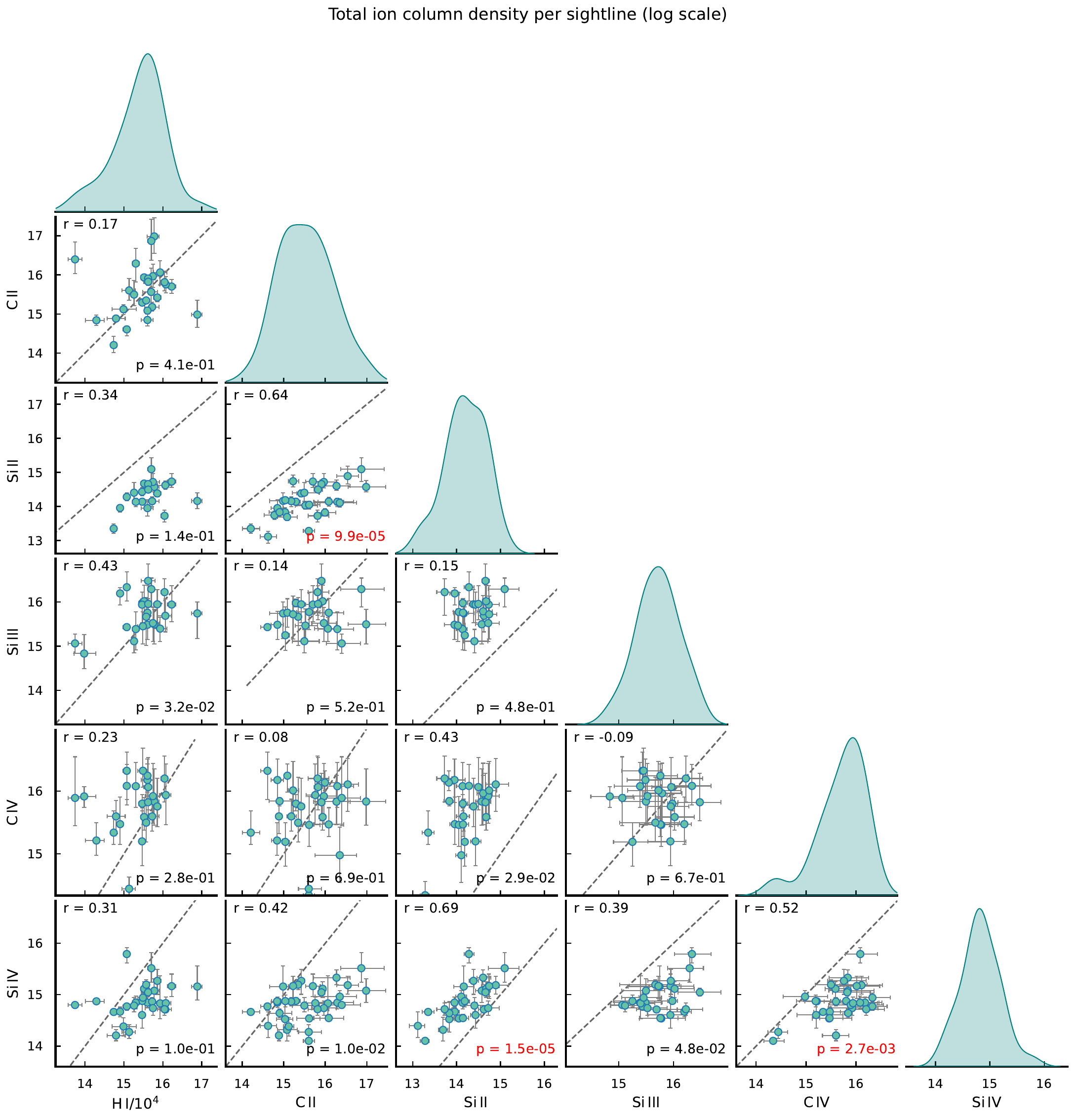}
    \caption{\textbf{Cross-transition comparison of total ion column densities per sightline derived from RT modeling for six individual transitions.} Each panel shows the correlation between the inferred total column densities of two ions, with Pearson $r$ and $p$ values labeled; $>3\sigma$ significant correlations are highlighted in red. \CII--\SiII\ exhibits the tightest relation, consistent with both tracing the same cool, low-ionization gas. Neutral hydrogen (scaled by $10^{-4}$ for comparability) shows no significant correlation with the metals. Overall, the column-density correlations are weaker and more diverse than the velocity-based ones, indicating that ion column densities may reflect local thermal and ionization structure rather than a globally coherent multiphase flow.
    \label{fig:compare_ion_columns}}
\end{figure*}

The total clump turbulent velocity, defined as $v_{\rm turb} = \sqrt{b_{\rm D,\,cl}^2 + \sigma_{\rm cl}^2}$,  captures the combined contribution of both the internal (microscopic) turbulence within individual clumps and the macroscopic random motions among clumps. Figure~\ref{fig:compare_turbulence} presents a cross-transition comparison of $v_{\rm turb}$ inferred independently from six transitions. As before, each subplot reports the Spearman rank correlation coefficient and $p$-value, with correlations exceeding the $3\sigma$ threshold highlighted in red.

As with the clump outflow velocity comparison, all five metal ions exhibit exceptionally tight and monotonic correlations, with typical coefficients of $r \approx 0.6$–$0.8$ and $p < 1\times10^{-3}$. The only notable exception is the \SiIII–\CIV\ pair, whose correlation falls just below the $3\sigma$ significance threshold, driven primarily by a few outliers. Consistent with previous results, we do not find statistically significant correlations between neutral hydrogen and the metal ions.

The tight correlations in $v_{\rm turb}$ across all metal ion species indicate that, much like the outflow velocity, the turbulent velocities traced by both low- and high-ionization species reflect similar dynamical conditions throughout the CGM. In other words, the cool and warm phases of the CGM appear to be co-moving not only in their bulk outflow velocities but also in their turbulent motions. Such behavior is expected if the phases are dynamically coupled within a multiphase flow, where cooling and mixing cause different gas phases to co-move with one another (e.g., \citealt{Gronke2018,Schneider2020}). This dynamical coherence suggests that the different ionization phases are physically coupled — likely embedded within a single multiphase outflow regulated by a common source of feedback-driven energy injection. These results further provide evidence that turbulence permeates multiple ionization layers of the clumpy gas, playing a key role in mediating energy exchange between different phases and helping to establish their shared kinematic evolution.

\subsection{Ion Column Densities}\label{sec:ion_columns}

We next examine how the total ion column densities per sightline ($N_{\rm ion,\,LOS} = (4/3) f_{\rm cl}N_{\rm ion,\,cl}$, see Eq. \ref{eq:NionLOS}) vary across different species in Figure~\ref{fig:compare_ion_columns}. Unlike the kinematic quantities, which exhibited uniformly strong, nearly one-to-one correlations across most metal ions, the column densities display a more complex and less coherent pattern.

Among all parameter pairs, C~\textsc{ii} and Si~\textsc{ii} exhibit the tightest correlation in column density, consistent with both ions tracing the same cool ($T \sim 10^4$\,K), low-ionization gas phase. A similarly strong trend is seen between \SiII\ and \SiIV, reflecting in part their shared elemental abundance as well as the coupled ionization structure of multiphase clumps. We also find a significant correlation between C \textsc{iv} and Si \textsc{iv}, though the significance lies roughly at the 3$\sigma$ level. As with the velocity-based quantities, neutral hydrogen\footnote{Note that our \HI\ column densities include contributions from both the primary and secondary clump populations.} shows no statistically significant correlation with the metal ions.

Overall, the inferred total column densities span $\sim10^{14}$ – $10^{17}$~cm$^{-2}$ for C \textsc{ii} and C \textsc{iv}, and $\sim10^{13}$ – $10^{16.5}$~cm$^{-2}$ for Si \textsc{ii} and Si \textsc{iv}. These ranges are broadly consistent with near-solar carbon-to-silicon abundance ratios, in which carbon is roughly an order of magnitude more abundant than silicon \citep{Asplund2021}.

Taken together, the ion-to-ion correlations in column density suggest a chemically and thermally non-coherent multiphase medium whose ionization layers do not exhibit the same degree of coupling seen in the velocity-based parameters. Although the outflow velocity and turbulent motions reveal strong dynamical coherence across ionization states, the column densities encode local ionization and density structure, leading to a more diverse and less uniform set of correlations across different ions.

\section{From single-line to multi-line joint spectral fitting}\label{sec:multi_line}
Having examined the correlations among the physical parameters inferred from single-line fitting of multiple transitions, a natural question that arises is whether these lines can be modeled in a self-consistent manner. In principle, there are several possible options to achieve such self-consistency, but most prove difficult to implement physically. First, the column densities of different ions do not closely track one another, so there is little justification for imposing a shared column density or fixed column-density ratios across species. Second, although the total turbulent velocity inferred from individual transitions shows strong correlations, the two separate turbulence components --- namely $b_{\rm D,\,cl}$ and $\sigma_{\rm cl}$ --- do not exhibit similarly strong correlations. This is expected, especially since the Doppler parameter $b_{\rm D,\,cl}$ explicitly depends on the atomic mass of the ions. As a result, the turbulent velocities cannot be meaningfully shared across transitions either. Ultimately, the only parameters that can be self-consistently shared are those associated with the bulk outflow kinematics: $v_0$ and $R$, which govern the radial outflow profile of the clumps, and the systemic velocity shift $\Delta v$.

\begin{figure*}
\centering
\includegraphics[width=\textwidth]{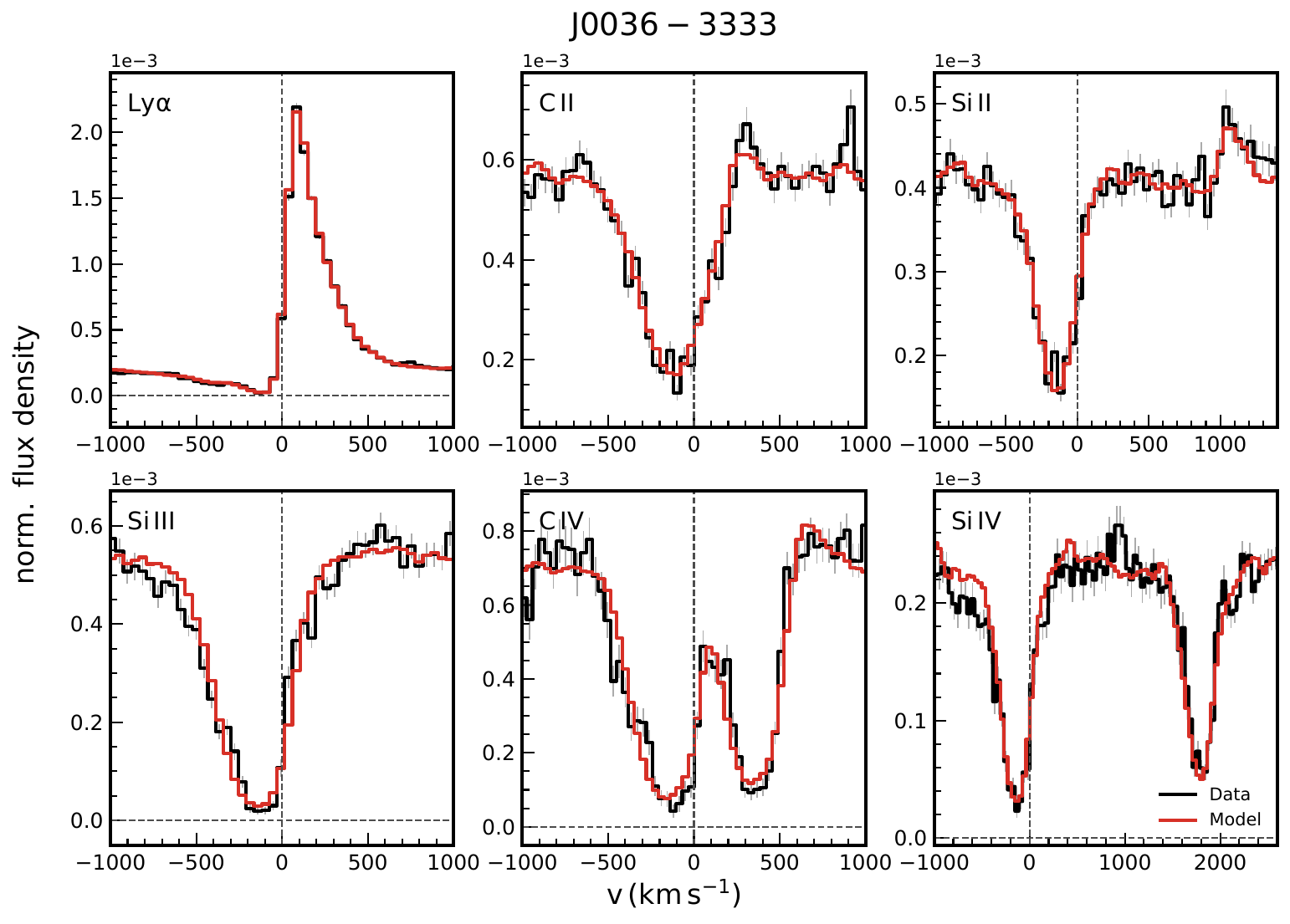}
    \caption{\textbf{Example of multi-line joint fitting for galaxy J0036–3333.} Shown are the observed line profiles (black) and best-fit models (red) for \lya\  and five metal lines (\CII, \SiII, \SiIII, \CIV, \SiIV). The five metal lines are simultaneously fitted with shared kinematic parameters using the joint-fitting framework, while \lya\ is modeled individually. Velocities are shown relative to the systemic redshift, indicated by the vertical dashed lines.}
    \label{fig:joint_example}
\end{figure*}

\begin{figure}
\centering
\includegraphics[width=0.47\textwidth]{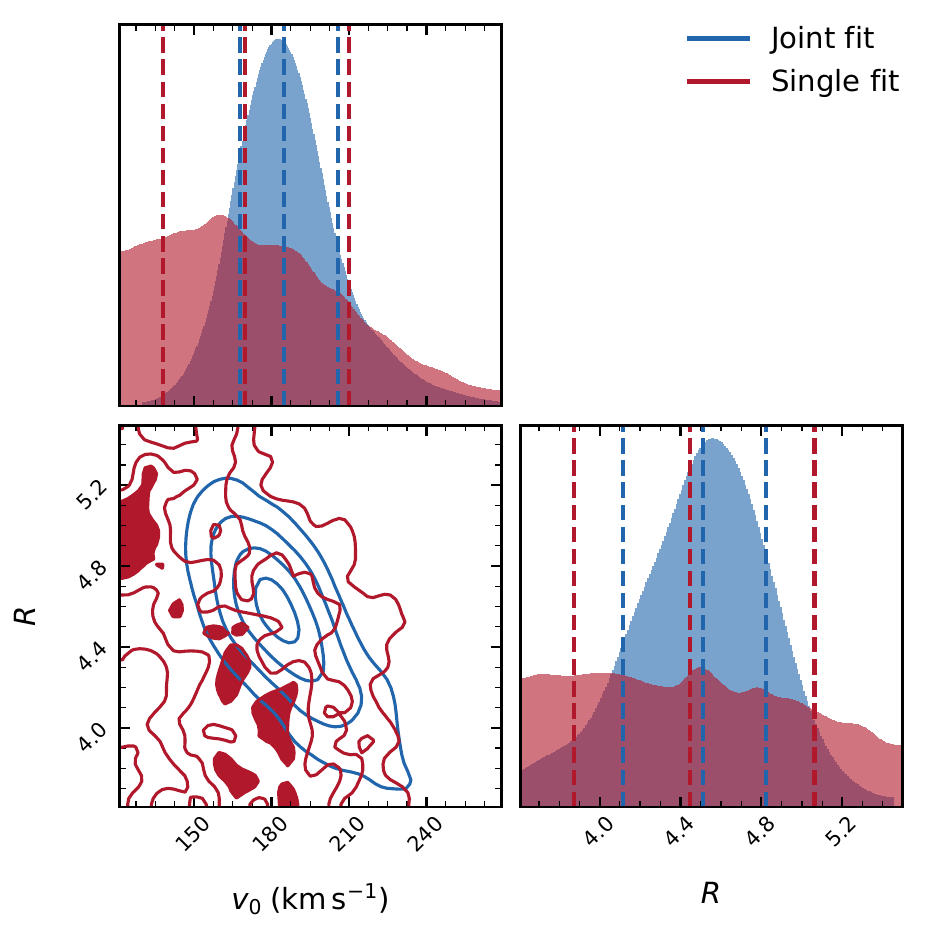}
    \caption{\textbf{Posterior distributions for the galaxy J0036$-$3333 from a single-line fit (\SiIII;} \textbf{red) and a joint fit using five metal lines (blue).} The red contours and histograms show the posterior distributions. The bottom left panel shows the joint posterior in the $v_0$–$R$ plane, and the top and right panels display the corresponding marginalized distributions. Dashed lines mark the 16th, 50th, and 84th percentiles. The joint fit produces significantly tighter and better-defined constraints than the single-line fit.
    \label{fig:joint_fitting_posterior}}
\end{figure}

Motivated by the consistency of the inferred $v_0$ and $R$ values across the five metal transitions (\CII, \SiII, \SiIII, \CIV, and \SiIV), we performed joint fits in which $v_0$, $R$, and $\Delta v$ were treated as shared parameters, while all remaining parameters for each transition were allowed to vary independently. In this framework, the total model likelihood is naturally defined as the product of the likelihoods of the individual transitions.

This joint-fitting approach proved highly effective: for essentially all galaxies in the sample, the best-fit spectra from the joint fits are comparable in quality to those obtained from the individual fits. In Figure~\ref{fig:joint_example}, we present galaxy J0036–3333 as an illustrative example, where the five metal lines are simultaneously fit using our joint-fitting framework, while \lya\ is modeled individually. More importantly, the joint method yields substantially tighter constraints on $v_0$, $R$, and $\Delta v$ -- particularly for galaxies with noisy spectra where single-line fits provide only weak constraints -- while leaving other parameters (e.g., clump volume filling factor, ion column density, and turbulent velocities) largely unaffected. Figure~\ref{fig:joint_fitting_posterior} illustrates this improvement for the galaxy J0036$-$3333. We compare the posterior distributions obtained from fitting only the \SiIII\ transition with those derived from a joint fit to all five metal lines. The joint fit significantly reduces parameter uncertainties (with the 1$\sigma$ width typically reduced by a factor of $\sim$ 2), and produces more compact and well-defined posteriors in the $v_0$–$R$ parameter space.

The fact that a single radial velocity profile can simultaneously reproduce the observed features of multiple transitions — spanning both low- and high-ionization species — strongly suggests that the Si$^+$, C$^+$, Si$^{2+}$, Si$^{3+}$, and C$^{3+}$ ions generally co-outflow within a coherent multiphase wind. This result indicates that the different ionic phases share a common large-scale kinematic structure rather than arising from dynamically independent components. We therefore adopt the joint-fit values of $v_0$, $R$, and $\Delta v$ for all subsequent analyses in Paper II, where they will be used to connect the inferred wind properties to global galaxy properties and to quantify the energy and momentum budget of the galactic wind.

In addition to the metal absorption lines, we also attempted incorporating the \lya\ emission line into our joint fitting to test whether the same $v_0$ and $R$ values that reproduce the metal transitions can also account for \lya. We find that for galaxies with both \lya\ spectra and metal lines, the $v_0$ and $R$ values required to reproduce the \lya\ profiles generally differ from those inferred from the metal transitions. A plausible explanation is that, at sufficiently low column densities, the \HI\ gas may remain optically thick to \lya\ even when the corresponding metal columns are too low to produce appreciable absorption. Alternatively, metals may be non-uniformly mixed with neutral hydrogen. This is consistent with the fact that \lya\ often requires a secondary clump population to reproduce the spectra, whereas the metal lines do not. Such differential mixing can lead to different scattering behaviors across transitions and therefore discrepant outflow velocity inferences. A comparison of the \lya-inferred and metal-inferred radial outflow velocity profiles for each galaxy is presented in the Appendix \ref{sec:best_fits}.

\section{Discussion}\label{sec:discussion}

In this section, we first summarize the key advances and physical insights enabled by the RT approach, and then examine the main assumptions and limitations that may affect the interpretation of the inferred parameters.

\subsection{Key Advances of this work}\label{sec:key_advances}

We highlight three major advances of this work that have broad implications for future studies of galactic winds in the CGM:

\begin{enumerate}
\item \textbf{Unified RT modeling framework:} We have shown that a single, physically motivated RT model can \emph{quantitatively} reproduce the diverse morphologies of multiple resonant and fluorescent emission and absorption lines, providing a coherent physical interpretation of line profiles spanning pure absorption, P-Cygni--like, and emission-dominated line shapes.

\item \textbf{Computational efficiency:}
Our fitting pipeline is computationally fast and scalable, making it feasible to perform joint, multi-transition inference across statistically significant galaxy samples. This capability transforms RT modeling from a case-by-case analysis to a population-level diagnostic of CGM kinematics and feedback energetics.

\item \textbf{Robust parameter constraints:}
Our RT framework constrains multiple physical parameters with
minimal degeneracy, including the gas covering factor, ionic column densities, and the kinematic structure of the outflow. In particular, it can \emph{disentangle} turbulent motions from coherent bulk outflows, yielding independent constraints on the turbulent velocity dispersion and the large-scale velocity field. This separation allows the energy and momentum associated with turbulence to be distinguished from those carried by bulk flows, enabling a clearer physical interpretation of CGM kinematics and feedback processes. A more detailed analysis is presented in Paper II.

\end{enumerate}

\subsection{Caveats and Limitations}\label{sec:caveats}

Our modeling adopts several simplifying assumptions that enable efficient exploration of parameter space but may not fully capture the physical complexity of multiphase outflows. In particular, we assume a constant clump size and column density  (and therefore a fixed clump mass) throughout the outflow. In reality, clumps are expected to evolve as they propagate: decreasing ambient pressure can drive expansion, while hydrodynamic instabilities and turbulence may induce fragmentation (e.g., \citealt{GronkeOh2018, GronkeOh2020, Li2020, Fielding2020, Fielding2022}).

In addition, the Doppler parameter $b_{\rm D,\,cl}$ in the model is generally much larger than the thermal velocity, especially for metal ions. Individual clumps are unlikely to be monolithic and may contain internal velocity gradients, unresolved turbulence, or substructure. Therefore, $b_{\rm D,\,cl}$ should be interpreted as an effective line-broadening parameter that incorporates both thermal motions and unresolved internal kinematics.

We also describe the random motions of clumps using a Gaussian velocity distribution characterized by a dispersion $\sigma_{\rm cl}$, effectively treating macroscopic turbulence as isotropic and statistically homogeneous. Real astrophysical turbulence is unlikely to be strictly Gaussian or scale-independent, so this prescription should be regarded as a phenomenological representation of the bulk velocity dispersion rather than a literal model of the turbulent cascade. Accordingly, $\sigma_{\rm cl}$ is best interpreted as an effective velocity dispersion that reproduces the observed line widths.

Lastly, we assume a spherically symmetric outflow geometry throughout this work. This approximation is likely reasonable for our sample, which is dominated by compact dwarf starbursts where feedback can drive quasi-isotropic winds. However, more massive disk galaxies often exhibit collimated or biconical outflows (e.g., \citealt{Veilleux2005, Xu2023b}). In such systems, the spherical geometry adopted here may not fully capture the directional escape of radiation or gas, and extending the framework to anisotropic geometries will be necessary when applying the model to those galaxies.

Taken together, these assumptions imply that the inferred parameters should be interpreted as effective, large-scale properties of the outflow that capture its dominant kinematic and structural behavior, rather than detailed descriptions of the underlying microphysics. Future spatially resolved observations and simulations that explicitly follow cloud evolution and wind geometry will therefore be essential for testing and refining these approximations.

\section{Conclusions}\label{sec:conclusion}

In this work, we have developed and applied \texttt{PEACOCK}, a three-dimensional Monte Carlo RT framework designed to jointly model rest-frame ultraviolet emission and absorption lines arising from multiphase, clumpy galactic winds, to systematically analyze a sample of 50 nearby star-forming galaxies. Our main conclusions are summarized as follows:

\begin{enumerate}

\item \textbf{A self-consistent multi-line RT framework reproduces the full diversity of observed UV line morphologies.} Using a single, physically motivated multiphase, clumpy CGM model, we reproduce 220 observed line profiles of \lya, \SiII, \CII, \SiIII, \SiIV, and \CIV\ that span pure absorption, pure emission, and P-Cygni--like morphologies. This is achieved without invoking ion-specific geometries or ad hoc emission prescriptions, establishing multi-line RT modeling as a robust and internally consistent interpretive framework.

\item \textbf{Macroscopic velocity dispersion is a necessary physical component of galactic outflows.} Purely radial, accelerating outflows often fail to reproduce observed UV absorption profiles, instead producing highly asymmetric and unphysical features. Incorporating macroscopic random motions among clumps naturally generates the broad, asymmetric absorption troughs seen in the data, indicating that turbulence must be treated as an intrinsic part of wind kinematics in the CGM.

\item \textbf{Different CGM model parameters leave distinct imprints on line profiles with minimal degeneracy.} Systematic parameter experiments show that key CGM properties — including ion column densities, bulk outflow velocities, and turbulent velocities — affect the emergent line profiles in qualitatively different ways. Consequently, these parameters can be independently constrained within our comprehensive RT framework.

\item \textbf{Low- and high-ionization metal lines trace a common, coherent kinematic structure.}  
Metal species spanning a wide range of ionization states are consistent with a shared, radially varying outflow velocity profile and exhibit tightly correlated turbulent velocities. This coherence indicates strong dynamical coupling between different CGM phases, implying that cool and warm gas not only co-move in bulk outflow but also share similar turbulent motions. Together, these results support a picture in which multiphase material is entrained within a common feedback-driven outflow rather than arising from independent kinematic components.

\item \textbf{Neutral hydrogen exhibits distinct physical properties from metal-traced gas.} In contrast to the strong ion-to-ion coherence observed among metal lines, \HI\ shows weak or marginal correlations with metal-derived outflow velocities, turbulent velocities, and column densities. This suggests that a substantial fraction of the neutral gas may be less well mixed with metals and may not share the same detailed kinematic structure as the metal-enriched outflow, even if both phases participate in the same large-scale galactic wind.

\item \textbf{Joint multi-line fitting significantly tightens constraints on gas outflow kinematics.} Imposing a common radial velocity law across multiple metal transitions yields substantially tighter constraints on the bulk outflow parameters while preserving the quality of the individual line fits. This consistency suggests that the shared velocity structure inferred from different ions reflects a genuine physical property of the multiphase outflow.

\end{enumerate}

Overall, our results demonstrate that joint physically grounded RT modeling across multiple ions provides a powerful bridge between UV observations and theoretical models of galactic outflows. By unifying emission and absorption diagnostics across a wide range of ionization states, this approach enables a self-consistent interpretation of CGM kinematics and energetics that cannot be achieved with phenomenological models or single-line analyses. The \texttt{PEACOCK} RT modeling framework thus opens a new pathway toward statistically robust and physically interpretable studies of galactic winds in the CGM from the local universe to the epoch of reionization. In a companion paper (Paper II), we will extend the framework developed here by relating the inferred wind properties to global host-galaxy characteristics, quantifying the kinetic energy and pressure budget of the cool-to-warm CGM, and exploring the physical origin of the inferred turbulent motions, including the role of stellar feedback as a dominant energy source.

\begin{acknowledgments}
We acknowledge the contributions of the CLASSY team members, whose efforts made this project possible. This work was carried out using the Advanced Research Computing at Hopkins (ARCH) core facility (rockfish.jhu.edu), which is supported by the National Science Foundation under grant OAC-1920103. MG acknowledges support from the Max Planck Society through the Max Planck Research Group.
\end{acknowledgments}





%
\facilities{HST (COS)}




\appendix
\section{Photon Frequency Redistribution in Fluorescent Transitions}
\label{sec:fluorescent_frequency}
To properly model the RT of photons for spectral lines that are not strictly resonant and have fluorescent channels (such as \SiII\ and \CII) the RT algorithm must be modified.
Here we describe how the final photon frequency is computed following each scattering event for transitions that permit fluorescent decay. 

Throughout this work, photon frequencies are expressed in Doppler units
\begin{equation}
x \equiv \frac{\nu - \nu_0}{\Delta\nu_{\rm D}}
\end{equation}
where $\nu_0$ is the reference transition frequency and $\Delta\nu_{\rm D}$ is the Doppler width. The incoming and outgoing photon directions are denoted by $\hat{\boldsymbol{k}}_{\rm in}$ and $\hat{\boldsymbol{k}}_{\rm out}$, respectively, and the atomic velocity $\boldsymbol{u}$ is measured in units of the thermal velocity $v_{\rm th}$.

The frequency of the incoming photon in the atomic rest frame is
\begin{equation}
x_{\rm in}^{\rm atom}
= x_{\rm in}^{\rm gas}
- \boldsymbol{u}\cdot\hat{\boldsymbol{k}_{\rm in}}
\label{eq:xin_atom_app}
\end{equation}
Likewise, the transformation from the atomic rest frame to the gas frame for the outgoing photon is
\begin{equation}
x_{\rm out}^{\rm gas}
= x_{\rm out}^{\rm atom}
+ \boldsymbol{u}\cdot\hat{\boldsymbol{k}}_{\rm out}
\label{eq:xout_gas_app}
\end{equation}

For a resonant event, the photon frequency is conserved in the atomic rest frame 
\begin{equation}
x_{\rm out}^{\rm atom} = x_{\rm in}^{\rm atom}
\label{eq:xout_atom_res}
\end{equation}
Combining Eqs.~(\ref{eq:xin_atom_app})--(\ref{eq:xout_gas_app}) gives the usual Doppler redistribution in the gas frame
\begin{equation}
x_{\rm out}^{\rm gas}
= x_{\rm in}^{\rm gas}
- \boldsymbol{u}\cdot\hat{\boldsymbol{k}}_{\rm in}
+ \boldsymbol{u}\cdot\hat{\boldsymbol{k}}_{\rm out}
\label{eq:xout_res_gas}
\end{equation}

For a fluorescent event, however, the absorbed photon is re-emitted at the line center of a different transition in the atomic rest frame (see also \citealt{Michel-Dansac2020})
\begin{equation}
x_{\rm out}^{\rm atom} = x_{\rm fluo}
\label{eq:xout_atom_fluo}
\end{equation}
independent of the incoming photon frequency. Substituting Eq.~(\ref{eq:xout_atom_fluo}) into Eq.~(\ref{eq:xout_gas_app}) yields
\begin{equation}
x_{\rm out}^{\rm gas}
= x_{\rm fluo}
+ \boldsymbol{u}\cdot\hat{\boldsymbol{k}}_{\rm out}
\label{eq:xout_fluo_gas_basic}
\end{equation}

For numerical implementation, we decompose the atomic velocity into components parallel and perpendicular to the incoming direction,
\begin{equation}
\boldsymbol{u}
= u_{\parallel}\hat{\boldsymbol{k}}_{\rm in}
+ u_{\perp}\hat{\boldsymbol{e}}_{\perp}
\end{equation}
and define $\mu \equiv \hat{\boldsymbol{k}}_{\rm in}\cdot\hat{\boldsymbol{k}}_{\rm out}$. Then
\begin{equation}
\boldsymbol{u}\cdot\hat{\boldsymbol{k}}_{\rm out}
= u_{\parallel}\mu
+ u_{\perp}\sqrt{1-\mu^2}
\end{equation}
then in the resonant case, we have 
\begin{align}
x_{\rm out,\,res}^{\rm gas}
&= x_{\rm in}^{\rm gas}
+ u_{\parallel}(\mu-1)
+ u_{\perp}\sqrt{1-\mu^2}
\label{eq:xout_res_final}
\end{align}
whereas in the fluorescent case

\begin{align}
x_{\rm out,\,fluo}^{\rm gas}
= x_{\rm fluo}
+ u_{\parallel}\mu
+ u_{\perp}\sqrt{1-\mu^2}
\label{eq:xout_fluo_final}
\end{align}

Equations~(\ref{eq:xout_res_final}) and (\ref{eq:xout_fluo_final}) illustrate the fundamental distinction between resonant and fluorescent scattering.
In the resonant case, the photon retains memory of its incoming frequency through the explicit $x_{\rm in}$ dependence, with Doppler shifts scaling as $(\mu - 1)$.
In contrast, for fluorescent scattering the photon frequency is reset in the atomic rest frame to the fluorescent line center $x_{\rm fluo}$, erasing any memory of the incoming frequency; the outgoing frequency in the gas frame is then determined solely by Doppler shifts associated with the atomic velocity projected along the outgoing direction, as encoded by the scattering geometry through $\mu$.

Following absorption, the excited atom may decay through multiple radiative channels.
For each excited level, the decay path is selected according to the corresponding branching ratios derived from Einstein coefficients and statistical weights.
Below we derive the transition probabilities and branching ratios used in the simulation. All probabilities are computed directly from Einstein coefficients and statistical weights, ensuring consistency with detailed balance and atomic selection rules.

The probability that a photon is absorbed through a transition originating from a lower level $i$ is proportional to the corresponding Einstein $B_{iu}$ coefficient.
The relative probabilities for absorption from lower levels $i$ into upper levels $u$ and $v$ are therefore
\begin{equation}
\frac{P_{i\rightarrow u}}{P_{i\rightarrow v}}
= \frac{B_{iu}}{B_{iv}}
\end{equation}

Using the Einstein relations and the principle of detailed balance
\begin{equation}
g_i B_{iu} = g_u B_{ui}
\end{equation}
together with the proportionality between the Einstein $A$ and $B$ coefficients for a given transition, the above ratio can be rewritten as
\begin{equation}
\frac{P_{i\rightarrow u}}{P_{i\rightarrow v}}
= \frac{g_u A_{ui}}{g_v A_{vi}}
\label{eq:B_to_A}
\end{equation}
where $g = 2J + 1$ denotes the statistical weight of each atomic level. This relation allows the absorption probabilities to be computed using tabulated Einstein $A$ coefficients and level degeneracies alone, without introducing additional free parameters.

As an example, we consider absorption from the two fine-structure ground states of \SiII\,$\lambda 1260$. We label the four relevant energy levels as $1, 2, 3,$ and $4$, ordered by increasing energy. The relative probability for absorption through the $2 \rightarrow 4$ and $2 \rightarrow 3$ channels is then
\begin{equation}
\frac{P_{24}}{P_{23}}
= \frac{B_{24}}{B_{23}}
= \frac{g_4 A_{42}}{g_3 A_{32}}
\end{equation}
where $g_i$ denotes the statistical weight of level $i$.

Substituting the corresponding statistical weights and Einstein $A$ coefficients yields
\begin{equation}
\frac{P_{24}}{P_{23}}
\simeq \frac{6 \times A_{42}}{4 \times A_{32}}
\approx 9.6
\end{equation}
Normalizing the probabilities then gives
\begin{equation}
P_{24} \simeq 0.90, \qquad
P_{23} \simeq 0.10
\end{equation}

In the Monte Carlo implementation, individual scattering paths are sampled hierarchically using conditional probabilities. At each scattering event, the absorption channel is first selected by Monte Carlo sampling from the set of all allowed bound--bound transitions, with probabilities proportional to the frequency-dependent absorption cross sections
\begin{equation}
P(i\rightarrow u \mid x)=\frac{\sigma_{iu}(x)}{\sum\limits_{(j\rightarrow v)\in \mathcal{T}} \sigma_{jv}(x)}
\label{eq:P_abs_general}
\end{equation}
where $\mathcal{T}$ denotes the full set of transitions included in the calculation. 
Conditioned on the selected absorption channel, the subsequent radiative decay path is then sampled using the appropriate branching ratios.

For example, if the absorption event populates level 3 from level 2 (i.e., the $2\rightarrow3$ channel is selected), the probability of the photon following the path $2\rightarrow3\rightarrow1$ is

\begin{equation}
P(2\rightarrow3\rightarrow1 \mid x)
= \frac{\sigma_{23}(x)}{\sum\limits_{(j\rightarrow v)\in \mathcal{T}} \sigma_{jv}(x)} \times P_{31}
\end{equation}
where $P_{31}$ is the branching ratio for decay from level 3 to level 1. 
Similarly, the probability of the alternative decay path $2\rightarrow3\rightarrow2$ is
\begin{equation}
P(2\rightarrow3\rightarrow2 \mid x)
= \frac{\sigma_{23}(x)}{\sum\limits_{(j\rightarrow v)\in \mathcal{T}} \sigma_{jv}(x)} \times P_{32}
\end{equation}
Other absorption and decay channels are treated analogously. This hierarchical sampling ensures that all scattering paths are selected consistently with both the frequency-dependent absorption probabilities and the radiative branching ratios.

\section{Connecting the RT Aperture Parameter to the Physical Aperture}\label{sec:compare_impact_param}

In our RT modeling, the parameter $b_{\max}$, or its normalized form $b_{\max}/R_{\rm out}$, serves as an effective aperture weighting parameter. Specifically, it controls the fraction of photons that contribute to the emergent model spectrum and therefore mimics the fraction of UV radiation captured by the observing aperture.

To interpret $b_{\max}/R_{\rm out}$ observationally, we compare it to the physical size of the spectroscopic aperture relative to the galaxy scale. The \emph{HST}/COS aperture has an effective radius of $\sim1.1''$ (accounting for vignetting), which we convert to physical units at the galaxy redshift to obtain $r_{\rm COS}$. We adopt an outer radius $R_{\rm out} = 20\,r_{50}$ for the RT model, where $r_{50}$ is the half-light radius measured from imaging \citep{Xu2022}, and assign a typical 15\% uncertainty to $r_{50}$. The ratio $r_{\rm COS}/R_{\rm out}$ therefore characterizes the fraction of the outflow region that is directly covered by the spectroscopic aperture.

Under the assumption that the UV-emitting stellar population approximately traces the inner launching region of the outflow, a correspondence is expected between $r_{\rm COS}/R_{\rm out}$ and the effective aperture weighting parameter $b_{\max}/R_{\rm out}$. We therefore compare these two quantities for our sample. For each galaxy, $b_{\max}/R_{\rm out}$ is taken as the average of the values inferred from all available metal transitions used in the RT modeling. We exclude the high-ionization \CIV\ and \SiIV\ lines in cases where the profiles are dominated by emission since they may be more sensitive to extended resonant scattering. When only a single suitable transition is available, we adopt the corresponding measurement.

\begin{figure}
\centering
\includegraphics[width=0.5\textwidth]{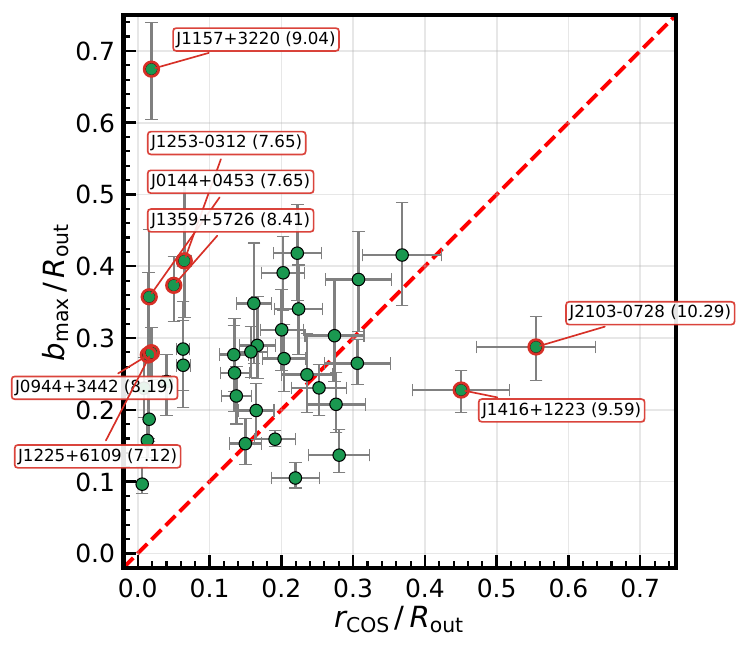}
    \caption{\textbf{Comparison between the RT–inferred aperture weighting parameter $b_{\max}/R_{\rm out}$ and the geometric aperture scale $r_{\rm COS}/R_{\rm out}$ for our sample.} Here $r_{\rm COS}$ is the physical radius corresponding to the $1.1''$ \emph{HST}/COS aperture at the galaxy redshift, and $R_{\rm out}=20\,r_{50}$ is the outer radius adopted in the RT model.  The dashed line indicates the one-to-one relation. Most galaxies lie close to the one-to-one relation, with differences typically within $\sim0.2$ dex, indicating that the RT modeling generally recovers an effective photon sampling region comparable to the geometric aperture size. The most significant deviations occur in a small subset of systems (highlighted with red circles and annotated with galaxy names and total stellar masses), which tend to exhibit complex or extended spatial morphologies where the effective photon sampling region does not simply follow the geometric aperture.
    \label{fig:bmax_rcos}}
\end{figure}

Figure~\ref{fig:bmax_rcos} compares $b_{\max}/R_{\rm out}$ with $r_{\rm COS}/R_{\rm out}$. Most galaxies lie broadly consistent with the one-to-one relation within moderate scatter, indicating that the RT modeling generally recovers an effective photon sampling region comparable to the geometric aperture size. Aside from a small number of outliers discussed below, the differences between $b_{\max}/R_{\rm out}$ and $r_{\rm COS}/R_{\rm out}$ are all within $\sim0.2$ dex. This agreement supports the interpretation of $b_{\max}$ as an aperture-weighting parameter that captures the fraction of the outflow contributing to the observed profile.

\begin{figure*}
\centering
\includegraphics[width=0.9\textwidth]{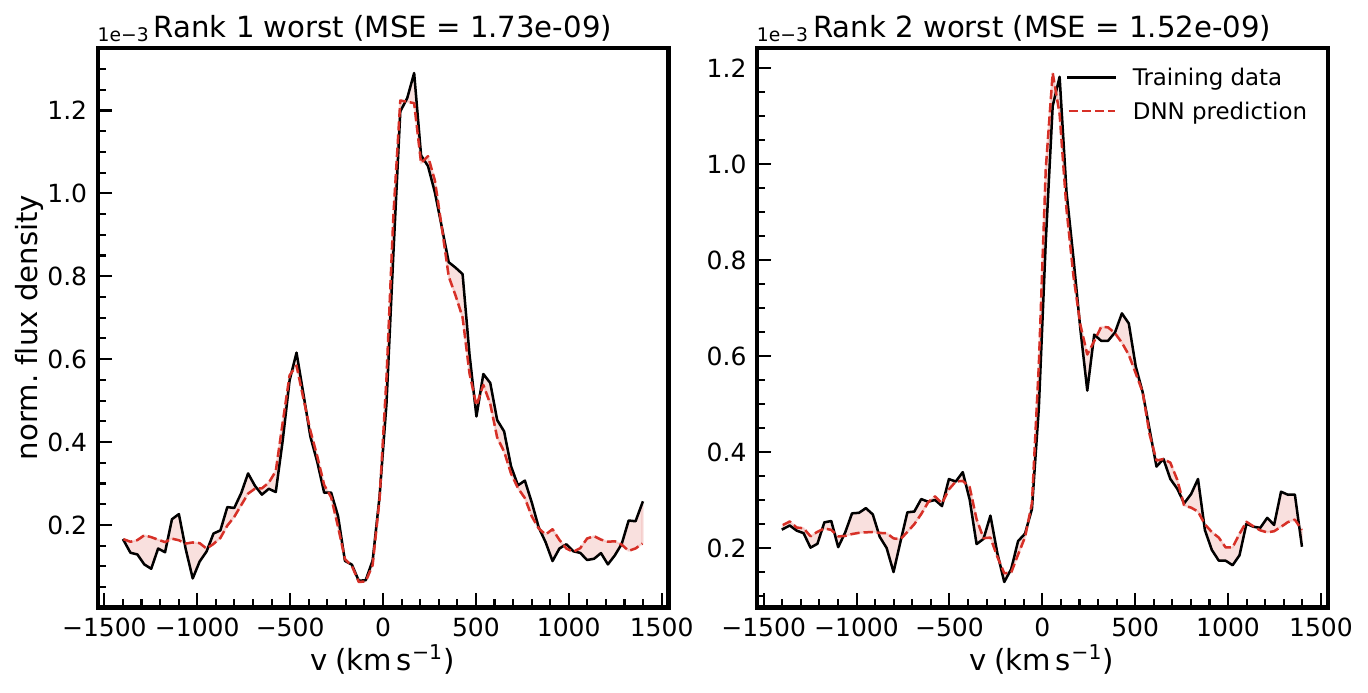}\\
    \caption{\textbf{Example validation of the neural network emulator used for \lya\ modeling of galaxy J0021+0052.}  Shown are the two spectra with the largest MSE across the entire dataset (training plus validation samples). The black curves denote the original RT model spectra and the red dashed curves the DNN predictions. The residual differences are shown as red shaded regions. The agreement demonstrates that even the worst-case errors are limited to small local fluctuations and do not alter the global line morphology. 
    \label{fig:dnn_validation}}
\end{figure*}

We highlight and discuss the largest deviations individually. Two of the most massive systems in the sample, J2103$-$0728 and J1416$+$1223 ($\log M_\star/M_\odot \sim 9.6$ -- 10.3), exhibit $b_{\max}/R_{\rm out}$ values systematically smaller than implied by $r_{\rm COS}/R_{\rm out}$. These galaxies show relatively regular and centrally concentrated morphologies in imaging, suggesting that the COS aperture encloses most of the UV-bright region. The smaller inferred $b_{\max}$ may indicate that the photons contributing most strongly to the observed profile originate from a more centrally concentrated region than implied by the geometric aperture size, possibly due to a compact starburst embedded within a more extended stellar component.

In contrast, J1157$+$3220 shows $b_{\max}/R_{\rm out}$ significantly larger than $r_{\rm COS}/R_{\rm out}$. This galaxy exhibits a strongly asymmetric and extended emission morphology, with substantial UV-bright structure outside the nominal aperture region. In such a configuration, photons originating at larger radii can still contribute to the observed absorption spectrum through scattering into the line of sight, leading to an effective sampling radius larger than the geometric aperture size and thus a larger inferred $b_{\max}$. Several intermediate-mass systems (J0144$+$0453, J1359$+$5726, J0944$+$3442, and J1225$+$6109; $\log M_\star/M_\odot \sim 7$ -- 8.5) also show similarly extended emission morphologies. For these galaxies, the spatial distribution of star-forming regions extends well beyond the COS aperture, naturally producing deviations from the one-to-one relation as the RT-inferred photon sampling region reflects contributions from gas located outside the central aperture.

One particularly interesting case is J1253$-$0312 (also known as SHOC~391), which shows a relatively small $r_{\rm COS}/R_{\rm out}$ despite its UV emission appearing largely enclosed within the aperture in imaging. The imaging suggests a compact, bright central star-forming region embedded within a more extended low-surface-brightness component, which may bias $r_{50}$ high relative to the size of the UV-bright core. In such a configuration, the geometric ratio $r_{\rm COS}/R_{\rm out}$ can be small even though the aperture captures most of the central UV light. 

Overall, the comparison demonstrates that $b_{\max}/R_{\rm out}$ and $r_{\rm COS}/R_{\rm out}$ are broadly consistent for the majority of the sample, supporting the physical interpretation of $b_{\max}$ as an effective aperture-weighting parameter. The remaining discrepancies arise primarily in galaxies with complex or extended spatial morphologies, where the effective photon sampling region no longer directly corresponds to the geometric aperture.

\section{Validation of the Neural Network Emulator}\label{sec:dnn_validation}

To evaluate the performance of the neural network emulator, we compared the network outputs against the original RT calculations. Each training dataset consists of 4000 synthetic spectra, of which 400 samples were reserved for validation and not used during training.

We use the DNN modeling of the \lya\ emission in the galaxy J0021+0052 as an illustrative example. Figure~\ref{fig:dnn_validation} shows the two spectra with the largest mean squared errors (MSE) across the entire dataset (training plus validation samples). Even for these worst-performing cases, the predicted spectra closely reproduce the overall line shapes of the true RT spectra. The residual differences (shown as red shaded regions) are small compared to the characteristic spectral features and do not alter the qualitative morphology of the line profiles.

The fact that the largest errors occur only in minor local fluctuations, rather than global profile mismatches, indicates that the neural network has successfully learned the mapping between physical parameters and spectral structure across the parameter space. We therefore conclude that the DNN emulator provides a reliable approximation to the full RT calculations and is sufficiently accurate for parameter inference within the uncertainties of the observational data.

\section{Line Profile Morphology Classification and Sample Subdivision}

Table \ref{tab:line_profiles} lists the morphology classification of the line profiles for all the galaxies in our combined sample. For each object, we report the assigned morphology class for the transitions \lya, \SiII, \CII, \SiIII, \SiIV, and \CIV\ when available; the morphology categories are defined in Section \ref{sec:individual_modeling}.

Based on the number of detected transitions, we divide the galaxies into three subsamples, highlighted using different colors in the table. Galaxies with all six transitions are shown in gold (the golden sample, 9 objects), those with four or five transitions in silver (the silver sample, 32 objects), and those with three or fewer transitions in bronze (the bronze sample, 9 objects).

{\renewcommand{\arraystretch}{0.97}
\begin{deluxetable*}{cccccccc}
\tablecaption{Line Morphology Classification for Galaxies in the Combined Sample\label{tab:line_profiles}}
\tablehead{
\colhead{Galaxy} & \colhead{$z$} 
& \colhead{\lya} 
& \colhead{\CII}  
& \colhead{\SiII}
& \colhead{\SiIII} 
& \colhead{\CIV}
& \colhead{\SiIV} \\
(1) & (2) & (3) & (4) & (5) & (6) & (7) & (8)
}
\startdata
\textcolor{gold}{\textbf{J0021+0052}} & 0.09839 & $a$ & $a$ & $b$ & $a$ & $a$ & $a$\\
\textcolor{gold}{\textbf{J0036--3333}} & 0.02060 & $b$ & $b$ & $b$ & $b$ & $a$ & $a$ \\
\textcolor{silver}{\textbf{J0055--0021}}$^\text{(H15)}$ & 0.16744 & $b$ & $b$ & $a$ & $a$ & -- & $a$ \\
\textcolor{silver}{\textbf{J0127--0619}} & 0.00550 & -- & $d$ & $c$ & -- & $b$ & $c$ \\
\textcolor{bronze}{\textbf{J0144+0453}} & 0.00532 & -- & $d$ & $b$ & -- & -- & $c$ \\
\textcolor{gold}{\textbf{J0150+1308}}$^\text{(H15)}$ & 0.14668 & $b$ & $f$ & $a$ & $a$ & $a$ & $a$ \\
\textcolor{bronze}{\textbf{J0337--0502}} & 0.01346 & -- & $e$ & $c$  & -- & $d$ & -- \\
\textcolor{bronze}{\textbf{J0405--3648}} & 0.00280 & -- & $d$ & $c$ & -- & -- & $c$ \\
\textcolor{silver}{\textbf{J0808+3948}} & 0.09123 & $c$ & Other & Other & Other & Other & Other \\
\textcolor{silver}{\textbf{J0823+2806}} & 0.04730 & -- & $c$ & $a$ & -- & $a$ & $a$ \\
\textcolor{silver}{\textbf{J0926+4427}} & 0.18030 & $c$ & $a$ & $a$  & $b$ & -- & $b$ \\
\textcolor{silver}{\textbf{J0934+5514}} & 0.00264 & -- & $e$ & $c$  & -- & $c$ & $a$ \\
\textcolor{gold}{\textbf{J0938+5428}} & 0.10210 & $f$ & $c$ & $a$ & $b$ & $a$ & $a$ \\
\textcolor{silver}{\textbf{J0940+2935}} & 0.00171 & -- & $e$ & $c$ & -- & -- & $a$ \\
\textcolor{silver}{\textbf{J0942+3547}} & 0.01483 & $e$ & $d$ & -- & $a$ & $b$ & $a$ \\
\textcolor{bronze}{\textbf{J0944+3442}} & 0.02005 & $a$ & $c$ & -- & -- & -- & -- \\
\textcolor{silver}{\textbf{J0944--0038}} & 0.00487 & $e$ & $d$ & $c$ & -- & $d$ & $a$ \\
\textcolor{silver}{\textbf{J1016+3754}} & 0.00391 & $c$ & $d$ & $a$ & -- & $c$ & $a$ \\
\textcolor{silver}{\textbf{J1024+0524}} & 0.03326 & $a$ & -- & $b$ & $b$ & $b$ & $a$ \\
\textcolor{gold}{\textbf{J1025+3622}} & 0.12720 & $a$ & $a$ & $b$ & $b$ & $a$ & $a$ \\
\textcolor{silver}{\textbf{J1044+0353}} & 0.01286 & $c$ & $d$ & -- & -- & $d$ & $d$ \\
\textcolor{silver}{\textbf{J1105+4444}} & 0.02148 & $b$ & -- & $b$ & $a$ & -- & $a$ \\
\textcolor{gold}{\textbf{J1112+5503}} & 0.13153 & $b$ & $b$ & $b$ & $b$ & $f$ & $a$ \\
\textcolor{bronze}{\textbf{J1113+2930}}$^\text{(H15)}$ & 0.17514 & -- & Other & Other & $b$ & -- & -- \\
\textcolor{bronze}{\textbf{J1119+5130}} & 0.00444 & $f$ & $a$ & -- & -- & -- & $a$ \\
\textcolor{silver}{\textbf{J1129+2034}} & 0.00466 & -- & $d$ & $c$ & -- & $b$ & $a$ \\
\textcolor{silver}{\textbf{J1132+1411}} & 0.01763 & $f$ & $d$ & -- & $b$ & -- & $a$ \\
\textcolor{bronze}{\textbf{J1132+5722}} & 0.00510 & -- & $d$ & $c$ & -- & -- & -- \\
\textcolor{gold}{\textbf{J1144+4012}} & 0.12695 & $b$ & $b$ & $a$ & $b$ & $a$ & $a$ \\
\textcolor{silver}{\textbf{J1148+2546}} & 0.04524 & $a$ & -- & $b$ & $a$ & $c$ & $c$ \\
\textcolor{silver}{\textbf{J1150+1501}} & 0.00250 & -- & $d$ & $c$ & $a$ & $b$ & $a$ \\
\textcolor{silver}{\textbf{J1157+3220}} & 0.01120 & $d$ & $a$ & -- & -- & $b$ & $a$ \\
\textcolor{silver}{\textbf{J1200+1343}} & 0.06690 & $a$ & -- & $a$ & $a$ & $b$ & $c$ \\
\textcolor{silver}{\textbf{J1225+6109}} & 0.00233 & -- & $e$ & $c$ & $b$ & $b$ & $a$ \\
\textcolor{silver}{\textbf{J1253--0312}} & 0.02267 & $e$ & $d$ & -- & $a$ & $b$ & -- \\
\textcolor{silver}{\textbf{J1314+3452}} & 0.00285 & -- & $d$ & $c$ & $b$ & $b$ & $a$ \\
\textcolor{bronze}{\textbf{J1323--0132}} & 0.02246 & $e$ & -- & -- & $a$ & $e$ & -- \\
\textcolor{silver}{\textbf{J1359+5726}} & 0.03390 & $c$ & $a$ & -- & $a$ & $b$ & $a$ \\
\textcolor{silver}{\textbf{J1414+0540}}$^\text{(H15)}$ & 0.08190 & $f$ & $c$ & -- & -- & $a$ & $a$ \\
\textcolor{gold}{\textbf{J1416+1223}} & 0.12316 & $f$ & $f$ & $b$ & $a$ & $f$ & $b$ \\
\textcolor{silver}{\textbf{J1418+2102}} & 0.00857 & $e$ & $e$ & -- & -- & $d$ & $a$ \\
\textcolor{silver}{\textbf{J1428+1653}} & 0.18170 & $a$ & $a$ & $a$ & $a$ & -- & $a$ \\
\textcolor{silver}{\textbf{J1429+0643}} & 0.17350 & $a$ & $b$ & $a$ & $a$ & -- & $b$ \\
\textcolor{silver}{\textbf{J1444+4237}} & 0.00219 & -- & $d$ & $c$ & -- & $d$ & $c$ \\
\textcolor{bronze}{\textbf{J1448--0110}} & 0.02738 & -- & -- & $a$ & $a$ & $b$ & -- \\
\textcolor{silver}{\textbf{J1521+0759}} & 0.09426 & $b$ & $a$ & -- & $a$ & $a$ & $a$ \\
\textcolor{silver}{\textbf{J1525+0757}} & 0.07579 & -- & $b$ & $a$ & $b$ & $a$ & $a$ \\
\textcolor{silver}{\textbf{J1545+0858}} & 0.03772 & $e$ & -- & -- & $a$ & $c$ & $c$ \\
\textcolor{silver}{\textbf{J1612+0817}} & 0.14914 & $b$ & $b$ & $b$ & $a$ & $a$ & -- \\
\textcolor{gold}{\textbf{J2103--0728}}$^\text{(H15)}$ & 0.13689 & $a$ & $b$ & $b$ & $a$ & $a$ & $a$ \\
		\hline
\enddata
\tablecomments{(1) Line morphology classification for 45 galaxies from the CLASSY sample and 5 galaxies from \citet{Heckman2015} (marked as H15). The morphology categories for each transition are defined in Section \ref{sec:individual_modeling}.}
\end{deluxetable*}}

\section{Best-fit Line Profiles for the Full Galaxy Sample}\label{sec:best_fits}

\begin{figure*}
\centering
\includegraphics[width=0.93\textwidth]{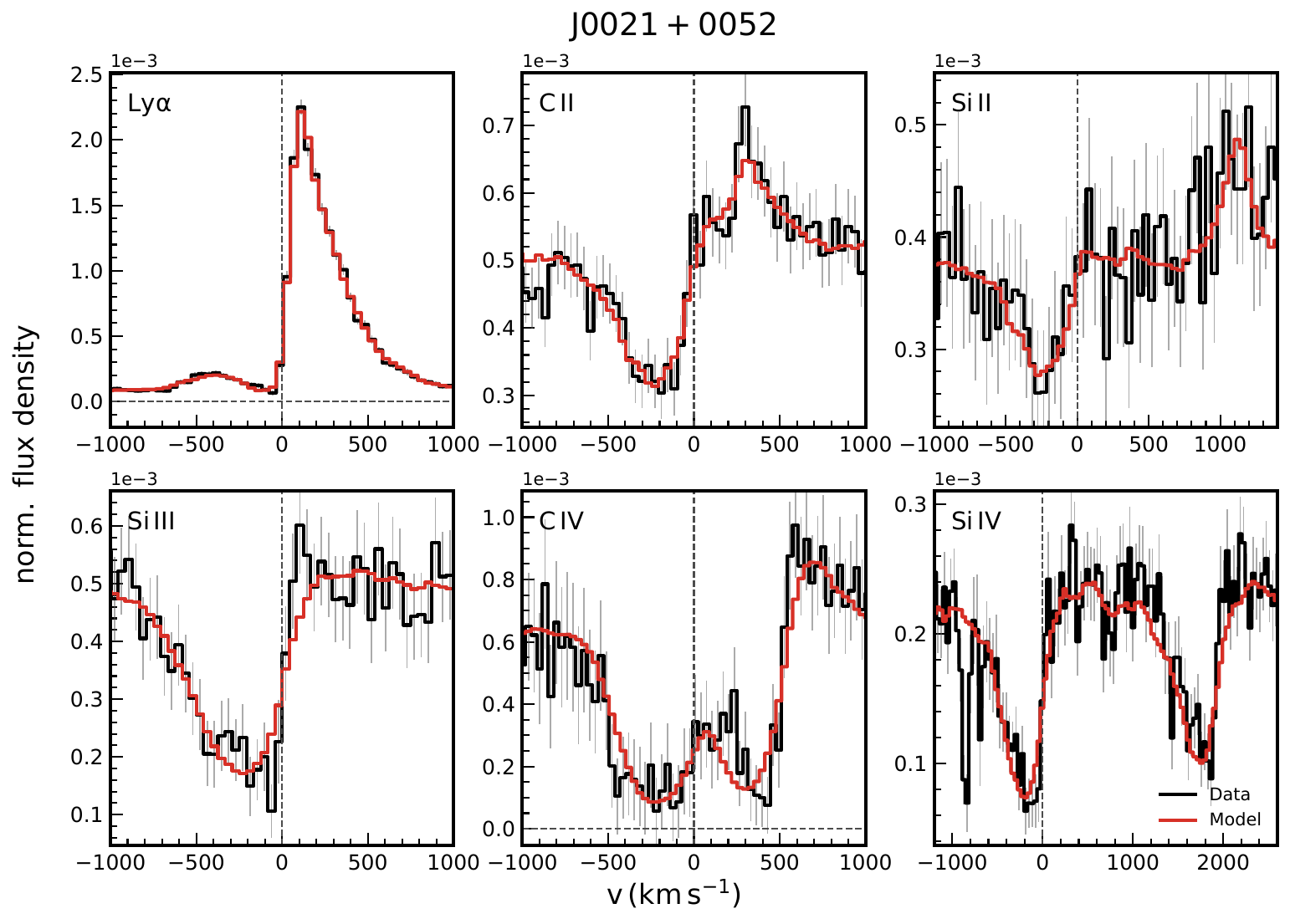}\\
\includegraphics[width=0.93\textwidth]{J0036-3333_new_joint.pdf}
    \caption{\textbf{Best-fit RT models (red) and the observed line profiles (black) for the galaxies in our sample.} 
    \label{fig:joint_fits1}}
\end{figure*}

\begin{figure*}
\centering
\includegraphics[width=0.9\textwidth]{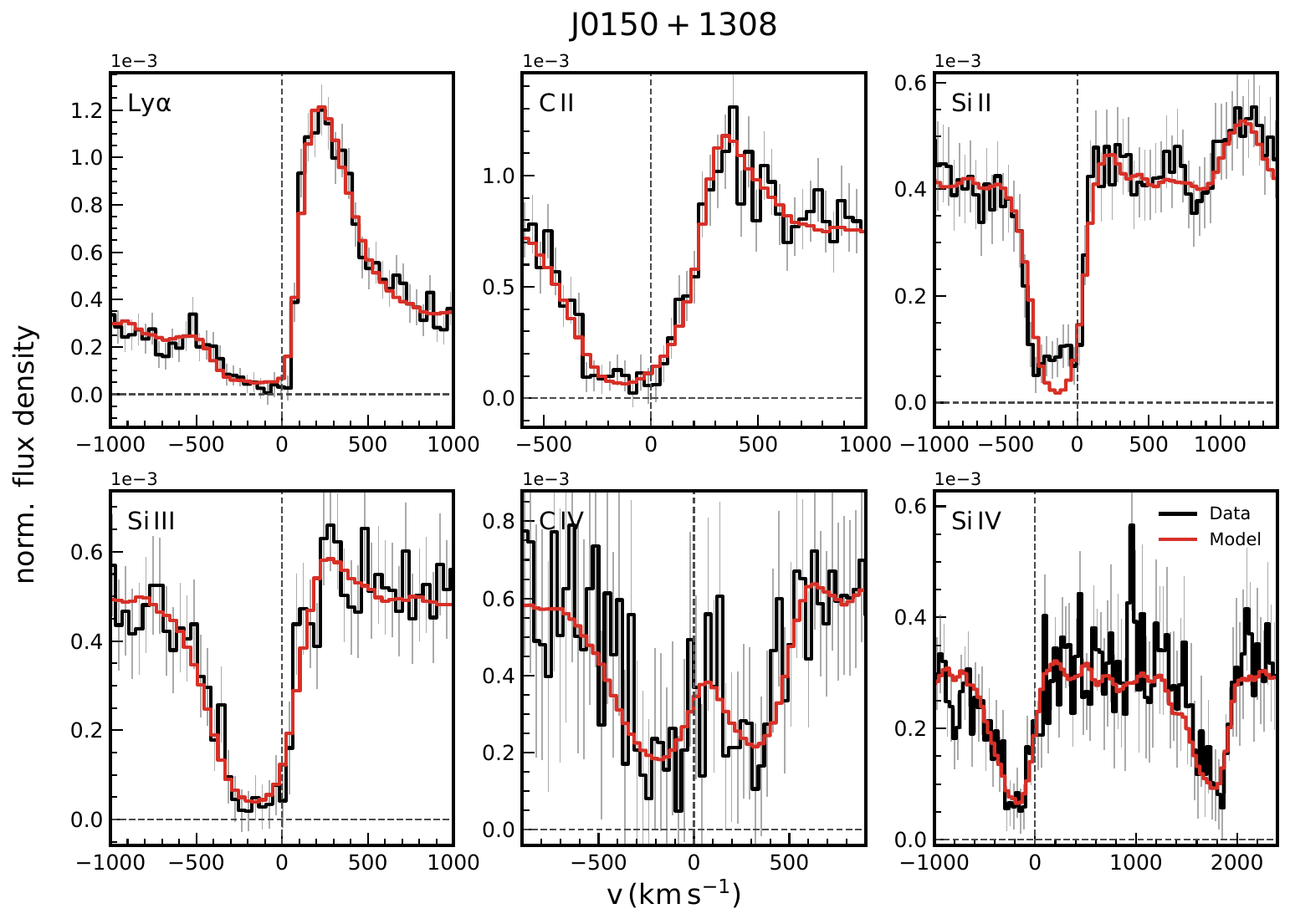}\\
\includegraphics[width=0.9\textwidth]{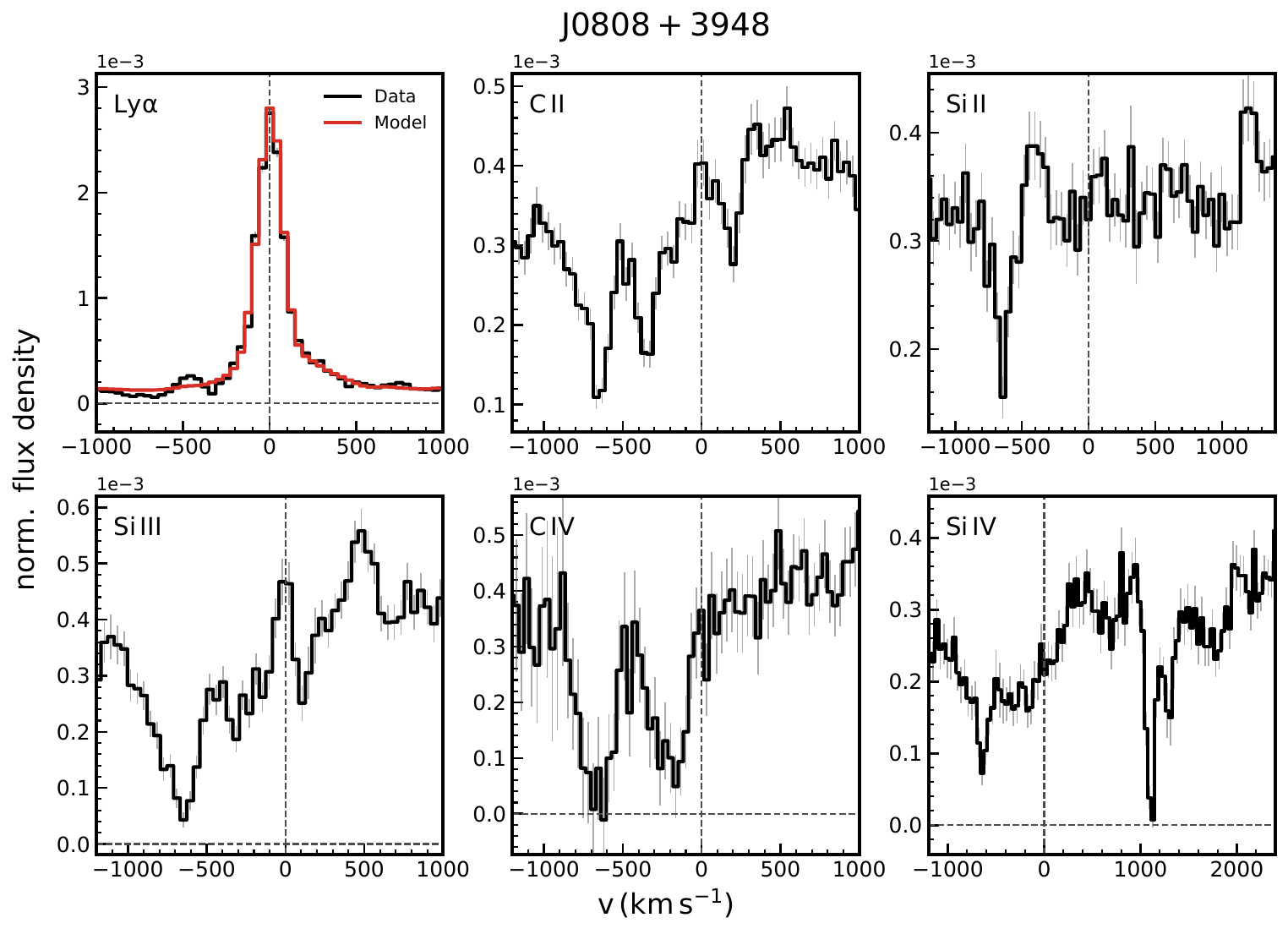}\\
    \caption{\textbf{Best-fit RT models (red) and the observed line profiles (black) for the galaxies in our sample.}  
    \label{fig:joint_fits2}}
\end{figure*}

\begin{figure*}
\centering
\includegraphics[width=0.9\textwidth]{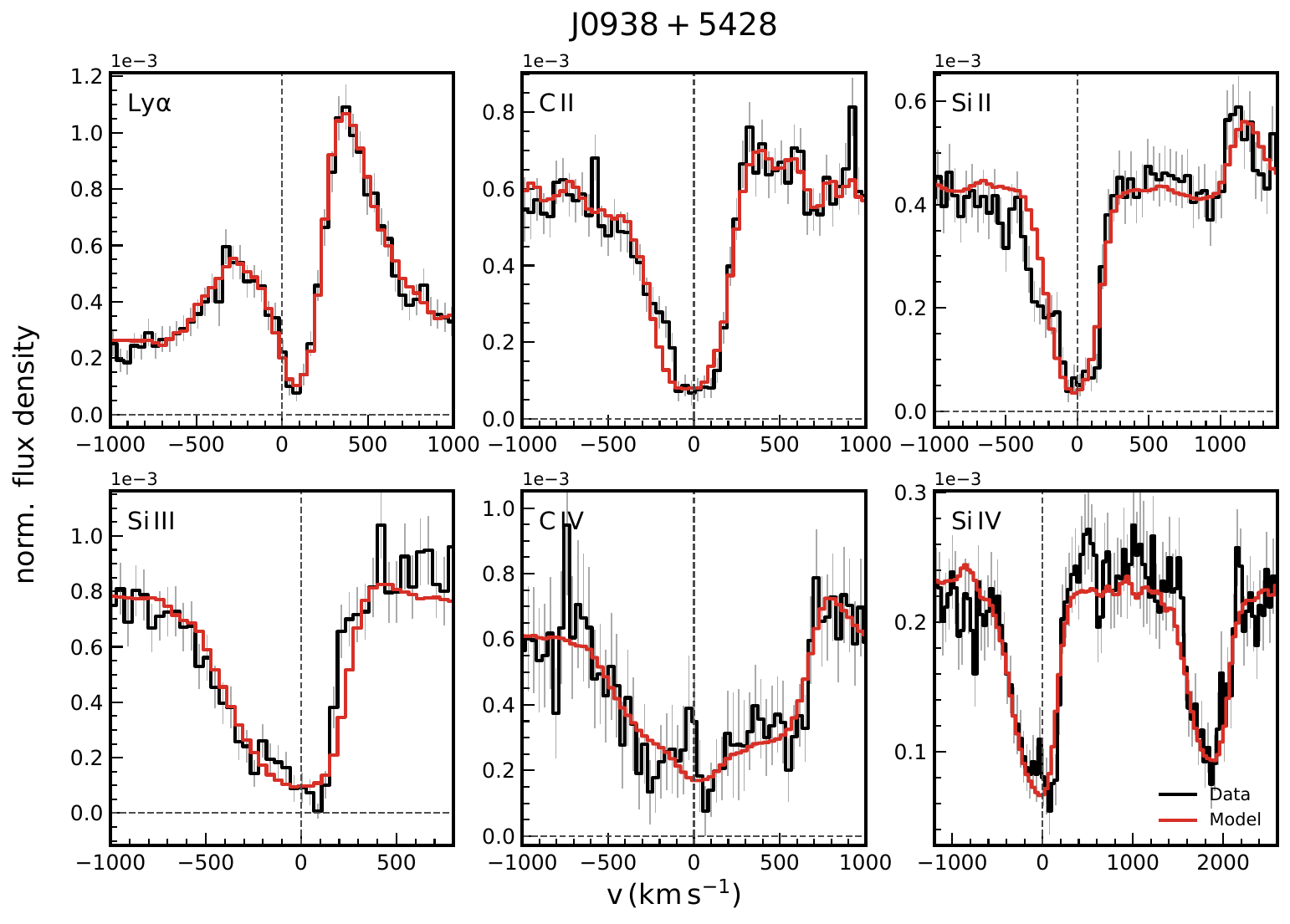}
\includegraphics[width=0.9\textwidth]{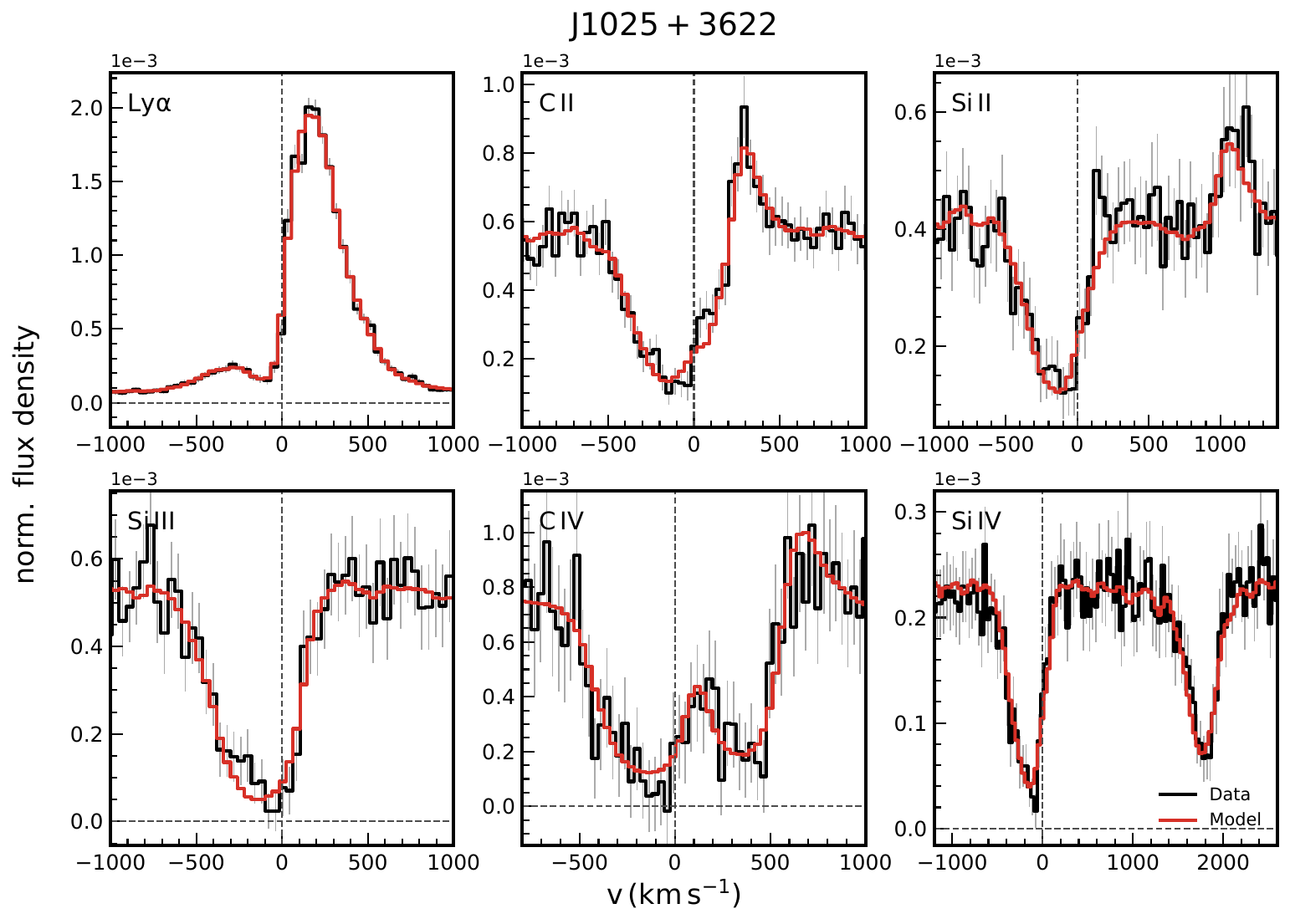}\\
    \caption{\textbf{Best-fit RT models (red) and the observed line profiles (black) for the galaxies in our sample.} 
    \label{fig:joint_fits3}}
\end{figure*}

\begin{figure*}
\centering
\includegraphics[width=0.9\textwidth]{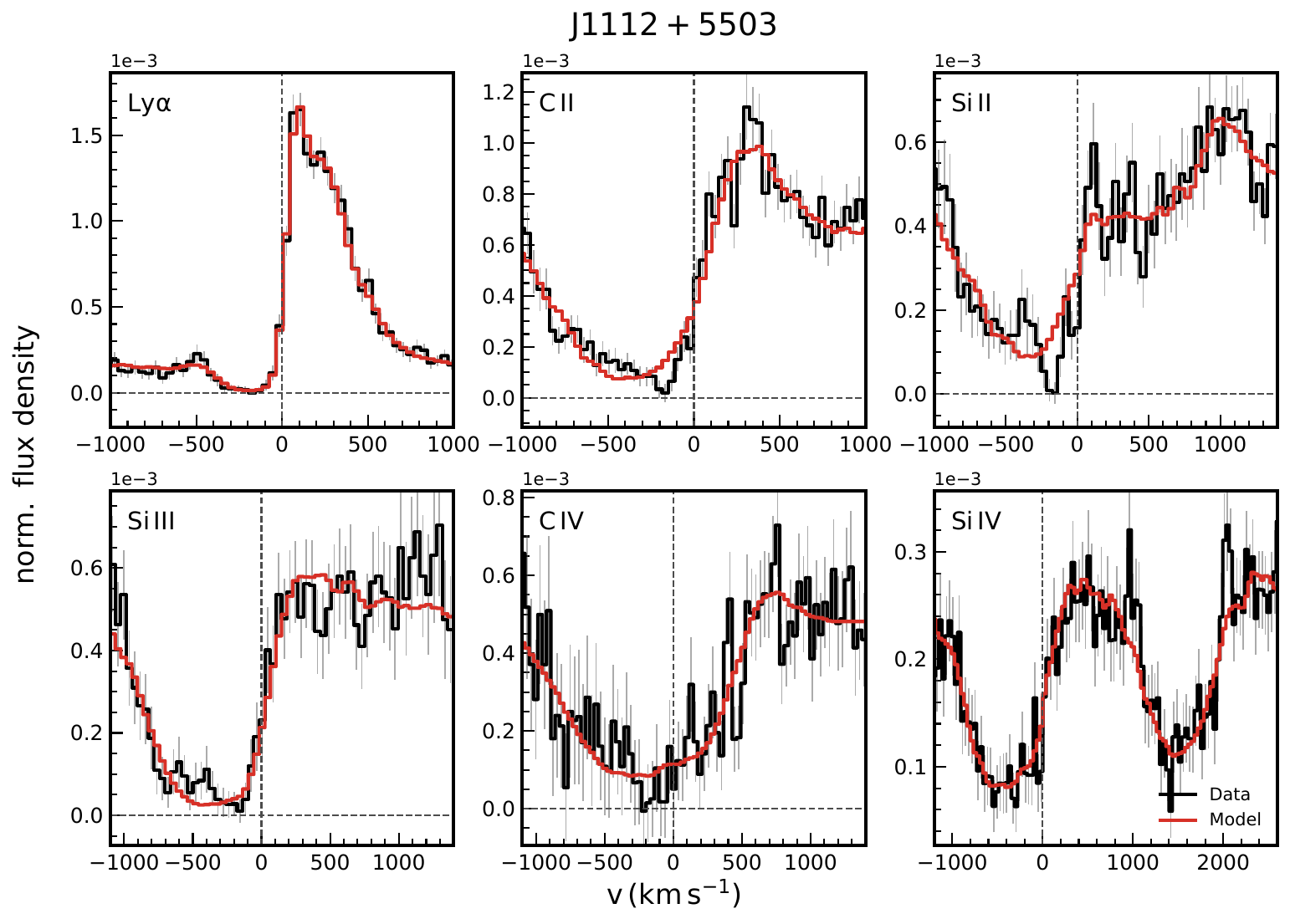}\\
\includegraphics[width=0.9\textwidth]{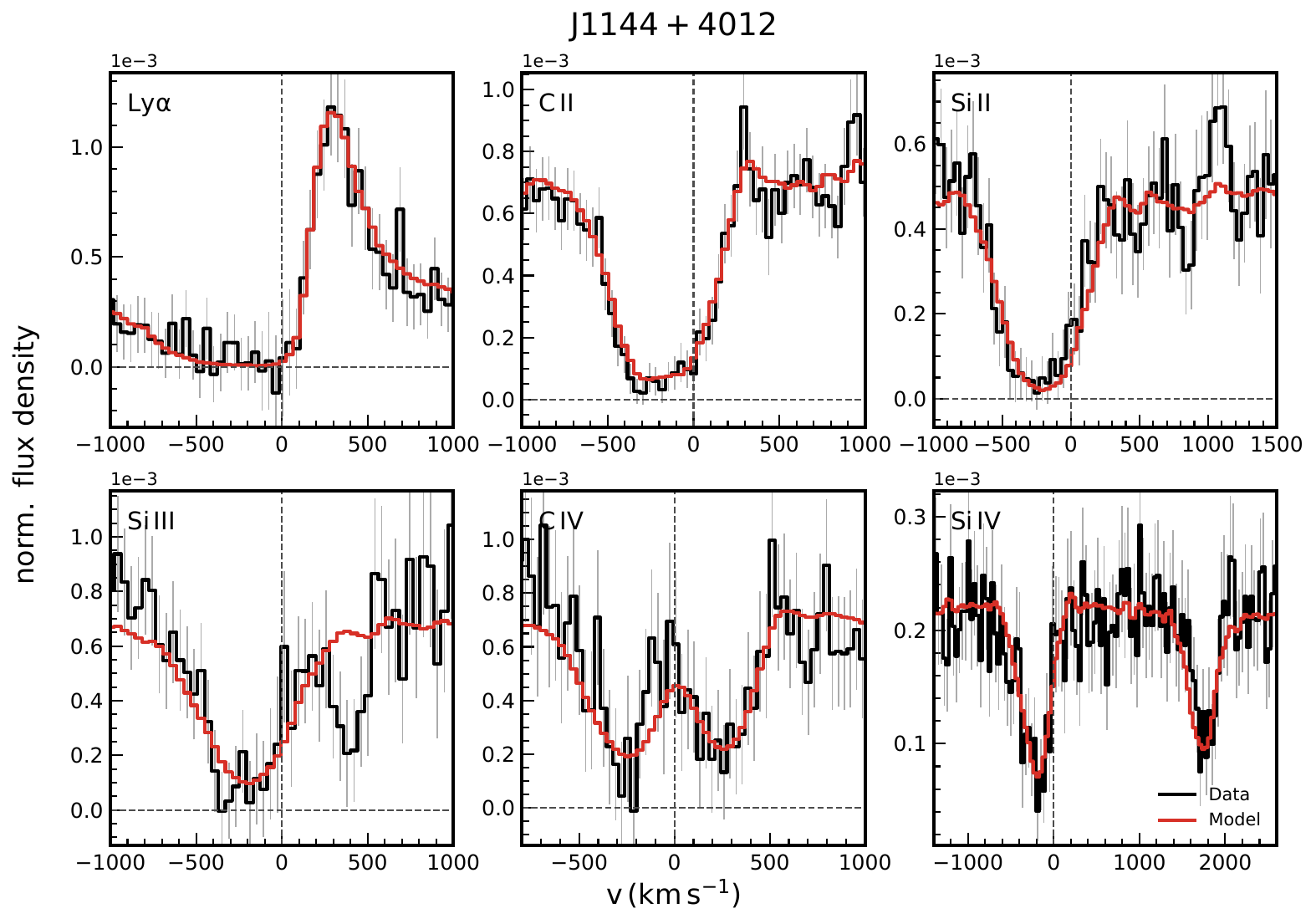}\\
    \caption{\textbf{Best-fit RT models (red) and the observed line profiles (black) for the galaxies in our sample.} 
    \label{fig:joint_fits4}}
\end{figure*}

\begin{figure*}
\centering
\includegraphics[width=0.9\textwidth]{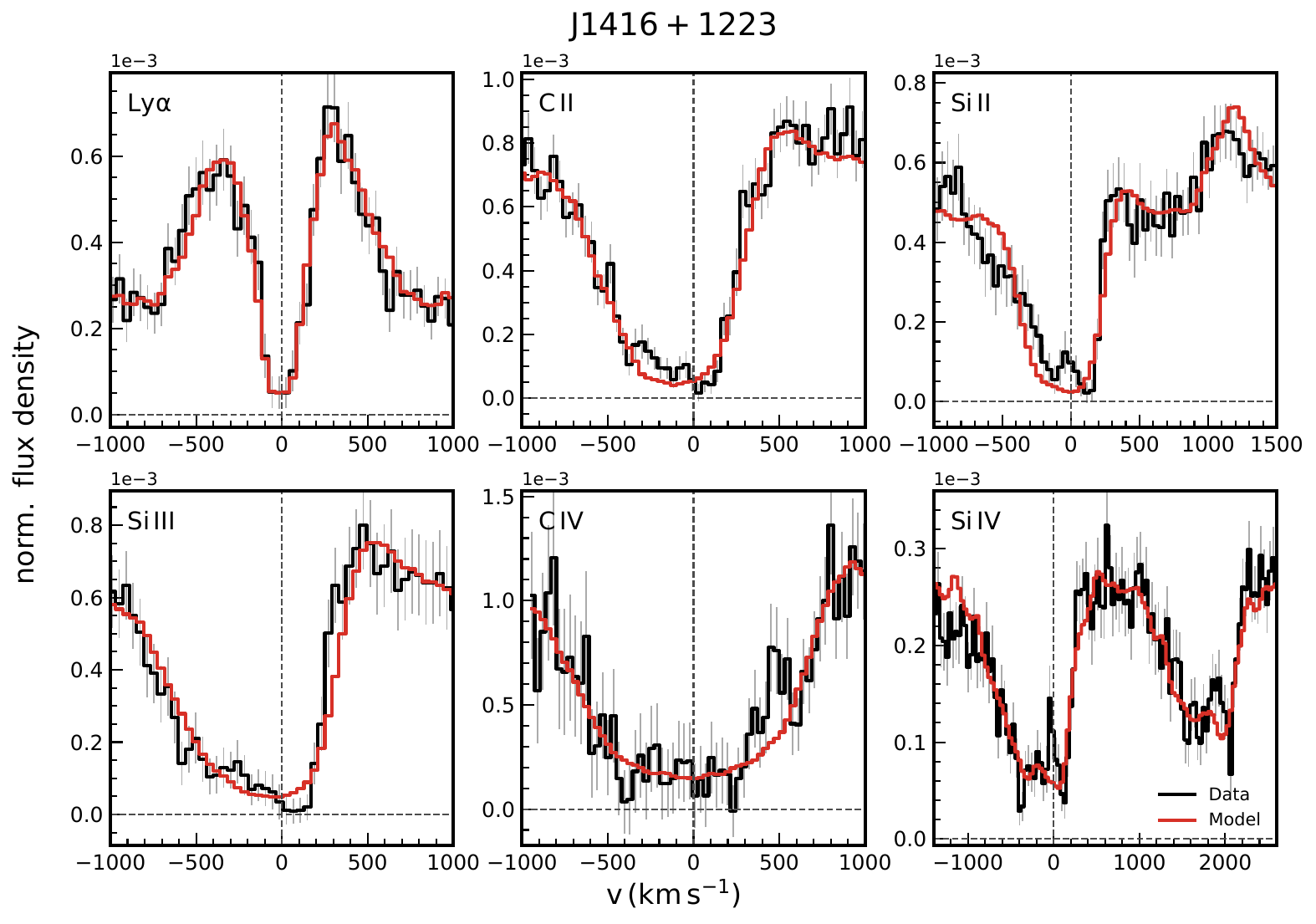}\\
\includegraphics[width=0.9\textwidth]{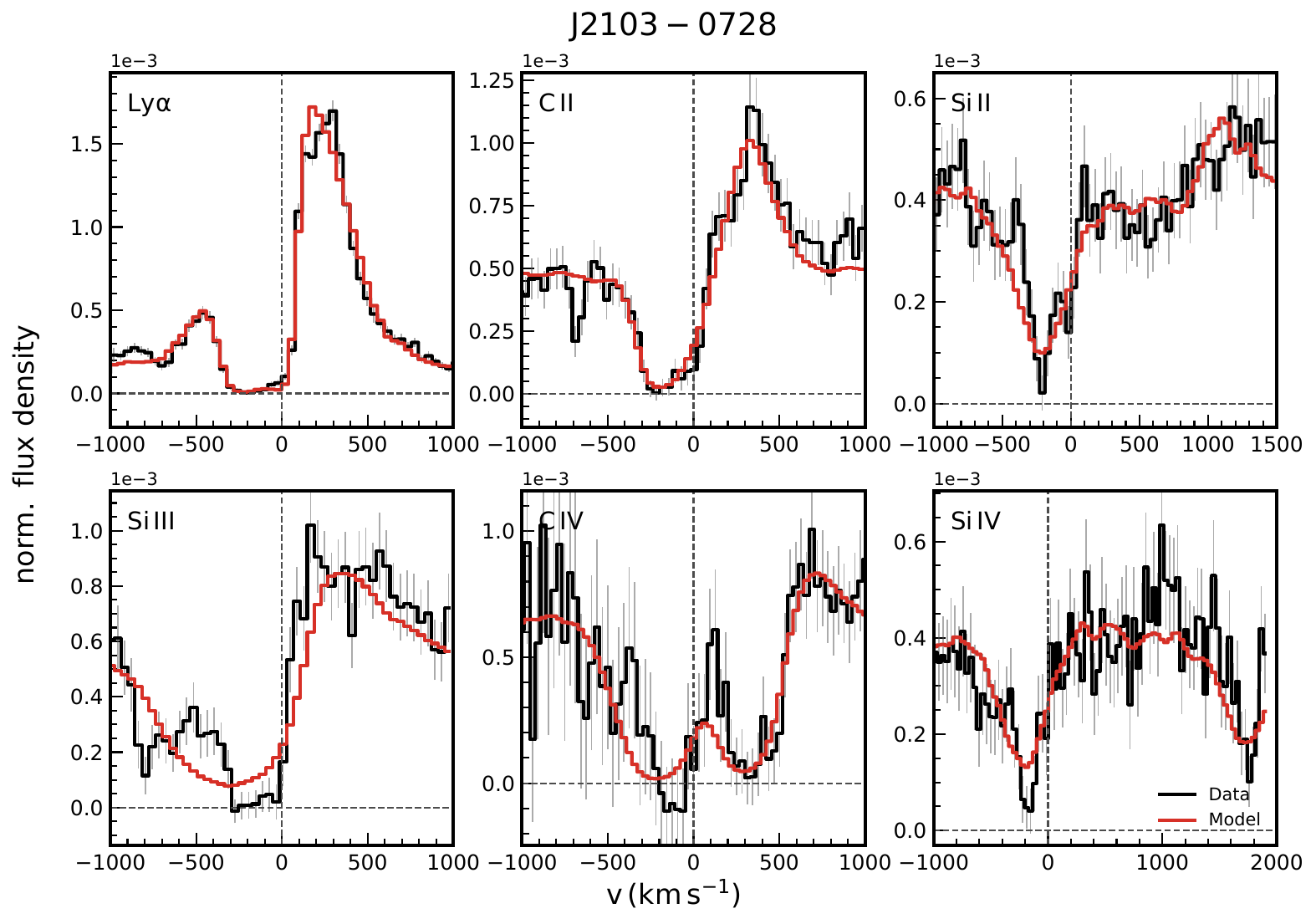}\\
    \caption{\textbf{Best-fit RT models (red) and the observed line profiles (black) for the galaxies in our sample.} 
    \label{fig:joint_fits5}}
\end{figure*}

\begin{figure*}
\centering
\includegraphics[width=0.9\textwidth]{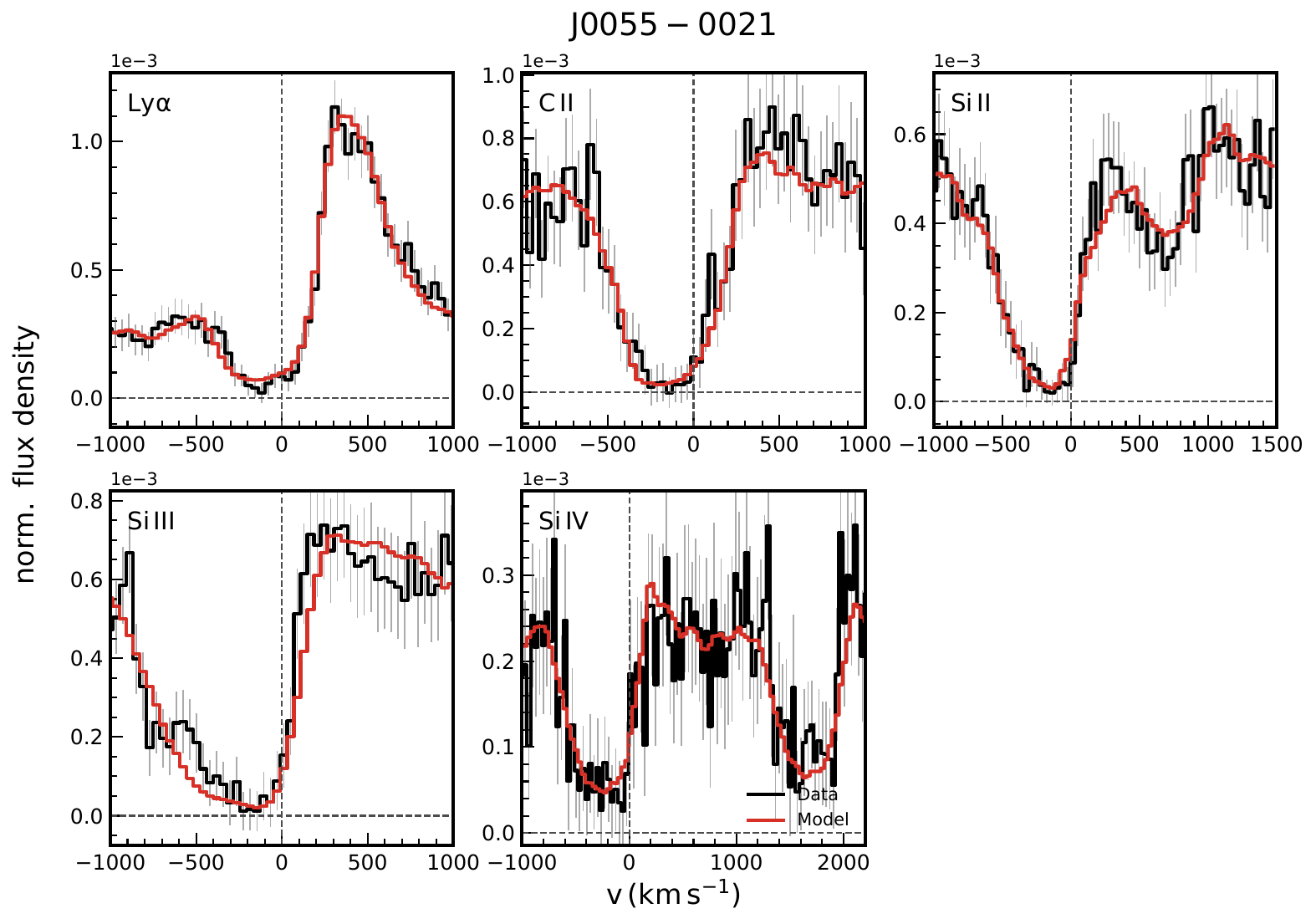}\\
\includegraphics[width=0.9\textwidth]{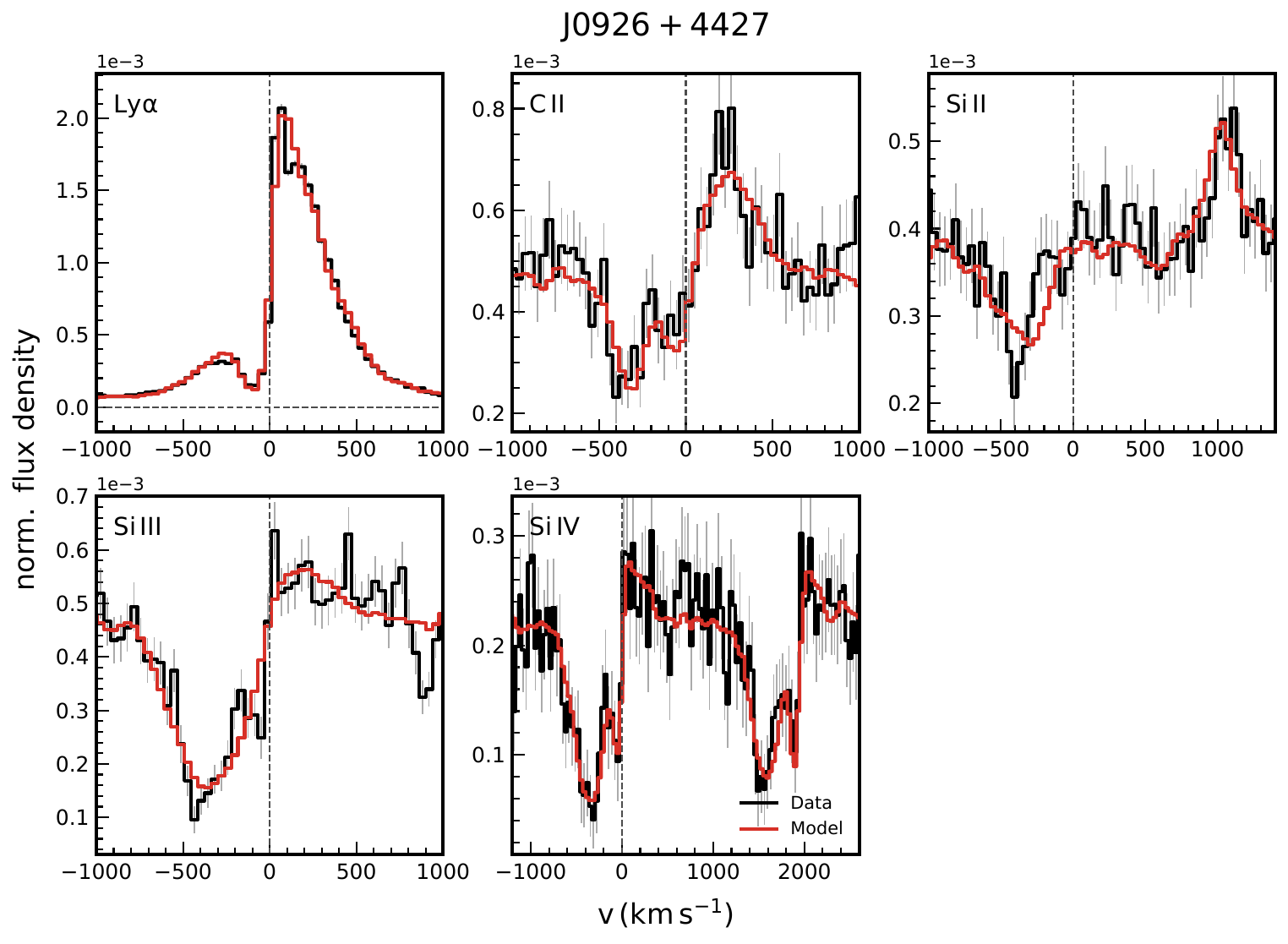}\\
    \caption{\textbf{Best-fit RT models (red) and the observed line profiles (black) for the galaxies in our sample.} 
    \label{fig:joint_fits6}}
\end{figure*}

\begin{figure*}
\centering
\includegraphics[width=0.9\textwidth]{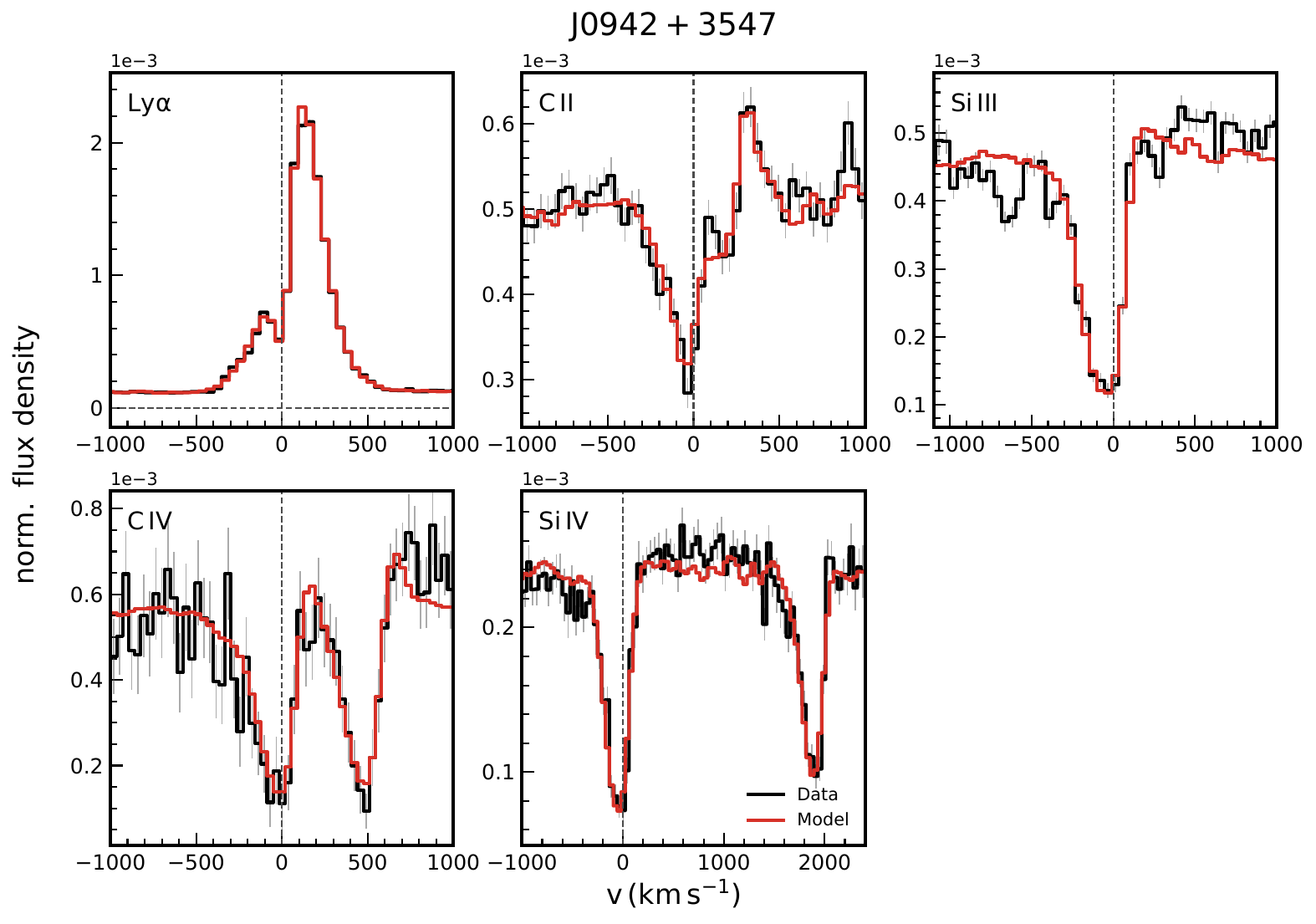}\\
\includegraphics[width=0.9\textwidth]{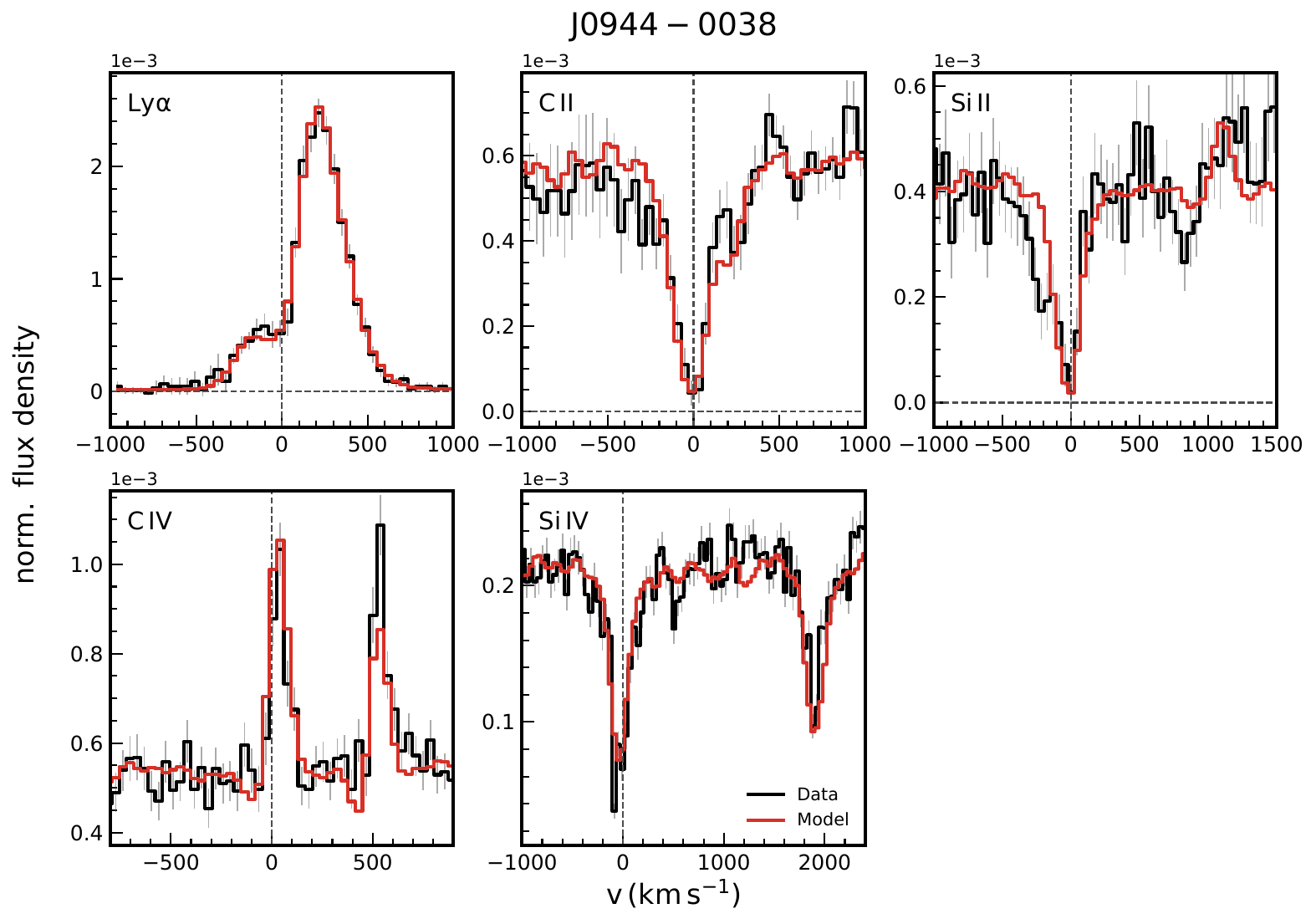}\\
    \caption{\textbf{Best-fit RT models (red) and the observed line profiles (black) for the galaxies in our sample.} 
    \label{fig:joint_fits7}}
\end{figure*}

\begin{figure*}
\centering
\includegraphics[width=0.9\textwidth]{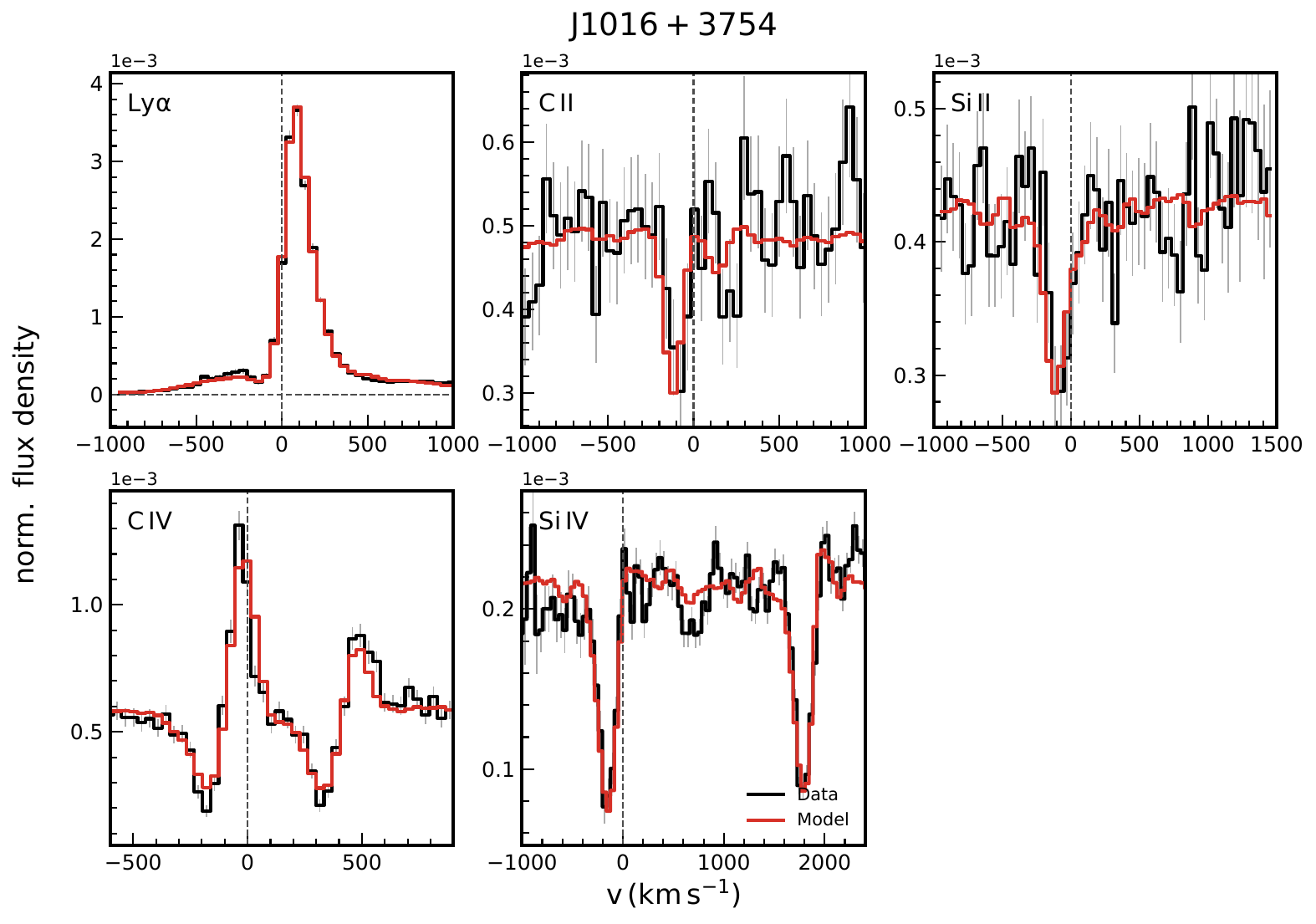}\\
\includegraphics[width=0.9\textwidth]{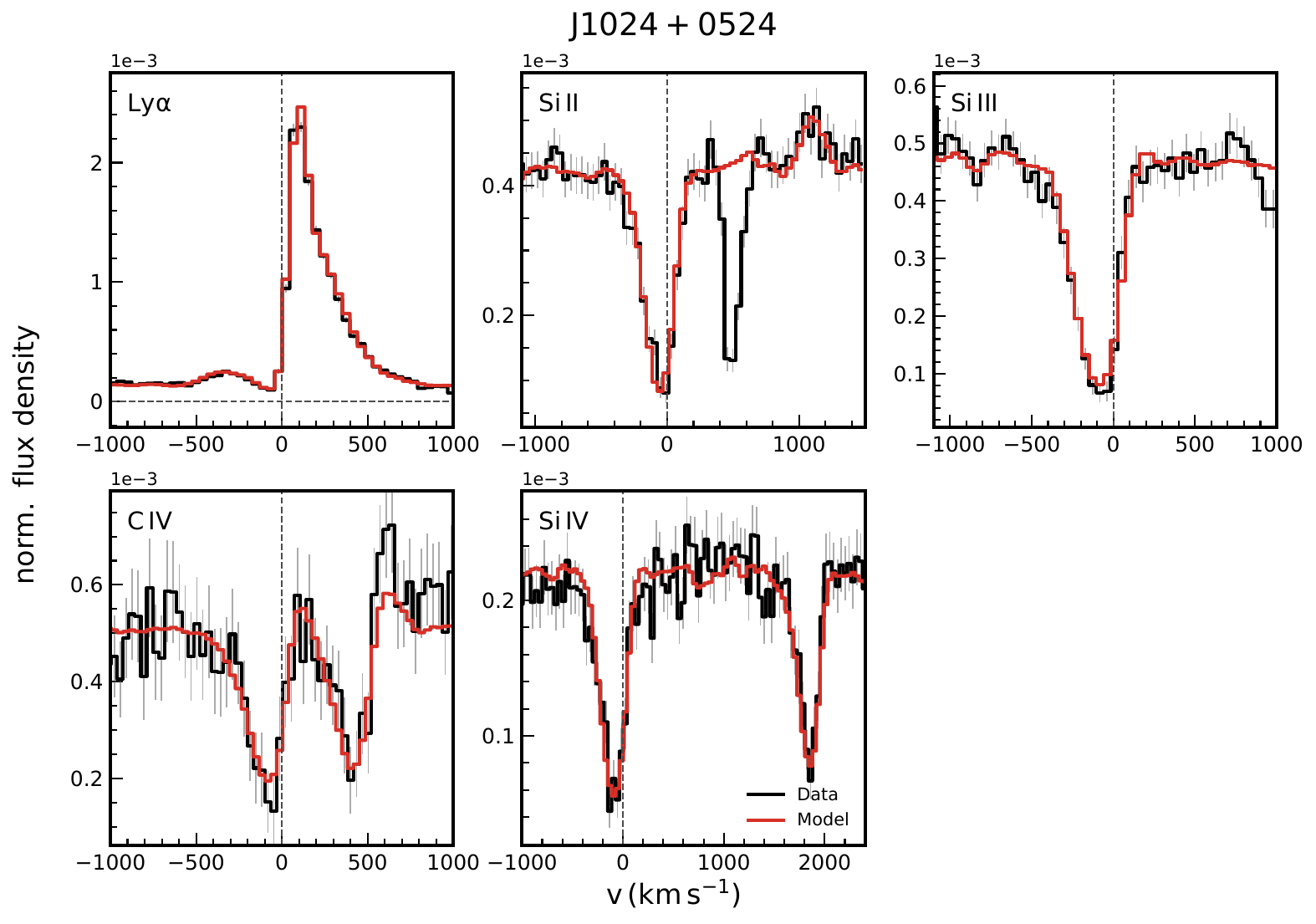}\\
    \caption{\textbf{Best-fit RT models (red) and the observed line profiles (black) for the galaxies in our sample.} 
    \label{fig:joint_fits8}}
\end{figure*}

\begin{figure*}
\centering
\includegraphics[width=0.9\textwidth]{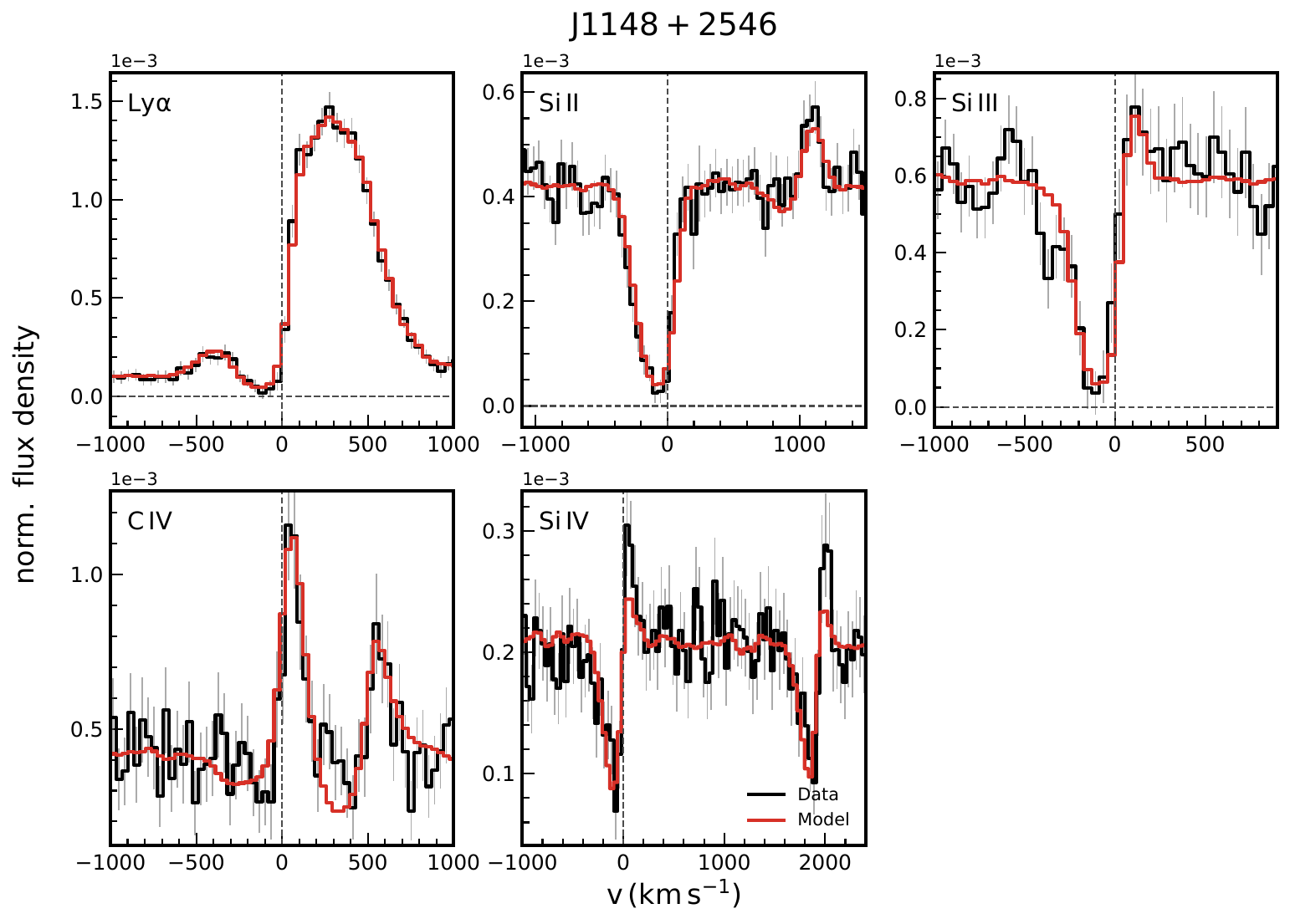}\\
\includegraphics[width=0.9\textwidth]{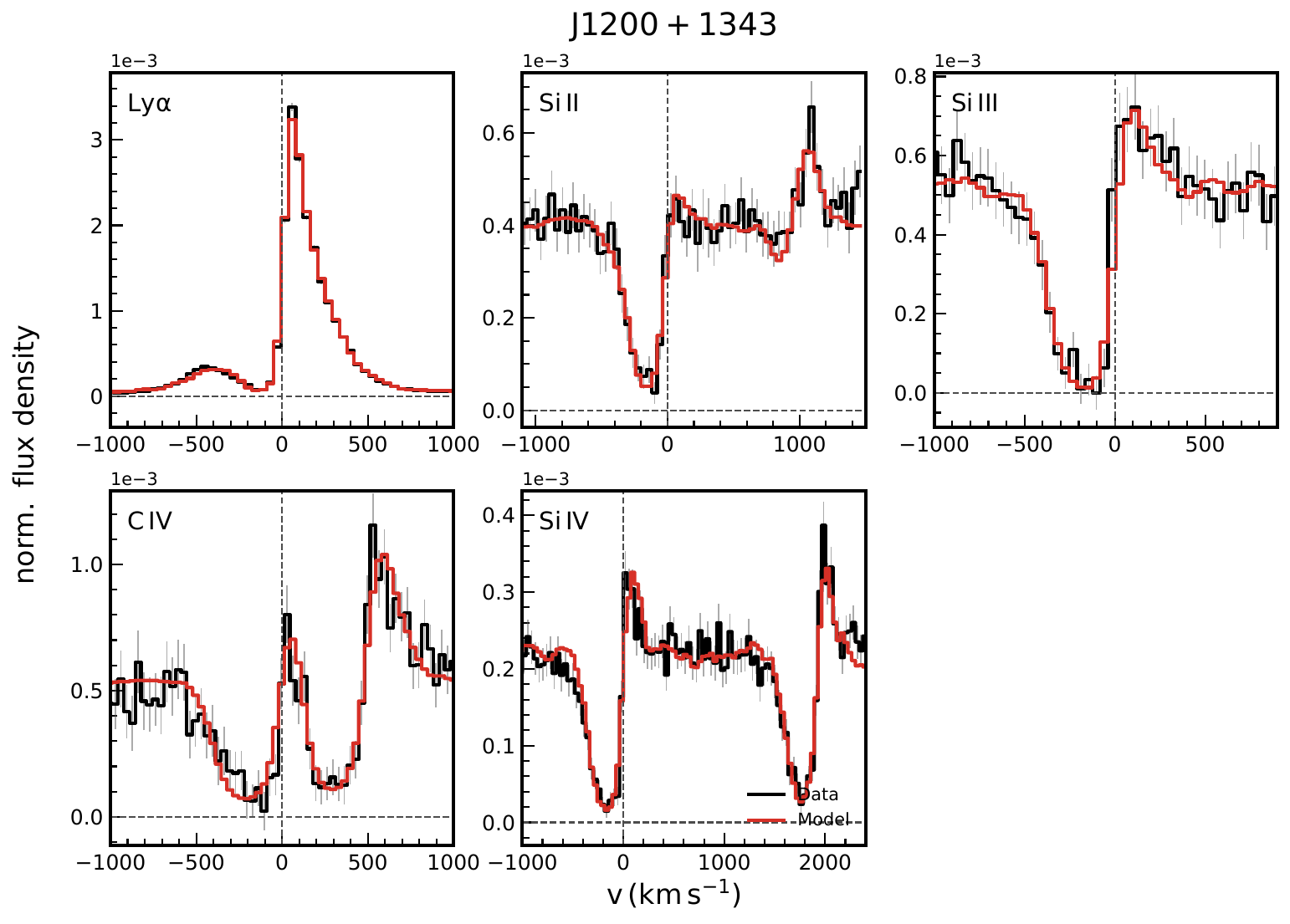}\\
    \caption{\textbf{Best-fit RT models (red) and the observed line profiles (black) for the galaxies in our sample.} 
    \label{fig:joint_fits9}}
\end{figure*}

\begin{figure*}
\centering
\includegraphics[width=0.9\textwidth]{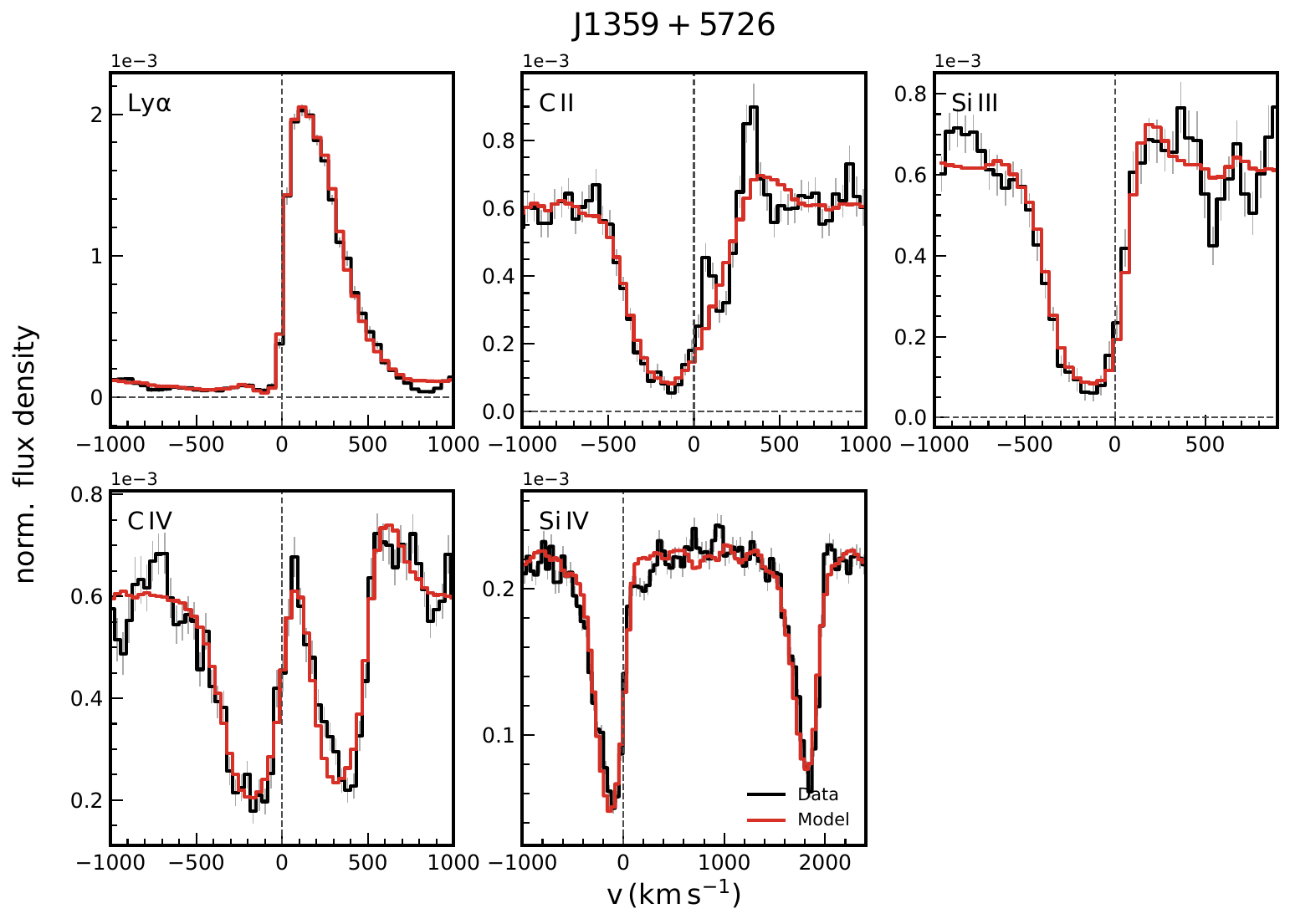}\\
\includegraphics[width=0.9\textwidth]{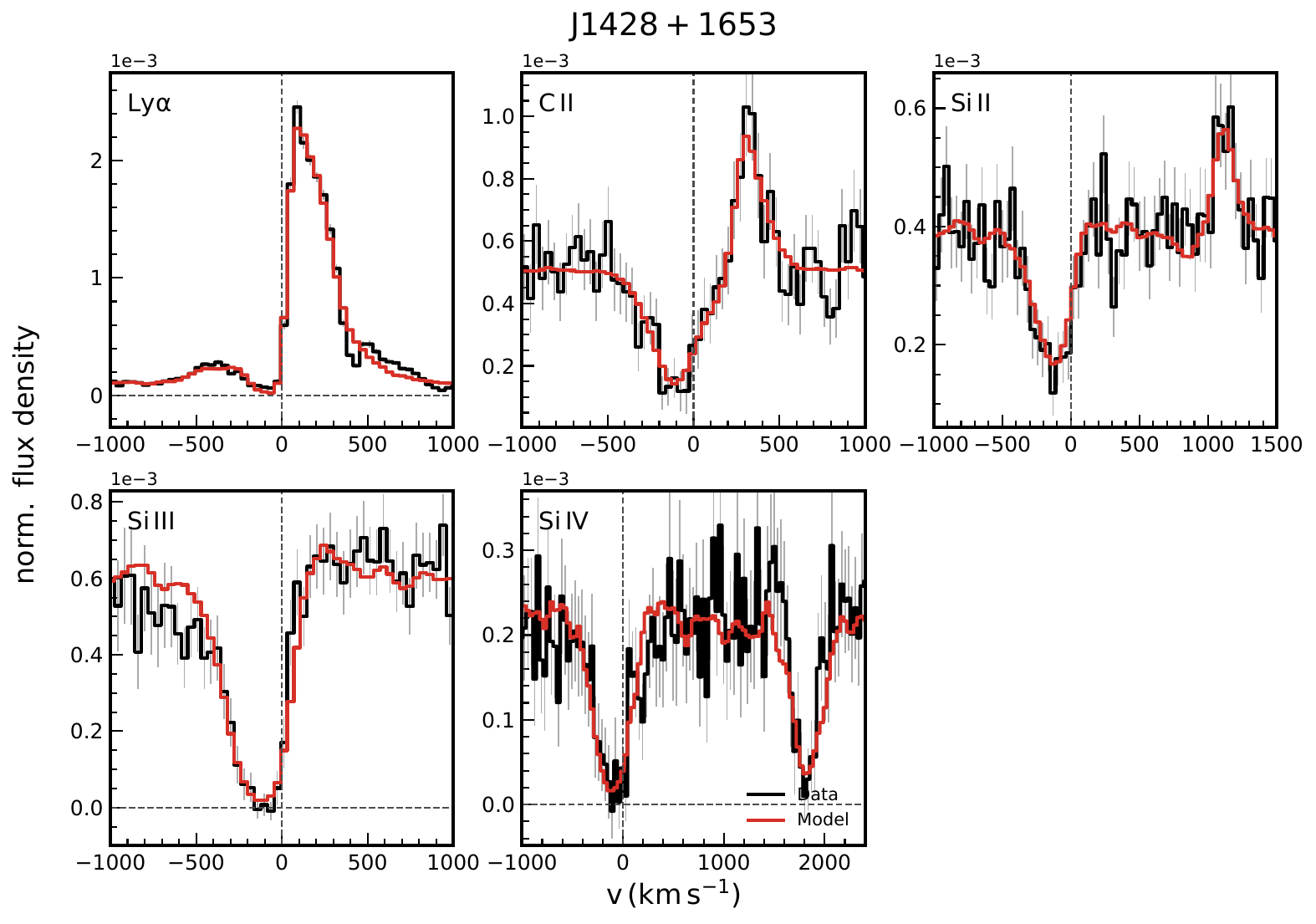}\\
    \caption{\textbf{Best-fit RT models (red) and the observed line profiles (black) for the galaxies in our sample.}  
    \label{fig:joint_fits10}}
\end{figure*}

\begin{figure*}
\centering
\includegraphics[width=0.9\textwidth]{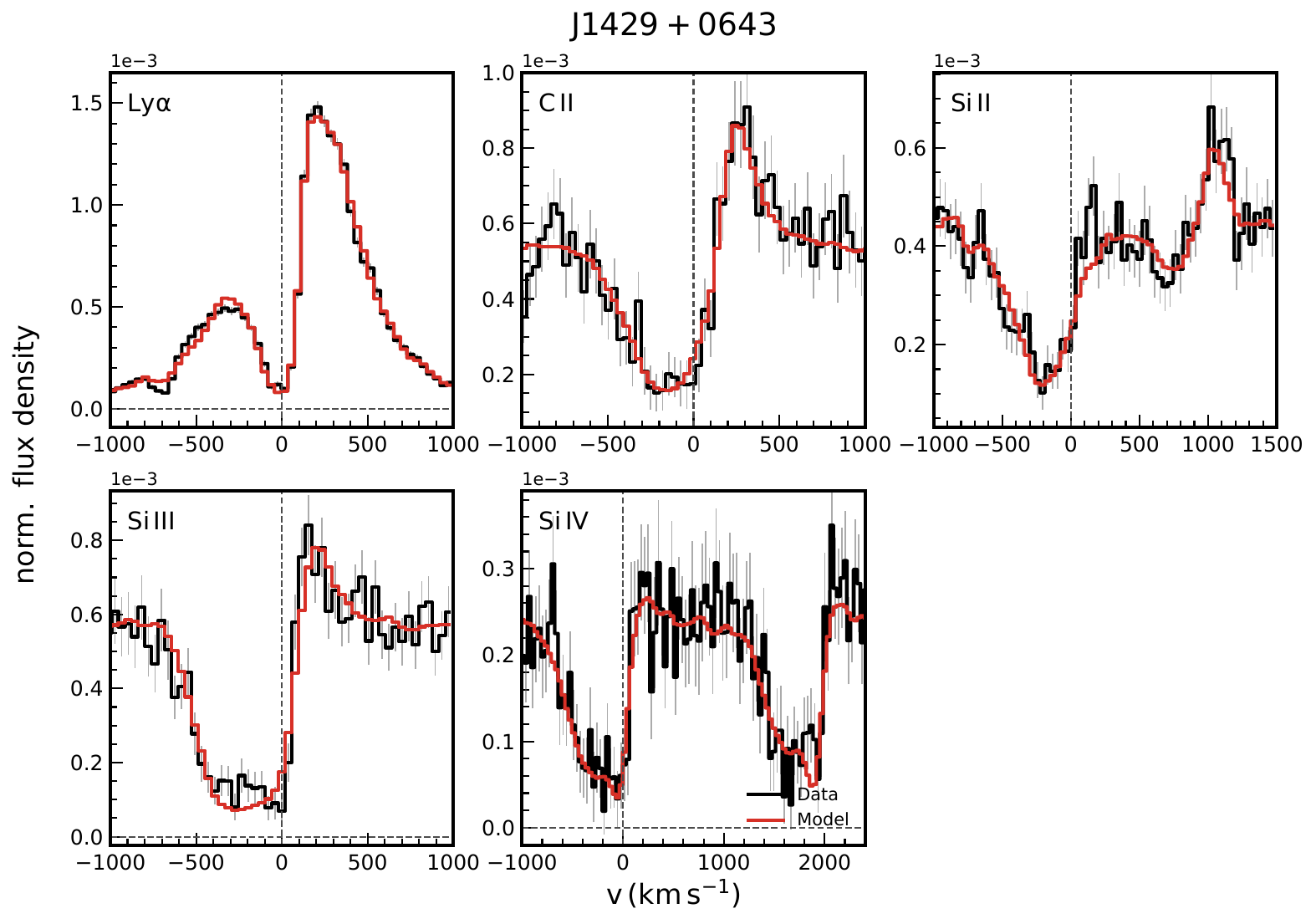}\\
\includegraphics[width=0.9\textwidth]{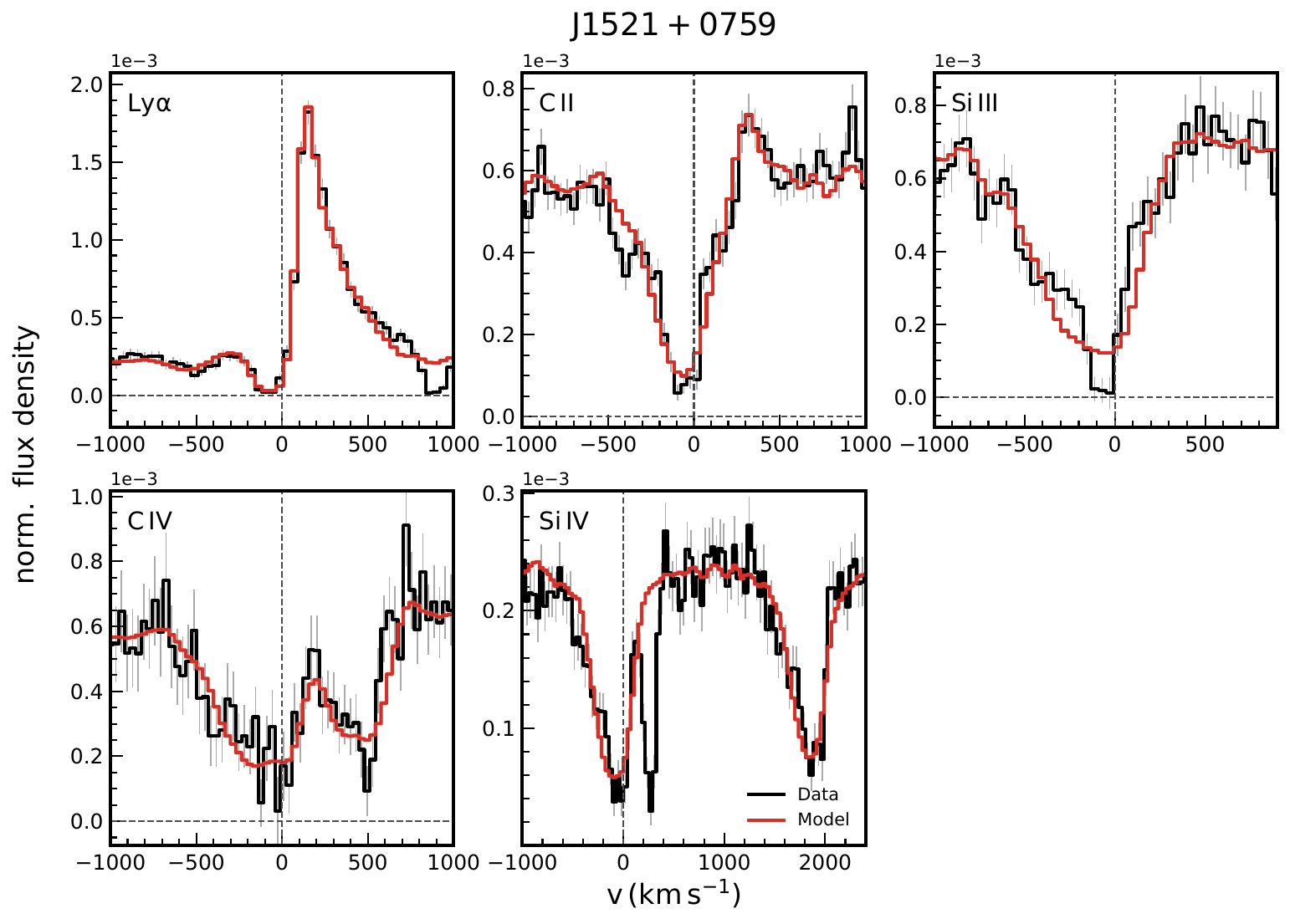}\\
    \caption{\textbf{Best-fit RT models (red) and the observed line profiles (black) for the galaxies in our sample.} 
    \label{fig:joint_fits11}}
\end{figure*}

\begin{figure*}
\centering
\includegraphics[width=0.9\textwidth]{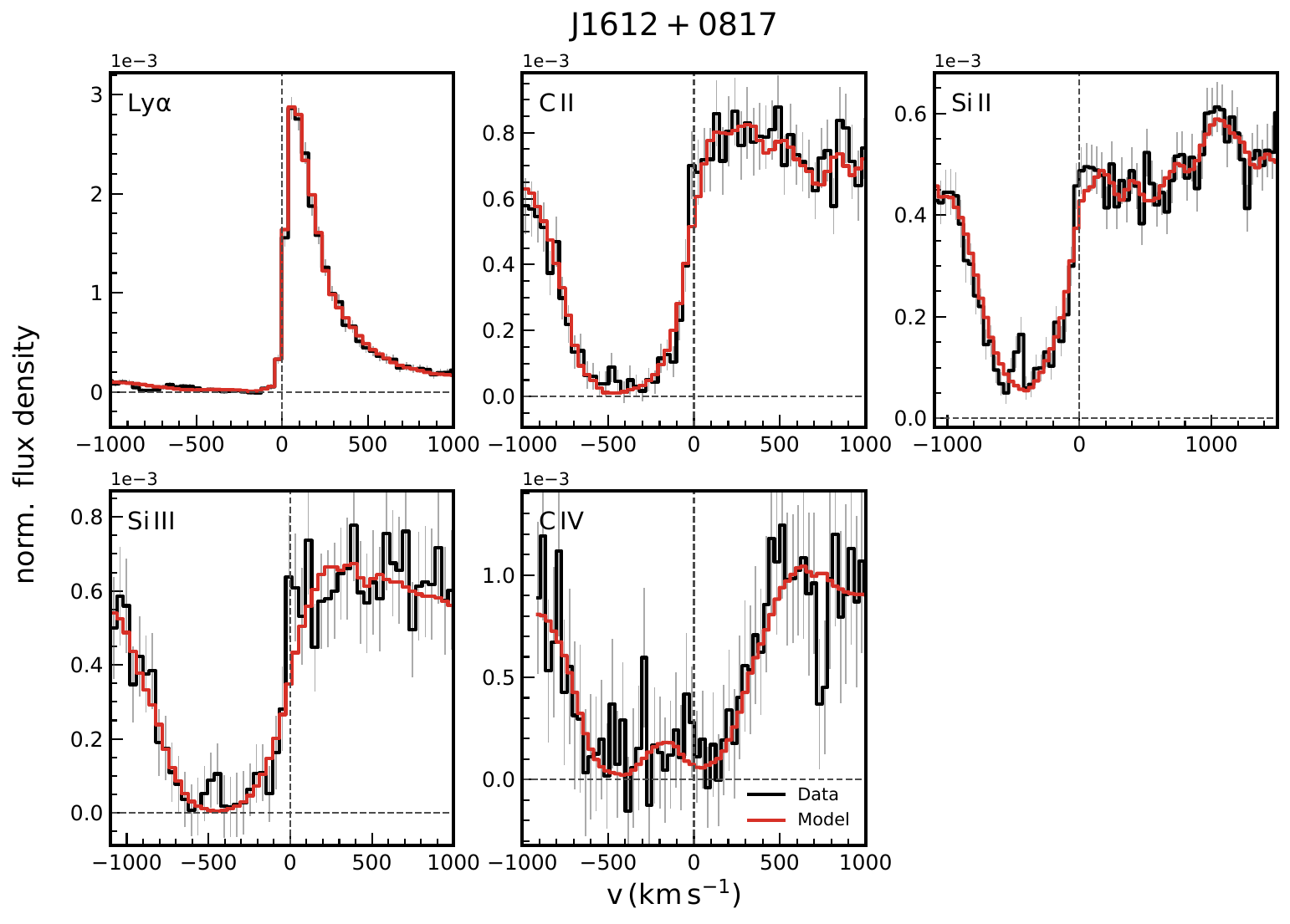}\\
\includegraphics[width=0.9\textwidth]{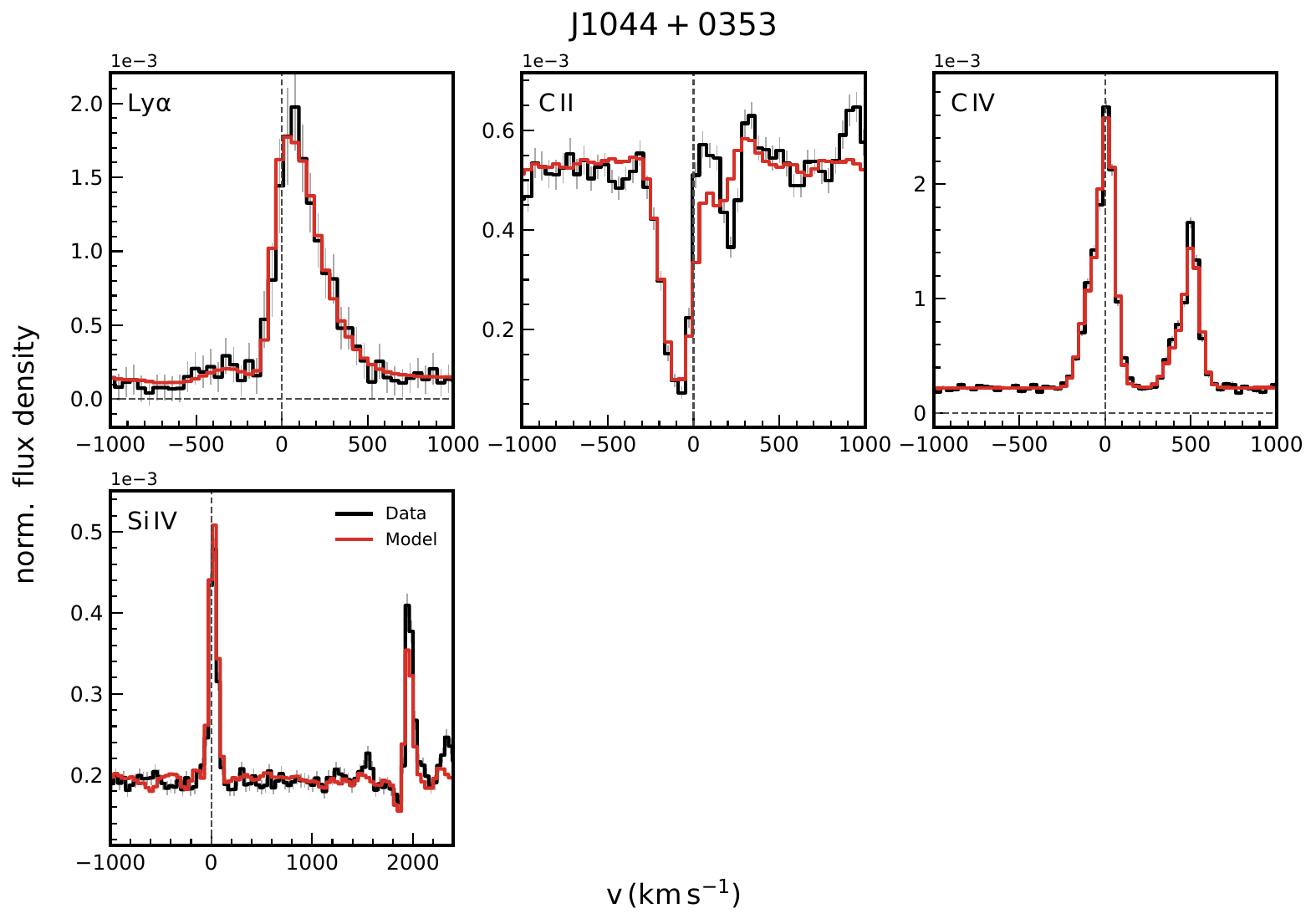}\\
    \caption{\textbf{Best-fit RT models (red) and the observed line profiles (black) for the galaxies in our sample.}  
    \label{fig:joint_fits12}}
\end{figure*}

\begin{figure*}
\centering
\includegraphics[width=0.9\textwidth]{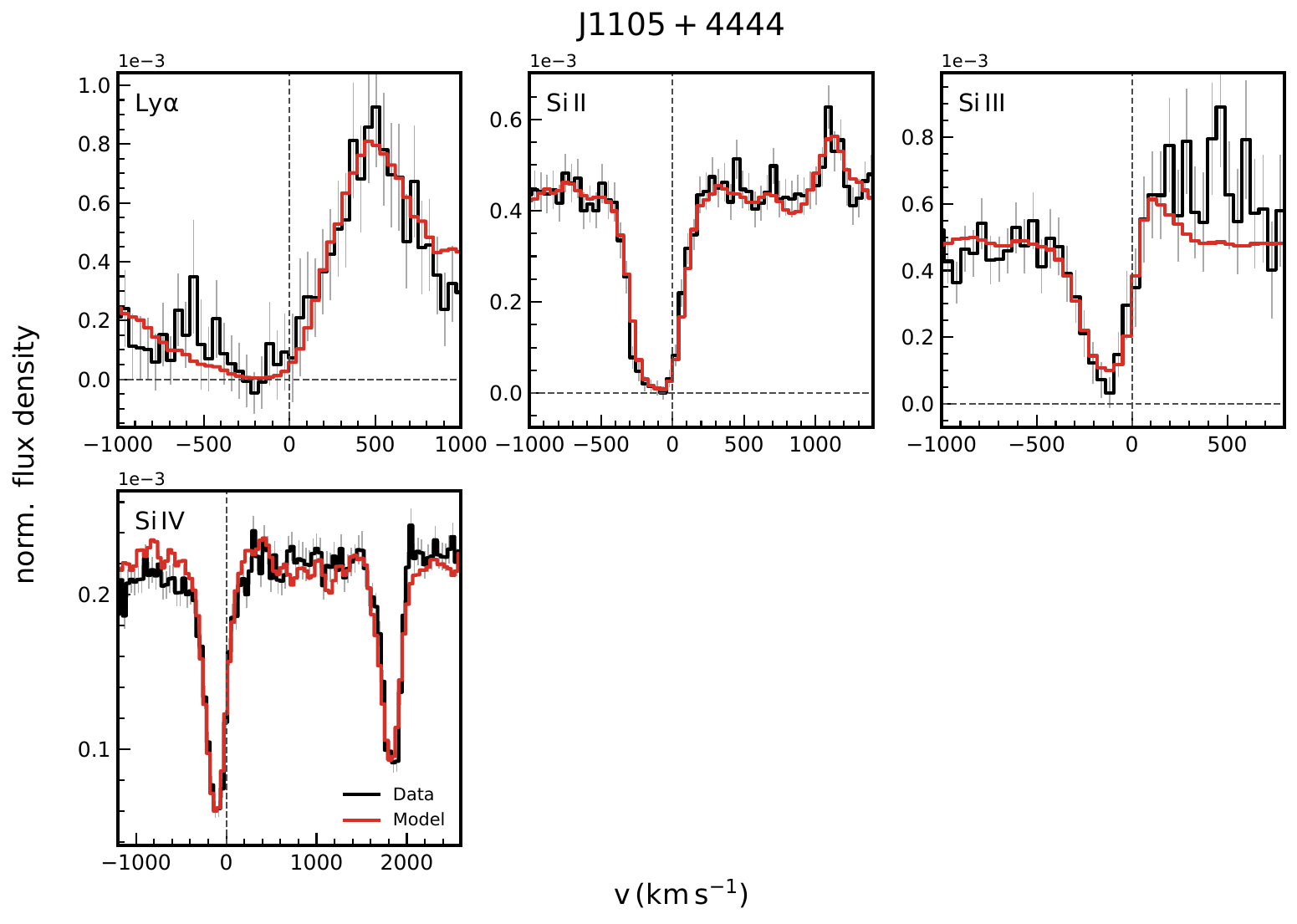}\\
\includegraphics[width=0.9\textwidth]{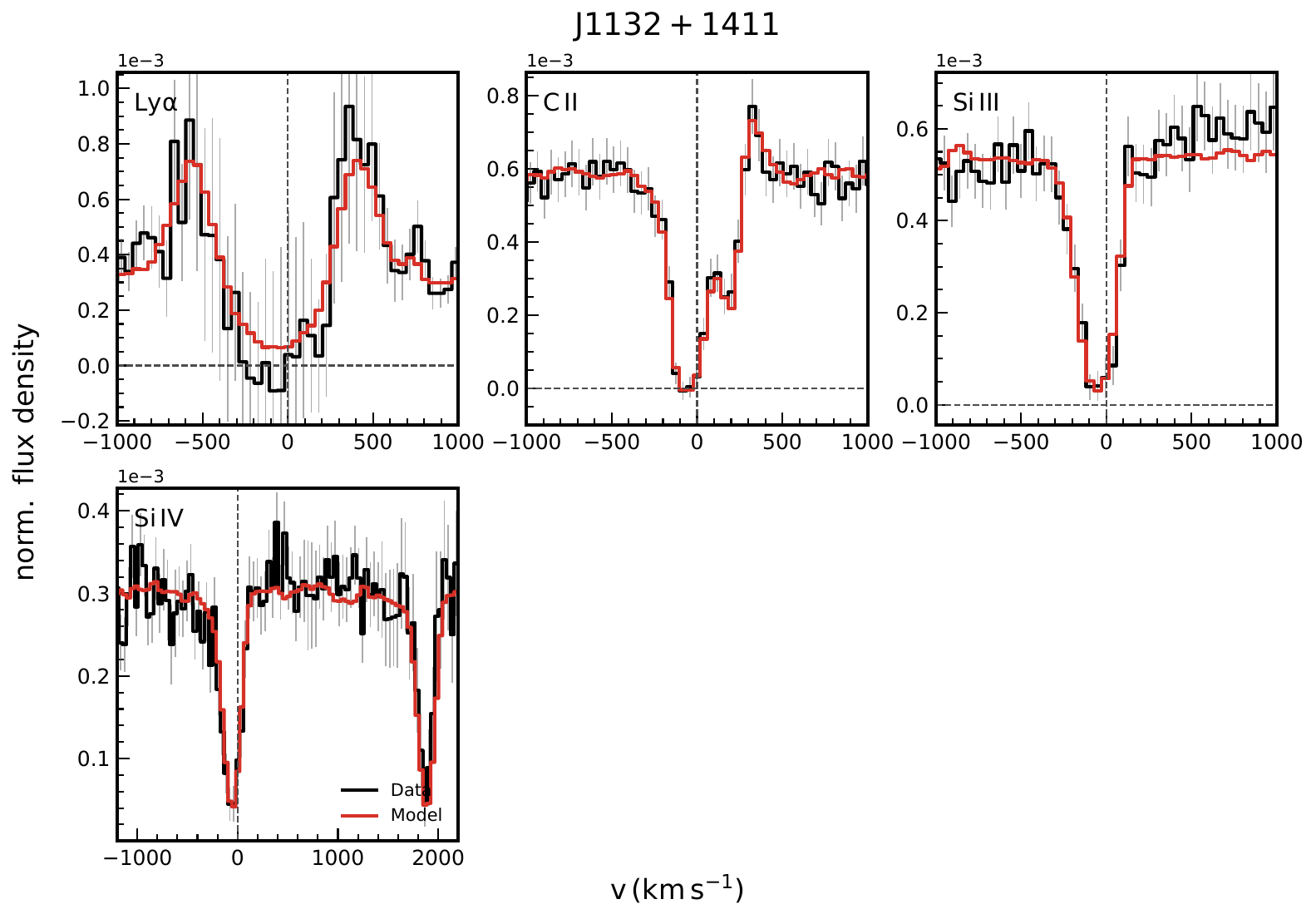}\\
    \caption{\textbf{Best-fit RT models (red) and the observed line profiles (black) for the galaxies in our sample.} 
    \label{fig:joint_fits130}}
\end{figure*}

\begin{figure*}
\centering
\includegraphics[width=0.9\textwidth]{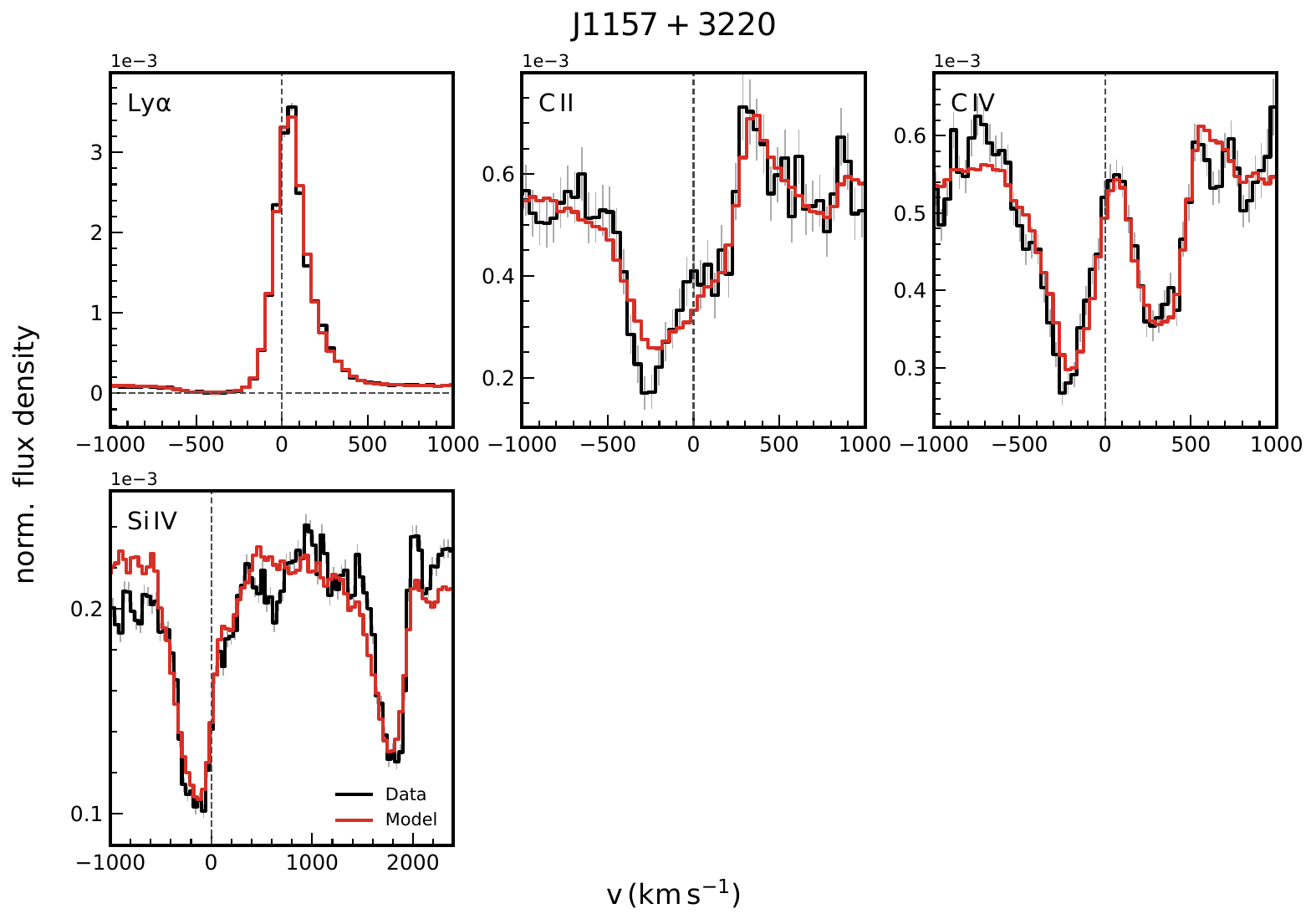}\\
\includegraphics[width=0.9\textwidth]{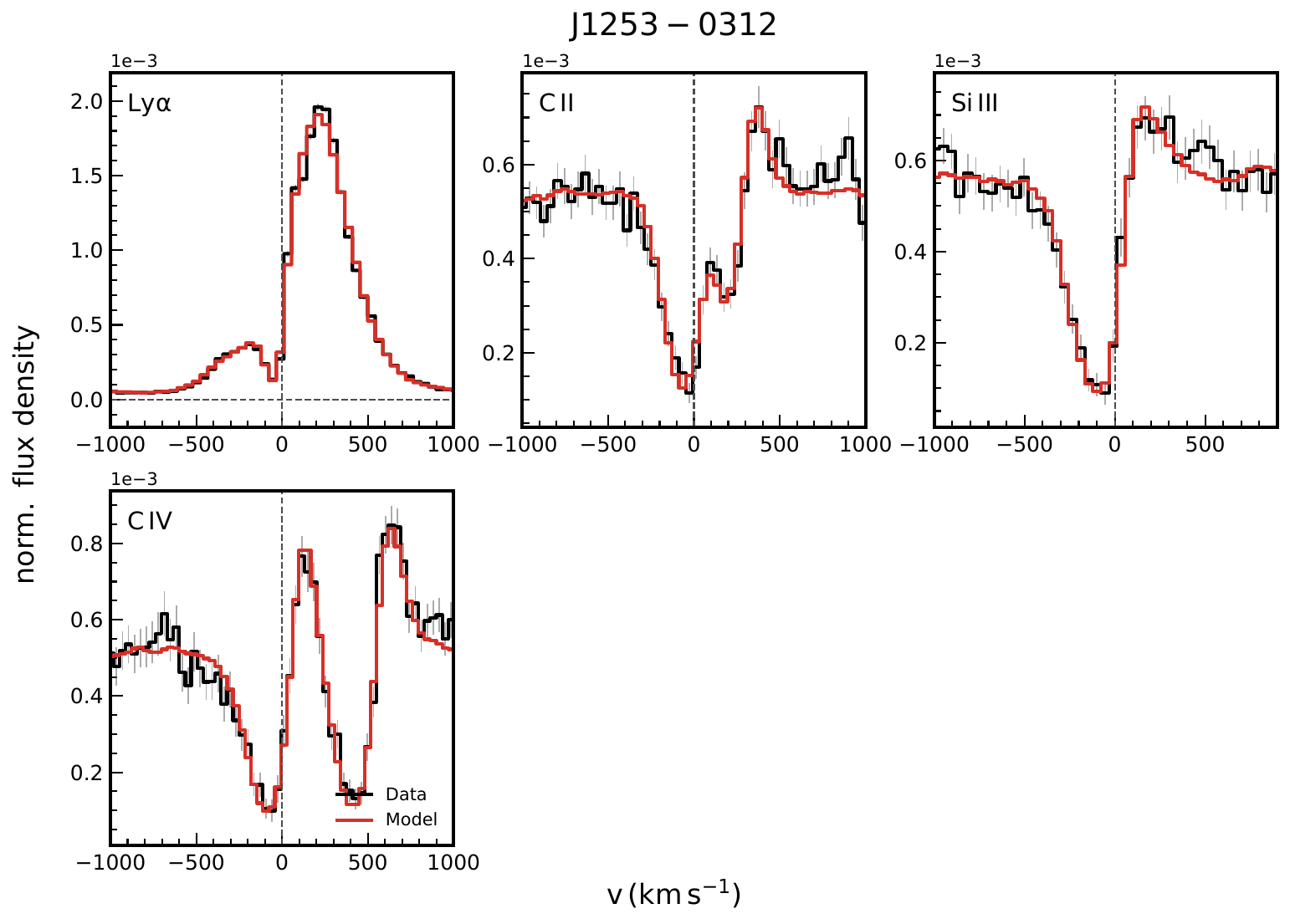}\\
    \caption{\textbf{Best-fit RT models (red) and the observed line profiles (black) for the galaxies in our sample.} 
    \label{fig:joint_fits13}}
\end{figure*}

\begin{figure*}
\centering
\includegraphics[width=0.9\textwidth]{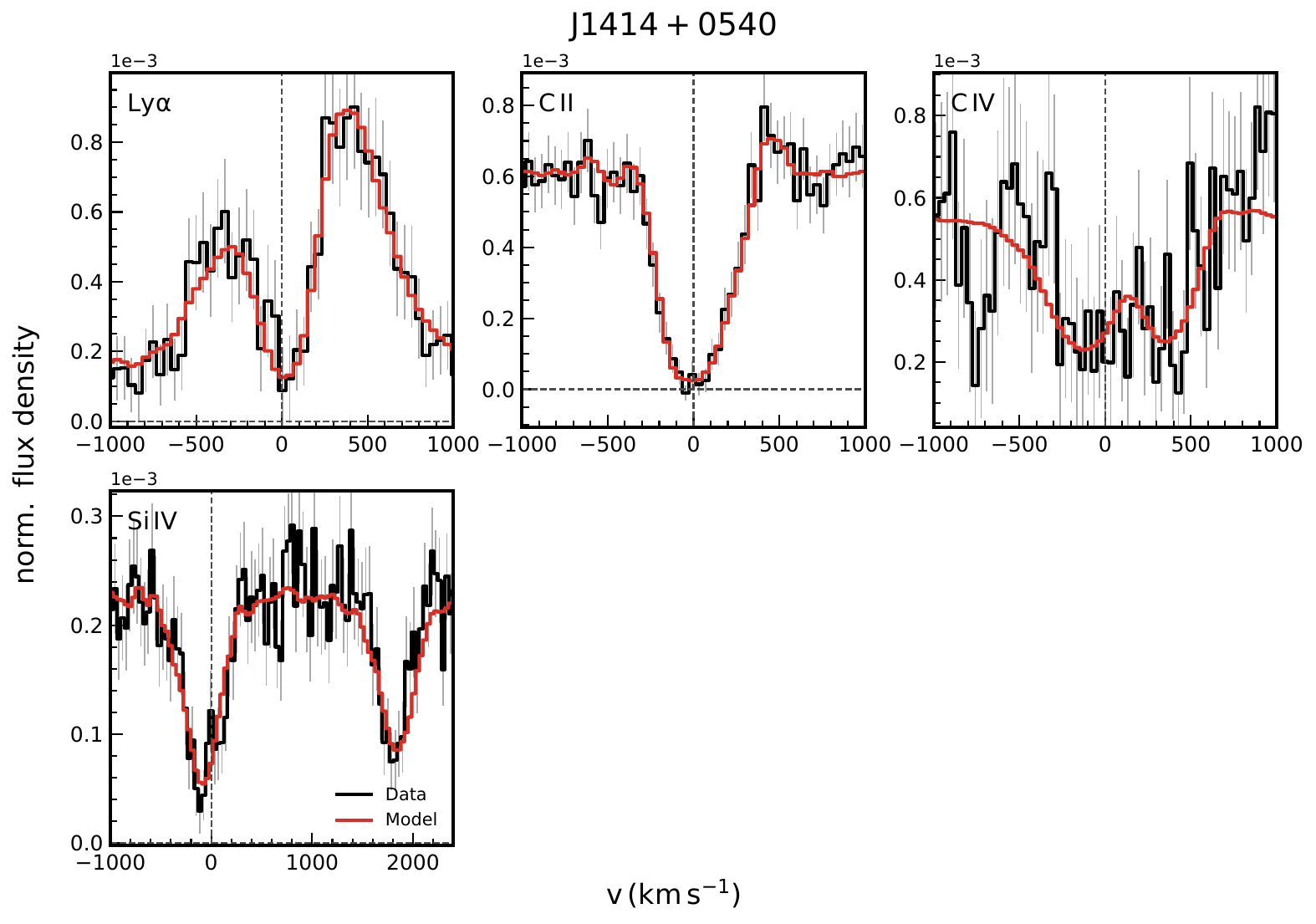}\\
\includegraphics[width=0.9\textwidth]{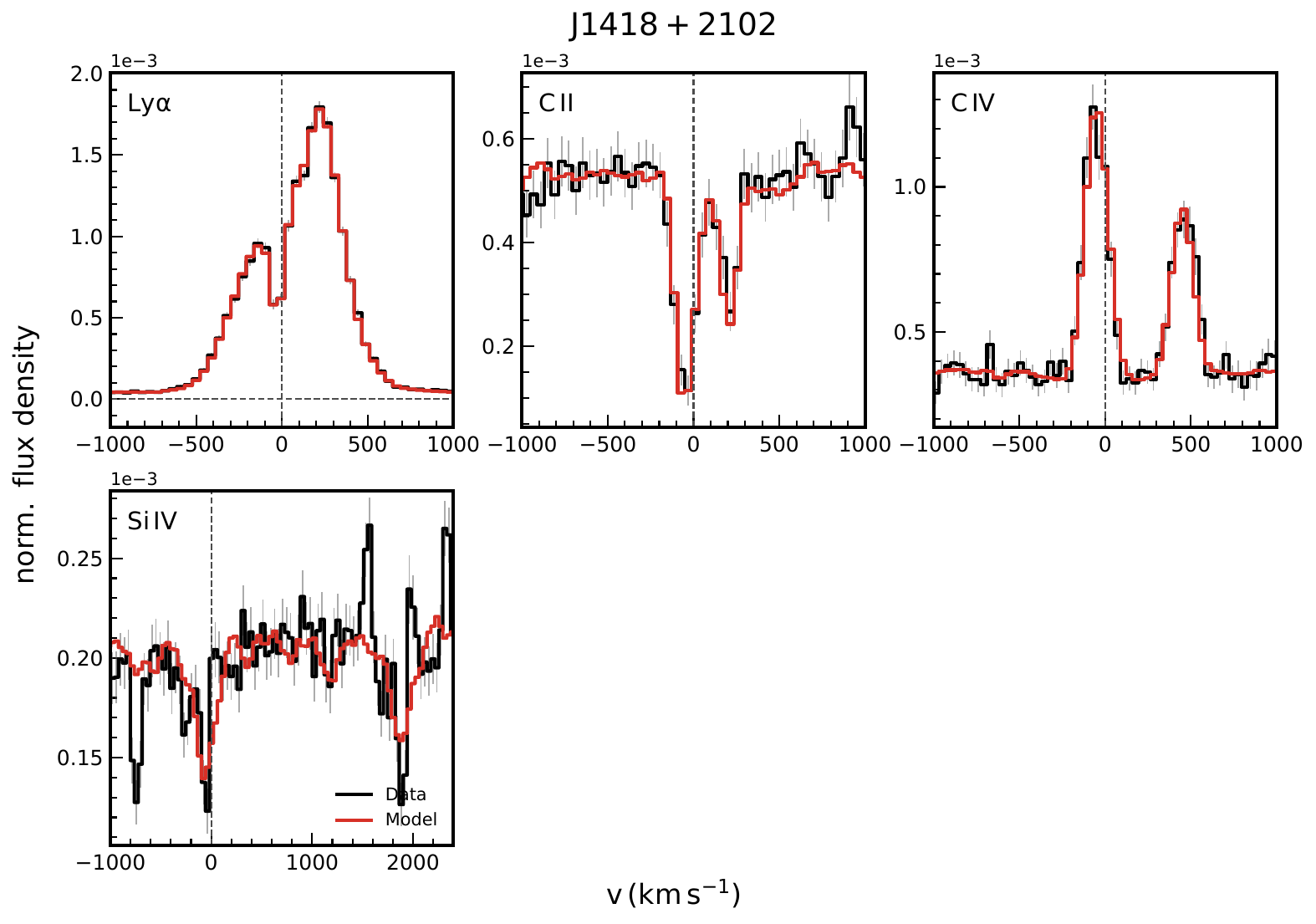}\\
    \caption{\textbf{Best-fit RT models (red) and the observed line profiles (black) for the galaxies in our sample.} 
    \label{fig:joint_fits14}}
\end{figure*}

\begin{figure*}
\centering
\includegraphics[width=0.9\textwidth]{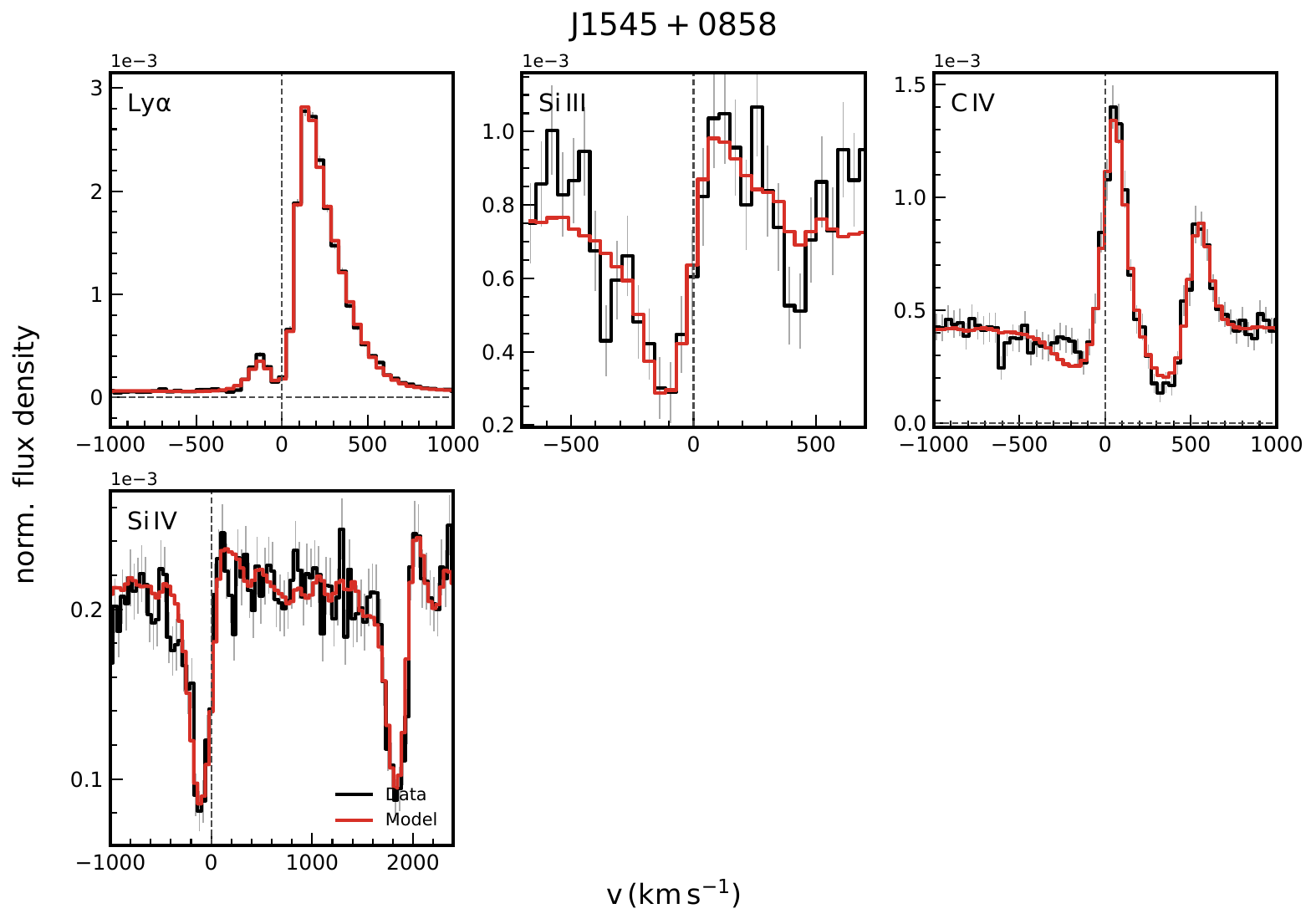}\\
\includegraphics[width=0.9\textwidth]{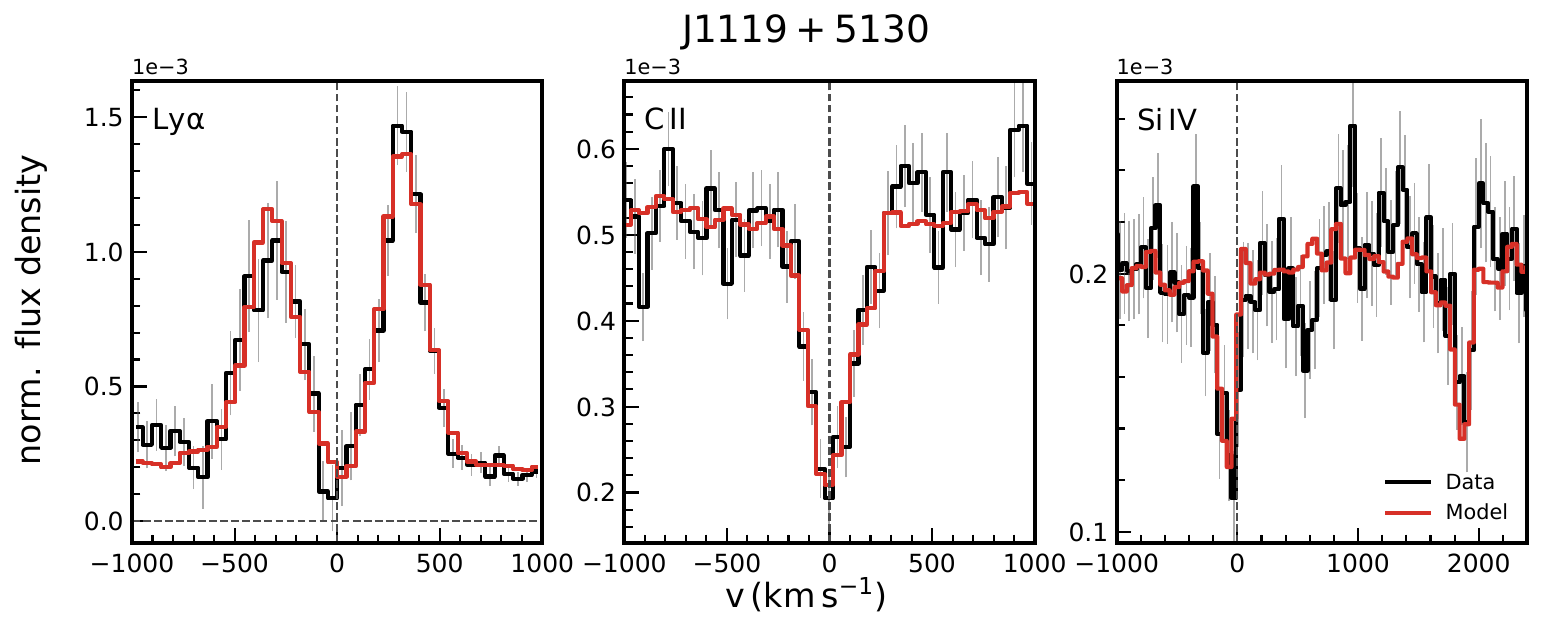}\\
\includegraphics[width=0.92\textwidth]{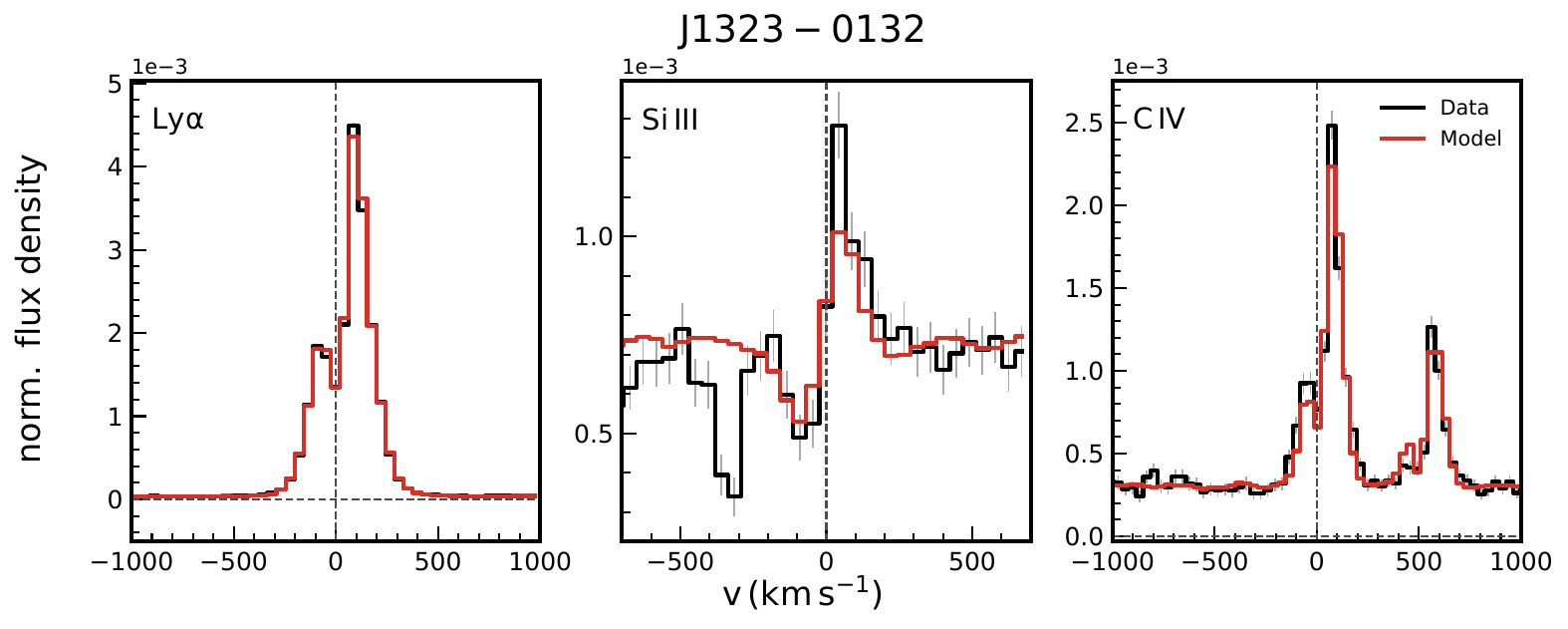}\\
    \caption{\textbf{Best-fit RT models (red) and the observed line profiles (black) for the galaxies in our sample.} 
    \label{fig:joint_fits15}}
\end{figure*}

\begin{figure*}
\centering
\includegraphics[width=0.9\textwidth]{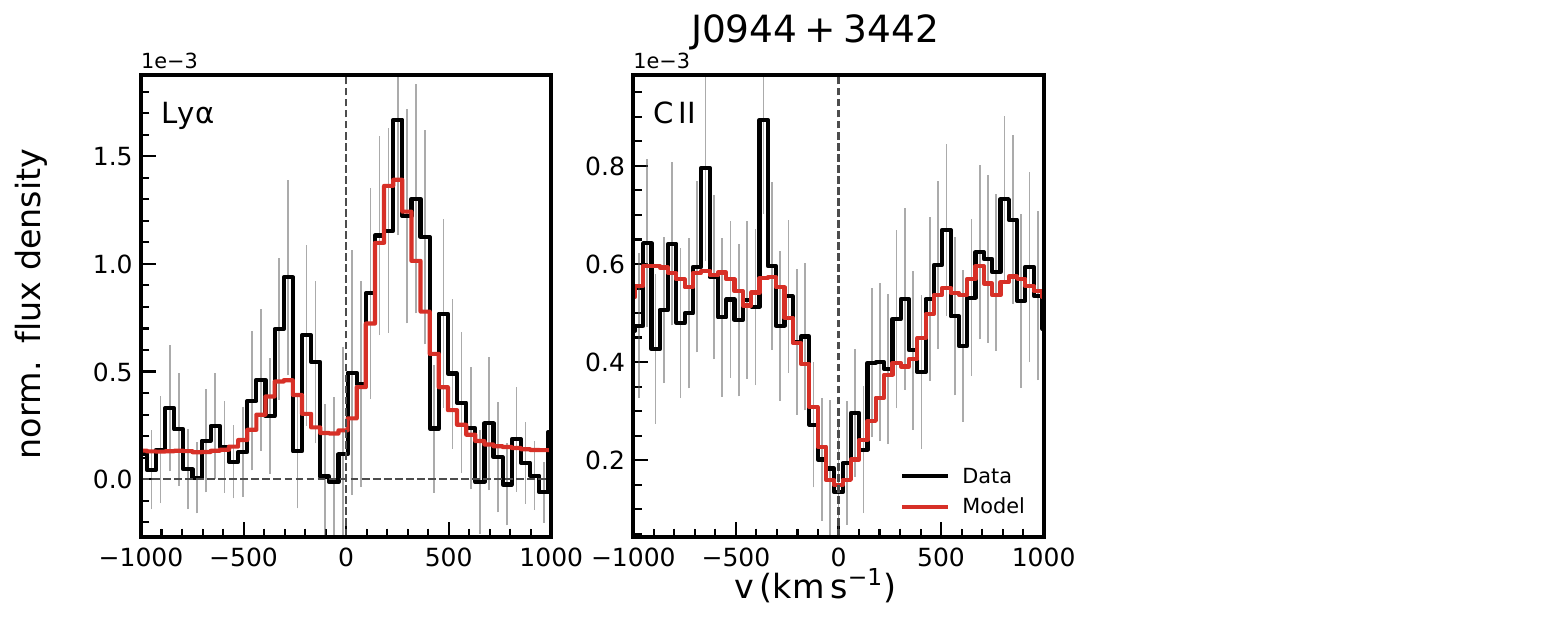}\\
\includegraphics[width=0.9\textwidth]{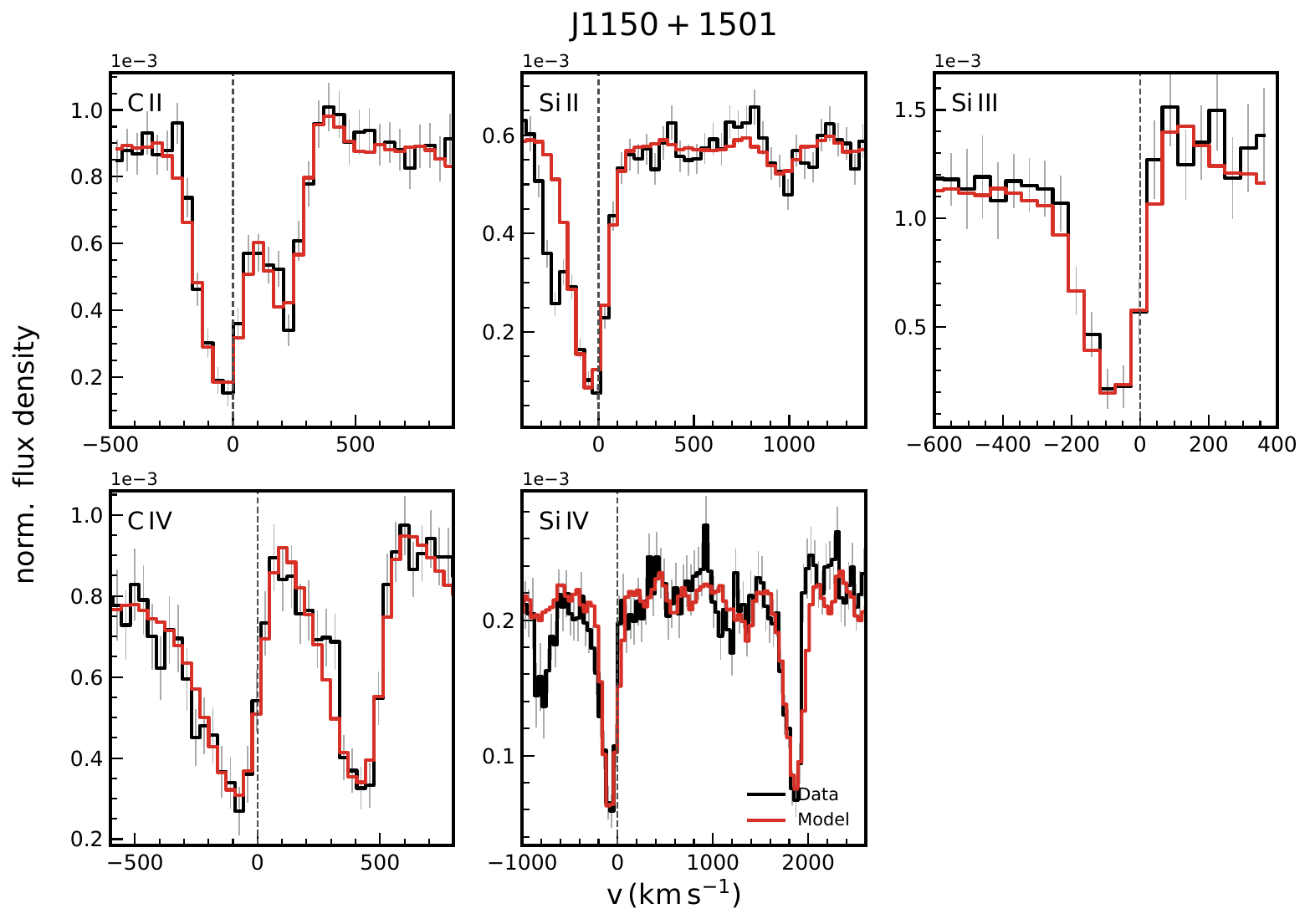}\\
    \caption{\textbf{Best-fit RT models (red) and the observed line profiles (black) for the galaxies in our sample.} 
    \label{fig:joint_fits16}}
\end{figure*}

\begin{figure*}
\centering
\includegraphics[width=0.9\textwidth]{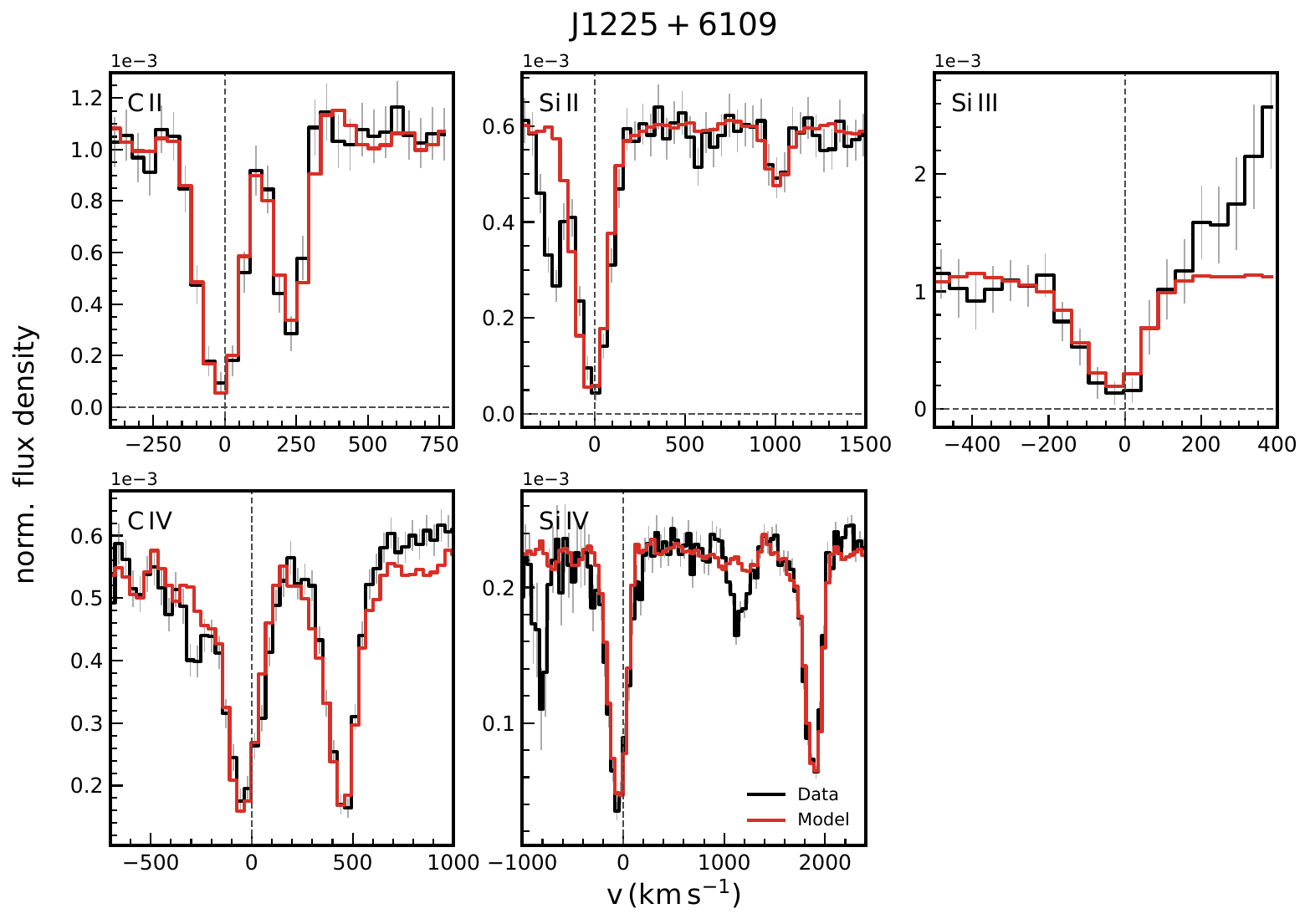}\\
\includegraphics[width=0.9\textwidth]{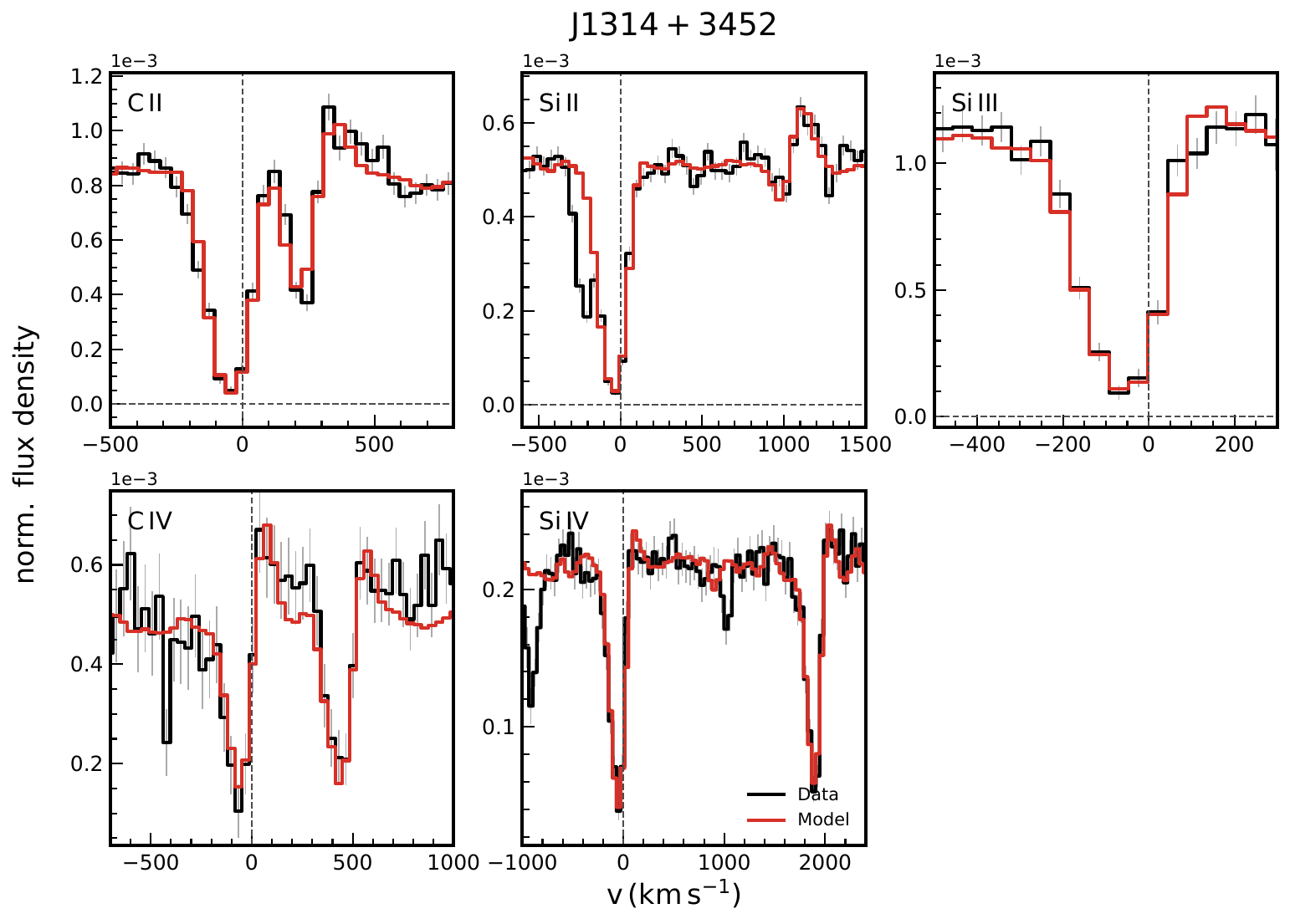}\\
    \caption{\textbf{Best-fit RT models (red) and the observed line profiles (black) for the galaxies in our sample.}  
    \label{fig:joint_fits17}}
\end{figure*}

\begin{figure*}
\centering
\includegraphics[width=0.9\textwidth]{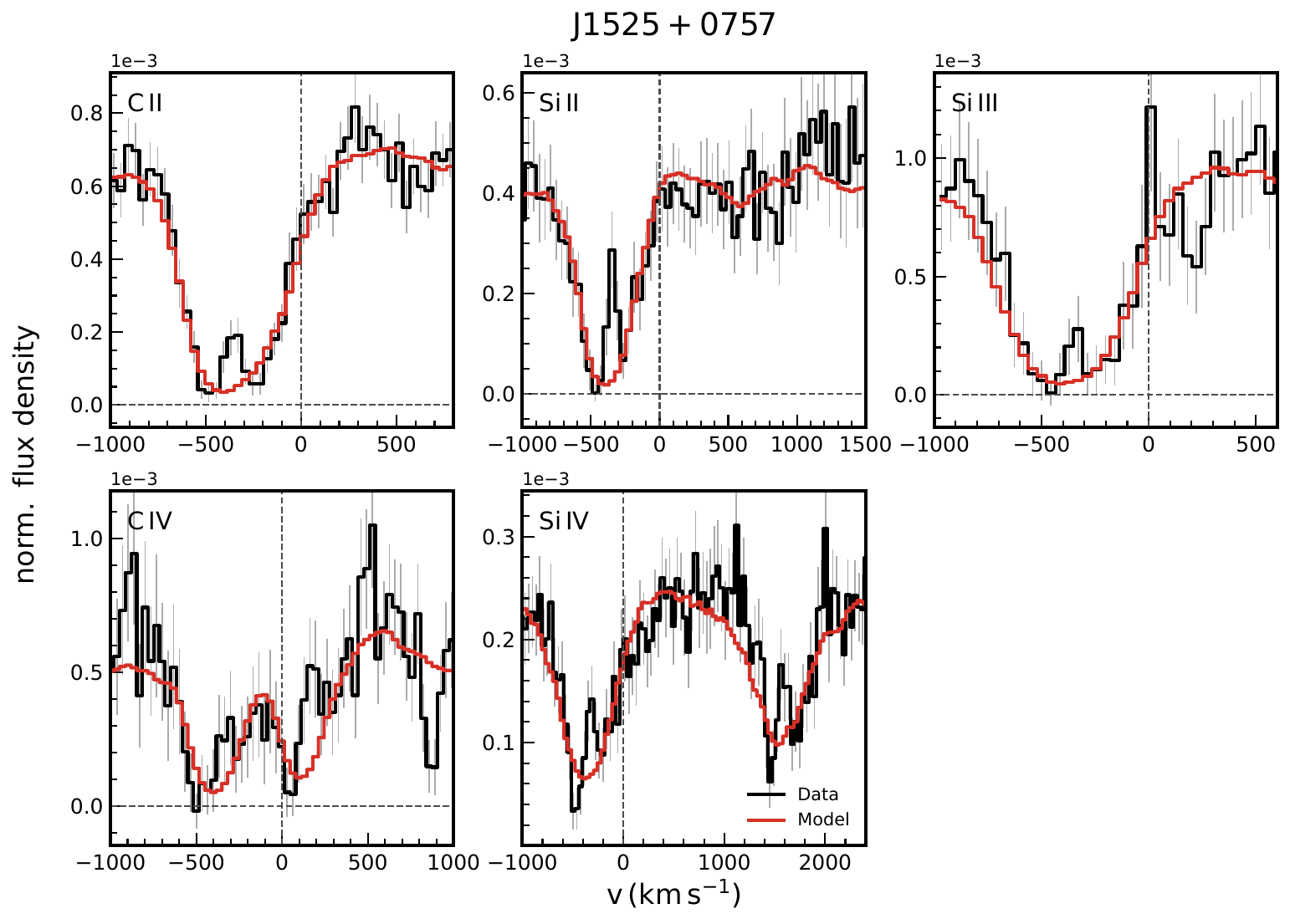}\\
\includegraphics[width=0.9\textwidth]{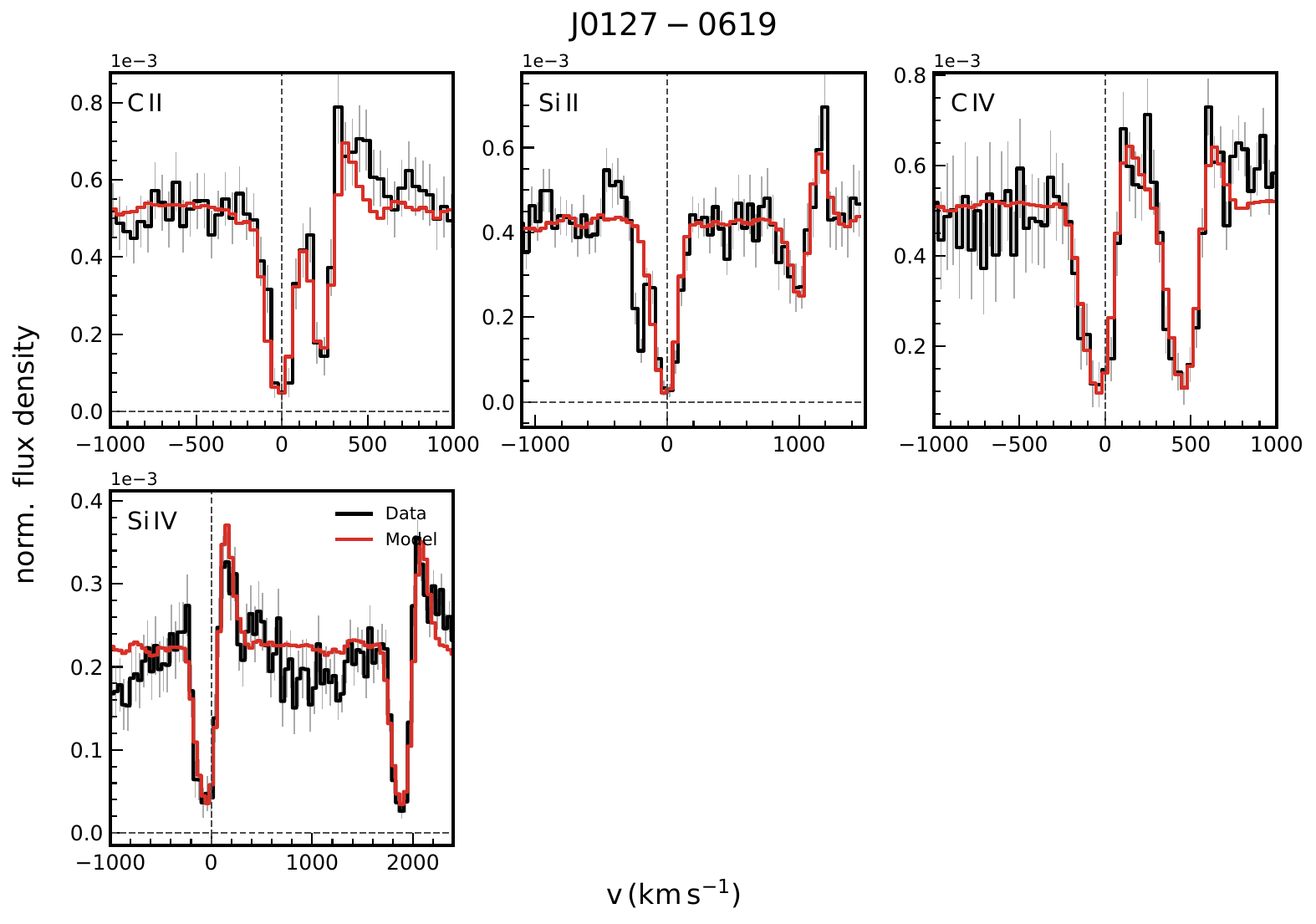}\\
    \caption{\textbf{Best-fit RT models (red) and the observed line profiles (black) for the galaxies in our sample.} 
    \label{fig:joint_fits18}}
\end{figure*}

\begin{figure*}
\centering
\includegraphics[width=0.9\textwidth]{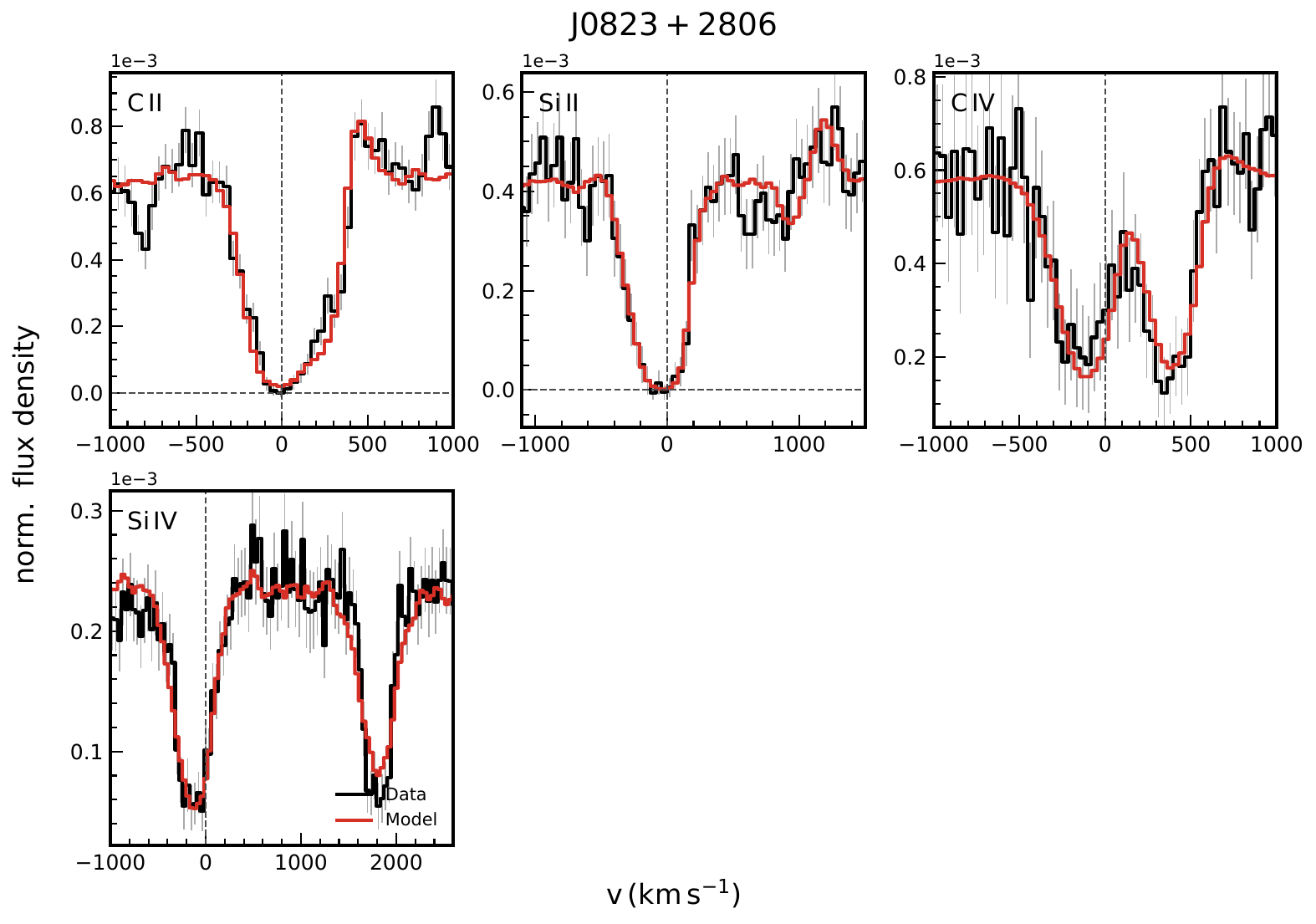}\\
\includegraphics[width=0.9\textwidth]{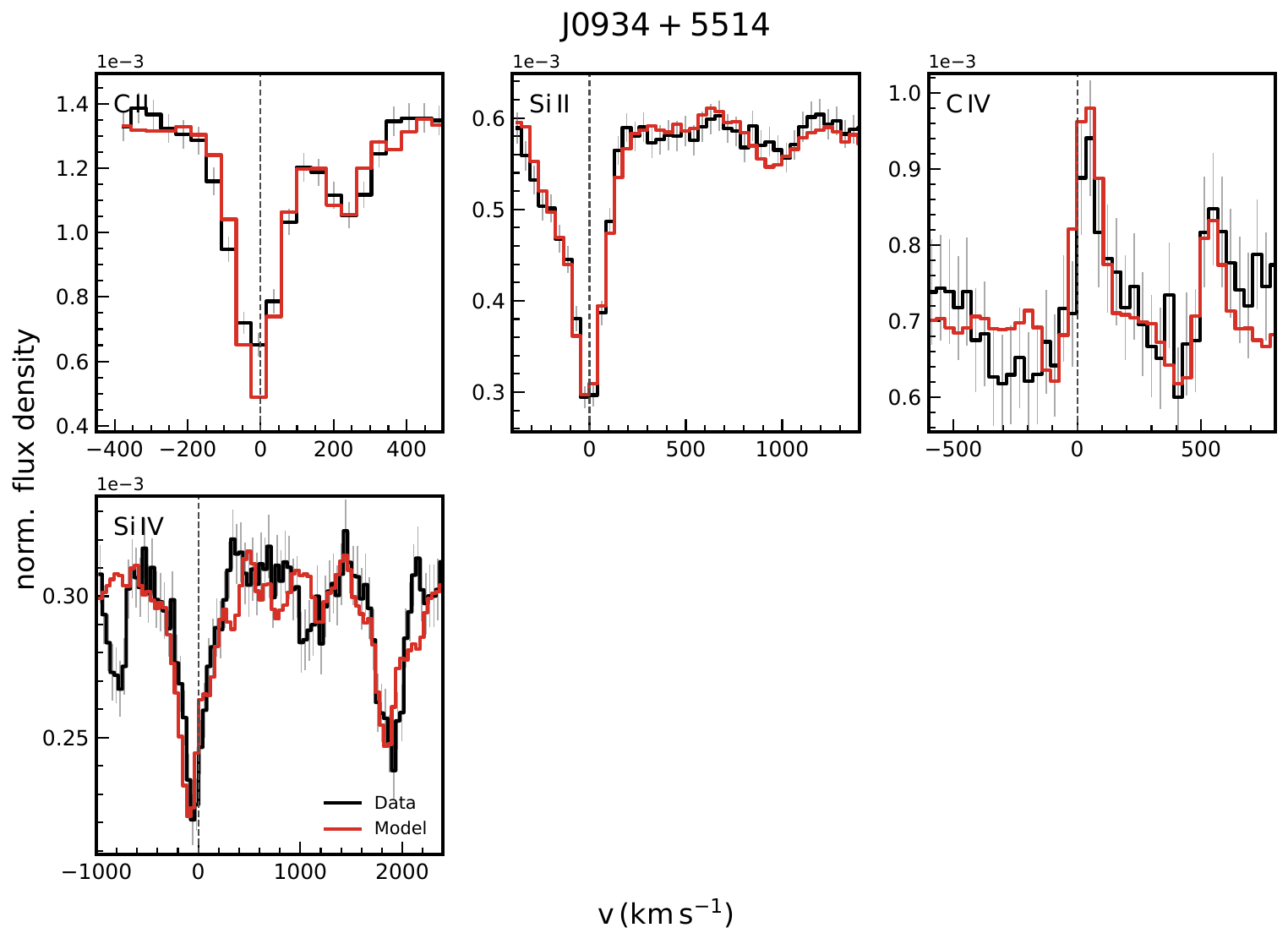}\\
    \caption{\textbf{Best-fit RT models (red) and the observed line profiles (black) for the galaxies in our sample.} 
    \label{fig:joint_fits19}}
\end{figure*}

\begin{figure*}
\centering
\includegraphics[width=0.9\textwidth]{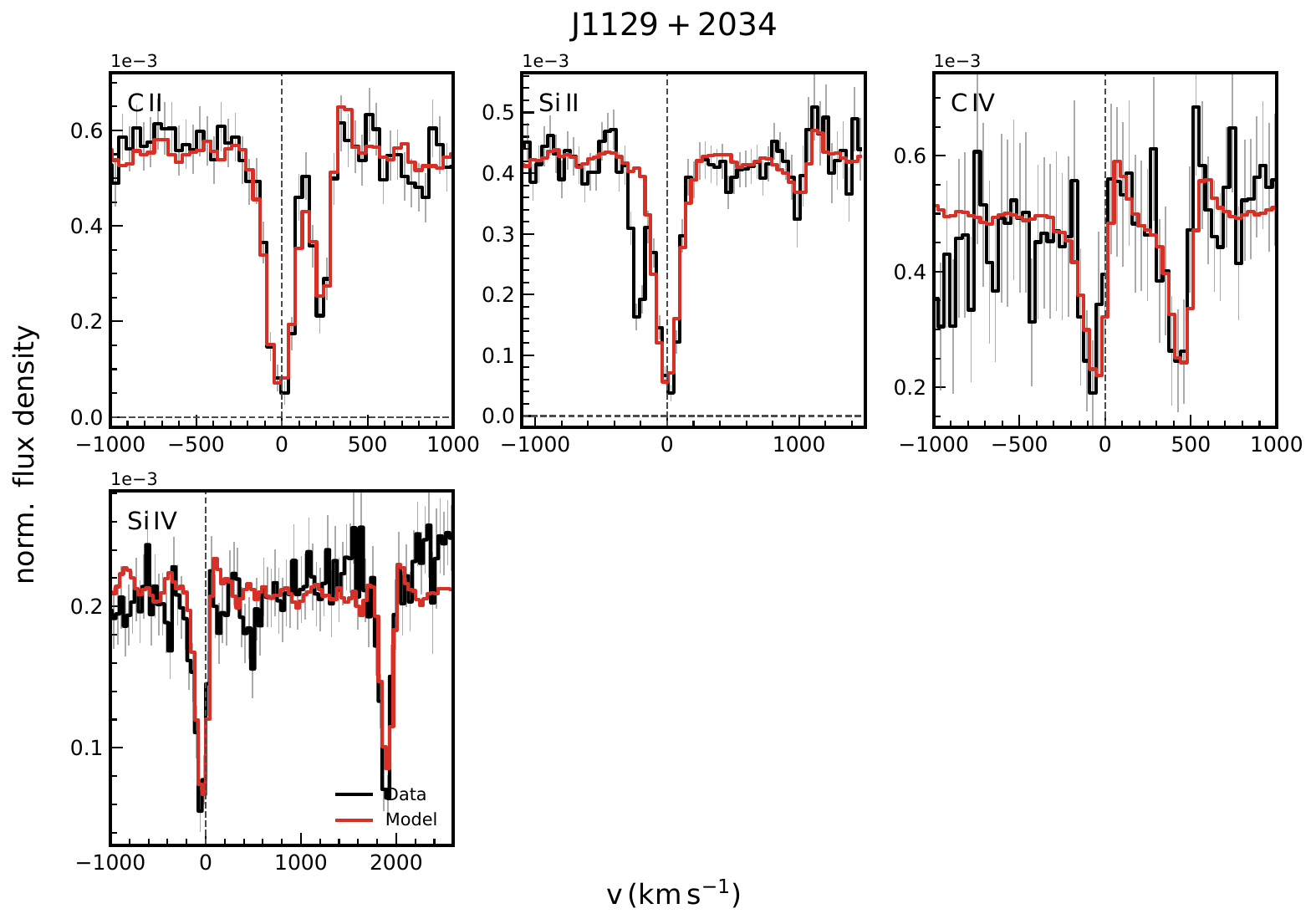}\\
\includegraphics[width=0.9\textwidth]{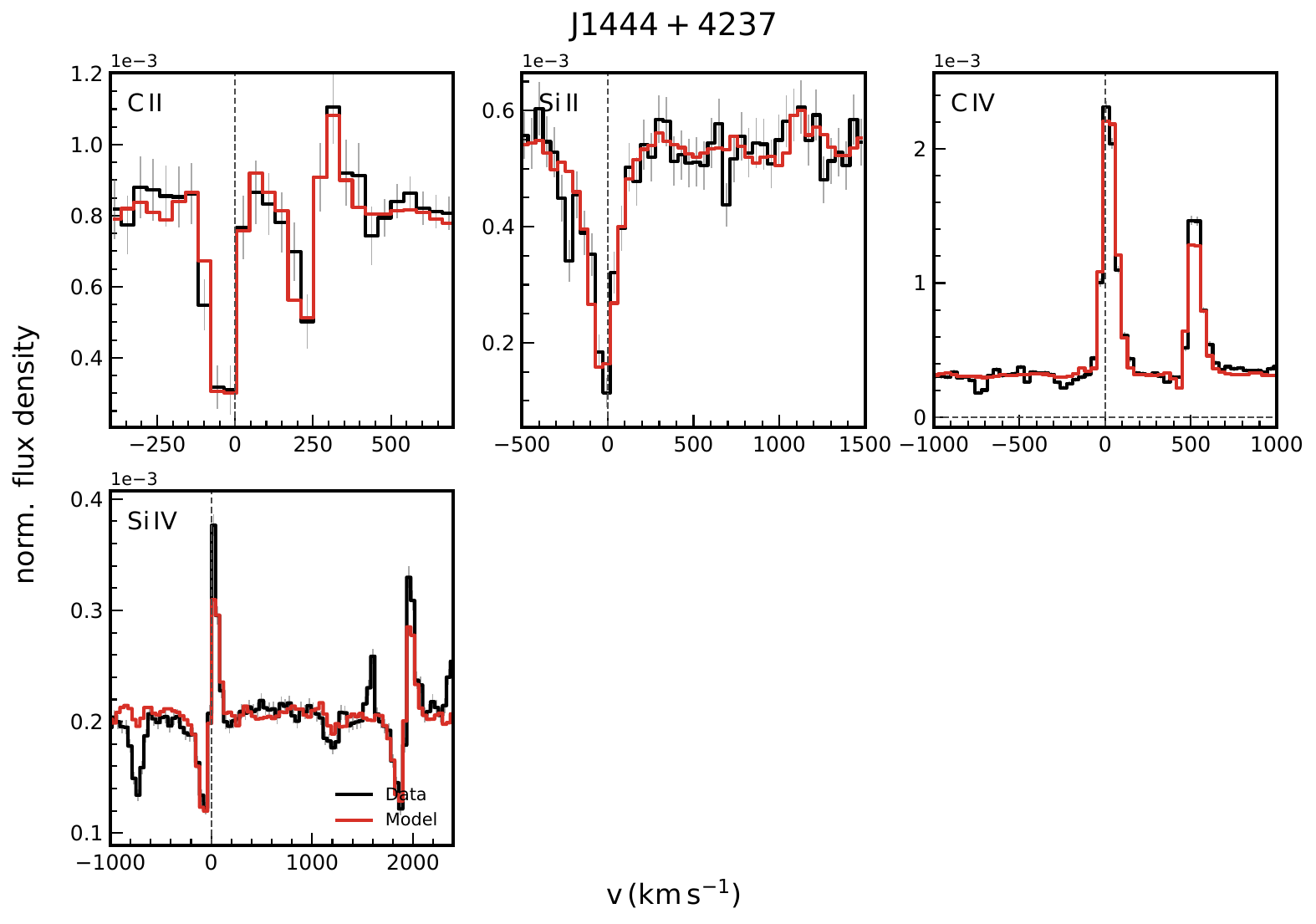}\\
    \caption{\textbf{Best-fit RT models (red) and the observed line profiles (black) for the galaxies in our sample.} 
    \label{fig:joint_fits20}}
\end{figure*}

\begin{figure*}
\centering
\includegraphics[width=0.84\textwidth]{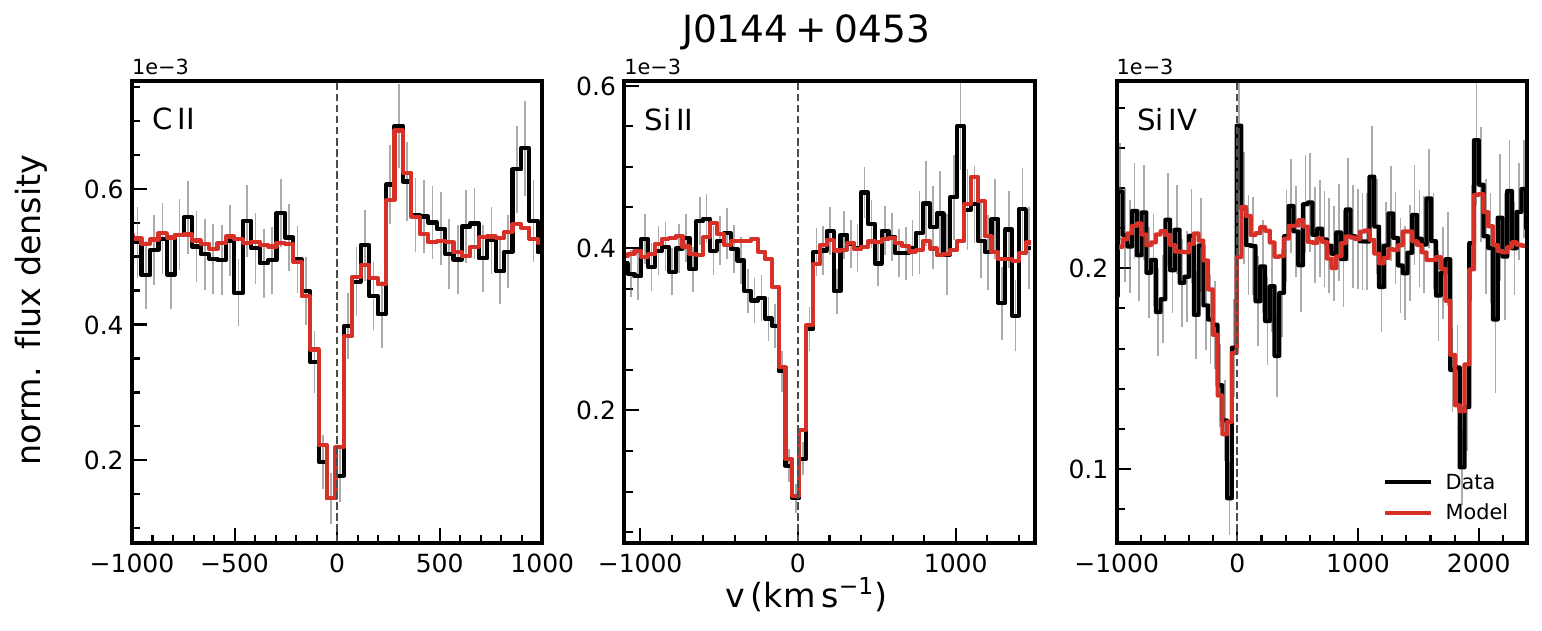}\\
\includegraphics[width=0.84\textwidth]{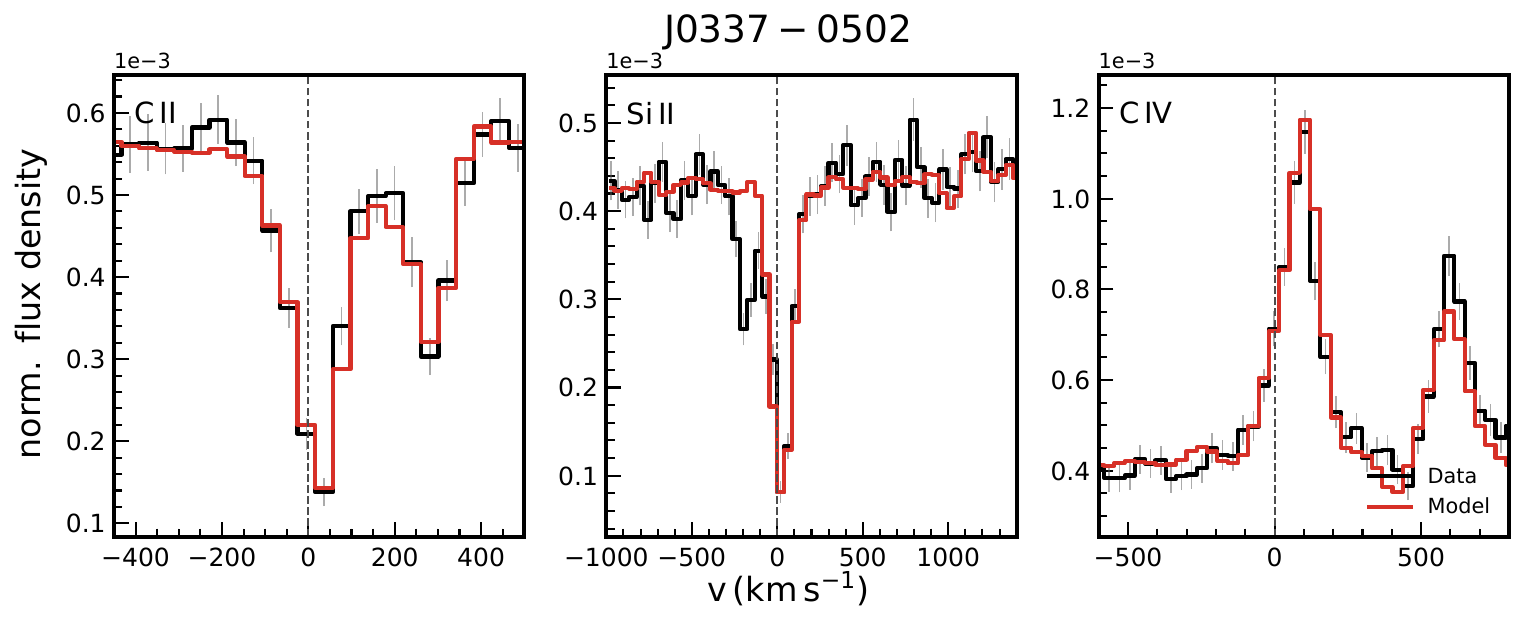}\\
\includegraphics[width=0.84\textwidth]{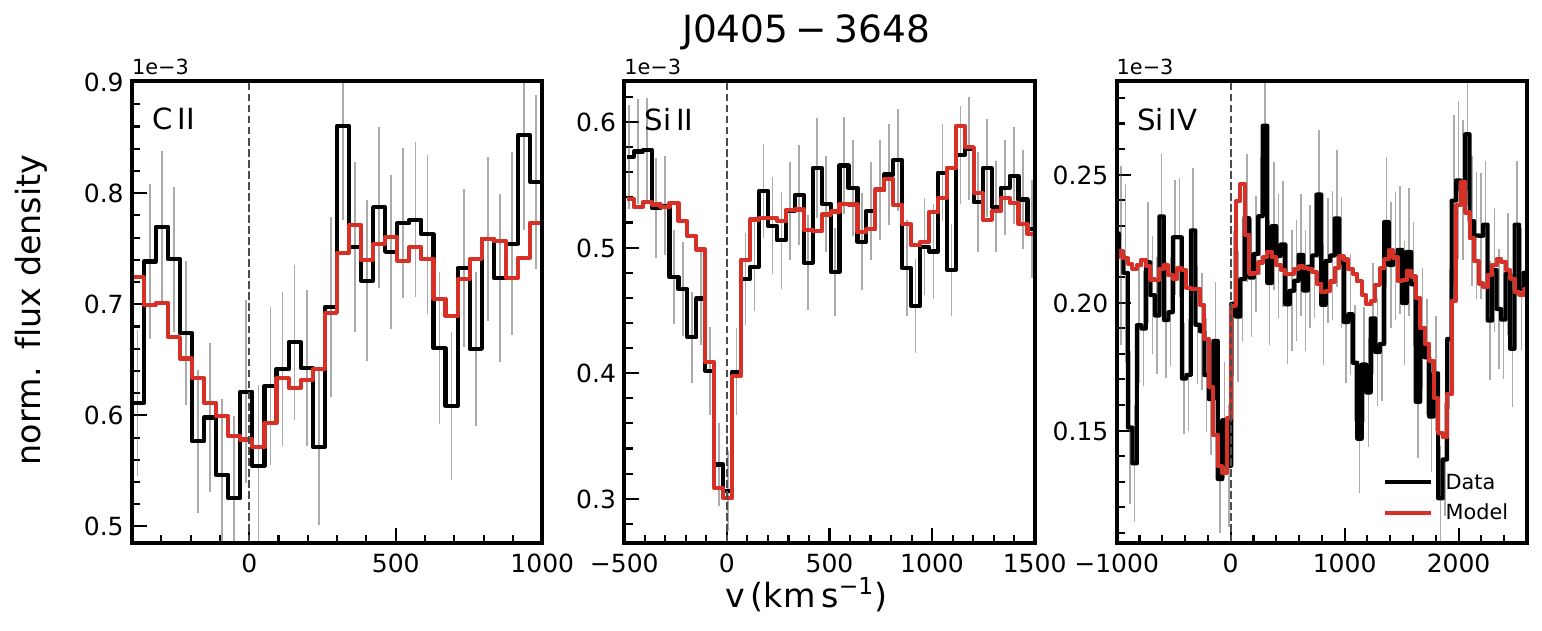}\\
\includegraphics[width=0.84\textwidth]{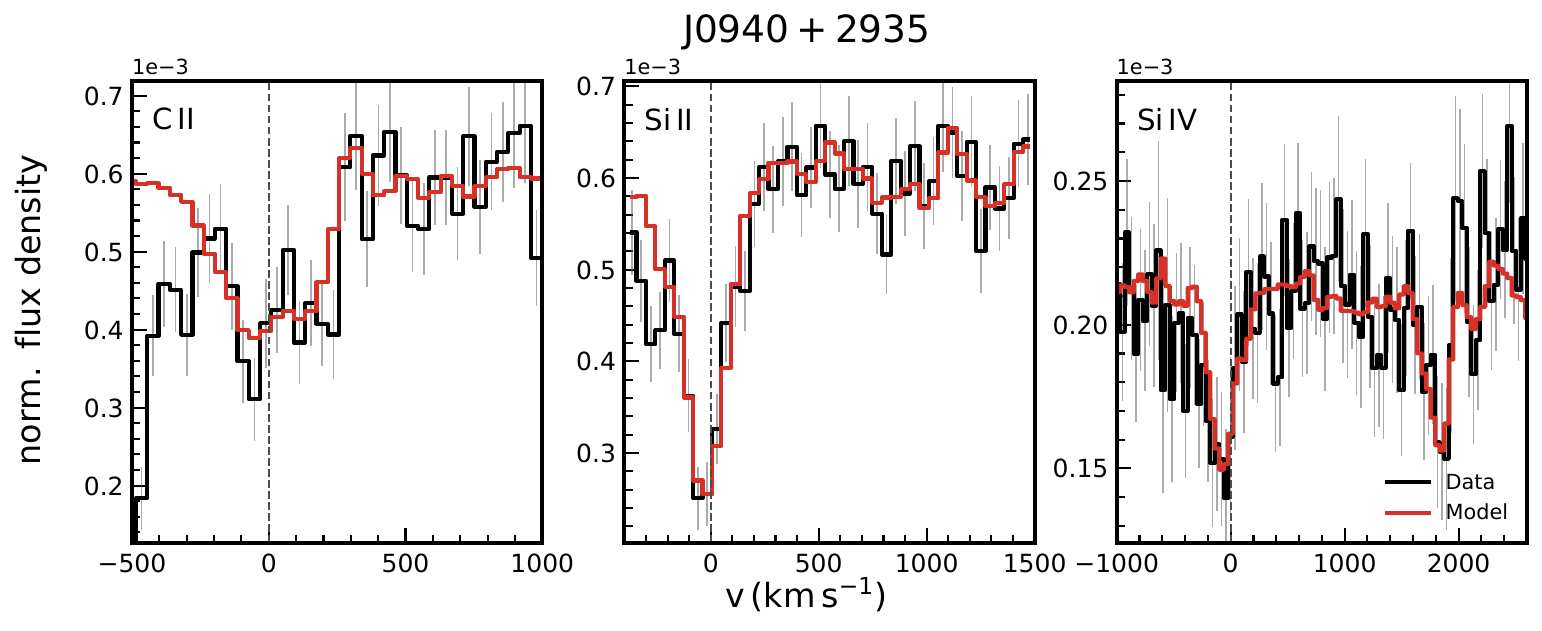}\\
    \caption{\textbf{Best-fit RT models (red) and the observed line profiles (black) for the galaxies in our sample.} 
    \label{fig:joint_fits21}}
\end{figure*}

\begin{figure*}
\centering
\includegraphics[width=0.9\textwidth]{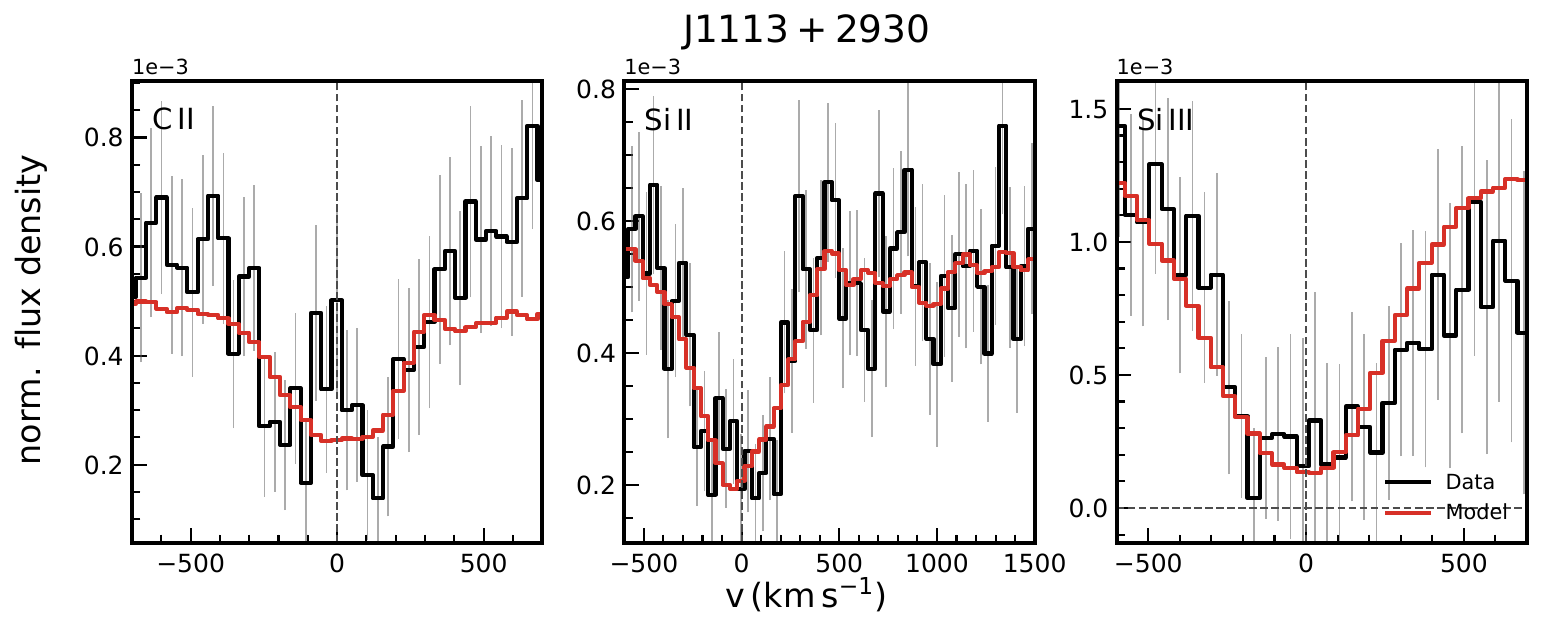}\\
\includegraphics[width=0.92\textwidth]{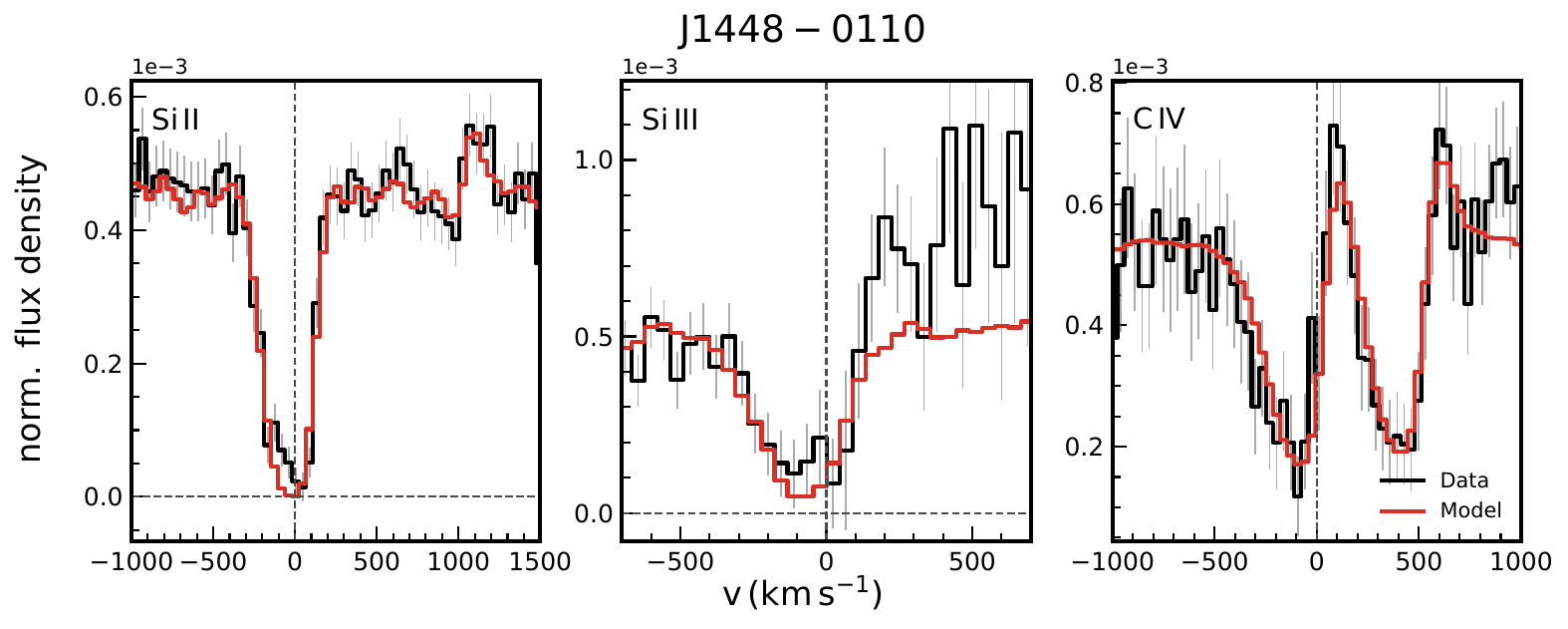}\\
\includegraphics[width=0.92\textwidth]{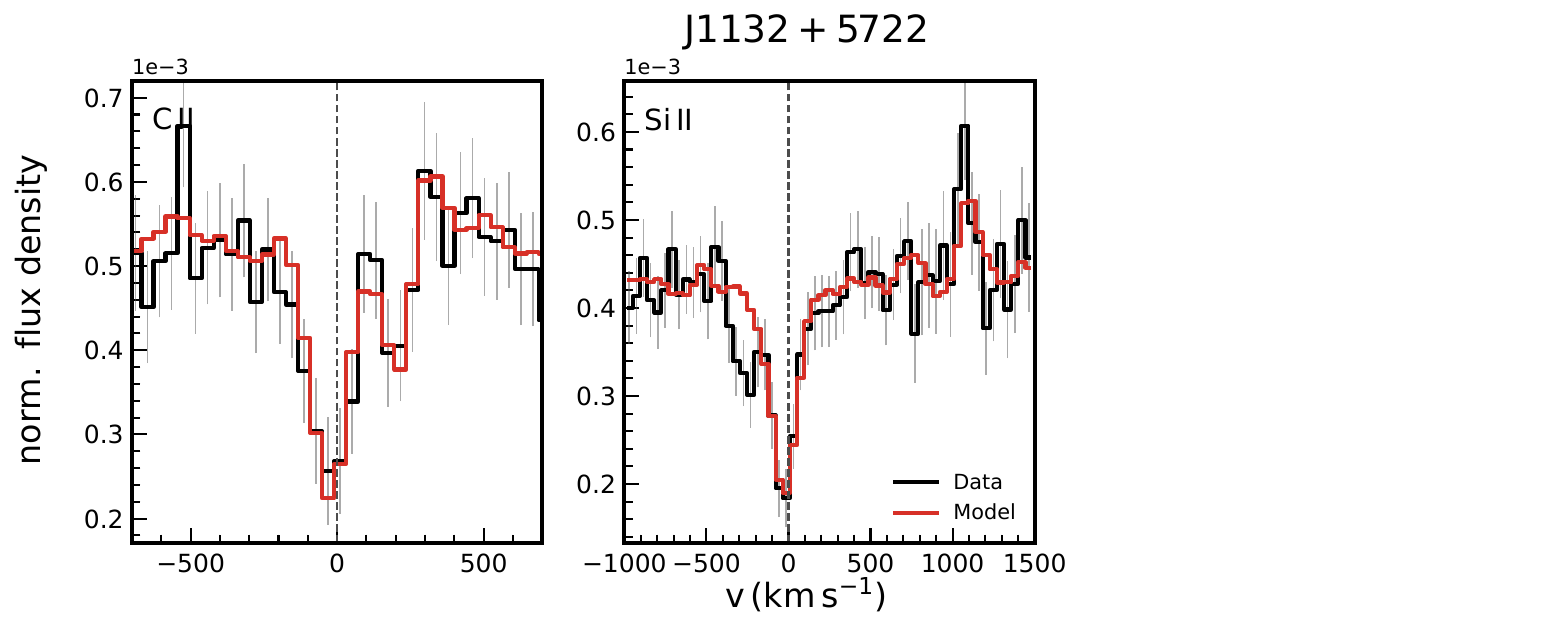}\\
    \caption{\textbf{Best-fit RT models (red) and the observed line profiles (black) for the galaxies in our sample.} 
    \label{fig:joint_fits22}}
\end{figure*}

Figures~\ref{fig:joint_fits1}--\ref{fig:joint_fits22} show the best-fit models for the individual line profiles of all galaxies in our sample, ordered by the number of available spectral lines. We begin with galaxies having complete coverage of \lya, \SiII, \CII, \SiIII, \SiIV, and \CIV, followed by systems with \lya\ plus four, three, and two metal transitions. We then show galaxies without detectable \lya\ emission but with metal absorption lines, ordered by the number of metal transitions available. The \lya\ emission line is fitted independently for each galaxy, whereas the metal lines are fitted jointly when multiple transitions are available, and the displayed models correspond to the joint best fits (see Section \ref{sec:multi_line}). The only exception is J0808+3948: while the \lya\ line is fitted, the metal-line profiles exhibit complex structures with multiple absorption and emission components that cannot be adequately reproduced by our model. For this system we therefore show only the observed spectra without model fits for the metal transitions.

\begin{figure*}
\centering
\includegraphics[width=\textwidth]{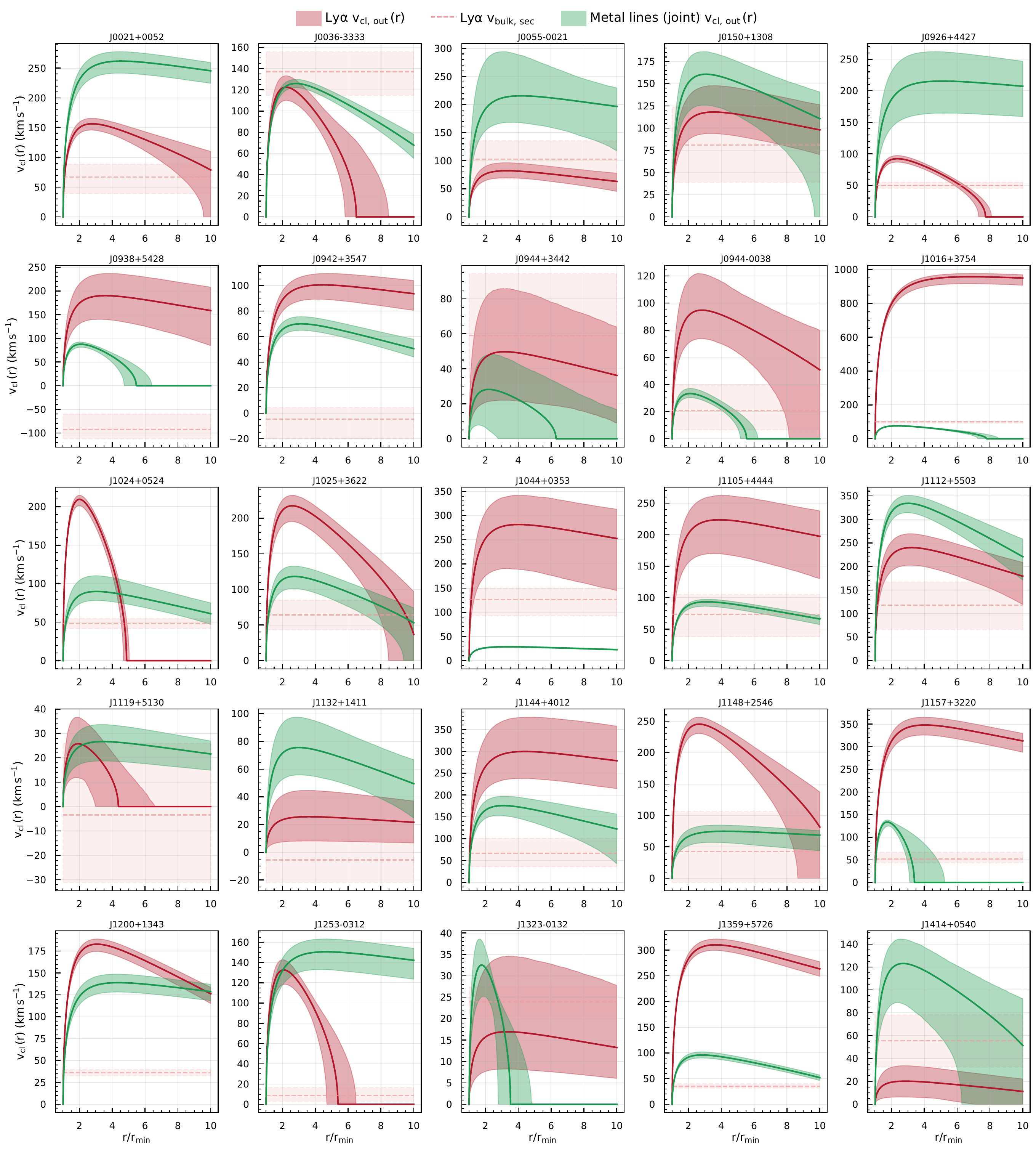}\\
   \caption{\textbf{Comparison of the inferred clump outflow velocity profiles for individual galaxies.} The red curves and shaded regions show the \lya-derived clump outflow velocity $v_{\rm cl,\,out}(r)$ and its 16th–84th percentile range, while the green curves and bands show the clump outflow velocity constrained from the joint fit to metal lines. The dashed horizontal lines mark the secondary bulk velocity component $v_{\rm bulk,\,sec}$ derived from \lya. Each panel corresponds to a different galaxy, labeled at the top. 
    \label{fig:vr_comparison_1}}
\end{figure*}

\begin{figure*}
\centering
\includegraphics[width=\textwidth]{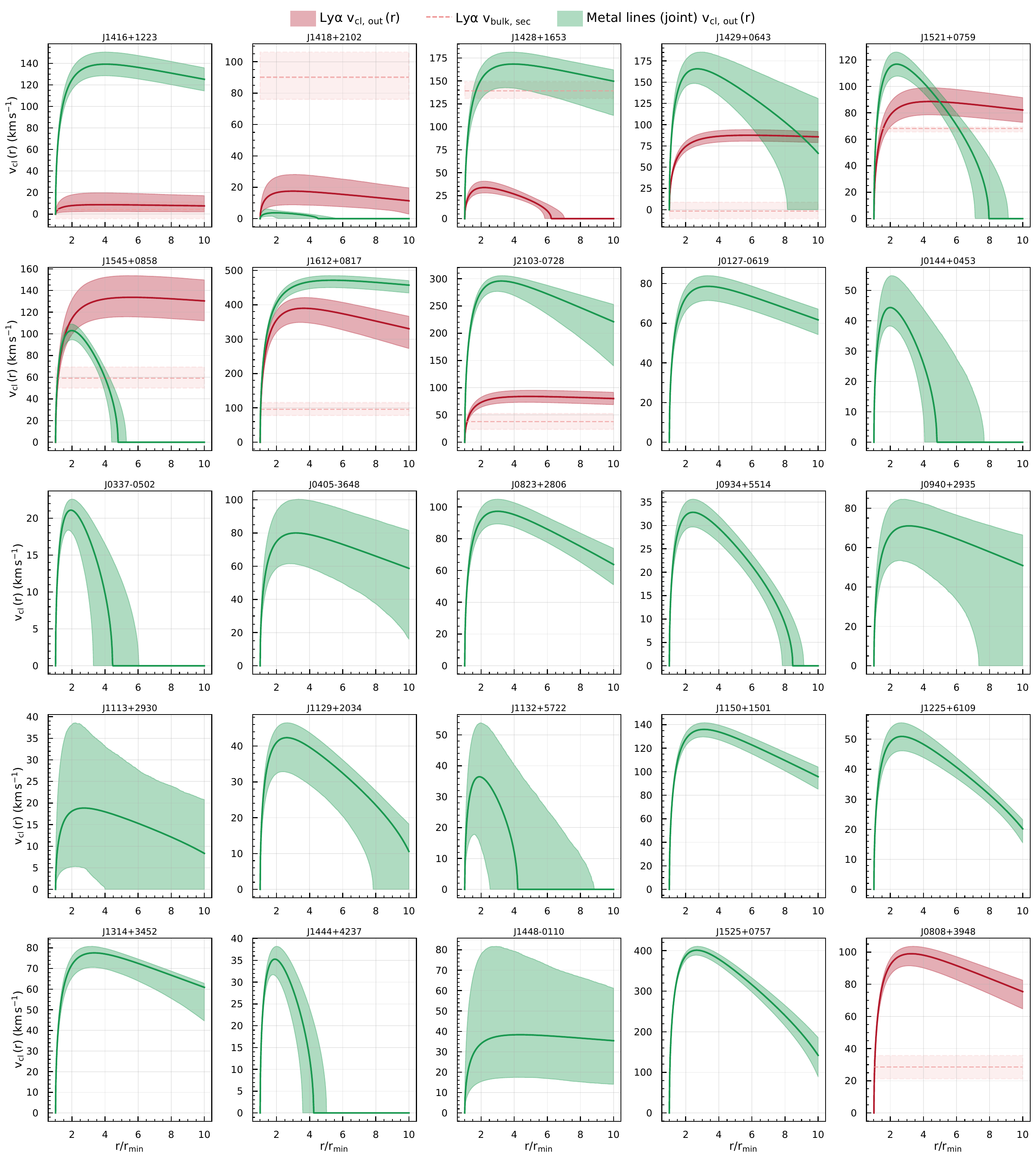}\\
   \caption{\textbf{Comparison of the inferred clump outflow velocity profiles for individual galaxies.} The red curves and shaded regions show the \lya-derived clump outflow velocity $v_{\rm cl,\,out}(r)$ and its 16th–84th percentile range, while the green curves and bands show the clump outflow velocity constrained from the joint fit to metal lines. The dashed horizontal lines mark the secondary bulk velocity component $v_{\rm bulk,\,sec}$ derived from \lya. Each panel corresponds to a different galaxy, labeled at the top. 
    \label{fig:vr_comparison_2}}
\end{figure*}

In addition to the line profile best-fits, we present a comparison of the inferred clump outflow velocity profiles. Figures~\ref{fig:vr_comparison_1} and \ref{fig:vr_comparison_2} show $v_{\rm cl,,out}(r)$ derived from \lya\ (when available) and from the joint metal-line fits for all galaxies, together with their associated uncertainties. The bulk velocity of the secondary clump population inferred from \lya, $v_{\rm bulk,,sec}$, is also indicated, allowing a direct comparison between the gas kinematics constrained by \lya\ and the metal transitions.

\section{Examples of Posterior Probability Distributions from the RT Modeling}\label{sec:posterior_examples}

\begin{figure*}
\centering
\includegraphics[width=\textwidth]{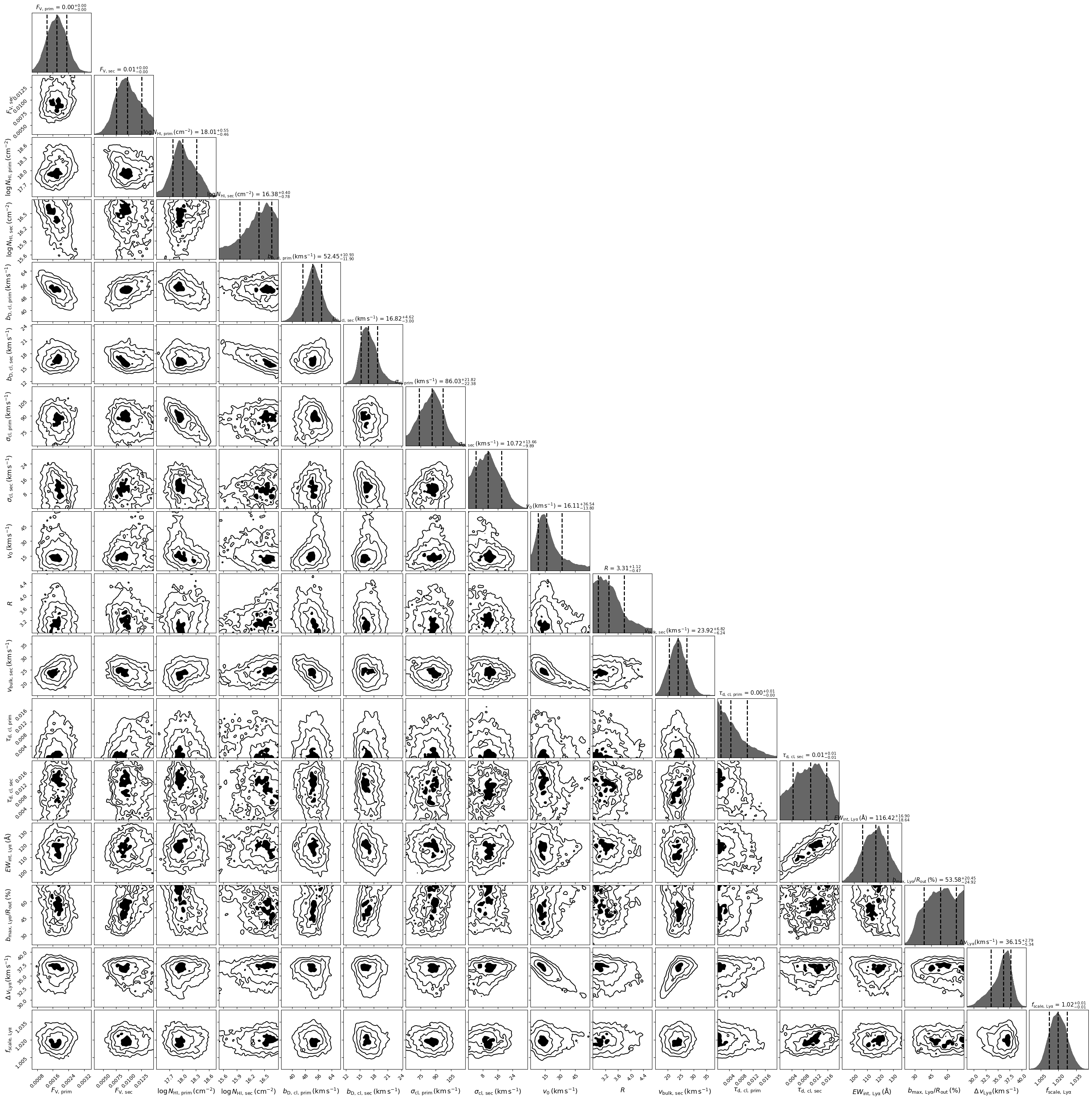}\\
\caption{\textbf{Posterior probability distributions from the fitting of the \lya\ emission line profile for J1323-0132.} The diagonal panels show the marginalized one-dimensional posteriors for each parameter, with the dashed vertical lines indicating the 16th, 50th, and 84th percentiles. The model consists of a primary outflowing component and a secondary semi-static component. For each component, the fitted parameters are the clump volume filling factor $F_{\rm V}$, the clump \HI\ column density $\log N_{\rm HI,\,cl}$, the clump Doppler parameter $b_{\rm D,\,cl}$, the clump velocity dispersion $\sigma_{\rm cl}$, and the clump dust optical depth $\tau_{\rm d,\,cl}$. The outflow velocity profile of the primary component is characterized by the outflow velocity normalization $v_0$ and the acceleration parameter $R$. The secondary component is characterized by a bulk velocity $v_{\rm bulk,\,sec}$. Additional fitted parameters include the intrinsic \lya\ equivalent width $\rm EW_{\rm int,\,Ly\alpha}$, the normalized maximum impact parameter $b_{\rm max,\,Ly\alpha}/R_{\rm out}$, the velocity offset $\Delta v_{\rm Ly\alpha}$, and the continuum scaling factor $f_{\rm scale,\,Ly\alpha}$.
    \label{fig:posterior_lya}}
\end{figure*}

\begin{figure*}
\centering
\includegraphics[width=\textwidth]{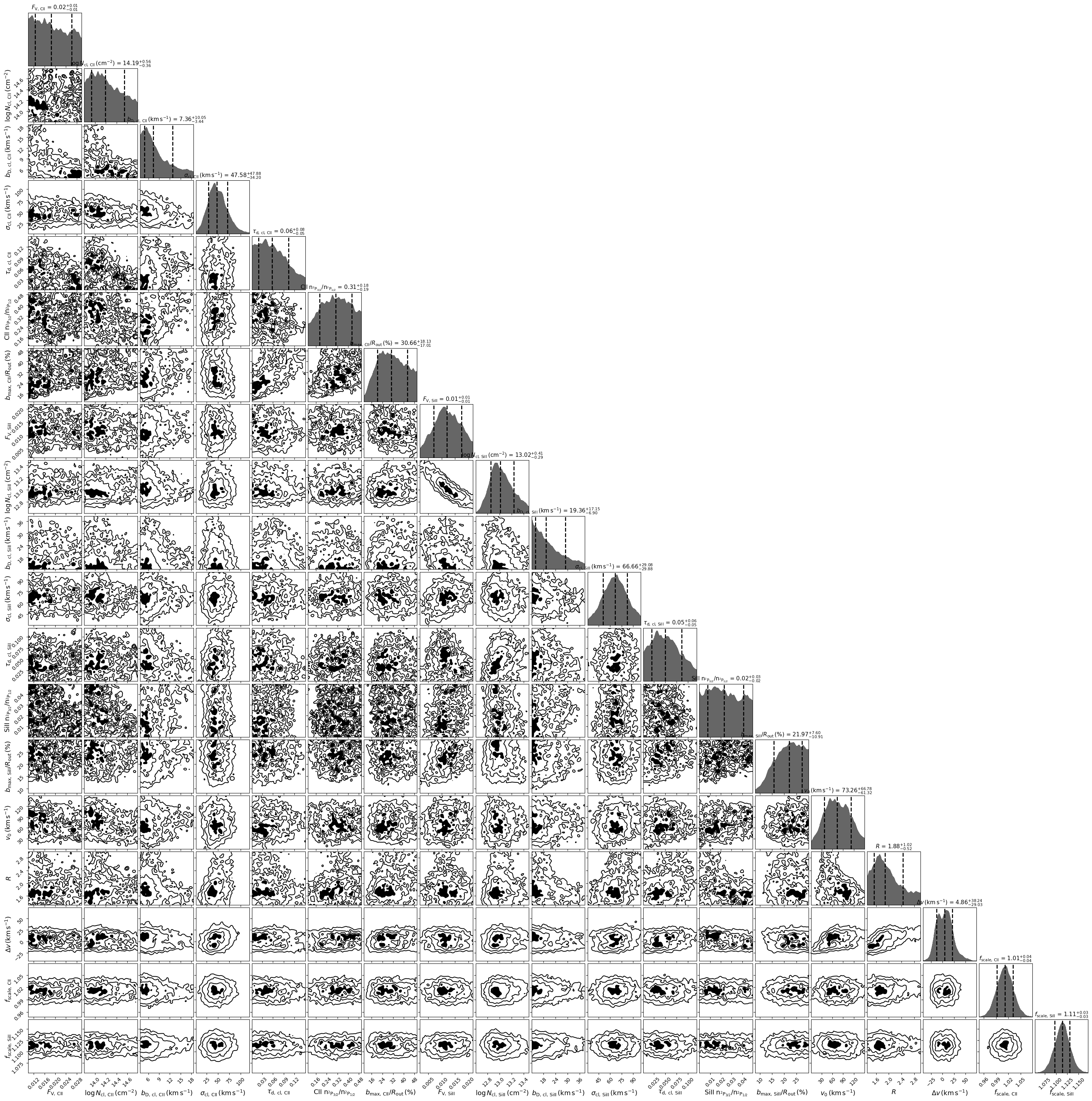}\\
   \caption{\textbf{Posterior probability distributions from the joint fitting of the \CII\ and \SiII} \textbf{absorption line profiles for J1132+5722.} The diagonal panels show the marginalized one-dimensional posteriors for each parameter, with the dashed vertical lines indicating the 16th, 50th, and 84th percentiles. For each ion, the fitted parameters are the clump volume filling factor $F_{\rm V}$, the clump ion column density $\log N_{\rm cl}$, the clump Doppler parameter $b_{\rm D,\,cl}$, the clump velocity dispersion $\sigma_{\rm cl}$, the clump dust optical depth $\tau_{\rm d,\,cl}$, the fine-structure population ratio $r_{\rm pop} = n_{^2P_{3/2}}/n_{^2P_{1/2}}$, and the normalized maximum impact parameter $b_{\rm max}/R_{\rm out}$. The shared parameters across both ions are the outflow velocity normalization $v_0$, the acceleration parameter $R$, and the velocity offset $\Delta v$. The continuum scaling factors $f_{\rm scale}$ for each ion are also fitted independently.
    \label{fig:posterior_metal}}
\end{figure*}

We present two representative examples of the full posterior probability distributions of the model parameters inferred from the RT fitting in Figures~\ref{fig:posterior_lya} and \ref{fig:posterior_metal}. The diagonal panels show the marginalized one-dimensional posteriors, while the off-diagonal panels display the two-dimensional parameter covariances.

Figure~\ref{fig:posterior_lya} corresponds to the fit of the \lya\ emission line profile of J1323$-$0132 using the two-component clumpy model described in Section~\ref{sec:lya}. The primary component represents the accelerating outflow, whereas the secondary component represents a semi-static clump population near systemic velocity. Figure~\ref{fig:posterior_metal} presents the joint constraints from the \CII\ and \SiII\ absorption profiles of J1132+5722, following the multi-line fitting procedure described in Section~\ref{sec:multi_line}. The posteriors exhibit moderate covariances between certain parameters, but the major model parameters are generally well constrained, with unimodal marginalized distributions that lie well within the adopted prior ranges.

\FloatBarrier
\bibliography{sample701}{}
\bibliographystyle{aasjournalv7}



\end{document}